SAARLAND UNIVERSITY

**Department of Computer Science**

MASTER THESIS

---

# An Executable Structural Operational Formal Semantics for Python

---

## Maximilian A. Köhl

**Supervisor**

Prof. Dr. Jan Reineke

**Reviewers**

Prof. Dr. Jan Reineke

Prof. Dr. Holger Hermanns

December 15, 2020

# Abstract

Python is a popular high-level general-purpose programming language also heavily used by the scientific community. It supports a variety of different programming paradigms and is preferred by many for its ease of use. With the vision of harvesting static analysis techniques like abstract interpretation for Python, we develop a formal semantics for Python. A formal semantics is an important cornerstone for any sound static analysis technique. We base our efforts on the general framework of structural operational semantics yielding a small-step semantics in principle allowing for concurrency and interaction with an environment. The main contributions of this thesis are twofold: first, we develop a meta-theoretic framework for the formalization of structural operational semantics in tandem with the necessary tool support for the automated derivation of interpreters from such formal semantics, and, second, we validate the suitability of this approach for the formalization of modern programming languages developing a semantics for Python.

## Acknowledgements

First and foremost, I would like to thank Prof. Dr. Jan Reineke for his support and guidance. Your valuable feedback was always helpful and pushed me to continuously improve my thoughts and writing. I would also like to thank Prof. Dr. Holger Hermanns for his support and insights on process calculi. In addition, I would like to thank my brother Jannis for the fruitful discussions we had and Michaela Klauck for proofreading this thesis.

# Erklärung

Ich erkläre hiermit, dass ich die vorliegende Arbeit selbständig verfasst und keine anderen als die angegebenen Quellen und Hilfsmittel verwendet habe.

# Statement

I hereby confirm that I have written this thesis on my own and that I have not used any other media or materials than the ones referred to in this thesis.

# Einverständniserklärung

Ich bin damit einverstanden, dass meine (bestandene) Arbeit in beiden Versionen in die Bibliothek der Informatik aufgenommen und damit veröffentlicht wird.

# Declaration of Consent

I agree to make both versions of my thesis (with a passing grade) accessible to the public by having them added to the library of the Computer Science Department.

Saarbrücken, ______________________    ______________________________

(Datum/Date)                    (Unterschrift/Signature)

# Contents

# List of Figures

# List of Snippets

# List of Definitions

# List of Theorems

# Chapter 1

# Introduction

*Python is more concerned with making it easy to write*
*good programs than difficult to write bad ones.*

Steve Holden [30], June 2005

Python is a high-level general-purpose programming language supporting a variety of different programming paradigms [13]. It is the second most popular programming language on GitHub[1] and according to the TIOBE Index for November 2020 [11]. Python markets itself as an "easy to learn, powerful programming language" and an "ideal language for scripting and rapid application development in many areas on most platforms" [21]. It is used by organizations such as Google [19], Facebook [45], and NASA [52]. Despite all its popularity, however, Python largely lacks static guarantees.

Sound static analysis techniques such as abstract interpretation [15] have been successfully applied to a variety of different programming languages including C [16], Java [55], and recently also to a subset of Python [22, 39] promising to rectify Python's lack of static guarantees. A prerequisite for harvesting the potential of sound static analysis techniques for Python is a formal semantics for Python. In contrast to other languages, however, Python lacks an explicit and comprehensive specification of its semantics [1, 39]. While the *Python Language Reference* (PLR) [20] partially specifies the semantics of Python, it is far from complete and often inaccurate or vague. As a consequence, existing work on the formalization of Python's semantics [53, 44, 27, 22] is based on the PLR but also on extensive experimentation with the reference implementation *CPython*. Although a more rigorous and formal definition of Python's semantics has been called upon also from some within the Python community, there is no established formal or even semi-formal semantics yet [1]. With this thesis, we aim to further future advancements in that direction by developing a *Structural Operational Semantics* (SOS) [42] for a significant subset of Python. Our vision is to eventually arrive at a powerful static analysis tool for Python leveraging abstract interpretation based on the semantics we develop. To our knowledge, we are the first who develop a structural operational semantics for Python. Therefore, we explicitly aim to assesses whether the SOS framework is in general suitable for the formalization of Python and to identify opportunities for future work.

---

[1] https://octoverse.github.com/#top-languages (Accessed: 2020/11/24)

Structural operational semantics have been pioneered by Gordon Plotkin in 1981 [42]. According to Plotkin himself, his seminal notes on SOS "had deliberately not been written in a theoretical framework as [he] wanted not to be constrained but rather to work naturally with the various features" [43, p. 10]. Nevertheless, in subsequent work, the ideas of SOS underwent a more formal treatment from concurrency theorists whose results are condensed in what is known as *rule formats* [2, 24, 43]. Taking inspiration from their work, we develop the necessary meta-theoretical means to make an SOS semantics *executable*. Due to Python's lack of specification it is crucial to asses the adequacy of a formal semantics with test cases [53, 44, 27]. Defining our Python semantics in our meta-theoretic framework enables the automated derivation of an interpreter closely coupled with the semantics. Then, by executing a suite of test cases, the thereby derived interpreter allows us to asses the adequacy of our formal semantics with respect to CPython.

**Contributions**   The core contributions of this thesis are twofold. First, we develop a meta-theoretic framework for the definition of structural operational semantics which enables the automated derivation of interpreters from a formal semantics, and, second, based on this meta-theoretic framework, we develop an executable small-step semantics for a significant subset of Python which in principle allows for concurrency and interaction with an environment. The formal semantics for Python also serves to demonstrate the suitability of the meta-theoretic framework for the definition of formal semantics for modern programming languages involving a variety of control flow mechanisms. As part of this thesis, we provide a prototypical implementation[2] of the proposed techniques. The inference rules presented here have mostly been exported from definitions within our meta-theoretic framework. Furthermore, this implementation finally enables us to evaluate our semantics with respect to CPython and existing work.

**Overview**   Corresponding to the two core contributions, this thesis consists of two major parts. The first part develops the necessary meta-theoretic framework for the formalization and mechanization of structural operational semantics. As explaining and motivating this framework directly with the rather complex structural operational semantics for Python would unnecessarily complicate its introduction, we introduce the framework using the well-established process calculus *Calculus of Communicating Systems* (CCS) [38, 37] as our running example. In the second part, we then introduce our structural operational semantics for Python utilizing the earlier introduced meta theory. As we argued above, it is crucial to be able to execute test cases on the formal semantics to establish sufficient evidence that it is indeed adequate. Hence, a meta-theoretic treatment of structural operational semantics enabling mechanization is a prerequisite for the main aim of this thesis. However, its development is not particularly Python-specific.

---

[2] The implementation is publicly available: https://github.com/koehlma/rigorous

**Structure**   In Chapter 2, we introduce the necessary preliminaries regarding the Python programming language and the general framework of structural operational semantics. We briefly revisit existing work on rule formats and further present and argue for various adequacy criteria for formal semantics in general. The adequacy criteria will later guide the design of the formal semantics for Python and serve as a basis for the final assessment of our formal semantics. We also introduce CCS whose formal semantics are a paradigmatic example of structural operational semantics.

In Chapter 3, we develop a meta-theoretic framework by fully formalizing a variant of the SOS framework thereby making it mechanizable. This chapter is intentionally kept minimalistic and rather abstract in order to not obscure the meta theory with details about a concrete programming language. Nevertheless, the inference rules of CCS introduced in Chapter 2 will motivate the various features of the meta formalism. After introducing the meta formalism, we demonstrate and discuss the accompanying tool support developed for this thesis again using CCS as an example.

In Chapter 4, we present a formal small-step semantics for Python leveraging the meta-theoretic framework introduced earlier. To this end, we flesh out the meta theory with language-specific details such as Python's data types and values. This chapter is furthermore intended to demonstrate the capabilities of the meta theory and its suitability for the formalization of modern programming languages with a variety of control-flow mechanisms. We discuss the advantages and limitations of our semantics over existing work and evidence its adequacy by applying it to a set of test cases.

We conclude this thesis in Chapter 5 by summarizing our contributions and comparing them to prior work. In particular, we discuss the advantages and limitations of our meta-theoretic framework and assess whether our formal semantics for Python meets the adequacy criteria motivated in Chapter 2. Finally, we give an outlook on opportunities for future work.

# Chapter 2

# Preliminaries

In this chapter, we set the stage and introduce the necessary preliminaries for what is to come. In Section 2.1, we give a brief introduction to the Python programming language highlighting some of its distinguishing features and identifying an interesting and significant subset of Python we aim to provide a formal semantics for. In Section 2.2, we introduce the general framework of *Structural Operational Semantics* (SOS) pioneered by Gordon Plotkin [42] and discuss general adequacy criteria for formal semantics. These adequacy criteria will later guide our design decisions and further serve as the basis for assessing our semantics. Furthermore, we touch upon existing formalizations of the SOS framework originating in work on rule formats. Finally, in Section 2.3, we briefly introduce the *Calculus of Communicating Systems* (CCS) [37, 38] which serves as our running example.

## 2.1   Python 101

In this section, we give a brief introduction to the Python programming language. Note that a comprehensive introduction to the language is out of scope. Our main focus will be on the fragment of Python which we formalize in Chapter 4. We argue that this fragment covers anything conceptually interesting and leaves out only details which can faithfully be captured in future work. While we introduce the semantics of concrete constructs in detail in Chapter 4, this section merely gives an overview.

We base this thesis on Python 3.7 which was the most recent version of Python at the time we began writing. Note that the version number of Python evolves in sync with the version number of the reference implementation *CPython*. The *Python Language Reference* (PLR) [20] specifies the *syntax* of Python, i. e., how valid Python programs are formed from textual primitives. Unfortunately, there is no explicit and complete formal or even semi-formal specification of the *semantics*, i. e., the computational meaning of syntactically valid programs, of Python [1, 39]. While the PLR partially specifies the semantics of Python, it is far from complete and often inaccurate or vague. Hence, the reference for this work is both, the PLR but also, where the PLR is insufficient, the reference implementation CPython.

### 2.1.1 Python by Example

Before we dig into the details of the language, here are a few examples showcasing some of the more unique features of Python.

```
1  def fibonacci(n):
2      assert n >= 0, "fibonacci is undefined for negative integers"
3      if n < 2:
4          return n
5      else:
6          return fibonacci(n - 1) + fibonacci(n - 2)
```

Snippet 2.1: Recursive procedure computing the Fibonacci sequence.

Snippet 2.1 shows a recursive procedure computing the Fibonacci sequence. Note that there are no braces or other characters for the separation of logical scopes or blocks because in Python indentation carries syntactic meaning. In Python, integers are arbitrary in precision and only restricted by the available main memory. Hence, this procedure has the intended mathematical semantics because no integer overflows may happen.

**Generators** An interesting feature of Python which has received considerable attention in the literature on formal semantics for Python [22, 44, 27, 53] are *generators*. Generators allow the convenient enumeration of arbitrary objects. For example, Snippet 2.2 generates all the Fibonacci numbers starting with the provided index into the Fibonacci sequence.

```
1  def fibonacci_sequence(start=0):
2      while True:
3          yield fibonacci(start)
4          start += 1
```

Snippet 2.2: Generator for the Fibonacci sequence.

```
1  for x in fibonacci_sequence(6):
2      print(x)
3      if x > 420000:
4          break
```

Snippet 2.3: Printing Fibonacci numbers.

Generators may not terminate, in fact, Snippet 2.2 does not terminate and therefore describes the infinite Fibonacci sequence. Generators can be used in `for` loops to lazily iterate over the generated objects. For instance, the `for` loop in Snippet 2.3 iterates over the infinite sequence of Fibonacci numbers starting with the Fibonacci number 8 at index 6 until a Fibonacci number greater than $420\,000$ is produced. The generator lazily (but inefficiently) produces only those Fibonacci numbers needed.

These examples show the more straightforward use of generators to represent finite or infinite sequences that are computed as needed. Phrased differently, a generator *produces* values as needed until it is *exhausted* which may never be the case. However, generators can also *consume* values.

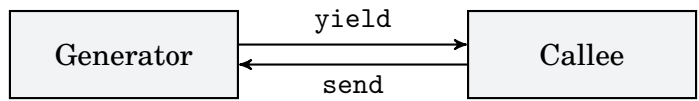

Figure 2.1: The generator protocol.

Figure 2.1 depicts how a callee interacts with a generator in general. Values can be *yielded* to the callee with the `yield` expression. When encountering a `yield`, the generator pauses execution and passes control back to the callee together with the produced value. Whenever the callee wants to retrieve the next value, it resumes the generator by *sending* a value to it—in case of Snippet 2.3 this value is implicitly `None`. One may think of this protocol as a simple synchronous communication channel between the callee and the generator. Snippet 2.4 shows a toy example using this functionality.

```
1  def fibonacci_channel(index=0):
2      while True:
3          index = yield fibonacci(index)
4
5
6  >>> generator = fibonacci_channel(6)
7  >>> generator.send(None)
8  8
9  >>> generator.send(8)
10 21
```

Snippet 2.4: Bi-directional Fibonacci generator.

After creating a generator by calling the procedure, the first value that has to be sent to the generator is always `None`. In this case, this leads the generator to produce the 6th Fibonacci number which is $8$ because $6$ has been passed as the initial index into the Fibonacci sequence as an argument to the procedure. Afterwards the generator consumes indices into the Fibonacci sequence and produces the respective Fibonacci numbers as a result.

While Snippet 2.4 is clearly not a realistic use case, the ability to send values to generators is heavily used for cooperative multitasking using generators as coroutines, e.g., in the popular network framework *Twisted* [60]. Only recently, with Python 3.5, explicit syntactic support for coroutines has been added to Python [51]. We do not support this explicit syntax because deep down there is no difference to how generators work.

Generators have been identified as challenging to formalize by Politz et al. [44] who present a translation of generators to classes. However, the correctness of this translation is at least not self-evident. In general, we believe that a formal semantics should be as perspicuous as possible and complex translations like the one presented by Politz et al. hamper the clarity of a semantics. The formal big-step semantics by Fromherz et al. ignores the possibility to send values to generators altogether [22]. With our semantics we explicitly aim to support generators in a way that does neither rely on complex translations nor sacrifices the possibility to send values to generators. After all, generators and the possibility to send values to them are heavily used in modern applications in the form of coroutines.

**Object Model**   In Python, everything is an object of a certain *class* or *type* including code, functions, and classes themselves. In contrast to other languages like Java [57] there are no non-object primitives. Classes in Python support multiple inheritance using the C3 linearization algorithm [7] and built-ins such as `int` and `str` can be subclassed. The creation of classes happens at runtime and can be customized in various ways via custom *meta classes*, i. e., the classes a class may be an object of.

In general, classes have special *dunder methods* such as `__add__`, `__call__`, or `__getitem__` recognizable by their double underscore naming schema. Among other things, these methods are used for operator overloading. For instance, `__add__` corresponds to the + operator. Snippet 2.5 shows an example where `__getitem__` is implemented to overload indexed access. Objects of the class `Fibonacci` support indexed access as if they were a sequence.

```
1  class Fibonacci:
2      def __getitem__(self, item):
3          return fibonacci(item)
4
5
6  >>> f = Fibonacci()
7  >>> f[6]
8  8
```

Snippet 2.5: Fibonacci class implementing item access.

By implementing dunder methods almost everything can be customized including attribute access or what happens when an object is called. This freedom means that Python's object model is a particular challenge to formalize especially because it is unclear what the primitives are. For example, neither calling an object nor accessing its attributes is a primitive because it might involve executing user-defined code and whether this is the case in general not statically decidable. For some reason this challenge has so far received only little attention in existing work. As a result, on closer inspection, most existing work seems to be inadequate in that respect as a result of an ad-hoc and oversimplified treatment of Python's object model. In contrast, we aim to base our semantics on a well-defined and parsimonious set of primitives. We come back to the details of the object model in Section 4.2.

**Decorators**  A convenient language feature of Python are *decorators*. Decorators allow modifying a function or a class after it has been created. They are functions, taking a class or a function and usually returning a class or function although they may also return something else.

```python
1   _cache = {}
2
3
4   def memoize(function):
5       def wrapper(*args, **kwargs):
6           try:
7               return _cache[(function, args, kwargs)]
8           except KeyError:
9               result = function(*args, **kwargs)
10              _cache[(function, args, kwargs)] = result
11              return result
12
13      return wrapper
14
15
16  @memoize
17  def fast_fibonacci(n):
18      assert n >= 0, "fibonacci is undefined for negative integers"
19      if n < 2:
20          return n
21      else:
22          return fast_fibonacci(n - 1) + fast_fibonacci(n - 2)
```

Snippet 2.6: Fibonacci computation with memoization using a decorator.

Snippet 2.6 shows an example of a decorator for memoization. The function `memoize` is a higher-order function taking a function, e.g., `fast_fibonacci`, and returning a function defined by `wrapper`, which wraps the original function and provides a cache for calls. The *decoration* of `fast_fibonacci` with `@memoize` has the following effect: After the function `fast_fibonacci` has been constructed, which happens at runtime, the function `memoize` is called with the function as an argument. The result returned by `memoize` is then bound to the name `fast_fibonacci` instead of the original function. Hence, the recursive calls to `fast_fibonacci` make use of the cache effectively turning the function into a memoized variant of the original procedure.

Decorators are extensively used and the standard library of Python provides useful general-purpose decorators for recurring problems. For instance, the `functools` module[1] provides a decorator for LRU caching of results of function invocations. Instantiating this decorator with an unbounded cache size, one essentially obtains the above `memoize` decorator.

**Properties**  Direct attribute access is discouraged[2] in many languages like Java because attributes are considered implementation details. Instead explicit getters and setters provide encapsulation decoupling the API of an ob-ject from its internal representation. In Python, however, the concept of *properties* allows defining getters and setters but using them with the ordinary syntax for attribute accesses thereby eliminating the need for explicit getters and setters. In particular, properties allow replacing attributes with getters and setters at any time without breaking an object's API.

---

[1] https://docs.python.org/3.7/library/functools.html (Accessed: 2020/12/14)
[2] https://stackoverflow.com/a/1568230 (Accessed: 2020/12/14)

```python
class X:
    @property
    def y(self):
        return random.randint(0, 10)

    @y.setter
    def y(self, value):
        print(f"y has been set to {value}")

>>> x = X()
>>> x.y
3
>>> x.y
5
>>> x.y = 42
y has been set to 42
```

Snippet 2.7: Usage of properties instead of explicit getters and setters.

Snippet 2.7 shows an example. On retrieving the attribute y the respective getter is executed which in this case just returns a random number. On setting the attribute y the respective setter is executed which in this case just prints a message containing the value the attribute is set to.

Note that properties are realized using decorators. The `property` decorator turns the respective method into a *descriptor*[3] whose getter is then the decorated method. The definition of a setter is optional. If omitted, the property is read-only and trying to set a value will raise an exception. A descriptor is an object implementing the dunder method `__get__` and optionally `__set__`. These methods are called as part of the attribute access. In general, descriptors allow implementing similar mechanisms with additional features. For instance, as of Python 3.8, the `functools` module provides a `cached_property` decorator providing a cache for the value of a property.

### 2.1.2 An Interesting Fragment

As we have seen, Python provides some interesting features posing a challenge for formalization such as generators and the highly customizable object model. For the scope of this thesis, we restrict ourself to a fragment of Python that we further define and argue for in the following. In particular, we argue that the fragment we are considering allows us to faithfully claim that it is possible to eventually support all of Python.

---

[3] https://docs.python.org/3.7/howto/descriptor.html (Accessed: 2020/12/14)

Python is famous for its extensive standard library and batteries-included mentality. It also offers a large amount of built-ins. We only support a subset of those built-ins, namely, *integers*, *strings*, *lists*, *dictionaries*, and *tuples*. The built-ins we are not supporting include *sets* and *bytes*. As we argue in Chapter 4, our semantics can be extended relatively easily with the missing built-ins in future work. For the built-ins we do support, we limit ourselves to a subset of the operations they are supporting.

$\langle expr \rangle$ ::= 'True' | 'False' | 'None' | 'Ellipsis'                    *(constants)*
    | $z \in \mathbb{Z}$ | $x \in Float$ | $s \in String$                    *(literals)*
    | '[' $\langle expr \rangle^*$ ']' | '(' $\langle expr \rangle^*$ ')'                    *(sequences)*
    | '{' ( $\langle expr \rangle$ ':' $\langle expr \rangle$ )* '}'                    *(dictionary)*
    | $\langle target \rangle$                    *(target)*
    | '+' $\langle expr \rangle$ | '-' $\langle expr \rangle$ | '~' $\langle expr \rangle$                    *(unary)*
    | 'not' $\langle expr \rangle$ | $\langle expr \rangle$ 'and' $\langle expr \rangle$ | $\langle expr \rangle$ 'or' $\langle expr \rangle$                    *(boolean)*
    | $\langle expr \rangle$ $\langle \diamond \rangle$ $\langle expr \rangle$                    *(binary)*
    | $\langle expr \rangle$ ( $\langle \bullet \rangle$ $\langle expr \rangle$ )$^+$                    *(comparison)*
    | $\langle expr \rangle$ '(' $\langle args \rangle$ ')'                    *(call)*
    | 'yield' $\langle expr \rangle$                    *(yield)*
    | 'lambda' $\langle params \rangle$ ':' $\langle expr \rangle$                    *(lambda)*

$\langle target \rangle$ ::= $\langle id \rangle$ | $\langle expr \rangle$ '.' $\langle id \rangle$ | $\langle expr \rangle$ '[' $\langle expr \rangle$ ']'                    *(target)*

$\langle \diamond \rangle$ ::= '+' | '-' | '*' | '/' | '//' | '%' | '**'                    *(binary operator)*
    | '>>' | '<<' | '|' | '&' | '^' | '@'

$\langle \bullet \rangle$ ::= '<' | '<=' | '=>' | '>' | '==' | '!='                    *(comparison operator)*
    | 'is' | 'is not' | 'in' | 'not in'

$\langle args \rangle$ ::= ([ '**' | '*' ] $\langle expr \rangle$)* ( $\langle id \rangle$ '=' $\langle expr \rangle$ )*                    *(arguments)*

$\langle stmt \rangle$ ::= $\langle expr \rangle$                    *(expression)*
    | $\langle target \rangle$ '=' $\langle expr \rangle$                    *(assignment)*
    | 'pass'                    *(nop)*
    | 'return' [ $\langle expr \rangle$ ]                    *(return)*
    | 'raise' [ $\langle expr \rangle$ ]                    *(raise)*
    | 'break' | 'continue'                    *(loop control)*
    | 'nonlocal' $\langle id \rangle$ | 'global' $\langle id \rangle$                    *(scope modifier)*
    | 'del' $\langle target \rangle$                    *(delete)*
    | 'assert' $\langle expr \rangle$ [ ',' $\langle expr \rangle$ ]                    *(assertion)*
    | 'if' $\langle expr \rangle$ ':' $\langle body \rangle$ 'else' ':' $\langle body \rangle$                    *(conditional)*
    | 'while' $\langle expr \rangle$ ':' $\langle body \rangle$ 'else' ':' $\langle body \rangle$                    *(while loop)*
    | 'for' $\langle id \rangle$ 'in' $\langle expr \rangle$ ':' $\langle body \rangle$ 'else' ':' $\langle body \rangle$                    *(for loop)*
    | 'try' ':' $\langle body \rangle$ 'finally' ':' $\langle body \rangle$                    *(try-finally)*
    | 'try' ':' $\langle body \rangle$ ( 'except' [ $\langle expr \rangle$ [ 'as' $\langle id \rangle$ ]] ':' $\langle body \rangle$ )$^+$ 'else' ':' $\langle body \rangle$                    *(try-except)*
    | ( '@' $\langle expr \rangle$ )* 'def' $\langle id \rangle$ '(' $\langle params \rangle$ ')' ':' $\langle body \rangle$                    *(function)*
    | ( '@' $\langle expr \rangle$ )* 'class' $\langle id \rangle$ '(' $\langle args \rangle$ ')' ':' $\langle body \rangle$                    *(class)*

$\langle params \rangle$ ::= ([ '**' | '*' ] $\langle id \rangle$ [ '=' $\langle expr \rangle$ ])*                    *(parameters)*

$\langle body \rangle$ ::= $\langle stmt \rangle^*$                    *(body)*

Figure 2.2: The abstract grammar of the Python fragment.

Figure 2.2 shows the abstract grammar of the fragment of Python we are supporting. For our purposes, we do not have to concern ourselves with the literal syntax of Python. The reference implementation provides a module `ast`[4] for parsing Python code which produces an abstract syntax tree corresponding to the abstract grammar defined in Figure 2.2.

---

[4] https://docs.python.org/3.7/library/ast.html (Accessed: 2020/12/14)

Considering the syntax of Python, we do not support coroutines with `async` and `await` syntax [51], `yield from` expressions and statements [18], comprehensions [20, §6.2.4.], formatted strings [20, §2.4.1.], `with` statements [20, §8.5.], and augmented assignments [20, §7.2.]. Those features do not add anything conceptually interesting that is not already covered by other language features we do support. Before there was explicit syntax for coroutines, coroutines have been implemented with generators, which we do support. The interesting aspects of `yield from` are already covered by `yield` and the remaining constructs are mostly syntactic sugar. Comprehensions are syntactic sugar for generators, formatted strings are syntactic sugar for string constructions, the `with` statement is syntactic sugar for `try` statements, and augmented assignments can be desugared by calling a dunder method on the assignment target followed by a normal assignment.

By excluding those features, we limit ourselves to an interesting fragment of Python without sacrificing the potential to faithfully capture the whole language eventually. Most of the syntactic features we exclude are mere syntactic sugar and built-ins can be added later.

### 2.1.3 Implementation Freedom

While there is no doubt about the syntax of Python, the PLR [20] leaves much room for implementation freedom with respect to its semantics. Naturally, when developing a formal semantics for Python, the question arises how one deals with this implementation freedom.

There are two entirely different ways how to deal with implementation freedom. One can try to capture the implementation freedom itself in the formal semantics, e. g., by means of non-determinism, or just concretize it. Existing work [53, 44, 27, 22] deals with implementation freedom by concretizing it in some way usually following the reference implementation CPython. We believe that trying to capture the implementation freedom and other vagueness of the PLR would be a futile endeavour and, hence, we follow the existing work in that regard. Note that this question is of particular importance with respect to static analysis because guarantees will only hold with respect to the concrete semantics and not necessarily for any implementation of Python that resolves the implementation freedom differently. When resolving implementation freedom we put an emphasis on efficient implementability, i. e., the formal semantics should be efficiently implementable, and staying close to the reference implementation of Python.

## 2.2 Structural Operational Semantics

In general, there exist different approaches to formal semantics. This thesis builds on top of the general framework of *Structural Operational Semantics* (SOS) as pioneered by Gordon Plotkin [42] in 1981. In this section, we briefly recapitulate the basic ideas and concepts of SOS. For a more comprehensive overview and introduction we refer to the paper by Plotkin. Afterwards we discuss general adequacy criteria for formal semantics and touch upon existing formalizations of the SOS framework.

The central ideas of SOS are to (i) cut the execution of a program into small pieces, coined *execution steps*, and (ii) define the execution steps for compound program structures in terms of the execution steps of their components. The resulting semantics are *structural* in the sense that a program's behavior is defined inductively based on its syntactic structure.

At the heart of SOS are inference rules of the form

$$\frac{P_1 \quad \cdots \quad P_k}{C} \; \textsc{r}$$

where $P_1$ to $P_k$ with $k \in \mathbb{N}_0$ are the *premises* and $C$ is the *conclusion* of the inference rule $\textsc{r}$. Inference rules without premises are called *axioms*. The general idea is that the conclusion can be derived from the premises or, phrased differently, if the premises are true then so is the conclusion. Inference rules are combined to *inference trees* representing a derivation starting with only axioms and involving the application of multiple rules.

In Chapter 3, we provide a full formal definition of inference rules and trees as they appear in this thesis. For our present purposes, however, a pre-theoretic and mere intuitive understanding shall suffice.

Let us introduce the following running example:

> **Example 1 (Binary Addition)**
> Consider a binary addition operator $+$ that is used to form *binary addition expressions* of the form $e_l + e_r$. Now, SOS inference rules for the evaluation of binary addition expressions may look like this:
>
> $$\frac{e_l \to u}{(e_l + e_r) \to (u + e_r)} \; \textsc{l}_+ \qquad \frac{e_r \to u}{(e_l + e_r) \to (e_l + u)} \; \textsc{r}_+ \qquad \frac{z_l, z_r \in \mathbb{Z} \quad z_l + z_r = z}{(z_l + z_r) \to z} \; \textsc{e}_+$$
>
> Here $\textsc{l}_+$, $\textsc{r}_+$, and $\textsc{e}_+$ are the *names* of the inference rules.

We introduce the meaning of the different colors in a second when discussing the formation of inference trees. Here, $\to$ should be read as *evaluates in an execution step to*. The rule $\textsc{l}_+$ then formalizes the idea that if an expression $e_l$ evaluates in an execution step to $u$ then the binary addition expression $e_l + e_r$ evaluates in an execution step to $u + e_r$. Analogously, the rule $\textsc{r}_+$ concerns the evaluation of the right operand. Using these rules, both operands can be evaluated stepwise until they become integers. If both operands are integers then the rule $\textsc{e}_+$ becomes applicable which formalizes the idea that if both operands are integers then $z_l + z_r$ evaluates to the sum of $z_l$ and $z_r$.

To prove an execution step $e \rightarrow e'$, e.g., $(2 + 13) + 42 \rightarrow 15 + 42$, an inference tree with the conclusion $e \rightarrow e'$ is constructed. For example:

$$\frac{\dfrac{2, 13 \in \mathbb{N} \quad 2 + 13 = 15}{2 + 13 \rightarrow 15} \ \text{\small E}_+}{(2 + 13) + 42 \rightarrow 15 + 42} \ \text{\small L}_+ \tag{2.1}$$

This inference tree consists of two *instances* of the inference rules provided in Example 1. These instances are:

$$\text{(a)} \ \frac{2 + 13 \rightarrow 15}{(2 + 13) + 42 \rightarrow 15 + 42} \ \text{\small L}_+ \qquad \text{(b)} \ \frac{2, 13 \in \mathbb{Z} \quad 2 + 13 = 15}{2 + 13 \rightarrow 15} \ \text{\small E}_+$$

Note that the premise of instance (a) is the conclusion of instance (b). This enables stacking the instance (b) on top of instance (a) and thereby form an inference tree as in (2.1). An instance is obtained by substituting the *meta variables* in the respective rule. For example, to obtain instance (a) from rule $\text{\small L}_+$, $e_l$ is substituted with $2 + 13$, $e_r$ with $42$, and $u$ with $15$. What appears to be premises in $\text{\small E}_+$ are technically constraints with respect to possible substitutions of the meta variables. Hence, $\text{\small E}_+$ is an axiom.

We color meta variables orange, symbols like $+$ or $\rightarrow$ turquoise, and values like $15$ or $42$ green. Later, in Chapter 3, we introduce these concept formally and define how inference rules are formed using them.

Inference rules and trees define and allow proving individual execution steps. Iteratively chaining individual steps then yields a program's complete behavior. For example, after evaluating the *program state* $(2 + 13) + 42$ to $15 + 42$, $15 + 42$ becomes the next program state and is evaluated to $57$:

$$\frac{15, 42 \in \mathbb{N} \quad 15 + 42 = 57}{15 + 42 \rightarrow 57} \ \text{\small E}_+$$

Technically speaking, the inference rules define an *evaluation relation* $\rightarrow$ over program states capturing the semantics of the language.

**Labeled Execution Steps** As we will see, it is often useful to assign *labels* to execution steps. For example, labels might be used to indicate side effects produced by an execution step:

$$\text{print 'Hello World!'} \xrightarrow{\text{\small PRINT! 'Hello World!'}} \bot$$

In this example, the program state print 'Hello World!' evaluates to $\bot$ with the side effect of printing 'Hello World!' as indicated by the label:

$$\text{PRINT! 'Hello World!'}$$

Formally, a structural operational semantics with labeled execution steps defines a ternary relation relating program states via labels with possible successor states. The semantics of some initial program state $\bar{s}$ is then captured by a *Labeled Transition System* (LTS) [42].

> **Definition 1 (Labeled Transition Systems)**
> A labeled transition system is a tuple $\mathfrak{T} = \langle S, \mathrm{Act}, \rightarrow, \bar{s} \rangle$ where $S$ is a set of *states*, $\mathrm{Act}$ is a set of *action labels*, $\rightarrow \subseteq S \times \mathrm{Act} \times S$ is the *transition relation*, and $\bar{s} \in S$ is the *initial state*.

The inference rules of a structural operational semantics define the transition relation $\rightarrow$ between program states, i. e., $\langle s, \alpha, s' \rangle \in \rightarrow$ if and only if $s \xrightarrow{\alpha} s'$ is derivable with the inference rules of a semantics.

## 2.2.1 Adequacy Criteria

Before developing a formal semantics, we should ask ourselves what we actually want from a formal semantics. What properties should a formal semantics ideally have? We now take the time to discuss some general adequacy criteria for formal semantics. Note that this discussion neither aims at being comprehensive nor at defining formal criteria.

Clearly, a formal semantics should adequately capture the intuitive semantics of a programming language. In case of Python, the PLR partially provides a natural language description of these intuitive semantics. A formal semantics is often the foundation for further techniques such as abstract interpretation [15] or symbolic execution [34]. The correctness proofs for these techniques always hold relative to the formal semantics of a language. If the actual implementation of a language and its formal semantics diverge, this renders the results of these techniques at least questionable if not invalid. Therefore, it is of uttermost importance that a formal semantics adequately captures the intuitive semantics of a language and to establish evidence that it also agrees with the implementation of a language.

To gain confidence that a formal semantics adequately captures the intuitive semantics one has in mind and also agrees with an implementation, it is useful to come up with some test cases based on the intuitive semantics one has in mind and examine how they execute according to the formal semantics and the implementation. To this end, a formal semantics should be *executable*, i. e., it should be possible to derive an interpreter from it in an automated fashion. This interpreter, in turn, can then be used to execute the test cases and establish evidence that the formal semantics indeed corresponds to the intuitive semantics as well as that it agrees with an implementation, in our case Python's reference implementation. We are well aware of the fact that testing is not proving, however, first, it is impossible to prove that a formal semantics captures the intuitive semantics one has in mind due to their informal nature, and, second, in practice it is infeasible to prove the equivalence of a formal semantics with an implementation like Python's reference implementation comprising hundreds of thousands lines of code.

Other important desiderata for formal semantics are *clarity*, *succinctness*,

and *intelligibility*. A formal semantics should not be an obscure blob of definitions. Instead, there should be a succinct set of clear definitions constituting a formal semantics. At best, there is a self-evident correspondence between individual definitions and intuitive language concepts. A succinct set of clear definitions self-evidently related to intuitive language concepts fosters comprehension of the formal semantics and thereby enables one to judge, beyond blind testing, whether the formal semantics actually adheres to the intuitive semantics one has in mind. It renders a semantics intelligible providing insights into the inner workings of a language.

Another enabler of intelligibility is parsimony. A formal semantics should be parsimonious with respect to the concepts and primitives it introduces. As it is impossible to formalize anything without appealing to intuition at some point, it is important that the concepts and primitives introduced by a formal semantics are well-understood and not too complex. For instance, in example Example 1 we used the addition of integers as a primitive. This primitive is well-understood and the operation it performs is rather basic.

In this thesis, we present a meta-theoretic formalism enabling the automated derivation of an interpreter from a formal semantics thereby making the semantics *executable*. This interpreter allows us to gain confidence in our semantics for Python by executing test cases as described above. In addition, we aim at a clear, succinct, intelligible, and parsimonious semantics. We assess in Chapter 5 whether our semantics meets these goals.

### 2.2.2 Rule Formats

According to Plotkin, his seminal notes on SOS "had deliberately not been written in a theoretical framework as [he] wanted not to be constrained but rather to work naturally with the various features" [43, p. 10]. As a result, SOS is more about the idea and there is no precise formal criterion on whether a semantics deserves the predicate *structural operational semantics* or not. This core idea is that "the rules should be syntax-directed [43, p. 7], hence the name *structural*. Nevertheless, in subsequent work, the ideas of SOS underwent a more formal treatment from concurrency theorists whose results are condensed in what is known as *rule formats* [2, 24, 43]. Rule formats restrict the form of inference rules such that all semantics adhering to a specific format provably share specific desirable properties.

Recall that we aim at a meta formalism enabling the automated derivation of interpreters from a formal semantics. To this end, we also need to formalize the concept of inference rules as done in work on rule formats. Taking inspiration from existing work, we aim to capture inference rules as they appear in Example 1 and elsewhere in the literature. We briefly recap the existing definitions found in work on rule formats and then state why these are not suitable for our purposes. Hence, we develop our own meta theory for SOS in Chapter 3 building on top of existing work.

Existing work in the area of rule formats shares as a common basis the usage of *terms* [24]. Let $\mathbb{X}$ be a set of *variables* and $\Sigma$ be a set of *function symbols*. Terms are defined inductively. Each variable $x \in \mathbb{X}$ is a term and for each $k$-ary function symbol $f \in \Sigma$, $f(t_1, \ldots, t_k)$ is a term where $t_i$ with $1 \leq i \leq k$ are terms [2]. Following Groote et al., a *Transition System Specification* (TSS) is a set of rules where each rule has the following form [24, p. 11]:

$$\frac{\left\{ P_i(L_i)t_i \;\; \text{or} \;\; t_i \xrightarrow{L_i} t_i' \;\middle|\; i \in I \right\} \qquad \left\{ \neg P_j(L_j)t_j \;\; \text{or} \;\; t_j \xslashed{\xrightarrow{L_j}} \;\middle|\; j \in J \right\}}{P(L)t \;\; \text{or} \;\; t \xrightarrow{L} t'} \tag{2.2}$$

Here, $t$, $t_x$, and $t_x'$ are terms, $L$ and $L_x$ are lists of terms, $I$ and $J$ are arbitrary and possibly infinite index sets, and $P$ and $P_x$ are predicate symbols over terms. By further restricting this general scheme one obtains different rule formats [24] such as the *de Simone* format [17], the *GSOS* format [9], or the *PATH* format [5]. Based on these restrictions one can then prove several meta theorems about all formal semantics, i.e., TSSs, comprising only rules adhering to a respective rule format [24].

The rules $\textsc{l}_+$ and $\textsc{r}_+$ of Example 1 can be captured in such a formalism assuming that $+$ is a binary function symbol written in infix notation and $e_l$, $e_r$, and $u$ are variables. To also capture the rule $\textsc{e}_+$, however, one has to somehow make sense of the addition occurring in the constraints of the rule. While rules in the literature [38, 42, 26, 50, 46, eg.] often contain such *operator terms* with a fixed computational meaning, they are not part of the formalisms introduced in the area of rule formats. Instead, constraints are seen as giving rise to a potentially infinite number of rules [5]. In case of $\textsc{e}_+$ this would mean that we would have a rule for each pair $z_l, z_r \in \mathbb{Z}$ such that for this rule $z$ is substituted with the sum of $z_l$ and $z_r$. Such *rule-generating* constraints can be almost arbitrary and are shifted to the meta level of specifying a TSS. In the next section, we will see that the inference rules for the well-established process calculus CCS also contain such operator terms.

While shifting operator terms to the meta level of specifying a TSS certainly is a smart move for theoretical purposes, we aim at a meta formalism enabling the automated derivation of interpreters from a formal semantics. Hence, we have to incorporate operator terms directly into our SOS framework and cannot leave them unformalized by shifting them to the meta level. We fully formalize operator terms and allow them to occur arbitrarily inside of terms. The resulting meta theory allows us to capture inference rules with operator terms—in particular, we can handle operator terms inside of premises and conclusions as found in the literature and do not have to desugar such usages to constraints. In terms of expressivity, we do not allow negative premises, i.e., premises requiring that a transition with a certain label is not possible as indexed by $J$ in the rule schema (2.2). Instead of forcing the premises and conclusions into a fixed layout tailored for SOS, we also push the transition relation $\rightarrow$ itself to the term level, i.e., the rules we allow do not necessarily adhere to the schema (2.2). In our framework, the premises and conclusions of inference rules can be almost arbitrary terms.

## 2.3 Calculus of Communicating Systems

The *Calculus of Communicating Systems* (CCS) has been introduced by Robin Milner in 1982 [37]. It is a paradigmatic example of process calculi and valuable for the investigation of various concurrency problems. Here, we use it as an example in order to demonstrate that we are able to capture inference rules as they appear in the literature and to give a intuition how concurrency may be dealt with in our Python semantics.

The process calculus CCS is used in the course *Nebenläufige Programmierung* at Saarland University which teaches students the concepts of concurrent programming. The inference rules we present here are taken from the script of that lecture [29] and the original papers by Milner [37, 38].

The intuitive idea of CCS is is to capture the behavior of *processes* which can perform certain *actions* by means of *process terms*. For instance, the process described by the process term (a . 0) does perform the action a and is unable to do anything else afterwards. Each process term $P$ induces a labeled transition system $\langle S, \mathrm{Act}, \rightarrow, P \rangle$ (cf. Definition 1) coined its *LTS semantics*. As for labeled transition systems, let Act be a set of actions. In addition, let **X** be a set of *process variables*. The set of process terms **P** over Act and **X** is inductively defined as follows

$$\mathbf{P} \ni P \ ::= \ 0 \ | \ (\alpha \ . \ P) \ | \ (P_1 + P_2) \ | \ (P_1 \parallel P_2) \ | \ (P \setminus H) \ | \ X \qquad (2.3)$$

where $\alpha \in \mathrm{Act}$ is an action, $H \subseteq Act$ is a set of actions, and $X \in \mathbf{X}$ is a *process variable*. Intuitively, the *prefix operator* $(\alpha \ . \ P)$ means that the process can perform an action $\alpha$ and afterwards behaves like the process $P$, the *choice operator* $(P_1 + P_2)$ means that the process behaves either like $P_1$ or like $P_2$, the *parallel operator* $(P_1 \parallel P_2)$ captures the concurrent execution of $P_1$ and $P_2$, and the *restriction operator* $(P \setminus H)$ means that the process behaves like the process $P$ but is unable to perform actions in the set $H$. Process variables are used to define recursive processes. To this end, they are bound to a process term by an environment $\Gamma$ mapping process variables to process terms. Intuitively, the process described by $X$ then behaves like $\Gamma(X)$. We now turn to the inference rules capturing these intuitive behaviors and defining the transition relation of a process term's LTS semantics.

The semantics of the prefix operator are captured by:

$$\frac{}{\Gamma \vDash (\alpha \ . \ P) \xrightarrow{\alpha} P} \text{ PREFIX}$$

Here, $\Gamma$ is the aforementioned environment in which the process executes in. Based on this inference rule one can for instance derive $\Gamma \vDash (\mathrm{a} \ . 0) \xrightarrow{\mathrm{a}} 0$, i. e., within the concrete environment $\Gamma : \mathbf{X} \rightarrow \mathbf{P}$ the process (a . 0) performs an action a and afterwards becomes the process 0.

The semantics of the choice operator + is captured by two rules:

$$\frac{\Gamma \vDash P \xrightarrow{\alpha} P'}{\Gamma \vDash (P + Q) \xrightarrow{\alpha} P'} \text{ CHOICE-L} \qquad \frac{\Gamma \vDash Q \xrightarrow{\alpha} Q'}{\Gamma \vDash (P + Q) \xrightarrow{\alpha} Q'} \text{ CHOICE-R}$$

Intuitively, if both process $P$ and $Q$ in $P + Q$ can perform some action respectively, then both rules are applicable and, hence, the process $P + Q$ non-deterministically behaves either like $P$ or like $Q$.

If two processes are executed in parallel, then either can perform an action on their own. This idea is captured by two rules:

$$\frac{\Gamma \vDash P \xrightarrow{\alpha} P'}{\Gamma \vDash (P \parallel Q) \xrightarrow{\alpha} (P' \parallel Q)} \text{ PAR-L} \qquad \frac{\Gamma \vDash Q \xrightarrow{\alpha} Q'}{\Gamma \vDash (P \parallel Q) \xrightarrow{\alpha} (P \parallel Q')} \text{ PAR-R}$$

Again, if both processes can perform an action this leads to non-determinism in terms of multiple *interleavings* of both processes.

Ignoring the environment, the presented inference rules fit into the schema (2.2). We now turn to inference rules that do not strictly fit into this schema, i.e., at least not without reverting to rule-gernerating side constraints which are left to the meta level of specifying TSS.

The rule REC looks up a process variable in the environment:

$$\frac{\Gamma \vDash P \xrightarrow{\alpha} P' \qquad P = \Gamma[X]}{\Gamma \vDash X \xrightarrow{\alpha} P'} \text{ REC}$$

As a result, a process $X \in \mathbf{X}$ can perform an action $\alpha$ within an execution environment $\Gamma$ if and only if $\Gamma(X)$ can perform an action $\alpha$. After performing $\alpha$ the process $X$ behaves just like $\Gamma(X)$ after performing $\alpha$. This rule does not fit the schema (2.2) because it utilizes the execution environment and contains the operator term $\Gamma[X]$. Here, $\Gamma[X]$ has the fixed computational meaning of applying whatever has been bound to the meta variable $X$ to whatever function has been bound to the meta variable $\Gamma$, i.e., if $\Gamma \mapsto \Gamma$ with $\Gamma : \mathbf{X} \to \mathbf{P}$ is a function and $X \mapsto X$, then it is the process term $\Gamma(X)$.

Milner later introduced a *fix-operator* replacing the environment [38]. To this end, the set of process terms as defined in (3.1) is extended with process terms of the form $(\texttt{fix } X = P)$ where $X$ is a process variable and $P$ is a process term. The semantics of this operator are defined as follows:

$$\frac{\Gamma \vDash P[X \mapsto (\texttt{fix } X = P)] \xrightarrow{\alpha} P'}{\Gamma \vDash (\texttt{fix } X = P) \xrightarrow{\alpha} P'} \text{ FIX}$$

Here, $P[X \mapsto (\texttt{fix } X = P)]$ is an operator term corresponding to a ternary replacement operator, i.e., the idea is to take whatever process term $P$ has been bound to $P$ and replace each occurrence of whatever process variable $X$ has been bound to $X$ with $P$ in $P$. The rule FIX again does not strictly fit the schema (2.2) because the ternary operator term occurs within a premise but the notion of terms introduced for (2.2) only knows function symbols which do not carry any fixed computational meaning.

As the name "Calculus of Communicating Systems" suggests, processes can communicate with each other. They do so by means of *synchronized actions*. At this point, we have to take a closer look at the set of actions Act. We assume that the set Act is partitioned into a set of *generative* actions of the form $(a\,!)$, a set of *reactive* actions of the form $(a\,?)$, and a set of *internal actions*. Two processes can synchronize if one process performs a generative action and the other performs a corresponding reactive action, i.e., one performs $(a\,!)$ and the other $(a\,?)$ for some $a$. For the corresponding inference rule, we introduce a *complement* operator defined as follows:

$$\overline{\alpha} := \begin{cases} (a\,!) & \text{if } \alpha = (a\,?) \\ (a\,?) & \text{if } \alpha = (a\,!) \\ \alpha & \text{otherwise} \end{cases}$$

The complement operator transforms a generative action in its reactive counterpart and vice versa. The synchronization rule is then:

$$\frac{\Gamma \vDash P \xrightarrow{\alpha} P' \quad \Gamma \vDash Q \xrightarrow{\overline{\alpha}} Q'}{\Gamma \vDash (P \parallel Q) \xrightarrow{\tau} (P' \parallel Q')} \text{ SYNC}$$

Intuitively, for two processes executing in parallel, if one performs a generative action $(a\,!)$ and the other performs the reactive counterpart $(a\,?)$, both processes synchronize and take an execution step together producing an internal $\tau$ action. Again, this rule does not strictly fit the schema (2.2) due to the operator term applying the complement operator.

Last but not least, the inference rule RES handles process terms of the form $(P \setminus H)$ by not allowing $P$ to perform actions $\alpha \in H$:

$$\frac{\Gamma \vDash P \xrightarrow{\alpha} P' \quad \alpha \notin H}{\Gamma \vDash (P \setminus H) \xrightarrow{\alpha} (P' \setminus H)} \text{ RES}$$

This rule fits the schema (2.2) again. However, seen as a part of a TSS definition it generates a rule for each subset $H \subseteq \text{Act}$.

**Outlook** The inference rules printed here match the rules found in the literature. They have been exported to LaTeX and rendered into this thesis utilizing the tool support we are going to develop in the next chapter. As they are identical, appart from the coloring which can easily be turned off, to the rules found in Milner's original papers [38, 37] and the script of the lecture Nebenläufige Programmierung [29] they are arguably intuitive to understand without knowing the formal meta theory behind them. Put differently, the meta theory we are about to develop allows defining structural operational semantics in a way that is intuitive and closely resembles how SOS are used in research as well as in introductory courses.

Figure 2.3 shows the interpreter for CCS automatically generated from the same definitions used to print the rules in this section. Provided with a process term and the inference rules, the tool derives execution steps by applying the inference rules and outputs a tree for each step.

```
PS rigorous> rigorous-ccs explore --print-trees "(a?.0 || a!.0) \ {a?, a!}"
Initial State: ((((a ?) . 0) ∥ ((a !) . 0)) \ {(a ?), (a !)})

Source: ((((a ?) . 0) ∥ ((a !) . 0)) \ {(a ?), (a !)})
Action: τ
Target: (0 ∥ 0) \ {(a ?), (a !)})
                prefix ──────────────────────────        prefix ──────────────────────────
                       ({} ⊢ ((a ?) . 0) = (a ?) ⇒ 0)           ({} ⊢ ((a !) . 0) = (a !) ⇒ 0)
          sync ──────────────────────────────────────────────────────────────────────────
                       ({} ⊢ (((a ?) . 0) ∥ ((a !) . 0)) = τ ⇒ (0 ∥ 0))                    τ ∉ {(a ?), (a !)}
          res ─────────────────────────────────────────────────────────────────────────────────────────────
                       ({} ⊢ ((((a ?) . 0) ∥ ((a !) . 0)) \ {(a ?), (a !)}) = τ ⇒ ((0 ∥ 0) \ {(a ?), (a !)}))
```

Figure 2.3: An automatically derived interpreter for CCS.

In the next chapter, we develop the theory and algorithms behind this tool and discuss its implementation. To this end, we introduce a notion of terms allowing for operator terms with a fixed computational meaning as they occur in the various rules we have encountered so far. Note that the corresponding operators sometimes operate on terms themselves like the replacement operator used in the rule FIX. The resulting meta theory and tool support is the basis for the Python semantics we develop in Chapter 4.

# Chapter 3

# Meta Theory

In this chapter, we formalize and mechanize a variant of SOS as introduced in
Section 2.2. To this end, we harvest results from the area of *term unification*
[4] and take inspiration from work on rule formats [24]. The fundamental
idea is that a *term*—which is the kind of mathematical object premises and
conclusions are formalized as—may be derived by an inference rule whose
conclusion somehow *matches* the term, i. e., the term and the conclusion are
*unifiable*. By recursion on the premises and backtracking, this then yields
an algorithm constructing inference trees.

Although developed explicitly for SOS, our *meta theory* is universal and may
serve as a foundation for other formalisms based on inference rules. We for-
mally define the necessary meta concepts such as inference rules, trees, and
derivation in inference rule systems, which we prove to be Turing-complete.
We devise a semi-algorithm—the *inference engine*—for derivation, which also
allows enumerating answers to *inference questions*. Building on the inference
engine, we demonstrate how to derive an interpreter from a formal semantics
expressed in our framework. Finally, we discuss our implementation of the
presented algorithms providing tool support and present a case-study around
the *Calculus of Communicating Systems* (CCS) [37].

The advantages of a fully formalized and mechanized meta theory for the def-
inition of formal semantics seem obvious: It renders the semantics tractable
by computer programs enabling the automatization of all kinds of tasks. For
instance, it allows deriving an interpreter for a language automatically which,
if done right, is closely coupled with the formal semantics and allows gaining
confidence in the semantics by executing test cases. In Chapter 4, we cash in
on these advantages and validate our formal semantics for Python by execut-
ing a set of test cases on a derived interpreter.

Note that this chapter is intentionally kept rather abstract and leaves open
the concrete instantiation of many parts of the meta theory to not obscure it
with details about a concrete programming language. In particular, almost
no assumptions about the kinds of data values are made. We instantiate the meta theory for Python in Chapter 4. By keeping the core abstract and small, we hope for a perspicuous and extensible implementation.

**Design Goals**   Our explicit design goals are: (a) a small language-agnostic core, (b) the ability to formalize semantics in a way that closely resembles how SOS are used in research, and (c) the ability to automatically derive an interpreter from a formal semantics. This interpreter is explicitly not required to be fast and efficient. Instead, a close coupling between the semantics and the interpreter should be ensured. Goal (b) is intended to ensure that a semantics defined within our framework is intuitive and easy to understand by everyone who is generally familiar with SOS even without reading this chapter and knowing the formal meta theory behind them.

**Structure**   As already mentioned, we build upon existing research from the area of term unification. In Section 3.1, we begin by introducing a notion of *terms*. Terms take the place of premises and conclusions in inference rules. We innovate by extending the traditional notion of terms with operator terms which carry semantic meaning and can be evaluated.

In Section 3.2, we pose the problem of *semantic term unification*. In contrast to traditional term unification, semantic term unification is not indifferent to the semantic meaning of operator terms. Therefore, the known algorithms for term unification have to be adapted to the new problem. We present the necessary changes to a traditional term unification algorithm and formally argue for their correctness. The resulting semantic unification algorithm then serves as the basis around which we build the inference engine.

In Section 3.3, we describe how inference rules and trees are formed out of terms and other objects. We subsequently introduce a corresponding notion of derivability in an inference rule system and prove it to be undecidable in general. Based on semantic term unification, we devise a semi-algorithm coined *inference engine* which allows us to ask and answer inference questions. We then describe how the generic inference engine is used in the context of SOS to mechanize the application of inference rules and thereby derive an interpreter based on the formal semantics of a language.

In Section 3.4, we finally discuss our implementation of the algorithms presented in the previous sections as an extensible Python library. Our implementation provides the tool support necessary to specify formal semantics and make them executable. We heavily rely on it in the next chapter when we introduce our formal semantics for Python. To round this section off, we demonstrate the meta theory and our tool on simple arithmetic addition (cf. Example 1) and on the process calculus CCS [37].

In Section 3.5, we conclude this chapter by summarizing our findings and briefly discussing alternative approaches and limitations.

Based on the insights gained in Chapter 4 by applying the theory put forward in this chapter to the Python programming language, we argue in Chapter 5 that we meet the aforementioned design goals. Furthermore, in Section 5.1, we compare our approach to formal semantics with related work such as the $\mathbb{K}$ *semantic framework* [49] and the *Process Algebra Compiler* [12].

## 3.1 Terms

When formalizing SOS, the first question arising is what kind of mathematical objects premises and conclusions are. Premises and conclusions are what we call *terms*. *Terms*—either in the form of an *atom* or a *composed term*—are the primary object of our meta-theoretic framework. In this section, we inductively define a set of terms $\mathcal{T}$. To this end, we expand upon existing notions from the area of term unification [4, 47].

> **Definition 2 (Term Atoms)**
> Let $\mathbb{X}$ be a countable set of *meta variables*, $\Sigma$ be a countable set of *symbols*, and $\mathbb{V}$ be a countable set of *data values* such that $\mathbb{X} \cap \Sigma = \emptyset$ and $\mathbb{X} \cap \mathbb{V} = \emptyset$. We call elements $\tau \in \mathbb{X} \cup (\Sigma \cup \mathbb{V})$ *term atoms*.

We already informally introduced these concepts in Chapter 2. As in Chapter 2, we visually distinguish the three different kinds of term atoms by color. We color symbols $s \in \Sigma$ turquoise, data values $\nu \in \mathbb{V}$ green, and meta variables $x \in \mathbb{X}$ orange. Note that a term atom may be both, a symbol and a data value, because we did not assume $\mathbb{V}$ and $\Sigma$ to be disjoint. In case a term is a symbol and also a data value we usually color it turquoise.

To build more complex compound terms from term atoms, we introduce two ways to combine terms into terms—*sequence terms* and *operator terms*. For instance, recapitulate the conclusion $(e_r + e_l) \rightarrow (u + e_l)$ of the inference rule $\text{R}_+$ of Example 1. Here, $e_r$, $e_l$, and $u$ are meta variables and $+$ and $\rightarrow$ are symbols that are combined into a sequence term.

> **Definition 3 (Sequence Terms)**
> Let $\tau_1, \ldots, \tau_n \in \mathcal{T}$ with $n \in \mathbb{N}_{>0}$ be terms, then $(\tau_1 \cdots \tau_n) \in \mathcal{T}$ is a *sequence term*. We call $\tau_i$ the *components* of the sequence $(\tau_1 \cdots \tau_n)$.

We use parenthesis to explicitly specify the structure of sequence terms and usually omit the outermost parentheses. For example, $\tau_1(\tau_2\tau_3)$ is a sequence that consists of $\tau_1$ and the inner sequence $\tau_2\tau_3$. Crucially, it is distinct from the flat sequence $\tau_1\tau_2\tau_3$, i.e., $(\tau_1(\tau_2\tau_3)) \neq (\tau_1\tau_2\tau_3)$, i.e., it is important not to understand $(\tau_1 \cdots \tau_n)$ as a mere concatenation of the components $\tau_1$ to $\tau_n$ but instead as a new and distinct object. Intuitively, Definition 3 then allows us to capture tree structures such as abstract syntax trees.

Let us turn back to Example 1 again. The rule $E_+$ also contains the constraint $z_l + z_r = z$ where $+$ is the arithmetic addition operator carrying a specific *semantic meaning*. As we already saw, most inference rules occurring in research make use of operators in some form or another. For instance, it is quite common that inference rules make use of partial functions or *mappings* to store context information such as type environments [50], register files [26], or program memory [46]. In all these cases, we may think of the rules as involving an operator taking a mapping and a *key* as arguments and evaluating to the value to which the key is mapped by the mapping. The rule REC of the CCS semantics (cf. Section 2.3) is a paradigmatic example of such rules. To capture these operator terms, we enrich our definitions as follows:

> **Definition 4 (Operator Terms)**
> Let $\mathfrak{O}$ be a set of variable-arity *operator symbols*, then $op(\tau_1, \ldots, \tau_k)$ is an *operator term* for each $k$-ary operator symbol $op \in \mathfrak{O}$. Each $k$-ary operator symbol $op \in \mathfrak{O}$ corresponds to a $k$-ary partial function $[\![op]\!] : \mathbb{V} \times \ldots \times \mathbb{V} \rightharpoonup \mathbb{V}$ capturing the semantic meaning of the operator symbol.

We intentionally stay largely agnostic on operators and the values they operate on. Contrary to the explicit aim of typical proof assistants such as Coq[1] to rely on a small but powerful set of primitives like the *Calculus of Inductive Constructions* [8], we aim at an extensible theory where it is easy to drop in primitives as needed. Ideally, it should be easy to extend our meta theory with new operators and values and thereby adapt it to a particular programming language or for a particular other purpose. For instance, for the rule SYNC of the CCS semantics (cf. Section 2.3) we introduced the complement operator into our framework. Introducing new data values and operators, i. e., primitives, into our framework is fairly easy because we make almost no assumptions about them. As a result, one can define whatever operators one needs for a formal semantics or other purpose.

Note that operators are not necessarily defined for all data values and may not even be defined for all values of a specific kind, like integers. For each operator symbol $op$, we denote the *domain* of the partial function $[\![op]\!]$, i. e., the set of arguments for which $op$ is defined, by $\mathrm{dom}([\![op]\!])$. For instance, an arithmetic division operator $\div$ is not defined if the divisor is zero. We come back to this in a second when introducing *term evaluation*.

Taking all these definitions together we arrive at the following definition:

> **Definition 5 (Terms)**
> Terms are syntactical objects inductively defined by
>
> $$\mathcal{T} \ni \tau \quad ::= \quad x \in \mathbb{X} \quad | \quad s \in \Sigma \quad | \quad \nu \in \mathbb{V} \quad | \quad (\tau_1 \cdots \tau_n) \quad | \quad op(\tau_1, \ldots, \tau_k)$$
>
> where $k, n \in \mathbb{N}_{>0}$ and $op \in \mathfrak{O}$.

---

[1] https://coq.inria.fr/ (Accessed: 2020/12/14)

With these definitions in place, we can already imagine how the premises and conclusions of inference rules may be constructed from terms. However, considering the inference rule $\textsc{r}_+$ of Example 1 again, we still owe a formalization of the equality constraint and membership condition. We address those later, in Section 3.3, when we introduce inference rules and trees formally. Let us for now further focus on terms themselves.

Traditionally, in the literature on term unification, terms are either *variable symbols*, i. e., our *meta variables*, or $k$-ary *function symbols* with $k \in \mathbb{N}_0$ that do not carry any semantic meaning [4]. Function symbols with $k = 0$ represent *constants*, i. e., our *symbols*. We deviate from the traditional definition by (a) introducing operator terms that carry semantic meaning and (b) the possibility to compose terms more generically to arbitrary sequences. The reason for the former deviation is that we aim to capture inference rules like $\textsc{e}_+$ of Example 1 that involve operators with a specific semantic meaning. In addition, the usage of arbitrary sequences allows for a bit more flexibility in the formation of terms. Most core concepts known from the area of term unification nevertheless translate easily to our notion of terms which allows us to harvest existing research from the area. Before going into further detail in Section 3.1.2, let us have a closer look at what renders our notion of terms different, viz. that some terms carry semantic meaning.

### 3.1.1 Term Evaluation

In contrast to the traditional definition of terms [4], our notion of terms includes operator terms that carry a specific semantic meaning: Intuitively, operator terms whose arguments are data values can be *evaluated* by applying their operators to those arguments. To capture this idea formally, we define an *evaluation function* which evaluates terms by evaluating operator terms whose arguments evaluate to data values:

> **Definition 6 (Evaluation Function)**
> Let $\mathrm{eval} : \mathcal{T} \to \mathcal{T} \cup \{\bot\}$ be the *evaluation function* where $\bot \notin \mathcal{T}$ is the *undefined indicator*. We define $\mathrm{eval}$ recursively by
>
> $$\mathrm{eval}(\tau) := \tau \quad \text{if } \tau \in \mathbb{X} \cup (\Sigma \cup \mathbb{V})$$
>
> $$\mathrm{eval}((\tau_1 \cdots \tau_n)) := (\mathrm{eval}(\tau_1) \cdots \mathrm{eval}(\tau_n)) \quad \text{if } \forall i : \mathrm{eval}(\tau_i) \neq \bot$$
>
> $$\mathrm{eval}(\mathsf{op}(\tau_1, \ldots, \tau_k)) := [\![\mathsf{op}]\!]\,(\nu_1, \ldots, \nu_k)$$
> $$\text{if } \forall i : \nu_i = \mathrm{eval}(\tau_i) \in \mathbb{V} \text{ and } \langle \nu_1, \ldots, \nu_k \rangle \in \mathrm{dom}([\![\mathsf{op}]\!])$$
>
> $$\mathrm{eval}(\mathsf{op}(\tau_1, \ldots, \tau_k)) := \mathsf{op}(\tau_1', \ldots, \tau_k')$$
> $$\text{if } \forall i : \tau_i' = \mathrm{eval}(\tau_i) \neq \bot \text{ and } \exists i : \tau_i' \notin \mathbb{V} \text{ and } \forall i : \tau_i' \notin \mathbb{V} \implies \mathrm{vars}(\tau_i') \neq \emptyset$$
>
> $$\mathrm{eval}(\tau) := \bot \quad \text{in all remaining cases}$$
>
> where $\mathrm{vars} : \mathcal{T} \to \mathscr{P}(\mathbb{X})$ is the *variable occurrence function* defined by:

$$\text{vars}(\mathtt{s} \in \Sigma) := \emptyset \qquad \text{vars}(\nu \in \mathbb{V}) := \emptyset \qquad \text{vars}(x \in \mathbb{X}) := \{\, x \,\}$$

$$\text{vars}(\text{op}(\tau_1, \ldots, \tau_k)) := \bigcup_{1 \leq i \leq k} \text{vars}(\tau_i) \qquad \text{vars}((\tau_1 \cdots \tau_n)) := \bigcup_{1 \leq i \leq n} \text{vars}(\tau_i)$$

We say that the evaluation function *is undefined* for a term $\tau$ if and only if $\text{eval}(\tau) = \bot$. We call a term $\tau$ *well-formed* if and only if the evaluation function is not undefined for that term. We call a term $\tau$ *fully-evaluated* if and only if $\text{eval}(\tau) = \tau$, i.e., evaluation has no effect on $\tau$.

If a term contains operator terms whose arguments evaluate to values outside of the domain of the operator, then $\text{eval}$ is undefined for that term. The evaluation function is also undefined on terms which contain an operator term such that there is an argument which does not evaluate to a value and does not contain any variables. If an argument does not evaluate to a value and does not contain any variables, there is no way how it will ever, in particular, after substitution, evaluate to a value. Here is an example:

**Example 2**
Let $\mathbb{V}$ be the rational numbers, i.e., $\mathbb{V} := \mathbb{Q}$. We define a binary *division operator* $\div$ with the partial function $[\![\div]\!] : \mathbb{V} \times \mathbb{V} \rightharpoonup \mathbb{V}$ being defined by:

$$[\![\div]\!] := \left\{\, \langle p, q \rangle \mapsto \frac{p}{q} \,\middle|\, p, q \in \mathbb{Q} \wedge q \neq 0 \,\right\}$$

Based on this definition we then have, for example:

$$\text{eval}(\div(3, 4)) = \frac{3}{4} \qquad \text{eval}(\div(x, 0)) = \div(x, 0) \qquad \text{eval}(\div(3, 0)) = \bot$$

Note that $\mathtt{s} \notin \mathbb{V}$ because $\mathbb{V} = \mathbb{Q}$. Hence, we also obtain:

$$\text{eval}(\div(3, \mathtt{s})) = \text{eval}(\div(x, \mathtt{s})) = \bot$$

As exemplified by the rule FIX of the CCS semantics (cf. Section 2.3), operators sometimes need to operate on compound terms. Imagine the following CCS process term: $\mathtt{fix}\ X = (a\ .\ X)$. Formally, this term is a nested sequence term consisting of symbols. Recall the usage of the replacement operator in the rule FIX: $P[X \mapsto (\mathtt{fix}\ X = P)]$. Instantiating this rule for the process term $\mathtt{fix}\ X = (a\ .\ X)$ gives us: $(a\ .\ X)[X \mapsto (\mathtt{fix}\ X = (a\ .\ X))]$. Here, the ternary replacement operator takes three terms as arguments and returns a term. The resulting operator term evaluates as follows:

$$\text{eval}((a\ .\ X)[X \mapsto (\mathtt{fix}\ X = (a\ .\ X))]) \; = \; a\ .\ (\mathtt{fix}\ X = (a\ .\ X))$$

Note that we did not made any assumptions about whether sequence terms or operator terms can be data values. Crucially, to make rules such as FIX work, we must not assume that $\mathbb{V}$ and $\mathcal{T}$ are disjoint. However, if one allows compound terms as values, special care has to be taken in order to not render the theory and introduced notions ill-defined. To this end and considering the interplay between substitutions and evaluation, we assume in the following that terms without variables, coined *closed terms*, that are fully evaluated are also data values but no other terms are. This restriction has the advantage that it is clear that an operator term can be evaluated only after the variables occurring in its arguments have been substituted and the resulting arguments have been evaluated. Furthermore, the evaluation of an operator term never introduces new variables, is always fully evaluated and, thus, eval is idempotent. We rely on these properties later for various proofs. Formally, we capture this assumption as follows:

**Assumption 1 (Term Reification)**
In the following we assume that:

(i) $\{\, \tau \in \mathcal{T} \mid \mathrm{vars}(\tau) \neq \emptyset \lor \mathrm{eval}(\tau) \neq \tau \,\} \cap \mathbb{V} = \emptyset$ and
(ii) $\{\, \tau \in \mathcal{T} \mid \mathrm{vars}(\tau) = \emptyset \land \mathrm{eval}(\tau) = \tau \,\} \subseteq \mathbb{V}$

Note that without the assumption (i) one faces various questions without clear answers. For instance, without the assumption (i) it is not guaranteed that eval is idempotent because operator terms may evaluate to terms that contain unevaluated operator terms again. For the same reason, it is not even guaranteed that a term becomes fully evaluated after applying eval finitely many times raising the question of how often eval should be applied. In contrast, with the assumption (i), eval is guaranteed to be idempotent because it cannot evaluate to a term which is not already fully evaluated. Furthermore, if terms with variables can be values then it is unclear whether one should first substitute a variable in an argument or evaluate the term without substitution. To cut a long story short: Assumption (i) obviates complications and is prima facie necessary to arrive at a sensible theory.

While the assumption (i) obviates complications, the assumption (ii) comes, as we will see, at the expense of complications. Nevertheless, these complications are worthwhile because assumption (ii) allows us to capture inference rules such as FIX of the CCS semantics that involve operators applied to compound terms. On closer inspection, the rules REC and SYNC also involve operators that take terms to be data values. A weaker version of (ii) including only those terms as values that are useful in a particular context could also be made. However, such a weaker version would not obviate the complications introduced by (ii) as long as it allows some compound terms to be values. For most purposes, however, we need operators to be able to operate on compound terms. Hence, we may just assume the stronger version.

### 3.1.2 Term Unification

The problem of term unification has first been posed by John A. Robinson [47] and is concerned with making terms syntactically identical by *substitut-* *ing* their variables. Recapitulate that in Section 2.2 we evoked the intuition that there are instances of inference rules that are obtained by substituting the occurring meta variables. Thus, let us begin by introducing the concept of *substitutions*. To capture substitutions formally, we closely follow the established notion from the area of term unification [4]:

**Definition 7 (Substitutions)**
A *substitution* is a finite-domain partial function $\sigma : \mathbb{X} \rightharpoonup \mathcal{T}$ mapping variables to terms. For each substitution $\sigma$ we define a function $[\sigma] : \mathcal{T} \to \mathcal{T}$:

$$[\sigma](\mathsf{s} \in \Sigma) := \mathsf{s} \qquad [\sigma](\mathsf{op}(\tau_1, \ldots, \tau_k)) := \mathsf{op}([\sigma](\tau_1), \ldots, [\sigma](\tau_k))$$

$$[\sigma](\nu \in \mathbb{V}) := \nu \qquad [\sigma]((\tau_1 \cdots \tau_n)) := ([\sigma](\tau_1) \cdots [\sigma](\tau_n))$$

$$[\sigma](x \in \mathbb{X}) := \begin{cases} \sigma(x) & \text{if } x \in \mathrm{dom}(\sigma) \\ x & \text{otherwise} \end{cases}$$

A substitution $\sigma$ is *idempotent* if and only if:

$$\forall \tau \in \mathcal{T} : [\sigma](\tau) = [\sigma]([\sigma](\tau))$$

We define the *variable range* $\mathrm{vrange}(\sigma)$ of a substitution $\sigma$ by:

$$\mathrm{vrange}(\sigma) := \bigcup_{x \in \mathrm{dom}(\sigma)} \mathrm{vars}(x)$$

Given a set $\{\tau_1 \doteq \tau_1', \ldots, \tau_m \doteq \tau_m'\}$ of *term equations*, the problem of *term unification* then is the problem of finding a substitution $\sigma$, coined a *unifier*, such that all equated terms become syntactically identical after substitution, i. e., $\forall i : [\sigma](\tau_i) = [\sigma](\tau_i')$, or determining that no such substitution exists. Term unification problems are known to be solvable in linear time [36]. It is also known that if there exists a solution then there exists an idempotent solution in terms of a *Most General Unifier* (MGU) which is uniquely defined up to variable renaming and subsumes all other solutions [4].

Traditionally, term unification is a purely syntactic notion and indifferent to the idea that certain terms carry semantic meaning. For instance, there is no unifier for $\{7 \doteq x + y\}$ simply because there is no substitution such that both terms become syntactically identical. For our purposes, however, we would like unification to care about semantic meaning.

Take again the rule FIX of the CCS semantics (cf. Section 2.3) and the process term $\mathtt{fix}\, X = (a \,.\, X)$ as an example. Now, for the rule FIX to work, we would like $(a \,.\, X)[X \mapsto (\mathtt{fix}\, X = (a \,.\, X))]$ to be unifiable with $\alpha \,.\, P$ such that we can infer the premise of the respective instance of FIX with $P \mapsto a \,.\, X$ and $X \mapsto X$ using the inference rule PREFIX. Figure 3.1 shows the inference tree one would obtain by intuitively applying the inference rules by hand. Recall that $(a \,.\, X)[X \mapsto (\mathtt{fix}\, X = (a \,.\, X))])$ evaluates to $a \,.\, (\mathtt{fix}\, X = (a \,.\, X))$. So, we need a theory of unification that is sensitive to evaluation.

$$\dfrac{\overline{\{\ \} \vDash (a \ . \ X)[X \mapsto (\texttt{fix}\ X = (a \ . \ X))] \xrightarrow{a} (\texttt{fix}\ X = (a \ . \ X))}}{\{\ \} \vDash (\texttt{fix}\ X = (a \ . \ X)) \xrightarrow{a} (\texttt{fix}\ X = (a \ . \ X))}\ \text{FIX}} \text{PREFIX}$$

Figure 3.1: An inference tree using FIX and PREFIX.

Term unification has been extended with equational theories giving rise to $E$-unification where syntactical equality is replaced by another equivalence relation [4]. Equational theories concerning algebraic properties such as commutativity, associativity, or distributivity of function symbols have been extensively studied [4]. Intuitively, for our purposes, we would like to treat those terms as equal that evaluate to the same term. With the appropriate equivalence relation in place, $\{\ 7 \doteq x + y\ \}$ then has for instance $\{\ x \mapsto 3, y \mapsto 4\ \}$ as a solution because $7$ and $3 + 4$ evaluate to the same term, namely $7$. In the next section, we introduce *semantic term unification* which replaces mere syntactical equality with equality modulo evaluation and in addition requires that evaluation is defined for the involved terms. Due to the latter requirement it is, however, not an instance of $E$-unification.

## 3.2 Semantic Term Unification

In Section 2.2, we evoked the intuition that instances of inference rules are obtained by substituting occurring meta variables. Mechanizing inference, hence, involves finding such substitutions. To this end, we just introduced the traditional and well-studied problem of term unification [47] which is the problem of finding a substitution such that certain terms become syntactically identical once their variables have been substituted. Recap that we deviated from the traditional notion of terms by introducing operator terms which carry semantic meaning. While traditional term unification is a pure syntactical notion and indifferent to the semantic meaning of terms, we generalize it to take into account the semantic meaning of terms:

> **Definition 8 (Semantic Term Unification)**
> A *semantic unification problem* is a finite set $\mathbf{E} = \{\ \tau_1 \doteq \tau_1', \ldots, \tau_m \doteq \tau_m'\ \}$ of term equations $\tau_i \doteq \tau_i'$ with $\tau_i, \tau_i' \in \mathcal{T}$. A *semantic unifier* for the problem $\mathbf{E}$ is a substitution $\sigma$ such that for all $1 \le i \le m$:
> $$\text{eval}([\sigma](\tau_i)) = \text{eval}\big([\sigma](\tau_i')\big) \ne \bot$$

In contrast to traditional term unification where terms do not contain operators with specific semantic meaning, we require that terms become syntactically identical after applying the substitution *and* evaluating the resulting terms. If there are no operators with semantic meaning present, then our def-inition agrees with the traditional notions. Phrased differently, we replaced mere syntactical equality with syntactical equality modulo evaluation which renders the problem more general but also more difficult to solve. In addition, we also require that the evaluation function is defined for all of the involved terms. Note that the relation $\{\langle \tau, \tau' \rangle \mid \mathrm{eval}([\sigma](\tau)) = \mathrm{eval}([\sigma](\tau')) \neq \bot\}$ is not an equivalence relation because it is not reflexive. Therefore, semantic term unification is not an instance of $E$-unification.

Having formulated the problem, we now investigate how to solve it, i.e., how to find a solution for a given semantic unification problem $\mathbf{E}$ or determine that none exists. To this end, we build upon the non-deterministic term unification algorithm due to Martelli and Montanari [36]. Before we present the algorithm, a few general words about solutions are in order.

Traditionally, unification problems have a unique solution up to variable renaming in terms of an MGU subsuming all other solutions [36]. It is quite obvious, however, that we run into cases like $\{7 \doteq x + y\}$ that do not only have an infinite amount of solutions over the integers but, in particular, no solution subsuming all other solutions. In general, a solution subsuming all other solutions may not exist for semantic term unification problems. Semantic term unification shares this property with $E$-unification [4]. Solving such cases then requires reasoning about the semantic meaning of operator terms and resolving non-determinism about which solution to choose.

The algorithm we present solves a semantic term unification problem as far as possible until a solution is obtained, it is known that no solution exists, or it reaches a point where reasoning about the semantic meaning of operator terms is required that goes beyond their mere evaluation. Naturally, solving the latter cases by reasoning about the semantic meaning of operator terms cannot be achieved by an algorithm agnostic with respect to the meaning of operator terms. Hence, in these cases, an external solver or other resolution strategies tailored to specific operator terms are required. We call the fragment the algorithm is guaranteed to outright solve the *agnostically solvable* fragment and show in Section 3.2.2 how the algorithm can be interfaced and extended with external solvers and other resolution strategies. We conjecture, however, that in the context of formal semantics, where variables are almost always bound by mere syntactical decomposition, problems outside of the agnostically solvable fragment rarely arise. In fact, neither the formal semantics for CCS nor for Python gives rise to such cases.

## 3.2.1 A Non-Deterministic Algorithm

The non-deterministic algorithm due to Martelli and Montanari [36] is based on transforming the set of term equations until it is *solved*. In doing so, transformations have to obey the invariant that if they transform a set of term equations into another set, then both sets must have the same solutions. We follow their approach and adapt it to semantic term unification.

Let us first define what it takes for a set to be solved:

**Definition 9 (Solved Sets)**
A set of term equations $\mathbf{E}$ is *semi-solved* if and only if it can be partitioned into a *solved part* $\mathbf{E}_s$ and a *remaining part* $\mathbf{E}_r$ such that: (i) all equations in $\mathbf{E}_s$ have the form $x \doteq \tau$ such that $x$ occurs only on the left-hand side of this equation in $\mathbf{E}$ and (ii) all equations in $\mathbf{E}_r$ have one of three forms: (1) the form $\mathsf{op}(\vec{\tau}) \doteq \tau$ or (2) the form $\tau \doteq \mathsf{op}(\vec{\tau})$ with $\tau \notin \mathbb{X}$, or (3) the form $x \doteq \tau$ such that $x$ occurs within an operator subterm of $\tau$ and, except for the left-hand side of $x \doteq \tau$, not outside of operator subterms of any of the equations in $\mathbf{E}$. Note that if a set of term equations is semi-solved, then the partition into $\mathbf{E}_s$ and $\mathbf{E}_r$ is uniquely defined because the form required by (i) is pairwise mutually exclusive with the term forms of (ii).

The set $\mathbf{E}$ is *solved* if and only if it is semi-solved and its remaining part is empty. A solved set $\mathbf{E}$ *induces* the semantic unifier $\sigma$:

$$\sigma := \{\, x \mapsto \tau \mid x \doteq \tau \in \mathbf{E} \,\}$$

For our purposes, we define warranted transformations by means of a binary relation $\Longrightarrow$ on sets of term equations and a special failure indicator $\mathsf{fail}$ which means that the problem does not have any solutions. In the following, we present and discuss the traditional *transformation rules* which define $\Longrightarrow$ as well as our changes to them. To this end, we use the following shorthand notation and denote $\langle X, Y \rangle \in \Longrightarrow$ by $X \Longrightarrow Y$. Our presentation is inspired by the Wikipedia page on term unification [14].

Adapting the rules to our problem changes the properties of the algorithm and the relation $\Longrightarrow$. In particular, we are not able to guarantee that a set of term equations becomes solved by transformation eventually if a solution exists because its solutions may depend on the semantic meaning of operators as discussed above. Nevertheless, we guarantee that after finitely many transformations no more transformations are applicable and that then either the failure indicator is reached or the set is semi-solved, which allows us to leverage an external solver and other strategies.

Before we present the rules let us quickly introduce some auxiliary notions:

**Definition 10 (Unguarded and Guarded Variables)**
We call an occurrence of a variable inside a term $\tau$ *unguarded* if and only if it appears outside of an operator subterm. We define the set of unguarded variables $\mathrm{uvars}(\tau)$ of a term $\tau$ recursively:

$$\mathrm{uvars}(\mathsf{s} \in \Sigma) := \emptyset \qquad \mathrm{uvars}(\nu \in \mathbb{V}) := \emptyset \qquad \mathrm{uvars}(x \in \mathbb{X}) := \{\, x \,\}$$

$$\mathrm{uvars}(\mathsf{op}(\tau_1, \ldots, \tau_k)) := \emptyset \qquad \mathrm{uvars}((\tau_1 \cdots \tau_n)) := \bigcup_{1 \leq i \leq n} \mathrm{uvars}(\tau_i)$$

Analogously, we call an occurrence of a variable inside a term $\tau$ *guarded* if and only if it appears inside of an operator subterm. We define the set of guarded variables $\mathrm{gvars}(\tau)$ of a term $\tau$ recursively:

$$\mathrm{gvars}(\mathtt{s} \in \Sigma) := \emptyset \qquad \mathrm{gvars}(\nu \in \mathbb{V}) := \emptyset \qquad \mathrm{gvars}(x \in \mathbb{X}) := \emptyset$$

$$\mathrm{gvars}(\mathrm{op}(\tau_1, \ldots, \tau_k)) := \bigcup_{1 \leq i \leq k} \mathrm{vars}(\tau_i) \qquad \mathrm{gvars}((\tau_1 \cdots \tau_n)) := \bigcup_{1 \leq i \leq n} \mathrm{gvars}(\tau_i)$$

Although each occurrence of a variable inside a term is either guarded or unguarded, the sets $\mathrm{gvars}(\tau)$ and $\mathrm{uvars}(\tau)$ may not be disjoint because a variable may appear guarded and unguarded in a term $\tau$.

Let us now turn to the transformation rules defining $\Longrightarrow$. The first transformation rule we are considering removes term equations from the set that refer to already syntactically identical terms:

$$\mathbf{E} \cup \{\, \tau \doteq \tau \,\} \implies \mathbf{E} \qquad\qquad\qquad \text{(TD\textsc{el})}$$

If terms are already syntactically identical, they contribute nothing to a solution because they remain syntactically identical independent of the substitution and they also always evaluate to the same result. Hence, (TD\textsc{el}) simply deletes such term equations from the set. However, while already syntactically identical terms contribute nothing to a solution, they may prevent a solution from existing. The rule (TD\textsc{el}) is perfectly adequate in the traditional setting but in our case it might eliminate terms for which evaluation is undefined which could result in there being solutions for $\mathbf{E}$ while there are none for $\mathbf{E} \cup \{\, \tau \doteq \tau \,\}$. Note that (TD\textsc{el}) is also always adequate for $E$-unification because an equational theory is required to define an equivalence relation which is per definition reflexive. For the purpose of semantic term unification, however, we have to adapt (TD\textsc{el}) as follows:

$$\mathbf{E} \cup \{\, \tau \doteq \tau \,\} \implies \mathbf{E} \quad \text{if } \mathrm{gvars}(\tau) = \emptyset \text{ and } \mathrm{eval}(\tau) \neq \bot \qquad \text{(D\textsc{el})}$$

In contrast to (TD\textsc{el}), (D\textsc{el}) respects our invariant. Again, already syntactically identical terms contribute nothing to a solution, however, they are only deleted by (D\textsc{el}) if they also do not prevent a solution from existing independently on how variables are substituted. This is ensured by requiring that there are no guarded occurrences of variables and that the terms are well-formed. Unguarded occurrences of variables cannot make the term to evaluate to $\bot$ independently on how they are substituted because they are outside of operators. Hence, (D\textsc{el}) respects the invariant.

While (D\textsc{el}) also handles symbols and values if they are identical, we need a rule that lets the algorithm fail in case there is a term equation with symbols or values that are not identical:

$$\mathbf{E} \cup \{\, \tau \doteq \tau' \,\} \implies \text{fail} \quad \text{if } \tau \neq \tau' \text{ and } \tau, \tau' \in \Sigma \cup \mathbb{V} \qquad \text{(AF\textsc{ail})}$$

Clearly, non-identical symbols and values cannot be made identical after substitution and evaluation independent on how variables are substituted. Hence, the set $\mathbf{E} \cup \{\, \tau \doteq \tau' \,\}$ does not have any solutions.

Taken together, the rules (DEL) and (AFAIL) handle all term equations that only involve term atoms. The next four rules we are considering concern sequence terms: Two sequence terms of the same length become syntactically identical after evaluation if and only if their components become syntactically identical after evaluation, therefore:

$$\mathbf{E} \cup \left\{ (\tau_1 \cdots \tau_n) \doteq (\tau_1' \cdots \tau_n') \right\} \implies \mathbf{E} \cup \left\{ \tau_i \doteq \tau_i' \mid 1 \leq i \leq n \right\} \qquad \text{(SEQ)}$$

Sequence terms of different length, however, cannot become syntactically identical after evaluation independently of the substitution, hence:

$$\mathbf{E} \cup \left\{ (\tau_1 \cdots \tau_n) \doteq (\tau_1' \cdots \tau_m') \right\} \implies \text{fail} \quad \text{if } n \neq m \qquad \text{(SFAIL)}$$

In addition, a sequence term can further never become syntactically identical after evaluation to symbols or non-sequence values:

$$\mathbf{E} \cup \left\{ (\tau_1 \cdots \tau_n) \doteq \tau \right\} \implies \text{fail} \quad \text{if } \tau \in \Sigma \cup \mathbb{V} \text{ and } \neg\text{isSeq}(\tau) \qquad \text{(SLFAIL)}$$

$$\mathbf{E} \cup \left\{ \tau \doteq (\tau_1 \cdots \tau_n) \right\} \implies \text{fail} \quad \text{if } \tau \in \Sigma \cup \mathbb{V} \text{ and } \neg\text{isSeq}(\tau) \qquad \text{(SRFAIL)}$$

Here, isSeq($\tau$) is a predicate that is true for a term $\tau$ if and only if $\tau$ is a sequence term. If $\tau \in \mathbb{V}$ and at the same time is a sequence term, then either (SEQ) or (SFAIL) applies dependent on the length of the involved sequences. Hence, the only case where neither of the four rules is applicable is if a sequence term is equated with an operator term or a variable.

Note that operator terms might evaluate to sequence terms only if some sequence terms are also data values (cf. Assumption 1). If operator terms cannot evaluate to sequence terms, we may also apply the rules (SLFAIL) and (SRFAIL) if a sequence term is equated with an operator term. In general, with knowledge about the meaning of operator terms and the set of available values those rules can be modified—for instance, if one knows that some operators never evaluate to sequence terms. However, our aim here is to provide rules that are agnostic to the meaning of operator terms under the previous assumption that sequence terms may be values. Hence, the rules we are presenting concern the general case where operators might evaluate to sequence terms. We will later see how knowledge about the semantic meaning of operator terms can be integrated into the algorithm in a general fashion.

Let us now turn to the transformation rules that deal with variables. If there is a solution for a variable, we have to substitute the variable with its solution in all term equations at some point in order for the remaining rules to further decompose the terms and find solutions for the remaining variables. Solutions for variables are term equations of the form $x \doteq \tau$ or $\tau \doteq x$. For simplicity, we canonicalize solutions as follows:

$$\mathbf{E} \cup \left\{ \tau \doteq x \right\} \implies \mathbf{E} \cup \left\{ x \doteq \tau \right\} \quad \text{if } \tau \notin \mathbb{X} \qquad \text{(SWAP)}$$

The condition $\tau \notin \mathbb{X}$ is important because we would like the transformations to eventually terminate. Without it a term equation of the form $x \doteq y$ could be transformed to $y \doteq x$ which in turn could be transformed to $x \doteq y$ and so on. In this case, we just accept $y$ as the solution of $x$.

If a solution for a variable is known and it still occurs in other term equations, then the variable is substituted with its solution:

$$\mathbf{E} \cup \{\, x \doteq \tau \,\} \implies \mathbf{E}[x \mapsto \tau] \cup \{\, x \doteq \tau \,\}$$
$$\text{if } x \in \mathrm{vars}(\mathbf{E}) \wedge x \notin \mathrm{vars}(\tau) \ \textbf{ or } \ x \in \mathrm{uvars}(\mathbf{E}) \wedge x \notin \mathrm{uvars}(\tau) \tag{\textsc{Elim}}$$

Here, $\mathrm{vars}(\mathbf{E})$, $\mathrm{uvars}(\mathbf{E})$, and $\mathbf{E}[x \mapsto \tau]$ are defined as follows:

$$\mathrm{vars}(\mathbf{E}) := \bigcup\nolimits_{\tau \doteq \tau' \in \mathbf{E}} \mathrm{vars}(\tau) \cup \mathrm{vars}(\tau')$$

$$\mathrm{uvars}(\mathbf{E}) := \bigcup\nolimits_{\tau \doteq \tau' \in \mathbf{E}} \mathrm{uvars}(\tau) \cup \mathrm{uvars}(\tau')$$

$$\mathbf{E}[x \mapsto \tau] := \{\, [\{\, x \mapsto \tau \,\}](\tau_1) \doteq [\{\, x \mapsto \tau \,\}](\tau_2) \mid \tau_1 \doteq \tau_2 \in \mathbf{E} \,\}$$

At first sight, it might seem like (Elim) would give rise to a potentially infinite cycle of transformations. However, this is prevented by requiring $x \in \mathrm{vars}(\mathbf{E})$ and $x \notin \mathrm{vars}(\tau)$ or $x \in \mathrm{uvars}(\mathbf{E})$ and $x \notin \mathrm{uvars}(\tau)$: After substituting the variable in $\mathbf{E}$ all its (unguarded) occurrences, except for the left-hand side of $x \doteq \tau$, are eliminated, i.e., $x \notin \mathrm{vars}(\mathbf{E}[x \mapsto \tau])$ or $x \notin \mathrm{uvars}(\mathbf{E}[x \mapsto \tau])$, respectively. Therefore, the rule does not apply again.

Traditionally, one would also add the following rule:

$$\mathbf{E} \cup \{\, x \doteq \tau \,\} \implies \text{fail} \quad \text{if } x \in \mathrm{vars}(\tau) \tag{\textsc{TCFail}}$$

However, in the case of semantic term unification, this rule is not solution preserving. The problem $\{\, x \doteq \mathsf{op}(x) \,\}$ may have semantic unifiers, for instance, if $\mathsf{op}$ is the identity function. We thus adapt (TCFail) as follows:

$$\mathbf{E} \cup \{\, x \doteq \tau \,\} \implies \text{fail} \quad \text{if } x \in \mathrm{uvars}(\tau) \wedge x \neq \tau \tag{\textsc{CFail}}$$

If $x$ occurs outside of an operator subterm in $\tau$ and $\tau \neq x$ then there is no solution. Note that if the rule (CFail) applies, then $\tau$ cannot be a variable, it also cannot be a symbol because $\mathrm{uvars}(\mathsf{s}) = \emptyset$ for all symbols, and it cannot be an operator term because then $\mathrm{uvars}(\tau)$ would be empty as well. Hence, $\tau$ must be a sequence term. Now, no matter how we substitute $x$ in that sequence term, that sequence will not become identical to this substitution after evaluation because the substitution remains modulo evaluation as a subterm of the sequence. Even for the set $\{\, x \doteq (x) \,\}$ there is no solution, because sequences are distinct objects and a sequence of length one containing a term is not considered identical to that term (cf. the remarks on Definition 3). In contrast, the set $\{\, x \doteq (\mathsf{s}\, \mathsf{op}(x)) \,\}$ has a solution if, for instance, the operator $\mathsf{op}$ evaluates to $\mathsf{s}$ provided with $(\mathsf{s}\, \mathsf{s})$ as an argument.

Hitherto, the presented transformation rules have not explicitly dealt with operator terms. Now, what remains are transformation rules for operator terms. Traditionally, one would introduce two rules for what is traditionally called function symbols [36]. Ignoring evaluation for now and treating operator symbols as function symbols those rules would be:

$$\mathbf{E} \cup \{\, \mathsf{op}(\tau_1, \ldots, \tau_n) \doteq \mathsf{op}'(\tau_1', \ldots, \tau_m') \,\} \implies \text{fail} \quad \text{if } \mathsf{op} \neq \mathsf{op}'$$

$$\mathbf{E} \cup \{\, \mathsf{op}(\tau_1, \ldots, \tau_n) \doteq \mathsf{op}(\tau_1', \ldots, \tau_n') \,\} \implies \mathbf{E} \cup \{\, \tau_i \doteq \tau_i' \mid 1 \leq i \leq n \,\}$$

However, neither of these rules adheres to our invariant that the sets before and after transformation are required to have the same solutions. It is possible that both, distinct operators and identical operators with different arguments, evaluate to the same result. Therefore, neither of those rules can be used in the context of semantic term unification.

Intuitively, without further knowledge about the semantic meaning of operator terms one cannot learn anything about the variables occurring in their arguments from their occurrence in a term equation. However, after establishing a solution for a variable, ($\textsc{Elim}$) ensures that this variable is substituted in the arguments of all operator terms. If all variables occurring in the arguments of an operator term in a term equation have been substituted, the operator term can be evaluated without changing the solutions:

$$\mathbf{E} \cup \{\, \tau \doteq \tau' \,\} \implies \mathbf{E} \cup \{\, \mathrm{eval}(\tau) \doteq \mathrm{eval}(\tau') \,\}$$
$$\text{if } \mathrm{eval}(\tau) \neq \tau \vee \mathrm{eval}(\tau') \neq \tau' \text{ and } \mathrm{eval}(\tau) \neq \bot \neq \mathrm{eval}(\tau') \qquad (\textsc{Eval})$$

The rule ($\textsc{Eval}$) is applicable only if at least one of the terms is not yet fully-evaluated and both terms are well-formed. Requiring that at least one of the terms is not yet fully-evaluated ensures that there will not be an infinite transformation cycle without progress being made because Assumption 1 guarantees that eval is idempotent. In case any of the involved terms is not well-formed, the set does not have any solutions, hence:

$$\mathbf{E} \cup \{\, \tau \doteq \tau' \,\} \implies \text{fail} \quad \text{if } \mathrm{eval}(\tau) = \bot \text{ or } \mathrm{eval}(\tau') = \bot \qquad (\textsc{UEval})$$

The rules we just provided uniquely define the *transformation relation* $\implies$ as the least relation satisfying the defining statements. Formally:

---

**Definition 11 (Transformation Relation)**
Let $\implies$ be the least binary relation defined by the statements:

$$(\textsc{Del}), (\textsc{AFail}), (\textsc{Seq}), (\textsc{SFail}), (\textsc{SLFail}), (\textsc{SRFail}),$$
$$(\textsc{Swap}), (\textsc{Elim}), (\textsc{CFail}), (\textsc{Eval}), \text{ and } (\textsc{UEval})$$

---

For instance, ($\textsc{Del}$) is the statement that $\langle \mathbf{E} \cup \{\, \tau \doteq \tau \,\}, \mathbf{E} \rangle \in \implies$ for all $\tau \in \mathcal{T}$ and $\mathbf{E}$ if $\mathrm{gvars}(\tau) = \emptyset$ and $\mathrm{eval}(\tau) \neq \bot$. As we argued, each rule fulfills the required invariant and transforms a set of term equations into another set only if both sets have the same solutions in terms of Definition 8. As a result we obtain Theorem 1 stating solution preservation.

---

**Theorem 1 (Solution Preservation)**
If $X \implies Y$ then $X$ and $Y$ have the same solutions.

---

A non-deterministic algorithm is obtained by iteratively applying $\implies$ to a semantic term unification problem until no transformations are applicable anymore. We now prove that (A) this process terminates no matter how the nondeterminism is resolved and (B) at the end either fail is reached or the set is semi-solved. We subsequently discuss how to proceed in case a semi-solved set with a non-empty remaining part is obtained.

**Transformation Terminates** While we already discussed termination for some of the rules, we now prove that after finitely many transformations no transformation rules are applicable anymore. Formally:

> **Theorem 2 (Transformation Terminates)**
> There are no infinite sequences $X_1 \Longrightarrow X_2 \Longrightarrow \ldots$ of transformations.

*Proof.* We assign a quintuple $N(X) = \langle n_x, n_{\mathsf{op}}, n_{\vec{\tau}}, n_s, n_{\doteq} \rangle \in \mathbb{N}^5$ of natural numbers to each relatum $X$ of $\Longrightarrow$ such that if $X \Longrightarrow Y$ then $N(X) > N(Y)$ where $>$ is the usual lexical order on tuples of natural numbers. As $\mathbb{N}^5$ is a well-founded and totally ordered set with respect to $>$ the existence of $N$ proves termination of iterated transformation. We define $N$ as follows: Let $\mathbf{E}$ be a set of term equations, then $N(\mathbf{E}) := \langle n_x, n_{\mathsf{op}}, n_{\vec{\tau}}, n_s, n_{\doteq} \rangle$, where $n_x$ is the number of variables in $\mathbf{E}$ that do not occur only once on the left-hand side of an equation in $\mathbf{E}$ plus the number of variables that occur more than once outside of an operator in a term equation, $n_{\mathsf{op}}$ is the total number of occurrences of operator symbols in $\mathbf{E}$, $n_{\vec{\tau}}$ is the total number of sequences in $\mathbf{E}$, $n_s$ is the total number of term equations of the form $\tau \doteq x$ in $\mathbf{E}$, and $n_{\doteq}$ is the total number of term equations in $\mathbf{E}$. We further define $N(\mathrm{fail}) := \langle 0, 0, 0, 0, 0 \rangle$. It is now easy to see that every rule reduces at least one component of the tuple while potentially increasing only components with lesser significance: (DEL) decreases $n_{\doteq}$ and potentially other components, (AFAIL) reduces each component to 0, (SEQ) reduces $n_{\vec{\tau}}$, i.e., the number of sequences, by two, (SFAIL), (SLFAIL), and (SRFAIL) all reduce each component to 0, (SWAP) reduces $n_s$ without increasing any other components, (ELIM) reduces $n_x$ by reducing at least one of its summands, (CFAIL) reduces each component to 0, (EVAL) reduces $n_{\mathsf{op}}$ while potentially increasing other components except $n_x$ because operators may only evaluate to fully evaluated terms which neither contain operators nor variables, and (UEVAL) again reduces all components to 0. ∎

Theorem 2 implies that any algorithm applying transformation rules terminates no matter which rules are applied in which order. Martelli and Montanari proved an analogous theorem for their term unification algorithm [36]. We followed their proof idea and adapted it to fit our changes to the rules. In conjunction with the invariant that transformations are solution preserving, we obtain that any algorithm applying transformation rules is guaranteed to terminate with a set of equations that has the same set of solutions as the original semantic term unification problem or with a failure fail and, it terminates with a failure only if there are no solutions.

**Transformation Yields a Semi-Solved Set**   It remains to show that if no rules are applicable to a set of equations, then it is semi-solved:

> **Theorem 3 (Transformation Yields a Semi-Solved Set)**
> Let $\mathbf{E}$ be a set of term equations. If no transformations apply to $\mathbf{E}$, i. e., there exists no $X$ such that $\mathbf{E} \Longrightarrow X$, then $\mathbf{E}$ is semi-solved.

*Proof.* Assume for the sake of contradiction that no rules are applicable but $\mathbf{E}$ cannot be partitioned into a *solved part* $\mathbf{E}_s$ and a *remaining part* $\mathbf{E}_r$ such that: (i) all equations in $\mathbf{E}_s$ have the form $x \doteq \tau$ such that $x$ occurs only on the left-hand side of this equation in $\mathbf{E}$ and (ii) all equations in $\mathbf{E}_r$ have one of three forms: (1) the form $\mathrm{op}(\vec{\tau}) \doteq \tau$ or (2) the form $\tau \doteq \mathrm{op}(\vec{\tau})$ with $\tau \notin \mathbb{X}$, or (3) the form $x \doteq \tau$ such that $x$ occurs within an operator subterm of $\tau$ and, except for the left-hand side of $x \doteq \tau$, not outside of operator subterms of any of the equation in $\mathbf{E}$. If $\mathbf{E}$ cannot be partitioned like that, then there exists a term equation in $\mathbf{E}$ that has neither of the four forms. However, if there exists a term equation in $\mathbf{E}$ that has neither of the four forms, then at least one transformation rule applies to $\mathbf{E}$. Thus, by contradiction, the set is semi-solved. It remains to show that if there exists a term equation in $\mathbf{E}$ that has neither of the four forms, then at least one transformation rule applies. Let $\tau \doteq \tau'$ be a term equation in $\mathbf{E}$ that has neither of the four forms. We do an exhaustive case distinction over $\tau$ and $\tau'$:

- If $\tau, \tau' \in \mathbb{V} \cup \Sigma$ then (DEL) or (AFAIL) apply.

- If $\tau$ and $\tau'$ are both sequences then (SEQ) or (SFAIL) apply.

- If one of the terms is a sequence and the other is a symbol or a non-sequence value, then (SLFAIL) or (SRFAIL) apply.

- If $\tau'$ is a variable and $\tau$ is not a variable then (SWAP) applies.

The remaining cases involve a variable on the left-hand side or an operator term on at least one of the sides while the other side is not a variable. If one of the terms is an operator and the other is not a variable, then the equation has one of the four forms and we do not consider it here.

It remains the case where $\tau$ is a variable that occurs not only on the left-hand side of this equation in $\mathbf{E}$ and (A) not within an operator subterm of $\tau'$ or (B) outside of an operator subterm, except the left-hand side of $\tau \doteq \tau'$, of some equation in $\mathbf{E}$. Assume without loss of generality $\tau = x$.

(A): If $x$ occurs not only on the left-hand side of this equation in $\mathbf{E}$ and not within an operator subterm of $\tau'$ then it occurs outside of an operator subterm of $\tau'$ or within some other term. If it occurs outside of an operator subterm of $\tau'$ then (CFAIL) applies if $\tau' \neq x$, or, otherwise, (DEL) applies. If it does not occur outside of an operator subterm of $\tau'$ and within some other term, then it does not occur in $\tau'$ at all and thus (ELIM) applies.

(B): If $x$ occurs not only on the left-hand side of this equation in **E** and outside of an operator subterm, except the left-hand side of $x \doteq \tau'$, of some equation in **E**, then it occurs outside of an operator subterm of $\tau'$ or outside of an operator subterm of some other term. If it occurs outside of an operator subterm of $\tau'$ then (CFₐₗₗ) applies if $\tau' \neq x$, or, otherwise, (Dᴇʟ) applies. If it does not occur outside of an operator subterm of $\tau'$ but outside of an operator subterm of some other term in **E**, then (Eʟɪᴍ) applies. ∎

Taken together, Theorem 1, Theorem 2, and Theorem 3, imply that the non-deterministic algorithm that iteratively applies transformation rules terminates, independent on how non-determinism is resolved, either with fail or a set of term equations that is semi-solved and has the same solutions as the original term unification problem. Unfortunately, if it terminates with a set of term equations it is not guaranteed that the set is also solved. As already hinted at above, we now characterize the *agnostically solvable* fragment of semantic term unification problems as the fragment for which the algorithm is guaranteed to terminate with a solved set or a failure fail.

In any case, if the algorithm terminates with a solved set, then we have found a solution which subsumes all other solutions because the solved set has the same solutions as the original set and, hence, any solution must adhere to the unifier induced by the set (cf. Definition 9).

The non-deterministic algorithm we just presented is unable to solve the complete class of semantic term unification problems, where *solving* means terminating with a failure fail or a solved set, because it runs into cases where reasoning about the semantic meaning of operator terms is required. For those cases, the algorithm is also unable to tell whether there exists a solution at all. Nevertheless, in the best case, it still reduces those problems, it is unable to outright solve, to small sets of remaining equations, i. e., the remaining part of the semi-solved set, which can then be quickly solved by an external solver or other resolution strategies.

### 3.2.2 Extending the Agnostic Algorithm

Recapitulate that it was an explicit aim of this chapter to stay largely agnostic on concrete data values and operators in order to enable extensibility. The meta theory should not be understood as a closed theory but as a formal framework which has to be adapted to a particular use case by introducing data values and operators *together* with additional reasoning strategies for those values and operators. Naturally, an algorithm that is agnostic is limited when it comes to cases where reasoning about the semantic meaning of operator terms is required. Let us now have a closer look at (A) the structure of problems the algorithm does not outright solve and (B) how to interface an external solver and other strategies with the algorithm as part of a concretization of the theory with values and operators.

**The Remaining Part**  The problems the algorithm does not outright solve are transformed into a set of term equations that is semi-solved and has a non-empty remaining part. All equations in that remaining part have the property that they involve an operator at the top of at least one of the terms or a cycle where a variable is equated with a term where the variable occurs again in an argument of an operator subterm. Furthermore, they do not contain any variables that have a solution as part of the solved part but at least some variables that do not yet have a solution because otherwise everything could be evaluated removing all the occurring operators and then dealt with by the algorithm. The problem now is to find solutions for those variables that do not have a solution yet. As those variables may also still occur inside arguments of operator terms in the solved part, their solution is, however, not always completely independent of the solved part. We now take the semi-solved set and segment it into independent subproblems.

Let $\mathbf{E}$ be a semi-solved set with a non-empty remaining part. The set of variables $\mathrm{vars}(\mathbf{E})$ still occurring in any of the equations of $\mathbf{E}$ can be uniquely partitioned as follows: We start with the initial partition $P_0 := \{\, \{\, x \,\} \mid x \in \mathrm{vars}(\mathbf{E}) \,\}$ and iteratively merge sets of the partion that respectively contain a variable such that both variables occur in some and the same term equation in $\mathbf{E}$. After finitely many iterations a fixed point is reached where no sets are merged anymore. Now, the thereby obtained partition traces dependencies between the variables. Let $P$ denote the partition obtained by this process. Each set of variables $\mathbf{X} \in P$ induces a set of term equations:

$$\mathbf{E}[\mathbf{X}] := \{\, \tau \doteq \tau' \in \mathbf{E} \mid \big(\mathrm{vars}(\tau) \cup \mathrm{vars}(\tau')\big) \cap \mathbf{X} \neq \emptyset \,\}$$

It is easy to see that the sets induced by different sets of variables are pairwise disjoint and that the induced set of term equations captures the term equations relevant to obtaining a solution for the variables in the set. Singleton sets of variables $\mathbf{X} = \{\, x \,\}$ such that $\mathbf{E}[\mathbf{X}] = \{\, x \doteq \tau \,\}$ with $x \notin \mathrm{vars}(\tau)$ can be ignored as they are already solved. Now, the remaining sets $\mathbf{E}[\mathbf{X}]$ can be solved independently of each other and if there is no solution for some of the sets then there is no solution for the overall problem.

We now consider three rather generic additional resolution strategies that may be useful to further tackle the obtained subproblems.

**Out of Range**  Recap the remarks on the transformation rules (SLF$_\textsc{ail}$) and (SRF$_\textsc{ail}$) where we deferred the incorporation of knowledge about the semantic meaning of operator terms to later. The idea was that there is no solution if an operator term is equated with a sequence term but sequences are not in the range of values the operator evaluates to. Here are three ways to incorporate knowledge about the range of operators.

Let $\mathbf{E}[\mathbf{X}]$ be a subproblem obtained from the semi-solved set. There exists no solution if there exists an equation $\tau \doteq \tau' \in \mathbf{E}[\mathbf{X}]$ that has one of the following forms: (1) the form $(\tau_1, \ldots, \tau_n) \doteq \mathrm{op}(\vec{\tau})$ or the form $\mathrm{op}(\vec{\tau}) = (\tau_1, \ldots, \tau_n)$ such that $\forall \nu \in \mathrm{range}(\llbracket \mathrm{op} \rrbracket) : \neg\mathrm{isSeq}(\nu)$, (2) the form $\nu \doteq \mathrm{op}(\vec{\tau})$ or the form $\mathrm{op}(\vec{\tau}) \doteq \nu$ such that $\nu \notin \mathrm{range}(\llbracket \mathsf{op} \rrbracket)$, and (3) the form $\mathsf{op}(\vec{\tau}) \doteq \mathsf{op}'(\vec{\tau}')$ with $\mathrm{range}(\llbracket \mathsf{op} \rrbracket) \cap \mathrm{range}(\llbracket \mathsf{op}' \rrbracket) = \emptyset$. Note that these forms are paradigmatic instances of the forms of equations that make up the remaining part.

**Bijective Operators**   In case an operator is bijective and its inverse is efficiently computable the following resolution strategy is applicable. Let $\mathbf{E}[\mathbf{X}]$ again be a subproblem obtained from the semi-solved set and $\mathsf{op}$ be a bijektive operator with the inverse $\mathsf{op}^{-1}$. Now, if there exists an equation of the form $\nu \doteq \mathsf{op}(\tau_1, \ldots, \tau_k)$ or the form $\mathsf{op}(\tau_1, \ldots, \tau_k) \doteq \nu$ in $\mathbf{E}[\mathbf{X}]$ such that $\nu \in \mathrm{range}(\llbracket \mathsf{op} \rrbracket)$, we can modify the set without changing its solutions as follows: We delete the original equation from the set and replace it with an equation $\tau_i \doteq \tau_i'$ for each argument such that $\mathsf{op}^{-1}(\nu) = \langle \tau_1', \ldots, \tau_k' \rangle$. On the resulting set, the semantic term unification algorithm is then invoked again and, if all goes well, yields further solutions or fail.

**Using an SMT-Solver**   If a subproblem is obtained that contains only equations that can be encoded as a *Satisfiability Modulo Theory* (SMT) problem [6] then an external off-the-shelf SMT solver can be used to obtain further solutions. This idea, however, begs the question why one does not encode semantic unification problems as SMT problems and uses an off-the-shelf solver in the first place. We address the advantages of our approach over an SMT encoding of semantic term unification problems in Section 3.5.

### 3.2.3   Final Remarks

Let us conclude this section on semantic term unification with a few final remarks. We introduced the problem of semantic term unification to be able to capture inference rules that contain operator terms that inter alia operate on terms themselves. In the remainder we will see various aspects discussed here in action and demonstrate their usefulness for our purposes. It should be noted, however, that semantic term unification is not fully developed yet and that an in depth comparison to existing techniques should be done in order to carve out more fundamental contributions.

Our focus here clearly is on extensibility of the theory with new data values and operators. During our formalization endeavors, it turned out advantageous that it was easy to introduce new operators and data values because, in the absense of not agnostically solvable problems, one does not have to provide more than an evaluation function for new operators.

There are also some open questions and problems: While the resolution of non-determinism in the algorithm is irrelevant for traditional term unification this is unfortunately not the case for semantic term unification. While it is by definition irrelevant for the agnostically solvable fragment because we require that the algorithm is *guaranteed* to solve those problem, there are cases where depending on the order in which the rules are applied one either obtains a semi-solved or a solved set. We are confident that these cases are practically irrelevant for our purposes because they seem to be rooted in cyclic equations which do not appear in our context. Here is an example:

$$\{ \, x \doteq (\mathtt{s}\ 3), x \doteq (\mathtt{s}\ \mathtt{op}(x)) \, \}$$

Depending on whether one applies (ELIM) with the first or second equation one obtains a solved or a semi-solved set assuming that $[\![\mathtt{op}]\!]((\mathtt{s}\ 3)) = 3$. What is missing so far, is an enlightening criterion for when a problem is agnostically solvable that goes beyond the mere characterization that it is a problem guaranteed to be solvable by our algorithm. It may also be desirable to change some of the transformation rules and, for instance, guarantee that the above problem becomes solved independent on the resolution of non-determinism. Another possibility would be to backtrack and try a different resolution in case one runs into a semi-solved set. As we do not expect such problems to occur in our context, we defer these questions to future work.

Starting from their non-deterministic algorithm, Martelli and Montanari devise a linear time algorithm which depends on preprocessing the set of equations [36]. We conjecture that their core insights carry over to semantic term unification. However, we leave it open to future work to investigate this further. For now, we stick to straightforward application of the presented transformation rules until a failure or semi-solved set is produced.

## 3.3   Inference Engine

Having established what kind of object premises and conclusions are, we now put the pieces together and formalize inference rules and trees. Intuitively, a term then is *derivable* in an *inference rule system*, i. e., from a set of inference rules, if and only if there exists a tree that has the term as a conclusion. We go on to prove that inference rule systems are Turing-complete and, hence, it is in general undecidable whether a term is derivable or not. With that result in mind, we present a semi-algorithm, called *inference engine*, that enumerates inference trees that have a conclusion of a particular form. Finally, we show how to leverage this algorithm to derive an interpreter based on a formal semantics of a language expressed in our framework.

Let us start by refining our intuitive notion of inference rules. In Section 2.2, we introduced inference rules informally as respectively comprising *premises* $P_1$ to $P_k$ with $k \in \mathbb{N}_0$ and a *conclusion* $C$. While a rule may have multiple premises it is not required to have premises. We called inference rules without premises *axioms* and observed that some rules, like E$_+$ of Example 1, may pose additional constraints on the binding of their meta variables. The distinguishing feature of those constraints is that, in contrast to premises, they they are not derived by an inference tree themselves. Put differently, while the premises are anchors for subtrees that have the respective premise as a conclusion, the additional constraints are more like conditions on the applicability of a rule. To capture such additional constraints, we refine the concept of inference rules: An inference rule has the form

$$\frac{P_1 \quad \cdots \quad P_k \qquad \tau_1 \doteq \tau_1' \quad \cdots \quad \tau_m \doteq \tau_m' \qquad B_1 \quad \cdots \quad B_n}{C} \; \textsc{r}$$

where $P_1$ to $P_k$ are *premises*, $\tau_i \doteq \tau_i'$ are term equations, and $B_1$ to $B_n$ are *boolean conditions*. Recapitulate the inference rule $\textsc{e}_+$ of Example 1:

$$\frac{z_l, z_r \in \mathbb{Z} \qquad z_l + z_r = z}{(z_l + z_r) \rightarrow z} \; \textsc{e}_+$$

The rule $\textsc{e}_+$ is an axiom, i. e., it does not have any premises. However, substitutions for the variables $z_l$, $z_r$, and $z$ are constraint as follows: The axiom is applicable if and only if $z_l$ and $z_r$ are substituted with integers and $z$ is substituted with the sum of both integers. Here, $z_l, z_r \in \mathbb{Z}$ is a boolean condition and $z = z_l + z_r$ is a term equation. Note that we usually omit the dot on top of the equals sign for term equations appearing in an inference rule in the following because this matches the notation found in the literature more closely not leaving a reader unfamiliar with our meta theory confused about the dot. We formalize these intuitions as follows:

**Definition 12 (Inference Rules)**
An *inference rule* $\textsc{r}$ is a tuple $\langle \mathbf{P}, \mathbf{E}, \mathbf{B}, C \rangle$ where $\mathbf{P} \in \mathcal{T}^*$ is a finite vector of *premises*, $\mathbf{E}$ is a finite set of term equations, $\mathbf{B}$ is a finite set of *boolean conditions*, and $C \in \mathcal{T}$ is the *conclusion*. A boolean condition is a decidable predicate $B$ over substitutions.

We define the set of variables occurring in a rule by:

$$\mathrm{vars}(\textsc{r}) := \mathrm{vars}(C) \cup \mathrm{vars}(\mathbf{E}) \cup \bigcup_{P \in \mathbf{P}} \mathrm{vars}(P)$$

In Section 2.2, we also evoked the intuition that an inference tree is obtained by stacking *rule instances* on-top of each other such that they form a tree and there is a subtree for each premise of a rule instance that has the respective premise as a conclusion. To this end, a rule instance was obtained by suitably substituting the meta variables occurring in the respective rule. We now have all the necessary formal tools to formalize the idea of rule instances:

**Definition 13 (Rule Instances)**
Let $\textsc{r} = \langle \mathbf{P}, \mathbf{E}, \mathbf{B}, C \rangle$ be an inference rule. An *instance* of $\textsc{r}$ is a pair $\langle \textsc{r}, \sigma \rangle$ where $\sigma$ is a substitution such that

(1) $\mathrm{dom}(\sigma) = \mathrm{vars}(\textsc{r})$ and $\mathrm{vrange}(\sigma) = \emptyset$,
(2) for all premises $P \in \mathbf{P}$, $\mathrm{eval}([\sigma](P)) \neq \bot$,
(3) for all term equations $\tau \doteq \tau' \in \mathbf{E}$, $\mathrm{eval}([\sigma](\tau)) = \mathrm{eval}([\sigma](\tau')) \neq \bot$,
(4) all boolean conditions are satisfied, i. e., $\forall B \in \mathbf{B} : B(\sigma)$, and
(5) $\mathrm{eval}([\sigma](C)) \neq \bot$.

The *conclusion* of an instance, denoted by $C(\langle \textsc{r}, \sigma \rangle)$, is the term $\mathrm{eval}([\sigma](C))$.

Before we formalize the composition of instances to inference trees, we have a closer look on the defining restrictions (1) to (5). The restriction (1) requires that all meta variables occurring in an inference rule are substituted such that no meta variables are left behind. Intuitively, meta variables are part of the meta theory and primarily a vehicle to define inference rules based on pattern matching. Hence, they should not *leak* outside the scope of the meta theory. The restriction (2) and (5) require that the variables are substituted in a way such that the evaluation function is defined for the premises and the conclusion, respectively. Finally, the restriction (3) requires that all equality constraints are *satisfied* analogously to semantic term unification, and restriction (4) requires that the boolean conditions are satisfied. Taking again $\mathtt{E}_+$ as an example, the substitution $\{\, z_l \mapsto 2, z_r \mapsto 13, z \mapsto 15 \,\}$ would yield the instance of $\mathtt{E}_+$ occurring in (2.1). It satisfies the condition that $z_l$ and $z_r$ are bound to integers and that $z$ is bound to the sum of these integers. The conclusion of that instance is $(2 + 13) \to 15$.

We now combine rule instances to inference trees:

> **Definition 14 (Inference Trees)**
> An *inference tree* is a pair $\mathtt{T} = \langle \langle \mathtt{R}, \sigma \rangle, \mathbf{T} \rangle$ where $\mathtt{R}$ is an inference rule and $\sigma$ is a substitution such that $\langle \mathtt{R}, \sigma \rangle$ is an instance of $\mathtt{R}$ and $\mathbf{T}$ is a vector of *subtrees*—one for each premise of $\mathtt{R}$. The *conclusion* of $\mathtt{T}$, denoted by $C(\mathtt{T})$, is the conclusion of the instance $\langle \mathtt{R}, \sigma \rangle$, i. e., $C(\mathtt{T}) := C(\langle \mathtt{R}, \sigma \rangle)$. We define the set of inference trees inductively as follows: Let $\mathtt{R} = \langle \langle P_1, \ldots, P_k \rangle, \mathbf{E}, \mathbf{B}, C \rangle$ be an inference rule. If $k = 0$, i. e., $\mathtt{R}$ is an axiom, then a pair $\langle \langle \mathtt{R}, \sigma \rangle, \langle \rangle \rangle$ is an inference tree if $\langle \mathtt{R}, \sigma \rangle$ is an instance of $\mathtt{R}$. Otherwise, if $k \neq 0$, then a pair $\langle \langle \mathtt{R}, \sigma \rangle, \langle \mathtt{T}_1, \ldots, \mathtt{T}_k \rangle \rangle$ is an inference tree if $\langle \mathtt{R}, \sigma \rangle$ is an instance of $\mathtt{R}$ and every $\mathtt{T}_i$ is an inference tree with $C(\mathtt{T}_i) = \mathrm{eval}([\sigma](P_i))$.

Definition 14 captures the intuitive idea that instances of inference rules are stacked on-top of each other to form an inference tree such that there exists a subtree for each premise of an instance that has the respective premise as a conclusion. Formally the inference tree (2.1) is then captured by:

$$\left\langle \underbrace{\langle \mathtt{L}_+, \{\, e_l \mapsto (2 + 13), e_r \mapsto 42 \,\} \rangle}_{\text{instance of } \mathtt{L}_+}, \underbrace{\left\langle \left\langle \underbrace{\langle \mathtt{E}_+, \{\, z_l \mapsto 2, z_r \mapsto 13, z \mapsto 15 \,\} \rangle}_{\text{instance of } \mathtt{E}_+}, \langle \rangle \right\rangle \right\rangle}_{\text{subtrees for } \mathtt{L}_+} \right\rangle$$

Inference trees are usually constructed from multiple instances of multiple rules that make up an *inference rule system*. Formally:

> **Definition 15 (Inference Rule Systems)**
> An *inference rule system* $\mathbf{R}$ is a finite set of inference rules.

The inference rule system is often implicit in the following. With these definitions in place, we formalize the concept of *derivability*:

> **Definition 16 (Derivability)**
> Let $\mathbf{R}$ be an inference rule system. A term $\tau$ is *derivable* in $\mathbf{R}$ if and only if there exists an inference tree $\mathsf{T}$ with $C(\mathsf{T}) = \tau$ such that $\mathsf{T}$ contains only rule instances of rules from $\mathbf{R}$. An inference rules system induces a set $\mathcal{L}(\mathbf{R}) \subseteq \mathcal{T}$ of derivable terms coined its *language*.

Intuitively, the execution steps defined by a formal semantics are constituted by the terms derivable in the corresponding inference rule system. For instance, the term $((2 + 13) + 42) \rightarrow (15 + 42)$ is derivable in the system specified in Example 1 and, hence, the corresponding execution step from $(2 + 13) + 42$ to $15 + 42$ is possible. Before we turn to the algorithmic side, we aim to classify the computational mightiness of the presented formalism.

### 3.3.1 Turing Completeness of Inference Rule Systems

The introduced formalism is Turing-complete and thus computationally universal. In the following, we prove Turing-completeness by showing that every Turing machine can be transformed into an inference rule system such that the terms derivable in that system encode precisely those inputs the Turing machine halts on. As a consequence thereof, there is in general no algorithm that yields a definite verdict about whether a particular term is derivable in a given inference rule system because this would solve the halting problem. Formally, we prove the following theorem:

> **Theorem 4 (Turing-Completeness)**
> For every turing machine TM there exists an inference rule system $\mathbf{R}_{\mathrm{TM}}$ such that for all words $\vec{x}$ over the input alphabet of TM, $\vec{x} \in \mathcal{L}(\mathrm{TM})$ if and only if $\mathrm{enc}(\vec{x}) \in \mathcal{L}(\mathbf{R}_{\mathrm{TM}})$. Here enc is an encoding of the word $\vec{x}$ as a term and $\mathcal{L}(\mathrm{TM})$ is the language induced by TM, i.e., the set of all input words for which the Turing machine halts in an accepting state.

We assume that the reader is already familiar with Turing machines and refer to common introductory texts on computability theory for further details. For our purposes, we use the following formal definition [31]:

> **Definition 17 (Turing Machines)**
> A *Turing machine* TM is a tuple $\langle Q, \mathfrak{T}, {\sqcup}, \mathfrak{I}, \bar{q}, F, \delta \rangle$ where $Q$ is a finite set of *states*, $\mathfrak{T}$ is a finite, non-empty *tape alphabet*, ${\sqcup} \in \mathfrak{T}$ is the *blank symbol*, $\mathfrak{I} \subseteq \mathfrak{T} \setminus \{{\sqcup}\}$ is a finite, non-empty *input alphabet*, $\bar{q}$ is the *initial state*, $F \subseteq Q$ is a set of *accepting states*, and $\delta : (Q \setminus F) \times \mathfrak{T} \rightarrow (Q \times \mathfrak{T} \times \{\triangleleft, \triangleright\}) \cup \{\bot\}$ is the *transition function*.

Let TM be an arbitrary but fixed Turing machine. We provide a transformation of TM into an inference rule system $\mathbf{R}_{TM}$ such that the terms derivable in that system encode precisely those inputs the Turing machine halts on. The fundamental idea is to simulate the tape of the Turing machine by two stacks, one for the part left of the current reading position and one for the part right of the current reading position, and then have inference rules move the tape head by pushing and poping from these stacks.

We assume that $\Sigma \supset (\mathfrak{T} \cup Q) \cup \{ \, \texttt{::}, \texttt{nil} \, \}$, i.e., the tape alphabet and states of the turing machine are symbols (cf. Definition 2) and there are two additional symbols, $\texttt{::}$ and $\texttt{nil}$, that are neither states nor part of the tape alphabet. An input $x_1 x_2 \cdots x_n \in \mathfrak{I}^+$, i.e., a finite word over the input alphabet, is then encoded by the nested sequence term $((\cdots (\texttt{nil} :: x_n) \cdots :: x_2) :: x_1)$. For instance, the input $a\,b\,c$ is encoded by $(((\texttt{nil} :: c) :: b) :: a)$. Note that the order of the input symbols has been reversed in the term. Intuitively, the input has been encoded as a stack where the first input symbol lies on the top. This encoding allows starting with the input as the stack representing the right part of the tape. Recapitulate that a *configuration* of a Turing machine comprises the current state, the tape, and the reading position. We capture configurations formally as sequence terms of the following form:

$$( \quad \overset{\text{current state}}{\overbrace{q}} \quad :: \quad \overset{\text{left part}}{\overbrace{xs}} \quad :: \quad \overset{\text{current character}}{\overbrace{y}} \quad :: \quad \overset{\text{right part}}{\overbrace{zs}} \quad )$$

Here, $q \in Q$ is the current state the Turing machine is in, $xs$ is the stack representing the left part of the tape, $y \in \mathfrak{T}$ is the currently read character, and $zs$ is the stack representing the right part of the tape. A transition then involves pushing the character that should be written on one of the two stacks and take the character to be read next from the other stack.

We now define inference rules for each pair $\langle q, y \rangle \in Q \times \mathfrak{T}$ depending on the value of $\delta(\langle q, y \rangle)$. Let us start with the case where the Turing machine moves left, i.e., $\delta(\langle q, y \rangle) = \langle q', y', \lhd \rangle$ for some $q' \in Q$ and $y' \in \mathfrak{T}$:

$$\frac{(q' :: xs :: x :: (zs :: y'))}{(q :: (xs :: x) :: y :: zs)} \; \text{\small LEFT}$$

The inference rule LEFT captures the following idea: Being in state $q$ and reading $y$ the Turing machine should move left, write $y'$ at the current position, and transition into state $q'$. To this end, the top character $x$ is popped from the stack representing the left part of the tape and put at the position of the currently read character in the successor configuration. The successor configuration appears as the premise of the inference rule. At the same time, the state is transitioned to $q'$ and $y'$ is pushed on the stack representing the right part of the tape. As the empty stack is represented by $\texttt{nil}$, the rule is applicable only if the left end of the tape has not been reached because otherwise the stack could not be decomposed in $xs$ and $x$.

It remains the case where the Turing machine moves right, i.e., $\delta(\langle q, y \rangle) = \langle q', y', \rhd \rangle$ for some $q' \in Q$ and $y' \in \mathfrak{T}$. This case is a bit more involved because the tape has to be extended on the right side with blank characters. Therefore, we need an additional rule extending the tape:

$$\frac{(q' :: (xs :: y') :: z :: zs)}{(q :: xs :: y :: (zs :: z))}\text{ RIGHT}\qquad\frac{(q' :: (xs :: y') :: {\sqcup} :: \mathtt{nil})}{(q :: xs :: y :: \mathtt{nil})}\text{ EXTEND}$$

The rule RIGHT is analogous to the rule LEFT. In particular, it is not applicable if the stack representing the right part of the tape is empty. However, if the stack representing the right part of the tape is empty, then EXTEND becomes applicable and the blank character is read.

Having introduced inference rules for each pair $\langle q, y\rangle \in Q \times \mathfrak{T}$, a configuration is derived from its successor. At the end, we would like to be able to derive configurations for which the Turing machine eventually halts. For this purpose, we introduce an axiom for each accepting state $q_f \in F$:

$$\frac{}{(q_f :: xs :: y :: zs)}\text{ HALT}$$

With these inference rules in place, it is easy to see that we obtained an inference rule system in which precisely those configurations are derivable for which the Turing machine eventually halts because configurations are derivable from their successors and halting configurations are derivable from axioms. It remains to introduce inference rules allowing us to derive encoded inputs for which the Turing machine halts:

$$\frac{(\bar{q} :: \mathtt{nil} :: x :: xs)}{(xs :: x)}\text{ INPUT}\qquad\frac{(\bar{q} :: \mathtt{nil} :: {\sqcup} :: \mathtt{nil})}{\mathtt{nil}}\text{ EMPTY}$$

The result is an inference rule system $\mathbf{R}_{\text{TM}}$ such that for each word $\vec{x}$ over the input alphabet, the respective encoding $\text{enc}(\vec{x})$ is derivable in the system if and only if the Turing machine TM halts on $\vec{x}$. As the transformation can be done for any Turing machine, we proved Turing-completeness and obtain undecidability of derivability as a corollary:

**Corollary 1 (Derivability is Undecidable)**
Let $\mathbf{R}$ be an inference system and $\tau$ be a term. It is in general undecidable whether $\tau \in \mathcal{L}(\mathbf{R})$, i.e., whether $\tau$ is derivable in $\mathbf{R}$.

This result is, of course, a double-edged sword. On the one hand, it proves that the introduced formalism is computationally universal and essentially a Turing-complete computational theory based on inference rules. On the other hand, it entails that no algorithm ultimately deciding derivability exists. This seems especially problematic as it is the explicit goal of this chapter to mechanize the application of inference rules as they appear in SOS. While no algorithm exists that ultimately decides derivability, nevertheless a semi-algorithm can be given that might yield a derivability verdict or just diverge without a result. Fortunately, the rules appearing in the context of SOS are structured such that this semi-algorithm terminates.

In the context of structural operational semantics, rule formats (cf. Section 2.2.2) restrict the form of inference rules such that all formal semantics adhering to those restrictions have desirable properties, for instance, that derivability becomes decidable or that bisimulation is a congruence relation [25]. Usually, it is required that the size of the terms is reduced by each rule, i. e., that the premises are smaller than the conclusion. Here "size" may mean the sum of the number of all atoms, sequences, and operators in a term. The proof for Turing-completeness relied on extending the tape and thereby giving rise to a premise that is larger than the conclusion.

As a final remark, note that we did not rely on operator terms with semantic meaning for that proof. In fact, the traditional notion of terms already suffices for Turing-completeness and undecidability.

### 3.3.2 A Semi-Algorithm for Derivation

Having established that derivability is in general undecidable, we now provide a semi-algorithm enumerating inference trees for terms of a given form. In contrast to an algorithm, a *semi-algorithm* is not guaranteed to terminate with a solution. As we just proved derivability to be undecidable in general, no algorithm exists that is guaranteed to find a definite answer to the question whether a certain term is derivable or not. Therefore, a semi-algorithm is the best we can do. Nevertheless, our algorithm is guaranteed to terminate if the inference rules do not give rise to *diverging trees*.

We first observe that, for the purpose of deriving an interpreter from a formal semantics, we need to solve a more general problem than just obtaining a derivability verdict for a given term. Imagine that we have a program state $2 + 13$. For an interpreter, we are now interested in the successor state or states of that state. Formally, we are interested in a substitution for $u$ in the term $(2 + 13) \to u$ such that the result is derivable. In general, we capture such *inference questions* by means of terms as follows:

**Definition 18 (Inference Questions)**
An *inference question* is a term $\tau$. An *answer* to an inference question in an inference rule system $\mathbf{R}$ is a tuple $\langle \sigma, \mathtt{T} \rangle$ where $\sigma$ is a substitution and $\mathtt{T}$ is an inference tree such that $\mathrm{dom}(\sigma) = \mathrm{vars}(\tau)$ and $\mathrm{eval}([\sigma](\tau)) = C(\mathtt{T})$, i. e., $\mathtt{T}$ is an inference tree witnessing that $\mathrm{eval}([\sigma](\tau))$ is derivable in the system $\mathbf{R}$ where $\sigma$ contains a solution for each variable.

Given an inference question, our semi-algorithm is supposed to enumerate all answers to that question. Note that the question of mere derivation is a special case of Definition 18. Hence, the semi-algorithm can also be used to enumerate the inference trees for a given closed term. By enumerating all answers to a given question, it is possible to explore all successor states of a given state or just a single successor state if one aborts the algorithm after an answer has been found. If the semi-algorithm terminates without an answer then it is impossible to derive a term of the given form.

Intuitively, enumerating answers to an inference question involves constructing inference trees and, thus, finding substitutions to obtain rule instances. The basic idea is to encode the constraints posed on these substitutions by Definition 18, Definition 14, and Definition 13 as a semantic term unification problem. However, not all constraints can be encoded as such, for instance, the boolean conditions. So, after obtaining a substitution with the semantic term unification algorithm, the remaining constraints have to be considered. If the binding of variables is completely determined by semantic term unification (cf. restriction (1) of Definition 13) then it suffices to merely check the boolean conditions on the obtained substitution. In case variables remain, a search for solutions for the remaining variables such that all restrictions of Definition 13 are satisfied is necessary. Let us first focus on the encoding of the constraints as a semantic term unification problem. To gain a better intuition for that process, we introduce the concept of *rule trees*:

> **Definition 19 (Rule Trees)**
> A *rule tree* is a pair $\text{TR} = \langle \text{R}, \mathbf{T} \rangle$ where $\text{R}$ is an inference rule and $\mathbf{T}$ is a vector of rule *subtrees*—one for each premise of $\text{R}$. The *conclusion* of $\text{TR}$, denoted by $C(\text{TR})$, is the conclusion of the rule, i.e., $C(\text{TR}) := C$. We define the set of rule trees inductively as follows: Let $\text{R} = \langle \langle P_1, \ldots, P_k \rangle, \mathbf{E}, \mathbf{B}, C \rangle$ be an inference rule. If $k = 0$, i.e., $\text{R}$ is an axiom, then a pair $\langle \text{R}, \langle \rangle \rangle$ is a rule tree. Otherwise, if $k \neq 0$, then a pair $\langle \text{R}, \langle \text{TR}_1, \ldots, \text{TR}_k \rangle \rangle$ is a rule tree if every $\text{TR}_i$ is a rule tree for each $1 \leq i \leq k$.

A rule tree is an inference tree but without substitutions. Given a rule tree $\text{TR}$, we may ask for a substitution for each of the involved subtrees such that the rule tree and the substitutions together form an inference tree. We encode the problem of finding such substitutions as a semantic term unification problem. As rules may appear multiple times in the same tree and we aim to construct a single semantic term unification problem for the whole rule tree, the encoding involves renaming the occurring variables. To this end, we construct a bijective substitution $\rho(\vec{n})$ for each subtree $\langle \text{R}, \mathbf{T} \rangle$ such that

$$\text{dom}(\rho(\vec{n})) = \text{vars}(\text{R}) \text{ and } \text{range}(\rho(\vec{n})) \subset \mathbb{X}$$

where $\vec{n}$ is the unique *identifer* of the respective subtree. The unique identifier of a subtree is a sequence of natural numbers obtained by tracing the path to the respective subtree. For instance, the empty sequence $\epsilon$ denotes the root tree, the singleton sequence $3$ denotes the tree corresponding to the third premise of the root tree, and the sequence $3\,2$ denotes the tree corresponding to the second premise of the third premise of the root tree. Formally, we define a function getTree taking a rule tree and an identifier and returning the corresponding subtree or $\perp$ if the identifier is invalid:

$$\mathrm{getTree}(\mathbf{TR}, \epsilon) := \mathbf{TR} \qquad \mathrm{getTree}(\langle \mathbf{R}, \langle \mathbf{TR}_1, \ldots, \mathbf{TR}_k \rangle \rangle, i \circ t) := \begin{cases} \mathbf{TR}_i & \text{if } 1 \le i \le k \\ \perp & \text{otherwise} \end{cases}$$

Here, $\circ$ is the usual concatenation operator on sequences. Further, we define an auxillary function returning the set of valid identifiers:

$$\mathrm{trees}(\mathbf{TR}) := \{\, \vec{n} \in \mathbb{N}^* \mid \mathrm{getTree}(\mathbf{TR}, \vec{n}) \ne \perp \,\}$$

With these definitions in place, we require that:

$$\forall \vec{n}, \vec{n}' \in \mathrm{trees}(\mathbf{TR}) : \vec{n} \ne \vec{n}' \implies \mathrm{range}(\rho(\vec{n})) \cap \mathrm{range}\big(\rho(\vec{n}')\big) = \emptyset$$

We impose this requirement such that each variable occurs only in the rule of a specific subtree after all variables have been renamed. As a result, all constraints can be encoded as a single large semantic term unification problem where each subtree of a rule tree contributes its own set of term equations encoding the constraints posed by that subtree on the substitution:

$$\begin{aligned} &\mathrm{uProb}(\langle \langle \langle P_1, \ldots, P_k \rangle, \mathbf{E}, \mathbf{B}, C \rangle, \langle \mathbf{TR}_1, \ldots, \mathbf{TR}_k \rangle \rangle, \vec{n}) \\ :=\quad & \{\, [\rho(\vec{n})](P_i) \doteq [\rho(\vec{n} \circ i)](C(\mathbf{TR}_i)) \mid 1 \le i \le k \,\} \\ \cup\;\; & \{\, [\rho(\vec{n})](\tau) \doteq [\rho(\vec{n})](\tau') \mid \tau \doteq \tau' \in \mathbf{E} \,\} \end{aligned}$$

The function $\mathrm{uProb}$ takes a subtree and its identifier and constructs a set of term equations where the variables are appropriately renamed such that all the term equations appearing in the set $\mathbf{E}$ of the rule occur in that set and for each premise of the rule, the premise and the conclusion of the corresponding subtree are equated. Let $\tau$ be an inference question such that w. l. o. g. there occur no variables in $\tau$ that appear in some renaming substitution $\rho(\vec{n})$. Given $\tau$ we obtain the following semantic term unification problem:

$$\{\, \tau \doteq [\rho(\epsilon)](C(\mathbf{TR})) \,\} \cup \bigcup_{\vec{n} \in \mathrm{trees}(\mathbf{TR})} \mathrm{uProb}(\mathrm{getTree}(\mathbf{TR}, \vec{n}), \vec{n})$$

Solving this problem means finding a substitution $\sigma$ such that the all premises are unified with the conclusion of the rules of the corresponding subtrees, the conclusion of the overall tree is unified with the inference question, and for each subtree the constraints imposed by the set $\mathbf{E}$ of the respective rule are satisfied. The substitution $\sigma$ induces a substitution $\sigma_{\vec{n}}$ for each subtree $\vec{n}$:

$$\sigma_{\vec{n}} := \{\, x \mapsto \sigma\big(x'\big) \mid x \in \mathrm{dom}(\rho(\vec{n})) \wedge \rho(\vec{n})(x) = x' \wedge x' \in \mathrm{dom}(\sigma) \,\}$$

Note that there may be multiple or even an infinite amount of substitutions that are solutions to the semantic term unification problem giving rise to multiple or even an infinite amount of answers to the inference question. Let us set aside this complication for a moment.

Now, if for each subtree $\vec{n}$ with $\mathrm{getTree}(\mathbf{TR}, \vec{n}) = \langle \mathbf{R}, \mathbf{T} \rangle$, $\mathrm{dom}(\sigma_{\vec{n}}) = \mathrm{vars}(\mathbf{R})$ and $\mathrm{vrange}(\sigma_{\vec{n}}) = \emptyset$ (cf. restriction (1) of Definition 13), it merely remains to check whether the boolean conditions are satisfied (cf. restriction (4) of Definition 13) in order to obtain an inference tree. Otherwise, again more involved reasoning about the boolean conditions and obtained substitutions is necessary. As it turns out, in practice, this is rarely necessary because the solutions for variables are almost always determined by syntactical decomposition in the context of formal semantics.

The actual semi-algorithm enumerating all answers to an inference question works in a similar fashion. However, it *incrementally* constructs rule trees together with corresponding semantic term unification problems for those rule trees. At its heart, the semi-algorithm performs a breadth-first search for inference trees. We now present and discuss pseudocode for the semi-algorithm. Note that the pseudocode only shows the computation of the substitution of an answer. The construction of the tree is straightforward bookkeeping and thus omitted in favor of clarity of presentation.

---

**Algorithm 1:** Inference Engine

**Data:** an inference question $\tau$ and an inference system $\mathbf{R}$

**1**   $pendingNodes := \text{newQueue}(\langle\langle\emptyset, \text{newQueue}(\langle\tau\rangle), \emptyset\rangle\rangle)$

**2**   **while** *pendingNodes is not empty* **do**

**3**     $\langle\mathbf{E}, queue, conditions\rangle := \text{pop}(pendingNodes)$

**4**     **if** *queue is empty* **then**

**5**       **foreach** $\sigma \in \text{solutions}(\mathbf{E}, conditions)$ **do**

**6**         $\text{emit}(\{\, x \mapsto \sigma(x) \mid x \in \text{vars}(\tau)\,\})$

**7**     **else**

**8**       $term := \text{pop}(queue)$

**9**       **foreach** $\langle\langle P_1, \ldots, P_k\rangle, \mathbf{E}, \mathbf{B}, C\rangle =_R r \in \mathbf{R}$ **do**

**10**         obtain a renaming substitution $\rho$ for the rule ʀ

**11**         $\mathbf{E}' := \mathbf{E} \cup \{\, term \doteq [\rho](C)\,\} \cup \{\,[\rho](\tau) \doteq [\rho](\tau') \mid \tau \doteq \tau' \in \mathbf{E}\,\}$

**12**         $conditions' := conditions \cup \{\,\langle\mathbf{B}, \rho\rangle\,\}$

**13**         **if** $\text{hasSolutions}(\mathbf{E}', conditions')$ **then**

**14**           $queue' := \text{copy}(queue)$

**15**           **foreach** $1 \le i \le k$ **do**

**16**             $\text{push}(queue', [\rho](P_i))$

**17**           $\text{push}(pendingNodes, \langle\mathbf{E}', queue', conditions'\rangle)$

---

Algorithm 1 performs a breadth-first search using a queue of nodes (Line 1) that have to be explored. A node in the search graph is a triple where the first component is a semantic term unification problem, the second component is a queue of terms, and the third component is a set of pairs $\langle\mathbf{B}, \rho\rangle$ where $\mathbf{B}$ is a set of boolean conditions and $\rho$ is a renaming substitution. The idea is that each node contains the semantic term unification problem constructed so far, a queue of terms for which subtrees have to be constructed, and the conditions with their respective renaming that have to be satisfied.

As long as there are nodes to be explored (Line 2) the algorithm pops a node from the queue of yet to be explored nodes (Line 3). It then checks whether there are terms for which subtrees have to be constructed (Line 4).

If the queue component of a node is empty, then no more subtrees have to be constructed. The algorithm thus iterates over the solutions of the semantic term unification problem taking into account the boolean conditions (Line 5) and emits a substitution (Line 6) as part of an answer to the inference question. The subtree for that substitution is constructed by straightforward book-keeping and, hence, omitted here in favor of clarity of presentation. For further details we refer to our implementation (cf. Appendix A). Note that the function solutions hides all the complications of taking a semantic term unification problem and the boolean conditions and computing a substitution that binds all variables and satisfies the boolean conditions given the respective renaming of variables. The details here depend on how boolean conditions are implemented and what data values and operators are available. Again there exists an agnostically solvable fragment which only requires an evaluation function for operator terms and a decider for boolean conditions. If the semantic term unification problem is agnostically solvable and determines the binding for all variables (cf. restriction (1) of Definition 13), then it merely remains to check whether the boolean conditions are satisfied.

If the queue component of a node is non-empty, then there are terms for which subtrees have to be constructed. To this end, a term is popped from the queue (Line 8) and a search for a rule whose conclusion is unifiable with the term is initiated (Line 9). In order to construct a subtree for a specific term a rule has to be found such that the conclusion of that rule and the term are semantically unifiable. If there exists no such rule then the node is a dead end because we have to construct a subtree for a term for which no tree exists as there is no rule that would yield that term as a conclusion. To apply a rule and construct a subtree, first of all, a renaming substitution has to be obtained (Line 10) such that the variables in the range of this substitution do not occur in any other renaming substitution or in the inference question. The term unification problem is then extended along the lines of uProb (Line 11) and the boolean conditions together with the renaming are recorded (Line 12). If the resulting term unification problem and set of conditions does not have a solution (Line 13) then the rule is not applicable. Otherwise, a successor node is constructed. To this end, the premises of the rule are pushed onto the queue of terms for which a subtree has to be constructed (Line 16). Note that the renaming is applied here and, hence, no renaming has to be applied to the term popped from the queue (Line 11). Finally, the resulting successor node is pushed onto the queue of pending nodes (Line 17).

Just like for semantic term unification, it is necessary to spell out the details of the functions solutions and hasSolutions when concretizing the theory with values and operators. Furthermore, there again is an agnostically solvable fragment for which it is sufficient to define an evaluation function for each operator symbol and a decider for each boolean condition. We conjecture that in the context of formal semantics, most inference questions and systems lie in that fragment because variables are almost always bound by mere syntactical decomposition. While we have no formal proof of that claim or even a better characterization of the agnostically solvable fragment, we demonstrate its sufficiency for the formal semantics we define in this thesis.

Intuitively, the algorithm works similar to how one would apply inference rules by hand. Starting with some term one would like to derive, one tries to apply rules. Each application of a rule that is not an axiom then yields new terms and the process is repeated. For some inference rule systems this process never terminates because there are *diverging trees* that keep on growing indefinitely. An example are inference rule systems of diverging Turing-machines. In principle, it is possible to augment the algorithm to detect some of these cases. However, we do not investigate this further and defer it to future work. In its current state, the algorithm terminates if and only if there are no such indefinitely growing trees because then the queue of remaining terms for which subtrees have to be constructed becomes empty eventually for every path. We do not provide a formal proof of that claim.

### 3.3.3 Deriving Interpreters

So far, we tackled the general problem of derivation in an inference rule system. A structural operational semantics is an inference rule system and syntactically valid programs manifest in terms.

We already hinted at the idea that successor states of a program state are explored by enumerating the answers to an appropriate inference question. An interpreter is obtained by computing a possible successor state for a program state, i.e., by aborting the semi-algorithm after an answer has been found. One can also explore the whole state space if a semantics is non-deterministic by simply considering all answers. Figure 3.2 depicts the process of program execution. Given a term representing a syntactically valid program, the *execution engine* constructs inference questions and passes them to the inference engine. The answers computed by the inference engine contain possible successor states based on which the execution enginge formulates new inference questions and so on. As a result, the execution engine computes a trace or traces of program states using the inference rules.

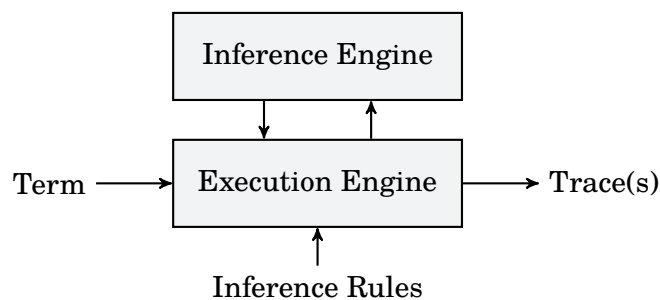

Figure 3.2: The process of program execution.

We exemplify this process in the next section with simple arithmetic expressions (cf. Example 1) and the process calculus CCS (cf. Section 2.3).

## 3.4   From Theory to Practice

In tandem with the meta theory, we developed the necessary tool support to specify inference systems, export them to LaTeX, and ask and answer inference questions. The complete implementation is available online[2] [35]. The core of our implementation is agnostic with respect to values and operators and, as we argue, easily extensible. The algorithmic core implementing semantic term unification and the inference engine is roughly 300 lines of Python code and available in the appendix of this thesis (see Appendix A). Including the data structures for terms, the complete core comprises roughly 600 lines of Python code. In this section, we showcase the practicability of the meta theory and accompanying tool support on two examples: A minimal example of arithmetic expressions based on Example 1 and a more involved example revolving around the process calculus CCS.

Instead of an explicit input language for the specification of inference rule systems, Python code is used directly to construct rules and inference rule systems. This has the advantage that Python can be used as a meta language to generate and manipulate rules. Furthermore, user-defined operators and data types can be easily added without needing to adapt an input language for rules. The implementation is agnostic with respect to data values and operators. New data types can be added as needed by sub-classing and are not part of the core. Our implementation, however, comes with a set of predefined generic values we deem useful such as integers and sets.

### 3.4.1   First Steps with Arithmetic Expressions

In order to get a feeling for the meta theory and its implementation, let us start by formalizing an extended variant of Example 1. Imagine a language of simple arithmetic expressions over the integers:

$$Expr \ni e \quad ::= \quad e_1 + e_2 \ \mid \ e_1 - e_2 \ \mid \ e_1 * e_2 \ \mid \ e_1 \ / \ e_2 \ \mid \ z \in \mathbb{Z} \qquad (3.1)$$

Analogously to Example 1, the semantics for this simple toy language comprises two rules, L-EVAL and R-EVAL, to evaluate the left and right operand of a binary expression, respectively:

$$\frac{e_l \xrightarrow{\alpha} u \quad \circ \in \{*, \ -, \ +, \ /\}}{(e_l \circ e_r) \xrightarrow{\alpha} (u \circ e_r)} \ \text{L-EVAL} \qquad \frac{e_r \xrightarrow{\alpha} u \quad \circ \in \{*, \ -, \ +, \ /\}}{(e_l \circ e_r) \xrightarrow{\alpha} (e_l \circ u)} \ \text{R-EVAL}$$

In addition there is a rule for each binary operator:

$$\frac{}{(z_l + z_r) \xrightarrow{\tau} z_l + z_r} \ \text{ADD-EVAL} \qquad \frac{}{(z_l - z_r) \xrightarrow{\tau} z_l - z_r} \ \text{SUB-EVAL}$$

$$\frac{}{(z_l * z_r) \xrightarrow{\tau} z_l \cdot z_r} \ \text{MUL-EVAL} \qquad \frac{}{(z_l \ / \ z_r) \xrightarrow{\tau} \lfloor z_l \div z_r \rfloor} \ \text{DIV-EVAL}$$

---

[2] https://github.com/koehlma/rigorous

Let us now have a closer look at the Python code defining that system.  See Appendix A.1 for the complete definition of the inference rule system as a Python program.  For readability, we start by defining auxillary functions for the creation of terms. For instance, the function `binary_expr` takes three terms, the left operand, the operator, and the right operand, and returns a sequence term representing a binary expression:

```python
def binary_expr(
    left: terms.Term, operator: terms.Term, right: terms.Term
) -> terms.Term:
    return terms.sequence(left, operator, right)
```

These auxillary functions are then used for the definition of rules:

```python
l_eval_rule = define.rule(
    name="l-eval",
    premises=(
        sos.transition(source=left_expr, action=sos.some_action, target=some_result),
    ),
    conditions=(is_binary_operator(some_operator),),
    conclusion=sos.transition(
        source=binary_expr(left_expr, some_operator, right_expr),
        action=sos.some_action,
        target=binary_expr(some_result, some_operator, right_expr),
    ),
)
```

This code snippet defines the rule L-EVAL.  In addition to the auxillary function `binary_expr` it uses the function `sos.transition` from the module `sos` which constructs a term representing a transition.  The module `sos` contains pre-defined auxillary functions for the definition of structural operational semantics within our framework.  Here, `left_expr`, `right_expr`, `some_result`, and `some_operator` are meta variables defined earlier:

```python
left_expr = define.variable("left_expr", text="el", math="e_l")
right_expr = define.variable("right_expr", text="er", math="e_r")

some_result = define.variable("result", text="u", math="u")

some_operator = define.variable("operator", text="◯", math="\\circ")
```

The function `is_binary_operator` constructs a boolean condition checking whether the operator is a binary operator.  To this end, it uses the module `sets` implementing a set data type and operators:

```python
def is_binary_operator(operator: terms.Term) -> inference.Condition:
    return booleans.check(sets.contains(BINARY_OPERATORS, operator))
```

The rule R-EVAL is defined analogously. In contrast to the rule $E_+$ of Example 1 the rule ADD-EVAL does not contain the condition $z_l, z_r \in \mathbb{Z}$. This condition is implicit as the addition operator $+$ is only defined for numbers.  Hence, substituting $z_l$ or $z_r$ with anything but a number, i. e., in this case an integer, will lead to the rule not being applicable for the evaluation function being undefined for such substitutions. The definition of ADD-EVAL uses the module numbers which provides a number data type and operators:

```
add_eval_rule = define.rule(
    name="add-eval",
    conclusion=sos.transition(
        source=binary_expr(left_int, BINARY_ADD, right_int),
        action=sos.ACTION_TAU,
        target=numbers.add(left_int, right_int),
    ),
)
```

Our implementation conveniently tracks the file and line number where each rule has been defined and outputs this information when the inference rule is printed on the console (see Figure 3.3).

```
PS rigorous> rigorous-arithmetic system print

Rule 'l-eval' (arithmetic/semantics.py:45):

         (el = α => u)    o ∈ {+, -, *, /}
l-eval  ───────────────────────────────────
            ((el o er) = α => (u o er))
```

Figure 3.3: Printing inference rules on the console.

With the inference rule system in place, we can now ask questions to the inference engine and thereby explore the evaluation of expressions. Let us take $((3 + 12) + (4 + 42))$ as an example. To iterate over the successor states, we formulate the following inference question $(((3 + 12) + (4 + 42)) \xrightarrow{\alpha} u)$. Asking the inference engine, we obtain two answers to that inference question. For both answers $\alpha \mapsto \tau$ but for $u$ we obtain two alternatives:

$$(15 + (4 + 42)) \qquad ((3 + 12) + 46)$$

The two answers reflect the fact that the semantics is non-deterministic because both L-EVAL and R-EVAL are applicable to the initial state. Hence, we obtain two successor states where the left and right operand has been evaluated, respectively. The inference engine does, however, not only provide us with the successor states but also with two inference trees (see Figure 3.4) witnessing that the initial state has those successors.

$$\frac{\dfrac{}{(3 + 12) \xrightarrow{\tau} 15} \text{ ADD-EVAL}}{((3 + 12) + (4 + 42)) \xrightarrow{\tau} (15 + (4 + 42))} \text{ L-EVAL}$$

$$\frac{\dfrac{}{(4 + 42) \xrightarrow{\tau} 46} \text{ ADD-EVAL}}{((3 + 12) + (4 + 42)) \xrightarrow{\tau} ((3 + 12) + 46)} \text{ R-EVAL}$$

Figure 3.4: Inference trees for adding integers.

By iterating this process we explore the whole state space. To this end, both obtained states are again transformed into a question, respectively:

(a) $((15 + (4 + 42)) \xrightarrow{\alpha} u)$      (b) $(((3 + 12) + 46) \xrightarrow{\alpha} u)$

The answers to both questions then yield $(15 + 46)$ as the successor whose successor finally is $61$. While both questions yield the same successor state, the inference trees witnessing the respective answers are different because for (a) EVAL-R is applied while for (b) EVAL-L is applied. As the inference engine exhaustively enumerates all answers to an inference question and it terminated we also know that no other states are reachable.

Note that all the inference rules and most terms depicted here have been generated and exported by our tool. Nevertheless, they look quite natural and exactly like the rules for arithmetic expressions you would expect in a textbook on structural operational semantics [42, cf.].

Before we come to the more involved semantics of the process calculus CCS, let us have a closer look on why no problems which are not agnostically solvable can arise from the semantics. When we explore successor states we always ask questions of the form $\tau \xrightarrow{\alpha} u$ where $\tau$ is a closed and fully-evaluated term. If we ask such questions, then all the variables occurring on the left-hand side of $\rightarrow$ in the conclusion of any inference rule are always bound by mere recursive syntactical decomposition of $\tau$. The rules ADD-EVAL, SUB-EVAL, MUL-EVAL, and DIV-EVAL have no other variables, hence, we also obtain instances of them by mere syntactical decomposition of $\tau$. Now, for the rules L-EVAL and R-EVAL the remaining variables $\alpha$ and $u$ are determined by syntactical decomposition of their premises. Intuitively, these considerations provide us with an order in which variables are bound to closed and fully-evaluated terms as an inference tree is constructed. First all variables except $u$ and $\alpha$ are bound in a rule by mere syntactical decomposition of the conclusion and then $\alpha$ and $u$ are bound by mere syntactical decomposition of a premise. Therefore, if we only ask questions of the form $\tau \rightarrow u$ then we never run into problems which are not agnostically solvable.

In contrast, if we would like to explore predecessor states we would ask questions of the form $u \xrightarrow{\alpha} \tau$ where $\tau$ is again a closed and fully-evaluated term. Now, for L-EVAL and R-EVAL an analogous reasoning applies but this time $u$ and $e_r$ respectively $e_l$ are bound by mere syntactical decomposition of the conclusion while $e_l$ respectively $e_r$ and $\alpha$ are bound by syntactical decomposition of the premises. However, for the remaining four rules, a binding for $z_l$ and $z_r$ can no longer be determined by mere syntactical decomposition. Instead reasoning about the operators is required, and, hence, exploring predecessors is not agnostically solvable. If we would ask the inference engine such questions, then it would throw an exception because we did not equip it with additional reasoning strategies suitable to resolve these situations. As our focus is on interpreters which compute successor states, such questions do however not pose a problem for the semantics we consider.

### 3.4.2 Calculus of Communicating Systems

In Section 2.3, we introduced the process calculus CCS as a paradigmatic example for structural operational semantics as they appear in literature. The presented inference rules for CCS look natural and like the rules in Milner's original papers [38, 37] and the script of the introductory lecture on concurrent programming at Saarland University [29]. Just like the rules for arithmetic expressions, they have been defined within a Python program and then exported to LaTeX and rendered into this thesis. The complete source code of this definition is available in Appendix A.2.

In this section, we briefly discuss some of the intricacies of putting the CCS semantics in our framework. The rules PREFIX, CHOICE-L, CHOICE-R, PAR-L, and PAR-R do not utilize any features we did not already encounter. Hence, we focus our discussion on the remaining four rules:

$$\frac{\Gamma \vDash P \xrightarrow{\alpha} P' \quad P = \Gamma[X]}{\Gamma \vDash X \xrightarrow{\alpha} P'} \text{ REC} \qquad \frac{\Gamma \vDash P \xrightarrow{\alpha} P' \quad \alpha \notin H}{\Gamma \vDash (P \setminus H) \xrightarrow{\alpha} (P' \setminus H)} \text{ RES}$$

$$\frac{\Gamma \vDash P[X \mapsto (\texttt{fix } X = P)] \xrightarrow{\alpha} P'}{\Gamma \vDash (\texttt{fix } X = P) \xrightarrow{\alpha} P'} \text{ FIX} \qquad \frac{\Gamma \vDash P \xrightarrow{\alpha} P' \quad \Gamma \vDash Q \xrightarrow{\overline{\alpha}} Q'}{\Gamma \vDash (P \parallel Q) \xrightarrow{\tau} (P' \parallel Q')} \text{ SYNC}$$

The rule REC contains the constraint $P = \Gamma[X]$ which uses the provided data type for *mappings*. A mapping is a partial function $\Gamma : \mathbb{V} \rightharpoonup \mathbb{V}$. An environment $\Gamma$ is formalized as such a mapping from process variables to process terms. The operator term $\Gamma[X]$ then evaluates to $\Gamma(X)$. The rule RES contains the boolean condition $\alpha \notin H$ which uses the provided data type for sets analogously to the rules L-EVAL and R-EVAL of the arithmetic expressions just that the set is obtained by syntactical decomposition. The rule FIX also just uses a replacement operator provided by our framework.

### Custom Operators and Data Types

The probably most interesting rule is SYNC which uses the CCS-specific complement operator. Thanks to the extensibility of our framework it is easy to add custom operators, assuming all problems are agnostically solvable, by providing nothing more than an evaluation function. Here is the definition of the complement operator by implementing it in Python:

```python
@terms.operator
@terms.check_arity(1)
def complement(arguments: terms.Arguments) -> t.Optional[terms.Term]:
    argument = arguments[0]
    if not isinstance(argument, terms.Sequence) or len(argument.elements) != 2:
        return None
    action, modifier = argument.elements
    if isinstance(modifier, terms.Symbol):
        if modifier.symbol == "!":
            return terms.sequence(action, terms.symbol("?"))
        elif modifier.symbol == "?":
            return terms.sequence(action, terms.symbol("!"))
    return None
```

The decorator `@terms.operator` tells our framework to add an operator symbol whose semantic meaning is captured by the decorated function (cf. Definition 4). The decorator `@terms.check_arity(1)` ensures that the operator has a fixed arity of one in this case, i. e., it accepts a single argument. Now, the function itself examines the first argument. If the first argument is a sequence term of the form $(a\,!)$ or $(a\,?)$ then the respective complement is returned. Otherwise, `None` is returned indicating that the operator is undefined for the provided arguments. Using the operator within a rule definition is achieved by calling the decorated function. The premises of the rule SYNC which use the complement operator are defined as follows:

```
premises=(
    sos.transition(
        environment=some_environment,
        source=some_process,
        action=sos.some_action,
        target=some_successor,
    ),
    sos.transition(
        environment=some_environment,
        source=other_process,
        action=complement(sos.some_action),
        target=other_successor,
    ),
)
```

To also tell our framework how to render the operator when exporting to LATEX another five lines of Python code are necessary. The operator is rendered using the macro `\overline` which puts a line on the operand:

```
@render.register_function_operator(semantics.complement)
def _render_complement(arguments: terms.Arguments, builder: render.BoxBuilder) -> None:
    builder.append_chunk("comp(", math="\\overline{")
    builder.append_term(arguments[0])
    builder.append_chunk(")", math="}")
```

This example demonstrates the extensibility of our framework. As long as all problems are agnostically solvable—which they usually are in the context of structural operational semantics where almost all variables are bound by mere syntactical decomposition—nothing more than an evaluation function has to be provided in order to define an operator for a particular purpose. By writing a few more lines the rules using this operator can then be exported to a paper or thesis looking natural and intuitive.

Adding custom data types is also straightforward. Imagine one would like a dedicated data type for process variables instead of using symbols. Such a data type is implemented by sub-classing `terms.Value`:

```
@d.dataclass(frozen=True)
class ProcessVariable(terms.Value):
    identifier: str
```

A value of this type is constructed by instantiating the class, e. g., to construct a process variable named X, `ProcessVariable("X")`. The dataclass decorator[3] `@d.dataclass` takes care of generating an equality and hash function for the custom datatype. These functions are used by our unification algorithm to determine whether two values are equal.

---

[3] https://docs.python.org/3.7/library/dataclasses.html (Accessed: 2020/12/14)

**A Command Line Tool**

By implementing a parser turning strings into terms and utilizing some of the convenience functions provided by our framework, we obtain a command line application from the inference rule system (Figure 3.5a). This command line application allows printing the system (Figure 3.5b) and exploring the state space of a CCS process (Figure 3.5c). The exploration happens via the inference engine and the inference trees can be printed as well.

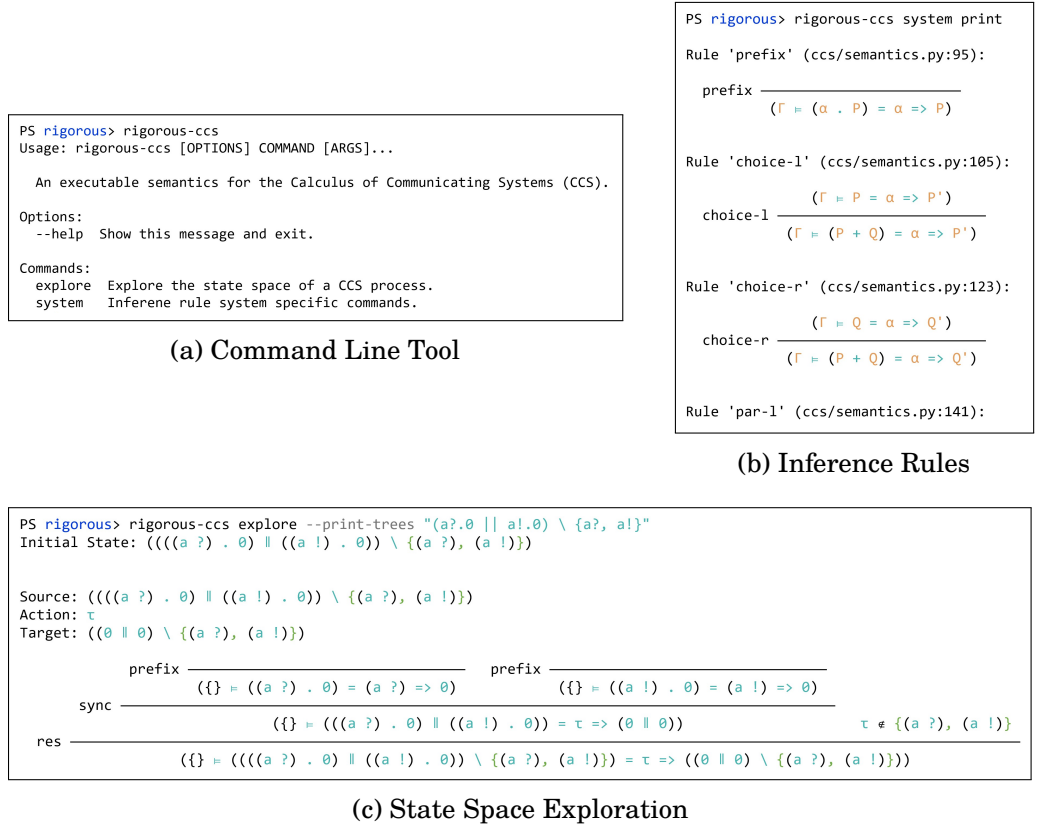

(a) Command Line Tool

(b) Inference Rules

(c) State Space Exploration

Figure 3.5: The tool obtained from the formal CCS semantics.

## 3.5 Summary and Discussion

We conclude this chapter with a brief summary and discussion of our contributions. Starting with the intuitive notion of inference rules motivated in Example 1, we presented a meta formalism for the formalization of inference rule systems. To this end, expanding upon existing research, we introduced the idea of semantic term unification which allowed us to capture inference rules like those for CCS which involve operator terms with a specific semantic meaning like lookups in an environment. We proved that derivation in an inference rule system, as we formalized it, is in general undecidable and that the presented formalism is Turing-complete. We presented a semi-algorithm coined the *inference engine* which enumerates answers to inference questions as a more general form of derivation. Based on the inference engine we demonstrated how an interpreter exploring the state space of a program can be obtained automatically from a formal semantics. Finally, we demonstrated the practicability of this approach on a semantics for simple arithmetic expressions and on the process calculus CCS.

While there are still open questions and further comparisons to related work are necessary to carve out more fundamental contributions, we believe that the presented formalism is a step forward and at least useful from a practical perspective for working with inference rule systems. In particular, its extensibility turned out useful while experimenting with our formal semantics for Python. Being able to add primitives as needed allows defining formal semantics on the desired level of abstraction.

In the next chapter, we build upon the results from this chapter and leverage the introduced meta theory to build a formal semantics for Python. However, before doing that, a few last remarks are in order.

**Why not SMT?** We promised earlier that we would compare our semantic term unification algorithm to the idea of encoding semantic term unification problems as SMT problems. Note that we did not investigate this in great detail and, thus, the following remarks are mostly preliminary. In any case, our semantic term unification algorithm is merely a piece of the overall theory we presented here. Hence, if such an encoding turns out beneficial, we just obtained another strategy for solving semantic term unification problems which we could plug into our broader framework.

If one likes to encode semantic term unification problems as SMT problems one first has to find an appropriate SMT fragment with equality that supports all the operators and data values one is interested in. In the context of formal semantics, this most likely includes operators that operate on compound terms. In case such a fragment exists and is supported by an off-the-shelf solver developing an encoding might actually be a good idea.

We see the advantage of our algorithm in its extensibility and flexibility especially with regard to the agnostically solvable fragment where nothing more than an evaluation function for an operator has to be provided. Within the agnostically solvable fragment, we allow almost arbitrary operators that would pose problems for an SMT-based approach such as arbitrary non-linear arithmetic. This flexibility is especially advantageous in the context of formal semantics where most problems seem to be in the agnostically solvable fragment as far as deriving interpreters is concerned. Furthermore, the algorithm can be interfaced with an SMT solver but most likely first reduces the problem size as we have seen in Section 3.2.2. In this regard, an SMT-based approach and our algorithm do not exclude each other but are complementary. Furthermore, recapitulate our conjecture that the core insights of Martelli and Montanari [36] carry over to our algorithm most likely leading to a more efficient version of the algorithm. Hence, it may also be an advantage that our algorithm is or can be made more efficient than an SMT encoding.

**Limitations**  Unfortunately, some rules in the literature make use of "…" where the reader is responsible for grasping the meaning of the three dots. Such rules cannot be directly formalized in our framework because the three dots do not have a clearly defined meaning. However, we are confident that it is possible to rewrite most of these inference rule systems without the dots. For future work, it might be interesting to investigate how certain instances of this notation can be integrated into our framework.

While the theory is not limited to the agnostically solvable fragment, our implementation is currently mostly limited in that regard. As we elaborated in Section 3.2.2, the semantic term unification algorithm has to be interfaced with external solvers and other resolution strategies in order to obtain solutions for problems outside of the agnostically solvable fragment. As we expect most problems that appear in the context of formal semantics to be agnostically solvable, we did not investigate this further.

**Open Problems**  The characterization of the agnostically solvable fragment is currently not enlightening. In particular, with regard to the inference engine a concise characterization of which inference questions are guaranteed to only give rise to agnostically solvable semantic term unification problems in a given inference rule system would be desirable. Furthermore, it would be desirable to obtain a restriction for inference rule systems that guarantees that derivation is decidable. For this purpose, further insights from the research on rule formats should be imported.

From a practical perspective, these problems do however not stand in the way of using the theory and its implementation to work with inference rules. In the worst case, the inference engine might just diverge or throw an exception if it encounters any problems it cannot solve.

**Beyond SOS**  While we developed the meta theory primarily for the purpose of formalizing the semantics of programming languages with an SOS-style formal semantics, it is quite general and most likely useful for other formalisms based on inference rules. Further exploration should be conducted in this regard in order to work out the limits of the framework and get a better idea of its weaknesses and strengths.

# Chapter 4

# Formal Semantics for Python

In this chapter, we develop a formal semantics for Python cashing in on the advantages of the meta-theoretic framework introduced earlier. This chapter demonstrates that our meta-theoretic framework is suitable for the formalization of modern programming languages involving a variety of control-flow mechanisms and interaction with an environment.

In general, there are multiple approaches to formal semantics. In Section 2.2, we introduced structural operational semantics as the approach we chose for this thesis. For now, we take this choice as the bedrock of our formal semantics and defer the discussion of its adequacy to Chapter 5.

Naturally, developing a formal semantics for a programming language involves design choices beyond the general semantic framework to use. In retrospect, one of the major design challenges of this work was to choose a parsimonious set of primitive data types and operations to base the formal semantics on. Historically, not much attention has been payed to having a parsimonious set of primitive building blocks. Instead, it has been an explicit design goal to build a language which is easy to extend:

> *I think my most innovative contribution to Python's success was making it easy to extend. That also came out of my frustration with ABC. ABC was a very monolithic design. […] You could write your own programs, but you couldn't easily add low-level stuff.*
>
> — Guido van Rossum, January 2003 [61]

Arguably, the ease with which "low-level stuff", i. e., primitives, can be added to Python and in particular its reference implementation played a huge role in its success. Unfortunately, it also lead to more or less mindlessly adding primitives as the need for them arose. As a result, there has never been a well-defined set of low-level primitives around which the language is built. This problem is aggravated further by a historically rather fuzzy distinction between the language itself and its reference implementation. Hence, this chapter devotes quite some effort to resolving this muddle.

**Design Goals**  We already discussed adequacy criteria for formal semantics in general in Section 2.2.1 and for Python in particular in Section 2.1. Before we dive into the details of the formal semantics for Python, let us explicitly phrase the main design goals for our formal semantics.

The three main design goals for our formal semantics are: (a) A parsimonious set of primitive data types and operations, (b) a semantics that concretizes implementation freedom and vagueness of the PLR in a way that permits an efficient implementation while staying close to the reference implementation, and (c) a semantics that is easy to extend with additional built-in types and operations we do not cover in this thesis.

A semantics that is not based on a well thought out set of primitives quickly becomes bloated and elusive. However, as we argued in Section 2.2.1, a semantics should be succinct and intelligible. Hence, we put considerable effort in designing the primitive data types and operations used by the semantics. As explained in Section 2.1.3, we believe that trying to formalize the implementation freedom and vagueness of the PLR is not worthwhile. Hence, we aim to concretize it in a way that permits an efficient implementation of the semantics while staying close to the reference implementation. Last but not least, as discussed above, Python owes its success in part to its extensibility. We like to retain this extensibility as far as possible in our formal semantics enabling the adaption to future versions of Python and allowing us to faithfully claim that the semantics can be made complete later by adding the parts we left aside for the scope of this thesis.

**Overview**  The formal semantics comprises four main parts: (i) An encoding of Python programs into terms so as to make them tractable by our meta theory, (ii) a formalization of a significant part of Python's data model capturing how Python objects are represented, (iii) an inference rule system constituting a structural operational semantics for a fragment of Python extended with a parsimonious set of primitives, and (iv) a *runtime* implementing built-ins and other operations using those primitives.

**Structure**  In Section 4.1 we describe how Python programs are translated into terms to be fed into the execution engine (cf. Section 3.3.3). Our focus will be on the general process of translation and the steps necessary to construct a term from a program but not on how individual expressions and statements are translated. We address the translation of individual expressions and statements later, in Section 4.3, when introducing the inference rules operating on the respective constructs.

In Section 4.2, we formalize Python's *data model*. To this end, we introduce a set of intuitive and natural primitives concretizing the set $\mathbb{V}$ of our meta theory. In Python, all data is represented by *objects* and relations between those objects. Using the earlier introduced primitives, we describe how Python objects are represented. Our focus here is on the structural representation of Python objects and not on the operations they support.

While Section 4.2 is all about how data is represented, we turn to the operational part of the semantics in Section 4.3 and Section 4.4. In Section 4.3, we present the inference rules for expressions and statements together with the terms—constructed from the respective Python constructs—they operate on. The inference rules capture a superset of the Python fragment defined in Section 2.1.2 which has been enriched with a parsimonious set of primitives as introduced in Section 4.2 and operations thereon. Most operations on Python objects are realized by *calling into* the *runtime*. The runtime is a collection of code capturing complex operations on Python objects that involve multiple steps of primitive operations. We exemplify parts of the runtime together with the respective syntactical constructs they cover. In Section 4.4, we present the inference rules that do not directly relate to expressions or statements. These inference rules concern interprocedural control flow, i. e., the management of frames for storing local variables and the call stack, and access to the memory storing Python objects.

In Section 4.5, we evaluate our formal semantics by harvesting a test suite from existing work and testing our semantics with that suite. Having formalized the semantics in the meta-theoretic framework presented in the previous chapter, we automatically obtain an interpreter from it. By running tests on this interpreter, we assess the completeness of the semantics and evidence its correctness. As we will see, using the standard algorithm for inference rule systems presented in Section 3.3 turns out to be very slow because even simple programs involve a large amount of execution steps. Hence, we also discuss some optimizations tailored to SOS we made to speed up the execution of programs. Furthermore, we compare our formal semantics to related work in the area of Python semantics [27, 44, 41, 22].

## 4.1 From Programs to Terms

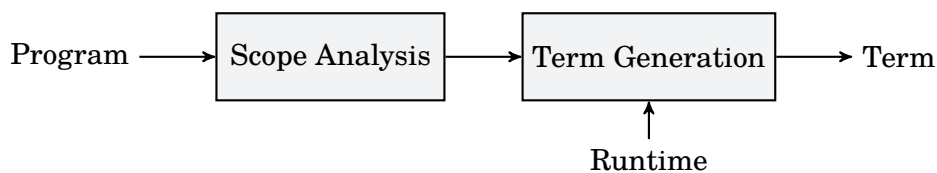

Figure 4.1: The translation pipeline.

To apply the techniques presented in Chapter 3 we first have to *translate* a Python program to a term (cf. Section 3.3.3). To this end, we assume that the Python program has been parsed into an *Abstract Syntax Tree* (AST), e. g., by the module ast[1] of the standard library. The translation itself is then a two-step process. First, a *scope analysis* is performed on the AST to determine how and where variables are bound and accessed. With the information of the scope analysis, the *term generation* subsequently generates a term from the AST. Figure 4.1 depicts this process.

---

[1] https://docs.python.org/3.7/library/ast.html (Accessed: 2020/12/14)

### 4.1.1 Scope Analysis

Python code is organized into *blocks*. A *block* is a syntactic group of code that executes *as a unit* [20, §4.1.]. Programs, function definitions, and class definitions constitute blocks. For each identifier used in a block, an *access mechanism* must be determined. For instance, an identifier used in a function may be *purely local* or refer to a *global variable*.

Snippet 4.1 shows an example of different access mechanisms. Calling `f` (line 16) constructs and returns a function g in which the identifiers x, y, and z are accessed with different mechanisms respectively. In this case, x comes from the global scope, i. e., the module where the function is defined in, y comes from the closure, more precisely the *cell* y of the closure, of the constructed function g, and z is a purely local identifier.

```python
1   x = "global"
2
3
4   def f():
5       y = "cell"
6
7       def g():
8           z = "local"
9           assert x == "global"  # mechanism GLOBAL
10          assert y == "cell"  # mechanism CELL
11          assert z == "local"  # mechanism LOCAL
12
13      return g
14
15
16  f()()
```

Snippet 4.1: Three different access mechanisms.

Python allows modifying access mechanisms with the keywords `global` and `nonlocal`. An identifier declared *global* with `global` will always come from the global scope no matter whether it is also *bound* in an enclosing lexical scope. An identifier declared *non-local* with `nonlocal` will always come from the nearest enclosing lexical scope where it is *bound*. Identifiers can be bound by various syntactical constructs. For instance, they are bound in a block if they occur on the left-hand side of an assignment. For further details on the binding of identifiers we refer to the PLR [20, §4.2.1.].

The scope analysis infers one of five access mechanisms for each identifier used within a block: `LOCAL`, `CELL`, `GLOBAL`, `CLASS_CELL`, and `CLASS_GLOBAL`. Depending on the access mechanism, accessing the identifier has a different semantics which we describe in more detail in Section 4.3.

The description of this name binding and resolution process in the PLR [20, §4.2.] is unfortunately not particularly clear. The scope analysis algorithm we extrapolated from the PLR is captured by the method `infer_mechanisms` of the `Block` class in the module `rigorous.semantics.python.syntax.blocks`

of the implementation [35]. We are not discussing it here and refer to the implementation for further details. Given that our semantics passes a variety of different tests (cf. Section 4.5) checking whether the name binding and resolution process works as it should, we are confident that the algorithm extrapolated from the PLR is indeed correct.

### 4.1.2 Term Generation

With the information of the scope analysis, a term is generated from a program. While the translation of expressions and statements into terms is, as we will see in Section 4.3, almost always straightforward, there is more to generating a term from a program than merely translating the expressions and statements it consists of. A term encoding a program state must also capture the whole state of program execution including the call stack and heap. To this end, we introduce the concept of *semantic layers* our semantics can be decomposed into. Figure 4.2 visualizes this idea.

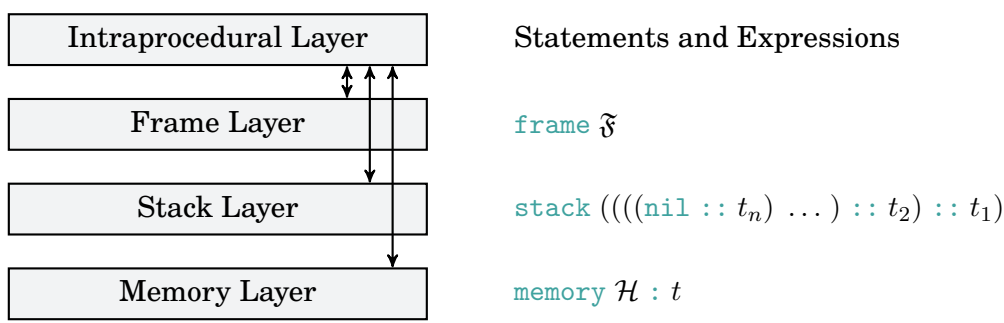

Figure 4.2: The five layers of our Python semantics.

The *intraprocedural layer* concerns the semantics of expressions and statements while the other three layers capture stack frames, the call stack, and the heap or memory. Every program state encoded into a term consists of all four layers, i. e., every term we are considering has the following form

$$\texttt{memory}\ \mathcal{H} : (\texttt{stack}\ ((((\texttt{nil} :: (\texttt{frame}\ \mathfrak{F}_n))\ \dots\ ) :: (\texttt{frame}\ \mathfrak{F}_2)) :: (\texttt{frame}\ \mathfrak{F}_1)))$$

where $\mathfrak{F}_1$ to $\mathfrak{F}_n$ are *frames* and $\mathcal{H}$ is the *heap*. A frame stores local variables and also the current state of execution of that frame as a term derived from statements and expressions. Intuitively, the inference rules will always execute the top-most frame until it returns at which point the control flows to its predecessor on the call stack. The heap is a partial function from *references* or addresses to values and must be pre-initialized with all the built-in functions, values, and types used by the program. Every built-in function, after all, is a Python object stored somewhere on the heap. In other words, the runtime and built-ins must be *linked* into the term (cf. Figure 4.1).

To translate a program, a term is generated which consists of one frame obtained from the program's code and a heap initialized with all built-ins and *constants* appearing in the program. For example, imagine a program using the string constant `"Hello World!"`. Now, when a term for that program is generated, a Python object for that string is constructed and put on the initialized heap. The initialized heap then contains all constants and built-ins and is put into the memory layer of the term to be generated. This process is analogous to how a compiler stores certain values in the data section of an object file[2] produced from a program.

We describe the layers and corresponding inference rules more formally in Section 4.4, for now a pre-theoretic intuition shall suffice.

**Built-Ins and the Runtime**

As mentioned earlier, we enrich the Python fragment defined in Section 2.1.2 with primitives. The term generator translates Python code to terms in one of two modes, the *user mode* or the *primitive mode*. Usual Python programs are translated in user mode while code that is part of the runtime, implementing, e. g., built-ins, is translated in primitive mode.

In primitive mode, certain expressions are translated into primitives directly. For instance, Snippet 4.2 shows a *runtime function* used to implement the process of accessing an item. Recapitulate Snippet 2.5 where we implemented `__getitem__` in order to overwrite item access. The semantics of a *subscript expression* $e_o[e_k]$ are essentially captured by a call `get_item(`$e_o$`, `$e_k$`)` to the runtime function `get_item`. This runtime function in turn does the following: Using the primitive `GET_SLOT` it retrieves the method `__getitem__` from the object. If the object does not have such a method, it raises a `TypeError` indicating that the object does not support item access and cannot be used in a subscript expression. If the object has such a method, then the result of the subscript expressions is obtained by invoking the method with the object as the first argument and the key as the second argument.

```python
def get_item(obj, key):
    slot = GET_SLOT(obj, "__getitem__")
    if slot is None:
        raise TypeError("object is not subscriptable")
    return slot(obj, key)
```

Snippet 4.2: Runtime function for accessing items.

Note that we use the ordinary syntax for calling also for primitives. During translation in primitive mode, the term generator identifies `GET_SLOT` as a primitive and constructs a respective term. In normal mode, the same code would be translated differently, namely to looking up the identifier `GET_SLOT` and then calling whatever object it is bound to.

We desugar much of Python's syntax to calls to runtime functions which themselves rely only on a parsimonious set of primitives. By doing this, we keep the semantics manageable and also separate actual primitives from functionality that involves multiple primitive steps. We discuss these primitive building blocks in more detail in Section 4.3. For more details on the translation process described here, we refer to our implementation [35].

---

[2] https://gcc.gnu.org/onlinedocs/gccint/Sections.html (Accessed: 2020/12/02)

## 4.2 Python's Data Model

Python's *data model* captures how the data a Python program interacts with is represented and what operations it supports [20, §3.]. From the perspective of a Python program, all data is represented by *objects* and their relations to other objects. This "everything is an object" philosophy includes functions and to some extent even the source code of a program itself.

Formalizing Python's data model has been one of the significant challenges of this work because it is unclear from the PLR what the actual primitives of the data model are. To get an idea of the complexity of the data model, the implementation of the basic built-in objects takes roughly 75 000 lines[3] of C code. This includes container types such as hash tables, called *dictionaries* in Python, and lists. Unfortunately, the mere execution of code already requires the implementation of most of these data types. Consider a simple program such as x = 4 which does nothing more than assigning 4 to the global variable x. However, the global namespace itself is a dictionary exposed via the function globals(). Hence, assigning 4 to x involves storing 4 within a dictionary using the string "x" as the key. The problem with this architecture is that it is impossible to implement most of these built-ins within Python because they need to be there before any code can run.

In this section, we formalize a significant part of Python's data model. Capturing everything is out of scope and not feasible in an academic context. Our focus will be on the structure of the objects and not on the operations they support. To this end, we define a set of *primitive data values* concretizing the set $\mathbb{V}$ of the meta-theoretic framework. These primitives are *integers* and *floats*, *booleans*, *unicode strings*, *undefined*, *vectors*, *records*, *mappings*, and *references*. Python's data model is formalized using these primitives by describing how Python objects are constructed from them.

### 4.2.1 Python Objects

In Python, every *object* has an *identity* and a *type* [20, §3.1.]. We capture objects formally by *references* referring to *object descriptors*, i. e., formally an object *is* a reference to an object descriptor.

References $r \in \mathbb{V}$ are values *referring* to other values by virtue of a *heap* mapping them to the values they refer to. References have a numeric or named

---

[3] Counted with cloc [59] on CPython version 3.7.9 in the directory Objects.

*identity*. For instance, $\mathrm{ref}(42)$ is a reference with the numeric identity $42$ and $\mathrm{ref(object)}$ is a reference with the named identity `object`. Two references are *identical* if their identity is the same, i. e., both have a numeric identity with the same numeric value or both have a named identity with the same name. Formally, a heap $\mathcal{H} : \mathbb{V} \rightharpoonup \mathbb{V}$ is a partial function from references to values. A reference $r \in \mathbb{V}$ refers to a value $\nu \in \mathbb{V}$ by virtue of a heap $\mathcal{H} \in \mathbb{V}$ if and only if $r \in \mathrm{dom}(\mathcal{H})$ and $\mathcal{H}(r) = \nu$. We usually use named references for built-in objects and numeric references for objects created while running a program. For instance, $\mathrm{ref(None)}$ is the object representing the constant `None` and $\mathrm{ref(TypeError)}$ is the built-in `TypeError` class. Naturally, two objects are identical if and only if they are the same reference.

Object descriptors are a special kind of *record*. A *record* $\mathrm{r} \in \mathbb{V}$ is a compound value associating *field names* to values much like a `struct` in C. Formally, a record is a finite-domain partial function from field names to values:

> **Definition 20 (Records)**
> Let $\mathbf{N}$ be a set of *field names*. A *record* is a partial function $\mathrm{r} : \mathbf{N} \rightharpoonup \mathbb{V}$ with a finite domain $\mathrm{dom}\,\mathrm{r}$. We use the following notation for records
>
> $$\langle \mathrm{n}_1 : \nu_1, \ldots, \mathrm{n}_n : \nu_n \rangle \tag{4.1}$$
>
> where $\mathrm{n}_i$ are field names and $\nu_i \in \mathbb{V}$ are values.
>
> The record corresponding to (4.1) is the partial function $\mathrm{r}$ with $\mathrm{dom}(\mathrm{r}) = \{\mathrm{n}_1, \ldots, \mathrm{n}_n\}$ and $\mathrm{r}(\mathrm{n}_i) = \nu_i$ for all $i$.

An object descriptor is a record with at least two fields, `cls` and `dict`, where `cls` contains a *type object* representing the type of the object and `dict` contains either $\mathrm{ref(None)}$ or a *dictionary-like object* storing the *instance attributes* of the object. Type and dictionary-like objects are references to special kinds of object descriptors with additional fields.

A typical example for an object is an integer object. Imagine a numerical computation whose result is the integer $31515$. This result would be represented by a reference referring to an object descriptor as follows:

$$\mathrm{ref}(42) \mapsto \langle \mathrm{cls} : \mathrm{ref(int)}, \mathrm{dict} : \mathrm{ref(None)}, \mathrm{value} : 31415 \rangle$$

We use this notation to indicate that the reference $\mathrm{ref}(42)$ refers to the object descriptor $\langle \mathrm{cls} : \mathrm{ref(int)}, \mathrm{dict} : \mathrm{ref(None)}, \mathrm{value} : 31415 \rangle$. The type of the object $\mathrm{ref}(42)$ is $\mathrm{ref(int)}$ which is the built-in `int` type. Integers do not have any instance attributes, hence, `dict` is set to $\mathrm{ref(None)}$. The integer object's value in terms of the primitive integer $31415$ is stored in the additional field `value` of the object descriptor. Note that there is a crucial difference between integer objects, which are references referring to object descriptors whose type is `int`, and primitive integers.

In the following, we present different kinds of built-in objects which differ in the additional fields of their respective object descriptors.

## 4.2.2 Types and Classes

The type of an object is the type object stored in the object descriptor's `cls` field. Type objects are central because every object has a type and the type of an object determines the operations it supports. Type objects refer to *type descriptors*. Type descriptors are object descriptors with six additional fields: `name`, `bases`, `mro`, `is_sealed`, `is_builtin`, and `layout`. The field `name` contains the *name* of the type as a primitive string, the field `bases` contains a possibly empty vector of type objects which are the *base types* of the type, the field `mro` contains a non-empty vector of type objects capturing the *Method Resolution Order* (MRO) of the type, the field `is_sealed` contains a boolean indicating whether the type can be subclassed, the field `is_builtin` contains a boolean indicating whether the type is a built-in, and the field `layout` contains a type object specifying the *layout* of the type. Before we discuss the role of those fields in more detail, here are two examples:

$$
\begin{array}{ll}
\mathrm{ref(object)} \mapsto \langle & \mathrm{ref(bool)} \mapsto \langle \\
\quad \mathrm{cls} : \mathrm{ref(type)}, & \quad \mathrm{cls} : \mathrm{ref(type)}, \\
\quad \mathrm{dict} : \mathrm{ref}(\ldots), & \quad \mathrm{dict} : \mathrm{ref}(\ldots), \\
\quad \mathrm{name} : \text{'object'}, & \quad \mathrm{name} : \text{'bool'}, \\
\quad \mathrm{bases} : [\,], & \quad \mathrm{bases} : [\,\mathrm{ref(int)}\,], \\
\quad \mathrm{mro} : [\,\mathrm{ref(object)}\,], & \quad \mathrm{mro} : [\,\mathrm{ref(bool)}, \mathrm{ref(int)}, \mathrm{ref(object)}\,], \\
\quad \mathrm{is\_sealed} : \mathrm{false}, & \quad \mathrm{is\_sealed} : \mathrm{true}, \\
\quad \mathrm{is\_builtin} : \mathrm{true}, & \quad \mathrm{is\_builtin} : \mathrm{true}, \\
\quad \mathrm{layout} : \mathrm{ref(object)} & \quad \mathrm{layout} : \mathrm{ref(int)} \\
\rangle & \rangle
\end{array}
$$

As the names suggests, $\mathrm{ref(object)}$ corresponds to the built-in type `object` and $\mathrm{ref(bool)}$ to the built-in type `bool`. One can not subclass $\mathrm{ref(bool)}$ and there are precisely two objects of this type:

$$
\begin{array}{l}
\mathrm{ref(True)} \mapsto \langle \mathrm{cls} : \mathrm{ref(bool)}, \mathrm{dict} : \mathrm{ref(None)}, \mathrm{value} : 1 \rangle \\
\mathrm{ref(False)} \mapsto \langle \mathrm{cls} : \mathrm{ref(bool)}, \mathrm{dict} : \mathrm{ref(None)}, \mathrm{value} : 0 \rangle
\end{array}
$$

### The Type Hierarchy

Types are organized hierarchically by a *subtype relationship* characterized by their base types. At the root of this type hierarchy is the type `object` which does not have any base types. Hence, the field `bases` of respective type descriptor is the empty vector $[\,]$. Formally, a vector $\vec{v} \in \mathbb{V}$ is a finite sequence of values. Every other type is a subtype of `object`.

Python supports multiple inheritance, i.e., a type may have multiple base types. A type's MRO is the order in which methods and *class attributes* are looked up in the type hierarchy. It is determined according to the C3 linearization algorithm proposed by Barrett et al. [7]. Class attributes of an object are instance attributes of the types in the MRO of its type. The type $\mathrm{ref(bool)}$ *inherits* form $\mathrm{ref(int)}$, hence, methods and class attributes are first looked up in $\mathrm{ref(bool)}$ then in $\mathrm{ref(int)}$ and finally in $\mathrm{ref(object)}$. As the MRO is a linearization of the inheritance hierarchy of a type, a type object is *a subtype of* another type object if and only if the latter type object appears in the MRO of the former type object's type descriptor. For instance, ref(`bool`) is a subtype of ref(`int`) because ref(`int`) appears in the MRO of ref(`bool`) or, more precisely, in the MRO of its type descriptor. For the same reason, ref(`bool`) is a subtype of ref(`object`) and itself. Note that the subtype relation is reflexive because the MRO always contains the type itself.

In addition to the subtype relationship there is an *instance relationship*. An object is *an instance of* a type object if and only if the type of the object is a subtype of the type object. For instance, ref(`True`) is an instance of ref(`int`) because its type, i. e., ref(`bool`), is a subtype of ref(`int`).

The *layout* of a type determines the fields of the object descriptors of its instances. For instance, the layout ref(`int`) of ref(`bool`) specifies that objects of the type ref(`bool`) have the same layout as integer objects, i. e., they have a field `value` containing a primitive integer. Types with incompatible layouts cannot be mixed within the inheritance hierarchy. For instance, trying to create a class which subclasses both ref(`int`) and ref(`str`), i. e., the built-in type for strings, will lead to a `TypeError` being thrown.

Recapitulate that types are also objects. Hence, they have instance attributes that can be set. However, due to how the built-ins are implemented within the reference implementation, setting attributes on built-in types is not allowed. The field `is_builtin` is used by our semantics to determine whether the instance attributes of a type can be set.

**Meta Classes**

Types are themselves objects whose descriptor's `cls` field contains a subtype of the `type` type, called their *meta class*. As a result, every type is an instance of `type`. The built-in type `type` is defined as follows:

$$
\begin{aligned}
\text{ref}(\texttt{type}) \mapsto \langle \\
\quad \texttt{cls} : \text{ref}(\texttt{type}), \\
\quad \texttt{dict} : \text{ref}(\dots), \\
\quad \texttt{name} : \texttt{'type'}, \\
\quad \texttt{bases} : [\,\text{ref}(\texttt{object})\,], \\
\quad \texttt{mro} : [\,\text{ref}(\texttt{type}), \text{ref}(\texttt{object})\,], \\
\quad \texttt{is\_sealed} : \texttt{false}, \\
\quad \texttt{is\_builtin} : \texttt{true}, \\
\quad \texttt{layout} : \text{ref}(\texttt{type}) \\
\rangle
\end{aligned}
$$

By sub-classing `type` the creation of classes can be customized. We describe how meta classes are used for the creation of classes when describing the semantics of class definitions in Section 4.3.

**Methods and Class Attributes**

Like with any other object-oriented language, the type of an object determines the operations it supports by providing *methods*. The methods a type provides are stored within its `dict` field. Calling a method on an object is an involved process roughly working as follows: First, the method is looked up using the normal mechanism for attribute lookups. For instance, calling `x.method()` invokes the mechanism for looking up the attribute `method` on the object `x`. This mechanism will first search for the attribute `method` within the instance attributes of `x`. If no such instance attribute is found, the mechanism for looking up class attributes is invoked. This mechanism traverses the MRO of the type of the object and stops as soon as it finds the attribute among the instance attributes of any of these types. If the respective instance attribute of the type is a descriptor (cf. Section 2.1.1) then the dunder method `__get__` on the descriptor is invoked and the result is returned. Otherwise, if the respective instance attribute of the type is no descriptor, the instance attribute is returned unmodified. Functions are descriptors and invoking their `__get__` method will return a *bound method*. A *bound method* is a callable object which internally stores the function and the object the attribute was accessed on [20, §3.2.]. For our example, looking up the attribute `method` on `x` returns a bound method which comprises a function implementing the method and the object `x`. After all this happened, the bound method is called. Calling a bound method calls the function implementing the method with the object stored within the bound method as first argument and passing through the arguments provided for the method call.

Notice how looking up a method involves looking up the method `__get__` on the function object implementing the method. To prevent an infinite recursion, the lookup process of dunder methods works differently and bypasses the normal mechanism [20, §3.3.10.].

The type descriptors we defined so far used ref(...) as a placeholder for some dictionary-like object containing the methods provided by the type. Usually, the field `dict` of an object descriptor contains an actual dictionary. For types, however, it contains a `mappingproxy` object. A `mappingproxy` object behaves like a dictionary just that it only accepts strings as keys. Formally, they are represented by means of object descriptors

$$\langle \texttt{cls} : \text{ref}(\texttt{mappingproxy}), \texttt{dict} : \text{ref}(\texttt{None}), \texttt{value} : \{\dots\} \rangle$$

where $\{\dots\} : \mathbb{V} \rightharpoonup \mathbb{V}$ is a finite-domain partial function from values to values such that $\text{dom}(\{\dots\})$ only contains primitive strings.

Like built-in functions, the methods of built-in types we support are put into function objects when the heap is initialized (cf. Section 4.1.2) which are then stored in the respective instance attributes of the type. The methods of built-in types are, like built-in functions, translated in primitive mode and, hence, have access to primitive operations.

There are two files, `rigorous/semantics/python/basis/source/builtins.py` and `rigorous/semantics/python/basis/source/runtime.py`, containing implementations of methods of built-in types using primitives [35]. Snippet 4.3 shows an excerpt of `builtins.py` implementing `__add__` for integer objects. Using the runtime function `lowlevel_isinstance` the function first checks whether the other object is an integer object. If this is the case, it creates a new integer object by adding the value of itself and the value of the other integer object using the primitive `number_add`. Otherwise, if the other object is not an integer object, it returns `NotImplemented`.

```
1  def __add__(self, other):
2      if lowlevel_isinstance(other, int):
3          return NEW_FROM_VALUE(
4              int,
5              number_add(VALUE_OF(self), VALUE_OF(other))
6          )
7      return NotImplemented
```

Snippet 4.3: Implementation of `__add__` for integer objects.

For a detailed exposition of the methods provided by built-in types, we refer to the two aforementioned files. Here, our focus is on the semantic primitives and not how the built-ins are implemented using them.

### 4.2.3 Built-In Types

Besides `int` and `bool` Python supports a large variety[4] of built-in types. Analogously to how ref(`bool`) is defined, our semantics defines a named reference and corresponding type descriptor for all built-in types we support. Let us now discuss some of them in more detail.

**Value Types**

Many built-in types are what we call *value types*. Objects of value types are references to *value descriptors*. A *value descriptor* is an object descriptor with an additional `value` field containing a primitive value. Examples for value types we already encountered are `int` and `bool`. Other value types are `float`, where the value is an IEEE 754 double-precision floating-point number [32], `str`, where the value is a Unicode string, i.e., a finite sequence of Unicode code points [58], `list` and `tuple`, where the value is a vector of objects, and `dict`, where the value is a vector of *entries*.

For instance, the floating-point number $3.1415$ is represented by

$$\langle \mathtt{cls} : \mathrm{ref}(\mathtt{float}), \mathtt{dict} : \mathrm{ref}(\mathtt{None}), \mathtt{value} : 3.1415 \rangle$$

and the string `"Hello World!"` is represented by:

$$\langle \mathtt{cls} : \mathrm{ref}(\mathtt{str}), \mathtt{dict} : \mathrm{ref}(\mathtt{None}), \mathtt{value} : \text{'Hello World!'} \rangle$$

Despite the fact that lists and tuples support different operations—lists are mutable while tuples are not—they are both represented using a vector of ob-jects as value. For example, a list containing two objects, ref(42) and ref(314) is represented by the object descriptor:

$$\langle \texttt{cls}:\text{ref}(\texttt{list}), \texttt{dict}:\text{ref}(\texttt{None}), \texttt{value}: [\,\text{ref}(42), \text{ref}(314)\,]\rangle$$

---

[4] https://docs.python.org/3.7/library/stdtypes.html (Accessed: 2020/12/14)

Dictionaries are Python's variant of hash tables. We represent them using a vector of *entries* as a value where an entry is a record with three fields, hash, key, and value, storing the hash of the key, the key itself, and the value associated with the key. Lookups, insertions and deletions happen by linear scanning over the entries. Of course, this representation of hash tables is highly inefficient for most practical purposes. However, our focus is on the semantics of dictionaries and not on their speed. The keys in a dictionary are ordered based on the insertion order and using a sequence of entries allows us to easily formalize this order.

**Singleton Objects**

The built-ins None, Ellipsis, and NotImplemented are all singleton objects. Singleton objects have in common that they are the only objects of their respective types and do not have any value associated with them. For instance, None is defined by the following object descriptor:

$$\text{ref}(\texttt{None}) \mapsto \langle \texttt{cls}:\text{ref}(\texttt{NoneType}), \texttt{dict}:\text{ref}(\texttt{None})\rangle$$

**Code Objects**

During the translation process (cf. Section 4.1), the term generated from a code block is placed into a *code object*. A code object is a reference to a *code descriptor*. A code descriptor is an object descriptor with the additional fields name, filename, cells, signature, doc, body, and is_generator. The field name contains the name of the code block (usually the name of the function or class), the field filename contains the name of the file the code originates from, the field cells specifies the *cells* the code introduces, the field signature contains the *signature* of the code, the field doc contains either ref(None) or a Python string object with the docstring [23] of the code, the field body contains the term the respective code block has been translated to, and the field is_generator contains a boolean indicating whether the code defines a generator (cf. Section 2.1.1).

Technically, code objects are defined as part of the PLR [20, §3.2.]. However, they are classified as internal object whose definition "may change with future versions of the [CPython] interpreter." Based on this quote from the PLR, we classify code objects as an implementation detail of CPython and represent them differently. Nevertheless, it would be quite easy to emulate CPython's code objects on-top of our formalization.

Every code object has a signature specifying which parameters it takes. A *parameter definition* is a record with two fields name and kind. The field name contains a primitive string obtained from the identifier of the parameter and the field `kind` contains one of four *parameter kinds*:

<div align="center">

'POSITIONAL_OR_KEYWORD'    'VARIABLE_POSITIONAL'

'KEYWORD_ONLY'    'VARIABLE_KEYWORD'

</div>

Our semantics does not support *positional-only* parameters. Positional-only parameters are mostly a relict of the standard library where certain built-in functions parse their arguments in a way that does not take into account the name of the parameter. Only recently, i.e., with Python 3.8, they have been properly added to the language itself [28]. However, recapitulate that we take Python 3.7 as the baseline for our semantics.

```python
1  def f(a, b, *c, d, **e):
2      print(a, b, c, d, e)
3
4
5  >>> f(1, 2, 3, 4, d=5, x=6, y=7)
6  1 2 (3, 4) 5 {'x': 6, 'y': 7}
```

Snippet 4.4: Example of various kinds of parameters.

Snippet 4.4 shows an example of a function definition with variadic positional and keyword parameters. The function requires two positional arguments, `a` and `b`, and accepts a variable number of additional positional arguments which are made available via the parameter `c`. As all additional positional arguments are collected into a tuple made available via the parameter `c`, the parameter `d` is required to be supplied as a keyword argument. Finally, all additional keyword arguments are collected into a dictionary made available vie tha parameter `d`. The signature of the code object obtained from that function definition would be the following vector:

$$
\begin{bmatrix}
\langle \texttt{name} : \text{`a'}, \texttt{kind} : \text{`POSITIONAL\_OR\_KEYWORD'} \rangle, \\
\langle \texttt{name} : \text{`b'}, \texttt{kind} : \text{`POSITIONAL\_OR\_KEYWORD'} \rangle, \\
\langle \texttt{name} : \text{`c'}, \texttt{kind} : \text{`VARIABLE\_POSITIONAL'} \rangle, \\
\langle \texttt{name} : \text{`d'}, \texttt{kind} : \text{`KEYWORD\_ONLY'} \rangle, \\
\langle \texttt{name} : \text{`e'}, \texttt{kind} : \text{`VARIABLE\_KEYWORD'} \rangle
\end{bmatrix}
$$

The field `cells` specifies the closure cells a piece of code introduces. Recapitulate Snippet 4.1 where within the inner function g the identifier y came from the function's closure. For this to work, the function f needs to create a *cell* for the variable y. A cell is essentially a Python object which stores another Python object. In case of Snippet 4.1, the code object created from the function definition f has the vector [ 'y' ] stored in its `cells` field indicating that a cell for the variable y should be constructed when executing the code of the function. Accesses to the variable y then go to the respective cell. For further details we refer to the PLR [20, §3.].

**Function Objects**

Code objects themselves just represent a piece of code but are not directly callable. Functions combine code objects with a global namespace and a closure in which the code can then be executed. In addition, they may also provide default values for some of the parameters. A *function object* is a reference to a *function descriptor*. A function descriptor is an object descriptor with five additional fields: `code`, `name`, `globals`, `cells`, and `defaults`. The field `code` contains the code object for the function, the field `name` contains the name of the function as a primitive string, the field `globals` contains a dictionary-like object representing the global namespace the function was defined in, the field `cells` contains the closure of the function, and the field `defaults` contains optional default values for some of the parameters.

The creation of code objects happens ahead-of-time when translating a program into a term. Likewise, the reference implementation CPython creates code objects when compiling a program to bytecode. Function objects are created when function definitions are executed and as part of the translation process for built-in functions and methods.

Take Snippet 4.1 as an example again. When this code executes, a function `f` is created. This function inherits its global namespace from the module it is defined in and has no closure because it is not enclosed in another function. When calling the function `f`, it creates another function g. This function inherits the global namespace from `f` and also the cells from `f`. In this case, there is a cell for the identifier `y`. Now, if the function `f` executes, it does so within the same global namespace as g and has the cells created around it available. When accessing an identifier, it uses the global namespace and the cells to access the identifiers `x` and `y`, respectively.

The name of the function is initially taken from the code object but can be changed independently afterwards. The cells and defaults are formally represented by *mappings*. In general, a mapping is a finite-domain partial function $m : \mathbb{V} \rightharpoonup \mathbb{V}$. We already encountered several instances of mappings, for instance, for `mappingproxy` objects. To store the cells and defaults mappings from primitive strings to objects are used.

In Section 4.3, we describe in more detail how we capture the semantics of function definitions and create function objects at runtime.

**Frame Objects**

Like code objects, *frame objects* are technically defined as part of the PLR but classified as internal [20, §3.2.]. We again take them to be implementation details and deviate from the definition within the PLR. Frames are exposed, for instance, as part of a *traceback*, i. e., a Python object representing a stack trace. Our semantics does not support such stack traces.

Another usage of frame objects are generators (cf. Section 2.1.1). While generator objects seem not to be specified as part of the PLR, the documentation of the `inspect` module[5] describes them. Generators allow pausing the execution of code by `yield` expressions. As part of this process, a frame is stored in the generator object so that it can be resumed later.

Frame objects are references to *frame descriptors*. Frame descriptors are object descriptors with two additional fields, `locals` and `body`. The field `locals` contains a mapping from primitive strings to values (mostly objects) and represents the local namespace of the frame. The field `body` contains a term representing the current state of execution.

The global namespace and cells, a piece of code executes in, are made available via two local identifiers, '`__globals__`' and '`__cells__`'.

In the upcoming section, we describe how frames are constructed and used by the semantics for generators and calling functions.

## 4.3 Intraprocedural Semantics

In this section, we present an inference rule system for an extended variant of the Python fragment we identified in Section 2.1.2. The inference rule system constitutes a structural operational semantics for Python and is formalized in our meta-theoretic framework (cf. Chapter 3) enabling the automatic derivation of an interpreter. The inference rules presented here have been exported to LaTeX from their definitions leveraging our tool.

Structural operational semantics operate on syntactic constructs. Within our framework, these constructs are closed terms. Like many other languages, Python has two central kinds of syntactic constructs, *expressions* and *statements*. While expressions evaluate to values, except for cases where an exception is thrown, statements do not evaluate to values. In this section, our focus is on the intraprocedural layer of the semantics (cf. Figure 4.2), i.e., on the semantics of expressions and statements.

For each kind of expression and statement we describe how it is translated into a closed term and present the inference rules operating on these terms. Furthermore, we describe the primitives added to the fragment defined in Section 2.1.2 and their semantics.

In total, the inference rule system comprises 98 inference rules.

### 4.3.1 The Semantics of Expressions

Expressions are syntactic constructs that evaluate to primitive values except for cases where an exception is thrown. We extend Python with a set of prim-itive expressions operating on primitive values. By translating built-in functions and methods in primitive mode (cf. Section 4.1), they have access to these primitive expressions, which allows us to program them using the superset of Python we are about to define.

---

[5] https://docs.python.org/3.7/library/inspect.html (Accessed: 2020/12/14)

**Primitive Building Blocks**

A recurring problem is that we need to evaluate multiple expressions one after another and then, after all the expressions have been evaluated, apply some kind of primitive operation to the results. Introducing dedicated rules for each of those primitives would be rather cumbersome and unnecessarily blow up the amount of rules. Instead, we introduce three primitive building blocks which handle primitive operations and runtime calls in a succinct way. First of all, we allow two expressions to be joined together with $::$ leading to terms of the form $(e_1 :: e_2)$ where the right expression is required to evaluate to a vector. The idea is that by nesting such expressions one can construct arbitrary vectors of values with the following rules:

$$\frac{e \xrightarrow{\alpha} u}{(e :: xs) \xrightarrow{\alpha} (u :: xs)} \text{ HEAD-EVAL} \qquad \frac{e \xrightarrow{\alpha} u \quad \texttt{is\_primitive}(x)}{(x :: e) \xrightarrow{\alpha} (x :: u)} \text{ TAIL-EVAL}$$

$$\frac{\texttt{is\_primitive}(v_l) \quad \texttt{is\_primitive}(v_r)}{(v_l :: v_r) \xrightarrow{\tau} \texttt{push\_left}(v_r, v_l)} \text{ PREPEND}$$

The rule HEAD-EVAL evaluates the *head*, i.e., the left subexpression of the expression. If the head has been evaluated to a primitive value, the rule TAIL-EVAL evaluates the *tail*, i.e., the right subexpression of the expression. As the tail is required to evaluate to a vector, the left part can be pushed to the front of that vector creating a new vector. This is done by the rule PREPEND. By chaining such expressions we can thereby generate arbitrary vectors of primitive values from expressions.

To *apply* primitive operations, we take terms of the form $(\texttt{apply } n \ e)$ to be expressions where n is a primitive string. The idea is to give each primitive operation a name and then apply it to a respective expression by first evaluating the expression and then applying the operation:

$$\frac{e \xrightarrow{\alpha} u}{(\texttt{apply n } e) \xrightarrow{\alpha} (\texttt{apply n } u)} \text{ APPLY-EVAL} \qquad \frac{\texttt{is\_primitive}(v)}{(\texttt{apply n } v) \xrightarrow{\tau} \texttt{apply}(n, \ v)} \text{ APPLY-EXEC}$$

The rule APPLY-EVAL evaluates the provided expression and APPLY-EXEC applies the primitive operation by using the apply operator which takes the name of the operation and the value the expression evaluated to. By using the earlier introduced way to generate arbitrary vectors from expressions, multiple arguments can be provided to a primitive operation.

In contrast to other primitive operators, apply is defined on all values but may evaluate to a closed term throwing an exception if the primitive is not defined for the provided values. This has the advantage that errors in the runtime, which uses these primitives, are reported.

Recapitulate Snippet 4.3. Here the primitive `number_add` has been used to add two integers. During translation in primitive mode, whenever a call expression $e_f(\vec{e})$ is encountered where the callee $e_f$ is an identifier that is a known named primitive, like `number_add`, the arguments of that call expression are translated and joined with `::` to obtain an expression evaluating the arguments from left to right and producing a vector. Finally, an expression (`apply` n $e$) is produced where n is the name of the primitive and $e$ is the expression obtained by joining the arguments. In case of Snippet 4.3, the expression `number_add(VALUE_OF(self), VALUE_OF(other))` is translated to

$$(\texttt{apply } `\texttt{number\_add}' \; (\mathit{self} :: (\mathit{other} :: [\;])))$$

where $\mathit{self}$ and $\mathit{other}$ are the expressions `VALUE_OF(self)` and `VALUE_OF(other)` translate to, respectively. `VALUE_OF(self)` and `VALUE_OF(other)` are *translation macros*. Translation macros are mere syntactic sugar transforming the provided expressions into another expression. In this case, they produce an expression loading the object descriptors of their arguments from memory and then extracting the field `value`.

Appendix B.1 contains a list of all named primitives used by the runtime with their signature and a brief documentation for each. Appendix B.3 contains a list of all translation macros with a brief documentation.

Note that `apply` is opaque and an atomic operation as far as the semantics is concerned. However, often it is necessary to apply multiple such operations to carry out a certain *logical operation*. For instance, accessing an item on an object involves looking up the method `__getitem__` on the object, raising a `TypeError` if that method does not exist, and calling the method otherwise (cf. Snippet 4.2). Such logical operations that involve multiple primitive operations are carried out by *runtime functions*. Unlike built-in functions, runtime functions are not Python function objects. Like ordinary functions, runtime functions encapsulate a piece of Python code which is, however, invoked via a dedicated mechanism similar to `apply`:

$$\frac{e \xrightarrow{\alpha} u}{(\texttt{runtime } \texttt{n } e) \xrightarrow{\alpha} (\texttt{runtime } \texttt{n } u)} \text{\textsc{runtime-eval}}$$

$$\frac{\texttt{is\_primitive}(v)}{(\texttt{runtime } \texttt{n } v) \xrightarrow{(\texttt{CALL make\_runtime\_frame}(\texttt{n}, v))} (\texttt{entry } .)} \text{\textsc{runtime-exec}}$$

While the mechanism is quite similar to the mechanism for `apply` presented above, instead of carrying out the operation by applying a primitive operator to the obtained values, the operation is carried out by constructing a frame and then calling into this frame. We will see how this call mechanism works from the side of the call stack in Section 4.4.2. The difference between calls into the runtime and Python function calls is how the stack frame is constructed. For calls to Python function objects, the frame is constructed by a runtime function taking a function object.

The primitive operator `make_runtime_frame` takes the name of a runtime function and a vector of arguments for that function and returns a frame de-scriptor. See Appendix B.2 for an overview of the available runtime functions with their signature and a brief documentation for each.

The expression (`entry .`) is an *entry point* to return to later. Depending on whether the operation succeeds and returns a result or throws an error, it is handled by the rule ENTRY-RESULT or ENTRY-ERROR:

$$\frac{}{(\texttt{entry .}) \xrightarrow{\text{(RESULT } v)} v} \text{ENTRY-RESULT} \qquad \frac{}{(\texttt{entry .}) \xrightarrow{\text{(ERROR } v)} (\texttt{raise } v)} \text{ENTRY-ERROR}$$

Note how the rules RUNTIME-EXEC, ENTRY-RESULT, and ENTRY-ERROR use actions (cf. Section 2.2) for the encapsulation of interprocedural control flow. Instead of modifying the call stack directly, the rule RUNTIME-EXEC merely creates a frame and communicates the intent that this frame should be executed next by means of a `CALL` action. The resulting program state (`entry .`) then allows returning from the call either by providing a result with a `RESULT` action or throwing an error with an `ERROR` action.

**Local Variables**

Local variables are identifiers whose access mechanism is `LOCAL` according to the scope analysis (cf. Section 4.1.1). The values bound to local variables are stored within the frame a piece of code executes in and can be retrieved via actions interacting with the frame layer (cf. Figure 4.2).

Identifiers occur in three different contexts: `LOAD`, `STORE`, or `DEL`. An identifier occurs in the `STORE` context if it is the left-hand side of an assignment and in the `DEL` context if it appears within a `del` statement. All other occurrences of identifiers are in the `LOAD` context.

A local variable within a `LOAD` context translates to (`load_local` $id$ $\bot$) where $id$ is the identifier as a primitive string. The rule LOAD-LOCAL handles these expressions via a `LOAD_LOCAL` action:

$$\frac{u = \texttt{ite}(v = \bot, (\texttt{runtime 'unbound\_local\_error'} \ (id :: [\,])), v)}{(\texttt{load\_local } id \ d) \xrightarrow{\text{(LOAD\_LOCAL } id \ v \ d)} u} \text{LOAD-LOCAL}$$

The `LOAD_LOCAL` action is handled by the frame layer (cf. Section 4.4.1). The primitive `load_local` expression allows defining a default value in case no value has been bound to the local variable. For translating local variables, this value is always set to $\bot$, i. e., a dedicated *undefined* value. The possibility to provide another default value is used by other parts of the semantics we discuss later. In case the value provided by the frame layer is $\bot$ the expression evaluates to a runtime call to `unbound_local_error`. This runtime call will raise an UnboundLocalError as required in case a local variable is not bound to a value [20, §4.2.2.]. Otherwise, if the value provided by the frame layer is not $\bot$, the expression evaluates to the provided value.

Technically, deleting variables [20, §7.5] and assigning to them [20, §7.2] is done in statements. Nevertheless, we allow for deletion and assignment ex-pressions which evaluate to the previously stored and the new value, respectively. This anticipates that starting with Python 3.8 there will be assignment expressions according to PEP 572 [3].

Assignments to local variables are translated to primitive expressions of the form $(\texttt{store\_local}\ id\ e)$ where $id$ is again the identifier as a primitive string and $e$ is the term the assigned expression translates to. The semantics are captured by two rules STORE-LOCAL-EVAL evaluating the expression and STORE-LOCAL-EXEC executing the assignment:

$$\frac{e \xrightarrow{a} u}{(\texttt{store\_local}\ id\ e) \xrightarrow{a} (\texttt{store\_local}\ id\ u)} \text{ STORE-LOCAL-EVAL}$$

$$\frac{\texttt{is\_primitive}(v)}{(\texttt{store\_local}\ id\ v) \xrightarrow{(\texttt{STORE\_LOCAL}\ id\ v)} v} \text{ STORE-LOCAL-EXEC}$$

Analogously to LOAD-LOCAL, a $\texttt{STORE\_LOCAL}$ action is used to instruct the frame layer to store the respective value.

Deleting a local variable returns the previously stored value or $\bot$ if no such value exists. Analogously to LOAD-LOCAL, an UnboundLocalError is produced if no value is bound to the respective identifier:

$$\frac{u = \texttt{ite}(v = \bot, (\texttt{runtime}\ \text{`unbound\_local\_error'}\ (id :: [\,])), v)}{(\texttt{delete\_local}\ id) \xrightarrow{(\texttt{DELETE\_LOCAL}\ id\ v)} u} \text{ DELETE-LOCAL}$$

**Non-Local Variables**

Non-local variables are identifiers whose access mechanism is not $\texttt{LOCAL}$ according to the scope analysis (cf. Section 4.1.1). All access mechanisms except $\texttt{LOCAL}$ are handled via calls to runtime functions. For instance, recapitulate that the local variable $\texttt{\_\_globals\_\_}$ of a frame is bound to the global namespace which is a dictionary-like object. Hence, loading a global variable is translated to the following runtime call

$$(\texttt{runtime}\ \text{`load\_global'}\ ((\texttt{load\_local}\ \text{`\_\_globals\_\_'}\ \bot) :: (id :: [\,])))$$

where $id$ is substituted with a string object representing the name of the global variable. Snippet 4.5 shows how this runtime function is implemented using primitives. Being a dictionary-like object, the function first tries to lookup the identifier in the namespace. If the identifier cannot be found within the namespace, a KeyError is thrown by the dictionary-like object. If this is the case, the function goes on to check whether there is a built-in with the respective name. If there is a built-in, the built-in is returned, otherwise, a NameError is thrown as required when accessing global variables that do not exist in the namespace and are not built-ins [20, §4.2.2.].

The access mechanisms CELL, CLASS_GLOBAL, and CLASS_CELL are handled analogously by a respective runtime function:

```
for CELL:
    (runtime 'load_cell' ((load_local '__cells__' ) :: (id :: [ ])))

for CLASS_GLOBAL:
    (runtime 'load_class_global' ((load_local '__dict__' ) :: ((load_local '__globals__' ) :: (id :: [ ]))))

for CLASS_CELL:
    (runtime 'load_class_cell' ((load_local '__dict__' ) :: ((load_local '__cells__' ) :: (id :: [ ]))))
```

The local variable `__dict__` contains the namespace of the class during the execution of a class definition. The runtime functions for `CLASS_GLOBAL` and `CLASS_CELL` first try to look up the variable within the namespace of the class and then revert to the global namespace and the cells, respectively.

Assignments to non-local variables are desugared as follows

```
for GLOBAL:  (runtime 'store_global' ((load_local '__globals__' ) :: (id :: (e_v :: [ ]))))
  for CELL:  (runtime 'store_cell' ((load_local '__cells__' ) :: (id :: (e_v :: [ ]))))
 for CLASS:  (runtime 'store_class' ((load_local '__dict__' ) :: (id :: (e_v :: [ ]))))
```

where $id$ is replaced with the name of the variable and $e_v$ is replaced with the term the assigned expression translates to. For assignments it makes no difference whether the access mechanism is `CLASS_GLOBAL` or `CLASS_CELL` because the value is always stored within the namespace of the class for both mechanisms. If the identifier is declared global or non-local within the body of the class, the access mechanism will be `GLOBAL` or `CELL`.

```
1  def load_global(namespace, identifier):
2      try:
3          return namespace[identifier]
4      except KeyError:
5          key = VALUE_OF(identifier)
6          builtins = LOAD(BUILTINS)
7          if mapping_contains(builtins, key):
8              return mapping_get(builtins, key)
9          else:
10             raise NameError()
```

Snippet 4.5: Runtime function for loading global variables.

### Constants and Literals

The constants `True`, `False`, `None`, and `Ellipsis` are replaced during the translation process with ref(True), ref(False), ref(None), and ref(Ellipsis), respectively. Integer, string, and float literals are translated by creating a new object on the initial heap (cf. Section 4.1).

### Sequence Literals

There are two kinds of sequence literals, one for lists and one for tuples (cf. Figure 2.2). We translate the expressions for the elements of these literals and then join them with :: to form an expression that evaluates to a vector of values.  The tuple and list objects are constructed by passing this vector of values to a respective runtime function

$$\text{for lists:}\quad (\texttt{runtime 'make\_list'} \ (\mathit{els} :: [\,]))$$
$$\text{for tuples:}\quad (\texttt{runtime 'make\_tuple'} \ (\mathit{els} :: [\,]))$$

where $\mathit{els}$ is substituted with the expression evaluating to the vector of values representing the elements of the sequence literal.

### Dictionay Literals

Dictionary literals are handled analogously to sequence literals.  With the help of $::$ and the inference rules HEAD-EVAL, TAIL-EVAL, and PREPEND a vector of key-value pairs is constructed and passed to a runtime function:

$$(\texttt{runtime 'make\_dict'} \ (\mathit{its} :: [\,]))$$

### Boolean Operators

Almost everywhere where a boolean is expected, one can use an arbitrary Python object which mimics a boolean by virtue of its `__bool__` method.  To this end, we add expressions of the form $(\texttt{bool } e)$ where $e$ is a term obtained by translating another expression.  Now, $(\texttt{bool } e)$ evaluates $e$ and then tries to convert it into a primitive boolean:

$$\frac{e \xrightarrow{\alpha} u}{(\texttt{bool } e) \xrightarrow{\alpha} (\texttt{bool } u)} \ \text{\small BOOL-EVAL}$$

$$\frac{\texttt{is\_reference}(v) \qquad v \notin \{\texttt{ref(None)}, \texttt{ref(False)}, \texttt{ref(True)}\}}{(\texttt{bool } v) \xrightarrow{\tau} (\texttt{bool } (\texttt{runtime 'convert\_bool'} \ (v :: [\,])))} \ \text{\small BOOL-EXEC}$$

$$\frac{b \in \{\texttt{true}, \texttt{ref(True)}\}}{(\texttt{bool } b) \xrightarrow{\tau} \texttt{true}} \ \text{\small BOOL-TRUE} \qquad \frac{b \in \{\texttt{false}, \texttt{ref(None)}, \texttt{ref(False)}\}}{(\texttt{bool } b) \xrightarrow{\tau} \texttt{false}} \ \text{\small BOOL-FALSE}$$

The runtime function `convert_bool` takes an arbitrary Python object and converts it into a boolean as specified by the PLR [20, §3.3.1.].

The basic build block for Python's boolean operators are *ternary expressions*:

$$\frac{c \xrightarrow{\alpha} u}{(c \ ? \ e_c : e_a) \xrightarrow{\alpha} (u \ ? \ e_c : e_a)} \ \text{\small TERNARY-EVAL}$$

$$\frac{}{(\texttt{true} \ ? \ e_c : e_a) \xrightarrow{\tau} e_c} \ \text{\small TERNARY-TRUE} \qquad \frac{}{(\texttt{false} \ ? \ e_c : e_a) \xrightarrow{\tau} e_a} \ \text{\small TERNARY-FALSE}$$

With these rules in place, a unary `not` is syntactic sugar for:

$$((\texttt{bool } e) \ ? \ \texttt{ref(False)} : \texttt{ref(True)})$$

The semantics of the boolean connectives `and` and `or` require *short circuiting*. The semantics of these operators are defined as follows: If the left operand of `and` is false, then it is returned. Otherwise, if the left operand of `and` is true, then the right operand is returned. Analogously, if the left operand of `or` is true, then it is returned. Otherwise, if the left operand of `or` is false, then the right operand is returned. Crucially, the left operand is evaluated only once, then its boolean value is checked and based on the boolean value a decision is taken. Using ternary expressions, these semantics of both boolean connectives are enshrined in the following inference rules:

$$\frac{e_l \xrightarrow{\alpha} u \quad \circ \in \{\texttt{and, or}\}}{(e_l \circ e_r) \xrightarrow{\alpha} (u \circ e_r)} \text{ BOOL-LEFT-EVAL}$$

$$\frac{\texttt{is\_primitive}(v_l)}{(v_l \texttt{ or } e_r) \xrightarrow{\tau} ((\texttt{bool } v_l) \texttt{ ? } v_l : e_r)} \text{ BOOL-OR-EXEC}$$

$$\frac{\texttt{is\_primitive}(v_l)}{(v_l \texttt{ and } e_r) \xrightarrow{\tau} ((\texttt{bool } v_l) \texttt{ ? } e_r : v_l)} \text{ BOOL-AND-EXEC}$$

**Unary and Binary Operators**

Unary operators—except `not` which is a boolean operator and handled as described above—are translated into runtime calls to `unary_operator`. The runtime function `unary_operator` takes two arguments, the operand and a primitive string with the name of the dunder method corresponding to the unary operator to be applied. The function checks whether the operand has the required method corresponding to the operator. If it does not have the required method, a `TypeError` is raised indicating that the operator is not supported on the operand. Otherwise, if it has the required method, the method is called and the result is returned. For instance, the unary minus expression `-e` is translated to the following runtime call

$$(\texttt{runtime 'unary\_operator'} (e :: (\texttt{'\_\_neg\_\_'} :: [\,])))$$

where $e$ is substituted with the term the expression $e$ translates to.

Analogously to unary operators, binary operators are translated into runtime calls to `binary_operator`. The runtime function `binary_operator` takes four arguments, the left and right operands, and two primitive strings with the names of the two dunder methods corresponding to the binary operator to be applied. Take the binary operator for substraction as an example. There are two dunder methods corresponding to this operator, `__sub__` and `__rsub__`. For this operator, the runtime function `binary_operator` first looks up the method `__sub__` on the left operand. If the method exists it is called with the left and right operand as arguments. In case the method does not return `NotImplemented`, its result is returned. Otherwise, if the method `__sub__` does not exist on the left operand or returns `NotImplemented`, the method `__rsub__` is called on the right operand with the right and left operand as arguments. In case this method also does not exist or returns `NotImplemented`, a `TypeError` is thrown indicating that the operands do not support the operator. Otherwise, if the method exists and does not return `NotImplemented` its result is returned. The process for the other binary operators is analogous using different method names [20, §3.3.8.].

An expression $e_l$ - $e_r$ is then translated to the following runtime call

$$(\texttt{runtime 'binary\_operator'} \ (e_l :: (e_r :: (\texttt{'\_\_sub\_\_'} :: (\texttt{'\_\_rsub\_\_'} :: [\ ])))))$$

where $e_l$ and $e_r$ are substituted with the terms the expressions $e_l$ and $e_r$ translate to, respectively.

**Comparison Operators**

Comparison operators in Python can be *chained*. For instance, the expression `3 < x < 6` requires x to be greater than $3$ but less than $6$. During translation, we wrap each chain of comparison operators within `cmp` in order to be able to distinguish `3 < x < 6` from `3 < (x < 6)`, both of which have a different semantics. The latter expression requires whatever result `x < 6` may return to be greater than $3$ and a binary comparison such as `x < 6` may return a non-boolean result [20, §3.3.1.]. Here are two examples:

$$e_1 < e_2 < e_3 \quad \rightsquigarrow \quad (\texttt{cmp} \ (e_1 < (e_2 < e_3))$$

$$e_1 < (e_2 < e_3) \quad \rightsquigarrow \quad (\texttt{cmp} \ (e_1 < (\texttt{cmp} \ (e_2 < e_3)))$$

The arrow $\rightsquigarrow$ should be read as "translates to". For the process of translation, each meta variable on the right side of $\rightsquigarrow$ is substituted with the term the respective expression on the left side of $\rightsquigarrow$ translates to.

Comparison expressions always evaluate from left to right. The rule CMP-LEFT-EVAL evaluates the left-most operand of a comparison expression:

$$\frac{e_l \xrightarrow{\alpha} u}{(\texttt{cmp} \ (e_l \bullet xs)) \xrightarrow{\alpha} (\texttt{cmp} \ (u \bullet xs))} \ \text{CMP-LEFT-EVAL}$$

As comparisons are wrapped within `cmp`, the following two rules are only applicable if the comparison is binary. If the comparison is binary, the right operand is evaluated after the left operand has been evaluated and, after both operands have been evaluated, the actual comparison is performed:

$$\frac{e_r \xrightarrow{\alpha} u \quad \texttt{is\_primitive}(v_l)}{(\texttt{cmp} \ (v_l \bullet e_r)) \xrightarrow{\alpha} (\texttt{cmp} \ (v_l \bullet u))} \ \text{CMP-RIGHT-EVAL}1$$

$$\frac{\texttt{is\_primitive}(v_l) \quad \texttt{is\_primitive}(v_r)}{(\texttt{cmp} \ (v_l \bullet v_r)) \xrightarrow{\tau} \texttt{compare}(\bullet, v_l, v_r)} \ \text{CMP-EXEC}1$$

The operator `compare` constructs a closed term which calls the appropriate runtime function for the comparison operator if both operands are Python objects and otherwise performs a primitive comparison returning a Python boolean. Neither of the rules requires us to check whether $\bullet$ is a comparison operator because this is ensured syntactically by requiring that $\bullet$ occurs inside of a `cmp` expression.

While the PLR explicitly states that "x < y calls x.__lt__(y)" [20, §3.3.1.] this is actually false. There are cases where x < y does not call x.__lt__(y) and there are other cases where x.__lt__(y) is called but more than that happens. The precise semantics of the comparison operators are complex and we had to take a look into the source code of the reference implementation to figure out what the semantics actually is. The result of this investigation is condensed in the runtime function rich_cmp [35].

If the comparison expression is non-binary, the next expression in the chain is evaluated after the left-most operand has been evaluated:

$$\frac{e_r \xrightarrow{o} u \quad \texttt{is\_primitive}(v_l) \quad \texttt{is\_comparison\_operator}(\bullet_i)}{(\texttt{cmp } (v_l \bullet_o (e_r \bullet_i xs))) \xrightarrow{o} (\texttt{cmp } (v_l \bullet_o (u \bullet_i xs)))} \text{ CMP-RIGHT-EVAL2}$$

After both operands have been evaluated, a ternary operator is constructed which compares the two left-most operands and based on the result evaluates to a comparison expression for the remainder or False:

$$\frac{\texttt{is\_primitive}(v_l) \quad \texttt{is\_primitive}(v_r) \quad \texttt{is\_comparison\_operator}(\bullet_i)}{(\texttt{cmp } (v_l \bullet_o (v_r \bullet_i xs))) \xrightarrow{\tau} ((\texttt{bool compare}(\bullet_o, v_l, v_r)) \ ? \ (\texttt{cmp } (v_r \bullet_i xs)) : \texttt{ref(False)})} \text{ CMP-EXEC2}$$

For the rules CMP-EXEC2 and CMP-RIGHT-EVAL2, we have to check whether the inner operator is a comparison operator because the right part could by any expression, in particular, some other binary operator expression.

**Call Expression**

*Call expressions* [20, §6.3.4] are used to call a *callable object* with a possibly empty sequence of *positional arguments* and a possibly empty set of *keyword arguments*. As arbitrary objects might be callable, if they implement the dunder method __call__, we formalize the semantics of call expressions by runtime calls. The runtime function call takes three arguments, a primitive vector of positional arguments, a primitive mapping of keyword arguments, and the call target. A call expression is then translated into a term

$$(\texttt{runtime 'call' } (\vec{a}_p :: (\vec{a}_k :: (f :: [\,]))))$$

where $\vec{a}_p$ is substituted with a term evaluating to a primitive vector corresponding to the positional arguments, $\vec{a}_k$ is substituted with a term evaluating to a primitive mapping corresponding to the keyword arguments, and $f$ is substituted with a term evaluating to the object to be called.

The vector of positional arguments is mostly constructed by joining the expressions with :: leveraging the rules HEAD-EVAL, TAIL-EVAL, and PREPEND. An exception are *starred arguments* [20, §6.3.4.] of the form *e. Starred arguments allow passing a variable number of arguments obtained from an iterable to a callable. To this end, $e$ needs to be evaluated. The resulting iterable is then turned into a vector which is concatenated with the vector for the other arguments again returning a vector:

$$(\texttt{apply 'sequence\_concat'}\ ((\texttt{runtime 'unpack\_iterable'}\ (e :: [\,])) :: (\vec{a}_p :: [\,])))$$

Keyword arguments are added with `mapping_set`:

$$(\texttt{apply 'mapping\_set'}\ (\vec{a}_k :: (id :: (e :: [\,]))))$$

Here, $\vec{a}_k$ is the empty mapping initially to which the keyword arguments are then added with `mapping_set` inductively.

Analogously to starred positional arguments, there are double starred arguments of the form $**e$. Here, $e$ needs to evaluate to a dictionary-like object, i. e., a key-value mapping, with string keys. Double starred arguments allow passing a variable number of keyword arguments obtained from a dictionary-like object to a callable. To this end, $e$ needs to be evaluated. The resulting dictionary-like object is then turned into a primitive mapping from primitive strings to objects using the runtime function `unpack_str_mapping` with which the primitive mapping of keyword arguments is extended:

$$(\texttt{apply 'mapping\_update'}\ (\vec{a}_k :: ((\texttt{runtime 'unpack\_str\_mapping'}\ (e :: [\,])) :: [\,])))$$

Internally, the runtime function `call` checks whether the object to be called is a function or some other object. In case the object is not a function, its `__call__` dunder method is invoked or a `TypeError` exception is thrown in case the object does not have a `__call__` method indicating that the object is not callable. In case the object is a function, the provided positional and keyword arguments are bound to variables based on the signature of the code of the function and the provided defaults for arguments (cf. Section 4.2.3). In addition, the cells introduced by the code object are created and initialized. Finally, a frame is constructed and called using `call`:

$$\frac{e \xrightarrow{\alpha} u}{(\texttt{call}\ e) \xrightarrow{\alpha} (\texttt{call}\ u)}\ \textsc{call-eval} \qquad \frac{\texttt{is\_primitive}(v)}{(\texttt{call}\ v) \xrightarrow{(\textsc{call}\ v)} (\texttt{entry .})}\ \textsc{call-exec}$$

The rule CALL-EVAL evaluates an expression required to evaluate to a frame descriptor. After the expression has evaluated to a frame descriptor, the frame descriptor is called by the rule CALL-EXEC analogously to how runtime functions are executed by RUNTIME-EXEC. The resulting entry point is handled by the rules ENTRY-RESULT and ENTRY-ERROR we already encountered.

In case the code object associated with the function is a generator code object, the constructed frame is not called with `call`. Instead, a generator object is constructed from the frame and returned to the caller.

**Generators and Yield**

Generators are one of the more unique features of Python (cf. Section 2.1). At the heart of generators is the `yield` expression. When encountering an expression of the form `yield` $e$, the execution of the current execution frame is paused after the expression $e$ has been evaluated. On pausing the execution and returning to the caller, the value, $e$ has evaluated to, is returned.

For handling `yield` expressions, we introduce two inference rules:

$$\frac{e \xrightarrow{\alpha} u}{(\texttt{yield } e) \xrightarrow{\alpha} (\texttt{yield } u)} \text{ YIELD-EVAL} \qquad \frac{\texttt{is\_primitive}(v)}{(\texttt{yield } v) \xrightarrow{\langle \text{YIELD } v \rangle} (\texttt{entry .})} \text{ YIELD-EXEC}$$

The rule YIELD-EVAL evaluates the expression and YIELD-EXEC instructs the stack layer (cf. Figure 4.2) to pass the control back to the caller and return the value. On the side of the generator, YIELD-EXEC leaves an entry point whose semantics is already captured by ENTRY-RESULT and ENTRY-ERROR. Hence, sending values to a generator or throwing an exception inside a generator (cf. Figure 2.1) can be achieved by `RESULT` and `ERROR` actions, respectively. When pausing a generator, its frame descriptor containing the current state of execution is stored within the respective generator object.

To send back a value to a generator, the frame descriptor is extracted from the generator object and a term of the form $(\texttt{send\_value } \mathfrak{F} \, v)$ is constructed by a primitive operation where $\mathfrak{F}$ is the frame descriptor to resume the execution from and $v$ is the value to send to the generator. The resulting `send_value` term is handled by the following inference rule:

$$\frac{}{(\texttt{send\_value } \widetilde{\mathfrak{F}} \, v) \xrightarrow{\langle \text{SEND\_VALUE } \widetilde{\mathfrak{F}} \, v \rangle} (\texttt{send .})} \text{ SEND-VALUE}$$

The rule SEND-VALUE instructs the stack layer (cf. Figure 4.2) to resume the execution of a generator from the given frame descriptor and send the provided value into the generator using a `RESULT` action.

Exceptions are thrown into a generator by an analogous mechanism:

$$\frac{}{(\texttt{send\_throw } \mathfrak{F} \, v) \xrightarrow{\langle \text{SEND\_THROW } \mathfrak{F} \, v \rangle} (\texttt{send .})} \text{ SEND-THROW}$$

Instead of sending a value with a `RESULT` action, a `SEND_THROW` action will instruct the stack layer to send an exception with an `ERROR` action. We discuss the complementary rules handling those actions in Section 4.4.2 when describing the stack layer in detail.

Both rules SEND-VALUE and SEND-THROW leave a *send handle* (`send .`). In case the generator threw an exception, send handles are handled equivalently to entry points by raising the exception in the caller's frame:

$$\frac{}{(\texttt{send .}) \xrightarrow{\langle \text{ERROR } v \rangle} (\texttt{raise } v)} \text{ SEND-RES-ERROR}$$

In case the generator returned a result and is exhausted, a `RESULT` action is generated by the stack layer. As a consequence, the send handle evaluates to a record with two fields, `frame` and `value`, where the field `frame` contains `ref(None)` indicating that there is no frame to resume to and the field `value` contains the value returned by the generator:

$$\frac{}{(\texttt{send .}) \xrightarrow{\langle \text{RESULT } v \rangle} \langle \texttt{'frame'} : \texttt{ref(None)}, \texttt{'value'} : v \rangle} \text{ SEND-RES-RESULT}$$

In case the generator yielded a value, a `VALUE` action is generated by the stack layer which contains the yielded value and a frame descriptor for resuming the execution of the generator. In this case, the send handle evaluates to
a record with the same fields as in SEND-RES-RESULT, however, with the field
`frame` containing the frame descriptor to be resumed:

$$\frac{}{(\texttt{send } .) \xrightarrow{(\text{VALUE } \mathfrak{F} \, v)} \langle\texttt{'frame'}: \mathfrak{F}, \texttt{'value'}: v \rangle} \text{ \small SEND-RES-VALUE}$$

The semantic primitives for generators presented here are used by the imple-
mentation of generator objects in `runtime.py`. For further details, how the
primitives are used, we refer to our implementation [35].

**Attributes and Items**

*Attribute expressions* [20, §6.3.1] and *subscript expressions* [20, §6.3.2] are
translated into runtime calls as follows:

$$e.\texttt{<ATTRIBUTE>} \quad \leadsto \quad (\texttt{runtime 'get\_attribute'} \; (e :: (attr :: [\,])))$$

$$e[e_k] \quad \leadsto \quad (\texttt{runtime 'get\_item'} \; (e :: (e_i :: [\,])))$$

Here, *attr* is the name of the attribute as a primitive string. The respective
runtime functions implement the mechanism for attribute and item lookups,
respectively. In particular, as described in Section 4.2.2, the mechanism for
looking up attributes is quite involved.

**Printing**

We promised earlier to address the interaction of Python programs with an
environment. To this end, we introduce print expressions of the form $(\texttt{print } e)$
that print primitive values. These primitive print expressions are used by our
implementation of the built-in function `print`.

The semantics of `print` expressions are captured by:

$$\frac{e \xrightarrow{\alpha} u}{(\texttt{print } e) \xrightarrow{\alpha} (\texttt{print } u)} \text{ \small PRINT-EVAL} \qquad \frac{\texttt{is\_primitive}(v)}{(\texttt{print } v) \xrightarrow{(\text{PRINT } v)} v} \text{ \small PRINT-EXEC}$$

The action PRINT is not handled by other rules leading the semantics to per-
form a step labeled with $(\texttt{print } v)$ where $v$ is the primitive value the respec-
tive expression evaluated to.

A full formalization of Python's input-output capabilities would involve a for-
malization of the necessary interaction with the operating system via system
calls. The primitive print expression is intended to demonstrate that this can
be done, however, doing so is out of scope for this thesis.

**Memory Operations**

Recapitulate that Python objects are references to object descriptors (cf. Section 4.2). As far as primitives are concerned, we need a way to *load* the object descriptor a reference refers to, a way to *store* or update the object descriptor a reference refers to, and a way to create *new* references referring to some object descriptor. So far, it has been a common theme to encapsulate side effects in terms of actions. Memory accesses are no exception here. Instead of putting the heap in some kind of execution context looped through every inference rule, memory accesses happen via actions.

Loading a reference happens via terms of the form $(\texttt{mem\_load}\ r)$ where $r$ is a term evaluating to a reference. Once $r$ has been evaluated to a reference, a $\texttt{MEM\_LOAD}$ action is used to retrieve the value the reference refers to:

$$\frac{e \xrightarrow{\alpha} u}{(\texttt{mem\_load}\ e) \xrightarrow{\alpha} (\texttt{mem\_load}\ u)}\ \text{\small MEM-LOAD-EVAL} \qquad \frac{\texttt{is\_primitive}(r)}{(\texttt{mem\_load}\ r) \xrightarrow{(\texttt{MEM\_LOAD}\ r\ v)} v}\ \text{\small MEM-LOAD-EXEC}$$

Creating a new reference referring to a specific object descriptor happens via terms of the form $(\texttt{mem\_new}\ e)$ where $e$ is a term evaluating to an object descriptor (or other primitive value) for which a new reference should be created. Once $e$ has been evaluated to a value, a $\texttt{MEM\_NEW}$ action is used to initialize and retrieve a fresh reference:

$$\frac{e \xrightarrow{\alpha} u}{(\texttt{mem\_new}\ e) \xrightarrow{\alpha} (\texttt{mem\_new}\ u)}\ \text{\small MEM-NEW-EVAL} \qquad \frac{\texttt{is\_primitive}(v)}{(\texttt{mem\_new}\ v) \xrightarrow{(\texttt{MEM\_NEW}\ r\ v)} r}\ \text{\small MEM-NEW-EXEC}$$

Finally, terms of the form $(\texttt{mem\_store}\ r\ e)$ are used for storing a value at a specific reference. Here, $r$ evaluates to a reference and $e$ evaluates to an object descriptor (or other primitive value) to be stored at the reference $r$ evaluates to. For $\texttt{mem\_store}$ three inference rules are necessary:

$$\frac{r \xrightarrow{\alpha} u}{(\texttt{mem\_store}\ r\ e) \xrightarrow{\alpha} (\texttt{mem\_store}\ u\ e)}\ \text{\small MEM-STORE-EVAL}1$$

$$\frac{e \xrightarrow{\alpha} u \quad \texttt{is\_primitive}(r)}{(\texttt{mem\_store}\ r\ e) \xrightarrow{\alpha} (\texttt{mem\_store}\ r\ u)}\ \text{\small MEM-STORE-EVAL}2$$

$$\frac{\texttt{is\_primitive}(r) \quad \texttt{is\_primitive}(v)}{(\texttt{mem\_store}\ r\ v) \xrightarrow{(\texttt{MEM\_STORE}\ r\ v)} r}\ \text{\small MEM-STORE-EXEC}$$

The rules MEM-STORE-EVAL1 and MEM-STORE-EVAL2 evaluate the reference and value, respectively. Once $r$ and $e$ have been evaluated, a $\texttt{MEM\_STORE}$ action is used to store the value at the provided reference.

We discuss the complementary rules handling the actions $\texttt{MEM\_LOAD}$, $\texttt{MEM\_NEW}$, and $\texttt{MEM\_STORE}$ in Section 4.4.3 when describing the memory layer.

### 4.3.2 The Semantics of Statements

*Statements* are syntactic constructs corresponding to operations that do not evaluate to values. To indicate that the corresponding operation has completed, we use the symbol $\bot$. Multiple statements can be chained to a *sequence*, e. g., the body of a function often is such a sequence of statements. To this end, two statements are concatenated with $;$ to a sequence term. There are two rules handling such sequences of statements:

$$\frac{s \xrightarrow{\alpha} t}{(s \; ; \; s') \xrightarrow{\alpha} (t \; ; \; s')} \text{ SEQUENCE-FIRST} \qquad \frac{}{(\bot \; ; \; s) \xrightarrow{\tau} s} \text{ SEQUENCE-ELIM}$$

The rule SEQUENCE-FIRST evaluates the first statement of a sequence and the rule SEQUENCE-ELIM eliminates the sequence once the first statement has completed leaving behind the second statement for execution.

Expressions can also be statements [20, §7.1]. To this end, they are wrapped within an `eval` statement which evaluates the expression and becomes $\bot$ once the expression has been evaluated to a primitive value:

$$\frac{e \xrightarrow{\alpha} u}{(\texttt{eval } e) \xrightarrow{\alpha} (\texttt{eval } u)} \text{ EVAL} \qquad \frac{\texttt{is\_primitive}(v)}{(\texttt{eval } v) \xrightarrow{\tau} \bot} \text{ EVAL-ELIM}$$

#### Assignment Statements

Python knows a variety of different kinds of assignment statements [20, §7.2]. We already described how accesses to variables are dealt with in the last section. To translate assignment statements to variables, these expressions are simply wrapped within an `eval` statement. The remaining assignments to attributes and items are translated as follows:

$$e.\texttt{<ATTRIBUTE>} = e_v \quad \rightsquigarrow \quad (\texttt{eval } (\texttt{runtime 'set\_attribute' } (e :: (attr :: (e_v :: [\,])))))$$

$$e[e_k] = e_v \quad \rightsquigarrow \quad (\texttt{eval } (\texttt{runtime 'set\_item' } (e :: (e_k :: (e_v :: [\,])))))$$

Here, $attr$ is the name of the attribute as a primitive string. The respective runtime functions implement the mechanism for setting attributes and items via the dunder methods `__setattr__` and `__setitem__`, respectively.

#### The `del` Statement

The `del` statement [20, §7.5] is used for deleting variables, attributes, and other kinds of assignment targets. We already defined how local variables are deleted (cf. Section 4.3.1). All other targets are deleted by calling into the runtime. For non-local identifiers, `del <IDENTIFIER>` is translated based on the access mechanism associated with the identifier

```
 for GLOBAL:  (eval (runtime 'delete_global' ((load_local '__globals__' ⊥) :: (id :: [ ])))))
   for CELL:  (eval (runtime 'delete_cell' ((load_local '__cells__' ⊥) :: (id :: [ ])))))
  for CLASS:  (eval (runtime 'delete_class' ((load_local '__dict__' ⊥) :: (id :: [ ])))))
```

where $id$ is replaced by a static reference to a string containing the name of the respective identifier (for GLOBAL and CLASS and for CELL to a primitive string). The deletion of attributes is translated as

$$\texttt{del } e.\texttt{<ATTRIBUTE>} \quad \leadsto \quad (\texttt{eval } (\texttt{runtime 'delete\_attribute'} (e :: (attr :: [\,]))))$$

where $attr$ is replaced by a static reference to a string containing the name of the respective attribute, and $\texttt{del } e[e_k]$ is translated as

$$\texttt{del } e[e_k] \quad \leadsto \quad (\texttt{eval } (\texttt{runtime 'delete\_item'} (e :: (e_k :: [\,]))))$$

where $e_k$ is substituted with the term $e_k$ translates to.

## The pass Statement

The pass statement "is a null operation" [20, §7.4.], i.e., it does nothing. We translate it to the symbol pass which is handled by the PASS rule:

$$\frac{}{\texttt{pass} \xrightarrow{\tau} \bot} \text{ PASS}$$

## The return Statement

The return statement [20, §7.6] is used to return a value from a function call and comes in two variants: return $e$ where $e$ is an expression and just return without an expression. In case the expression is omitted, it is simply instantiated to None, i.e., return is equivalent to return None. The translation of return $e$ into a corresponding term is then straightforward

$$\texttt{return } e \quad \leadsto \quad (\texttt{return } e)$$

where $e$ is substituted with the translation of $e$.

The semantics of return statements are captured by two rules: RETURN-EVAL which evaluates the expression and RETURN-EXEC which signals to the execution context that a value should be returned:

$$\frac{e \xrightarrow{\alpha} u}{(\texttt{return } e) \xrightarrow{\alpha} (\texttt{return } u)} \text{ RETURN-EVAL} \qquad \frac{\texttt{is\_primitive}(v)}{(\texttt{return } v) \xrightarrow{(\texttt{RETURN } v)} \bot} \text{ RETURN-EXEC}$$

Returning a value induces non-local or interprocedural control flow and, thus, is handled via actions. To this end, a RETURN $v$ action signals to the surrounding execution context that the value $v$ should be returned. The surrounding execution context can then react to this request, e. g., by executing the cleanup handler of a try-finally statement before returning.

**The `if` Statement**

The `if` statement [20, §8.1] is used for the conditional execution of code. While the concrete syntax of Python allows for having multiple alternatives with their respective own conditions using the `elif` keyword, Python's own parser module[6] treats them as syntactic sugar for nested `if` statements. We follow the same approach here and assume that all `if` statements have been desugared such that they can be translated as follows

$$\begin{array}{l} \texttt{if } e_c: \\ \quad s_c \\ \texttt{else}: \\ \quad s_a \end{array} \quad \rightsquigarrow \quad (\texttt{if } (\texttt{bool } e_c) \texttt{ then } s_c \texttt{ else } s_a)$$

where $e_c$, $s_c$, and $s_a$ are substituted with the terms $e_c$, $s_c$, and $s_a$ translate to, respectively. If an `if` statement lacks an `else` case, we simply add such a case with a single `pass` statement which does nothing. Note that the condition is wrapped into `bool` to convert it to a primitive boolean.

The semantics of `if` statements are captured by three rules: IF-EVAL which evaluates the condition until it becomes a primitive boolean, IF-TRUE which reduces an `if` statement to the *consequent body* $s_c$ if the condition has been evaluated to `true`, and IF-FALSE which reduces an `if` statement to the *alternate body* $s_a$ if the condition has been evaluated to `false`:

$$\frac{e \xrightarrow{\alpha} u}{(\texttt{if } e \texttt{ then } s_c \texttt{ else } s_a) \xrightarrow{\alpha} (\texttt{if } u \texttt{ then } s_c \texttt{ else } s_a)} \text{ IF-EVAL}$$

$$\frac{}{(\texttt{if true then } s_c \texttt{ else } s_a) \xrightarrow{\tau} s_c} \text{ IF-TRUE} \qquad \frac{}{(\texttt{if false then } s_c \texttt{ else } s_a) \xrightarrow{\tau} s_a} \text{ IF-FALSE}$$

**The `break` and `continue` Statements**

The `break` [20, §7.9] and `continue` [20, §7.10] statements are used to control the execution of `while` and `for` loops. They interact with the nearest enclosing loop by terminating it or making it continue with the next cycle, respectively. If they occur within `try-finally` statements, the respective cleanup handlers are executed before the loop is affected. The translation of both statements is straightforward. They are translated to the symbols `break` and `continue` whose semantics are defined by the following rules:

$$\frac{}{\texttt{break} \xrightarrow{\text{BREAK}} \bot} \text{ BREAK} \qquad \frac{}{\texttt{continue} \xrightarrow{\text{CONTINUE}} \bot} \text{ CONTINUE}$$

Both statements induce non-local control flow and may trigger the execution of multiple cleanup handlers before they affect the loop. Hence, their semantics need to signal the respective effect to the surrounding execution context via actions. These BREAK and CONTINUE actions are handled by corresponding rules for `try-finally`, `while`, and `for` statements.

---

[6] https://docs.python.org/3/library/ast.html

**The `while` Statement**

The `while` statement [20, §8.2] is used for the repeated execution of code as long as some condition is true. A special feature of Python's `while` statement is that it may have an `else` case which is executed once if the condition is or became false. Analogously to the `if` statement, we add an `else` case with a single `pass` statement if a `while` statements lacks the `else` case. We then translate `while` statements as follows

$$
\begin{array}{ll}
\texttt{while } e_c\texttt{:} & \\
\quad s_b & \\
\texttt{else:} & \rightsquigarrow \quad (\texttt{[ (bool } e_c\texttt{) ]c (while (bool } e_c\texttt{) do } s_b \texttt{ else } s_a\texttt{))} \\
\quad s_a &
\end{array}
$$

where $e_c$, $s_b$, and $s_a$ are substituted with the terms $e_c$, $s_b$, and $s_a$ translate to, respectively. Again, the condition is wrapped into `bool` to convert it to a primitive boolean. While the semantics of most statements are defined *destructively*, i. e., they modify the terms in place without retaining a copy of the original, we have to retain the original loop and condition for subsequent repetitions of the loop. To this end, we duplicate the condition and body before we begin with their evaluation. Hence, the condition appears twice in the translated term, once in the *loop part* starting with `while` and once in the *condition container* `[ · ]c`. The occurrence of the condition in the condition container is evaluated by the rule WHILE-EVAL:

$$
\frac{e \xrightarrow{\alpha} u}{(\texttt{[ } e \texttt{ ]c } L) \xrightarrow{\alpha} (\texttt{[ } u \texttt{ ]c } L)} \text{ WHILE-EVAL}
$$

If the condition has been evaluated to `false`, the loop terminates and can be discarded by replacing it with the `else` case:

$$
\frac{}{(\texttt{[ false ]c (while } e \texttt{ do } s_b \texttt{ else } s_a\texttt{))} \xrightarrow{\tau} s_a} \text{ WHILE-FALSE}
$$

Otherwise, if the condition has been evaluated to `true`, the body needs to be executed. This is done by duplicating the body from the loop part into a *body container* `[ · ]b` and thereby retaining it for later:

$$
\frac{}{(\texttt{[ true ]c (while } e \texttt{ do } s_b \texttt{ else } s_a\texttt{))} \xrightarrow{\tau} (\texttt{[ } s_b \texttt{ ]b (while } e \texttt{ do } s_b \texttt{ else } s_a\texttt{))}} \text{ WHILE-TRUE}
$$

The body container manages the execution of the body of the loop and in particular reacts to *loop actions*, i. e., `BREAK` and `CONTINUE` instructing the loop to abort and continue with the next cycle, respectively. To this end, the body is executed and any non-loop action is passed through:

$$
\frac{s \xrightarrow{\alpha} t \quad \neg\texttt{is\_loop}(\alpha)}{(\texttt{[ } s \texttt{ ]b } L) \xrightarrow{\alpha} (\texttt{[ } t \texttt{ ]b } L)} \text{ WHILE-BODY}
$$

If the execution of the body has terminated with $\bot$ a condition container is constructed using the retained condition from the loop part. As a result, the condition is evaluated again and the loop repeats until the condition becomes false or the loop is otherwise terminated:

$$\frac{}{(\lceil \bot \rceil \mathtt{b} \ (\mathtt{while} \ e \ \mathtt{do} \ s_b \ \mathtt{else} \ s_a)) \xrightarrow{\tau} (\lceil e \rceil \mathtt{c} \ (\mathtt{while} \ e \ \mathtt{do} \ s_b \ \mathtt{else} \ s_a))} \ \text{\small WHILE-REPEAT}$$

What remains are rules for when the body produces a `BREAK` or `CONTINUE` action in response to encountering a `break` or `continue` statement, respectively. In case a `BREAK` action is produced by the body, the loop terminates immediately, i. e., without executing the `else` case:

$$\frac{s \xrightarrow{\text{BREAK}} t}{(\lceil s \rceil \mathtt{b} \ L) \xrightarrow{\tau} \bot} \ \text{\small WHILE-BREAK}$$

In case a `CONTINUE` action is produced by the body, a condition container is constructed using the retained condition from the loop part. Analogously to WHILE-REPEAT this leads to the loop repeating:

$$\frac{s \xrightarrow{\text{CONTINUE}} t}{(\lceil s \rceil \mathtt{b} \ (\mathtt{while} \ e \ \mathtt{do} \ s_b \ \mathtt{else} \ s_a)) \xrightarrow{\tau} (\lceil e \rceil \mathtt{c} \ (\mathtt{while} \ e \ \mathtt{do} \ s_b \ \mathtt{else} \ s_a))} \ \text{\small WHILE-CONTINUE}$$

**The `raise` Statement**

The `raise` statement [20, §7.8] is used to raise exceptions and comes in two variants: `raise` $e$ where $e$ is an expression and just `raise` without an expression. We first consider the case where an expression is provided.

If an expression is provided, it is required to evaluate to either an instance or to a subclass of `BaseException` which, as the name suggests, is the base class of all exceptions. If the expression evaluates to an instance of `BaseException`, then this instance is thrown directly, otherwise, the instance to be thrown is obtained by instantiating the provided class without any arguments. If the expression evaluates to something that is neither an instance nor a subclass of `BaseException` a runtime error needs to be thrown. We encapsulate this behavior in the runtime function `ensure_exception`. Hence, a `raise` statement with an expression is translated as follows

$$\texttt{raise } e \quad \rightsquigarrow \quad (\texttt{raise } (\texttt{runtime 'ensure\_exception'} \ (e :: (\texttt{get\_active\_exc} :: \lceil \, \rceil))))$$

where $e$ is replaced by the term the expression $e$ translates to.

The semantics are captured by two rules: RAISE-EVAL which evaluates the call into the runtime and RAISE-EXEC which finally throws the exception once the call into the runtime has evaluated to a primitive value:

$$\frac{e \xrightarrow{\alpha} u}{(\texttt{raise } e) \xrightarrow{\alpha} (\texttt{raise } u)} \ \text{\small RAISE-EVAL} \qquad \frac{\texttt{is\_primitive}(v)}{(\texttt{raise } v) \xrightarrow{(\text{THROW } v)} \bot} \ \text{\small RAISE-EXEC}$$

Throwing an exception induces non-local control flow and, hence, like all non-local control flow, is handled via actions. To this end, a `THROW` $v$ action signals to the surrounding execution context that an exception $v$ has been thrown.

The surrounding execution context can then react to this exception, e. g., by executing the handler of a `try-except` statement.

If no expression is provided, the PLR states that "the last exception that was active in the current scope" should be re-raised. Unfortunately, the PLR stays unspecific on what it means for an exception to be "the last exception that was active in the current scope." Considering the other occurrences of the term "scope" in the PLR, one may very well conclude that in line six of Snippet 4.6 the `TypeError` should be re-raised. However, this is not the case. When one executes Snippet 4.6, a runtime error is raised due to there being no "last exception that was active in the current scope."

```
1  try:
2      raise TypeError()
3  except:
4      pass
5
6  raise  # produces a `RuntimeError`, not a `TypeError`
```

Snippet 4.6: No active exception resulting in a `RuntimeError`.

After further experiments with the reference implementation, it became evident that the formulation "last exception that was active in the current scope" is misleading at best. Instead, `raise` re-raises the exception that *is active* in the respective execution context. Code that executes in the context of an *active* exception executes either within a `finally` or an `except` context, i. e., is either part of a cleanup or an exception handler.

To capture the semantics of `raise` without an expression, we need a mechanism to retrieve the active exception from the surrounding execution context. As part of this mechanism, the rule GET-ACTIVE-EXC communicates with the surrounding context via `GET_ACTIVE_EXC v` actions and evaluates the symbol `get_active_exc` to the respective active exception:

$$\frac{}{\text{get\_active\_exc} \xrightarrow{\left(\text{GET\_ACTIVE\_EXC } v\right)} v} \text{ GET-ACTIVE-EXC}$$

Complementary rules to supply the active exception exist as part of the semantics of `try-finally` and `try-except` statements. We describe them later in their respective subsections. If there is no active exception in the current execution context, then `None` is supplied by the rules for execution frames (see Section 4.4.1). With this mechanism in place, we capture the semantics of `raise` without an expression by translating it as follows

$$\text{raise} \rightsquigarrow (\text{raise } (\text{runtime } \text{'check\_active\_exception'} (\text{get\_active\_exc} :: [\,])))$$

where the runtime function `check_active_exception` constructs and returns a runtime error if the provided object is `None`, i. e., if there is no active exception, and returns the object unmodified otherwise.

The active exception is also supplied to `ensure_exception` in case an expression is provided. The active exception is used to set the attribute `__context__` of the exception to be raised [20, §7.8.].

**The `try-finally` Statement**

In Python, the ideas behind having cleanup and exception handlers are amalgamated within the `try` statement [20, §8.4]. However, for our purposes, we separate them into `try-finally` which is used to define cleanup handlers and `try-except` which is used to define exception handlers. Hence, a `try` statement which has both, a cleanup handler and except handlers, is split into a `try-except` and a `try-finally` statement where the `try-except` statement is placed in the `try` body of the `try-finally` statement. The translation of `try-finally` statements to terms is straightforward

$$\begin{array}{l} \texttt{try:} \\ \quad s_b \\ \texttt{finally:} \\ \quad s_c \end{array} \quad \rightsquigarrow \quad (\texttt{try } s_b \texttt{ finally } s_c)$$

where $s_b$ and $s_c$ are substituted with the terms $s_b$ and $s_c$ translate to.

When executing a `try-finally` statement, the `try` body is executed until it *terminates*, either by executing successfully or by producing a *terminating* action inducing non-local control flow. Terminating actions are `BREAK`, `CONTINUE`, (`THROW` $v$), and (`RETURN` $v$). By inducing non-local control flow, these actions lead to the `try-finally` being discarded. Hence, these actions should trigger the execution of the *cleanup handler* $s_c$. The execution of the `try` body is handled by the rule FINALLY-BODY which passes all non-terminating actions through to the surrounding execution context:

$$\frac{s \xrightarrow{\alpha} t \quad \neg\texttt{is\_terminating}(\alpha)}{(\texttt{try } s \texttt{ finally } s_c) \xrightarrow{\alpha} (\texttt{try } t \texttt{ finally } s_c)} \text{ FINALLY-BODY}$$

In case the `try` body produces a terminating action, this action has to be saved and the cleanup handler should be executed. The saved action will be reproduced after the cleanup handler finished executing. To this end, the rule FINALLY-CLEANUP1 constructs a *finally container* ($[\,s_c\,]\texttt{f} \cdot \alpha$) containing the cleanup handler $s_c$ and terminating action $\alpha$:

$$\frac{s \xrightarrow{\alpha} t \quad \texttt{is\_terminating}(\alpha)}{(\texttt{try } s \texttt{ finally } s_c) \xrightarrow{\tau} ([\,s_c\,]\texttt{f} \cdot \alpha)} \text{ FINALLY-CLEANUP1}$$

In case the `try` body finishes successfully, the cleanup handler is executed directly and no action has to be stored and reproduced later:

$$\frac{}{(\texttt{try } \perp \texttt{ finally } s_c) \xrightarrow{\tau} s_c} \text{ FINALLY-CLEANUP2}$$

Remember that we introduced the concept of active exceptions to capture the semantics of `raise` statements. Whenever a cleanup handler executes due to an exception, then the inner code executes with that exception being active. Hence, the finally container must react appropriately to `GET_ACTIVE_EXC` actions requesting the currently active exception. The rule FINALLY-CONTAINER executes the finally container and passes through all actions as long as the ac-tion is not of the form $(\texttt{GET\_ACTIVE\_EXC}\ v)$ or no throw action has been saved, i. e., execution does not happen due to an exception:

$$\frac{s \xrightarrow{\alpha} t \quad \neg\texttt{is\_get\_active\_exc}(\alpha) \vee \neg\texttt{is\_throw}(\beta)}{([\,s\,]\texttt{f}\,.\,\beta) \xrightarrow{\alpha} ([\,t\,]\texttt{f}\,.\,\beta)} \text{ \small FINALLY-CONTAINER}$$

This rule is complemented by the rule FINALLY-GET-ACTIVE-EXC which reacts to $\texttt{GET\_ACTIVE\_EXC}$ actions if a throw action has been saved:

$$\frac{s \xrightarrow{(\texttt{GET\_ACTIVE\_EXC}\ v)} t}{([\,s\,]\texttt{f}\,.\,(\texttt{THROW}\ v)) \xrightarrow{\tau} ([\,t\,]\texttt{f}\,.\,(\texttt{THROW}\ v))} \text{ \small FINALLY-GET-ACTIVE-EXC}$$

Together both rules ensure that the cleanup handler is executed and a potentially active exception is provided to the inner execution context whenever requested. It remains to discard the finally container and reproduce the stored action if the cleanup handler finished executing:

$$\frac{}{([\,\bot\,]\texttt{f}\,.\,\alpha) \xrightarrow{\alpha} \bot} \text{ \small FINALLY-DONE}$$

## The try-except Statement

The try-except part of a try statement [20, §8.4] allows defining multiple exception handlers which potentially restrict the type of exceptions to be caught and bind them to a name. Furthermore, an else body can be specified which should be executed if and only if no exception is thrown in the try body. The translation of such try statements involves desugaring the except handlers into a single except handler. We explain this process in more detail later and first define the semantics for terms of the form $(\texttt{try}\ s\ \texttt{except}\ s_h\ \texttt{else}\ s_a)$ which represent try-except statements with an else body and a single unrestricted exception handler not binding the exception to a name.

As long as no exception is thrown, the rule EXCEPT-BODY executes the try body of a try-except statement by passing through actions:

$$\frac{s \xrightarrow{\alpha} t \quad \neg\texttt{is\_throw}(\alpha)}{(\texttt{try}\ s\ \texttt{except}\ s_h\ \texttt{else}\ s_a) \xrightarrow{\alpha} (\texttt{try}\ t\ \texttt{except}\ s_h\ \texttt{else}\ s_a)} \text{ \small EXCEPT-BODY}$$

If the try body finished executing, the else body is executed:

$$\frac{}{(\texttt{try}\ \bot\ \texttt{except}\ s_h\ \texttt{else}\ s_a) \xrightarrow{\tau} s_a} \text{ \small EXCEPT-DONE}$$

Otherwise, if an exception is thrown, an *except container* $([\,s_c\,]\texttt{e}\ v)$ is created containing the exception handler $s_c$ and the exception $v$:

$$\frac{s \xrightarrow{(\texttt{THROW}\ v)} t}{(\texttt{try}\ s\ \texttt{except}\ s_h\ \texttt{else}\ s_a) \xrightarrow{\tau} ([\,s_h\,]\texttt{e}\ v)} \text{ \small EXCEPT-HANDLE}$$

The except container manages the execution of the exception handler and, in particular, analogously to a finally container provides the inner execution context with the active exception via `GET_ACTIVE_EXC` actions

$$\frac{s \xrightarrow{\text{(GET\_ACTIVE\_EXC } v)} t}{(\texttt{[ } s \texttt{ ]e } v) \xrightarrow{\tau} (\texttt{[ } t \texttt{ ]e } v)} \text{ EXCEPT-GET-ACTIVE-EXC}$$

while other actions are passed through by the except container:

$$\frac{s \xrightarrow{\alpha} t \quad \neg\texttt{is\_get\_active\_exc}(\alpha)}{(\texttt{[ } s \texttt{ ]e } v) \xrightarrow{\alpha} (\texttt{[ } t \texttt{ ]e } v)} \text{ EXCEPT-EXEC}$$

If the exception handler finished executing, the except container is discarded:

$$\frac{}{(\texttt{[ } \bot \texttt{ ]e } v) \xrightarrow{\tau} \bot} \text{ EXCEPT-CAUGHT}$$

Now, with these rules in place, Python's `try` statements can be desugared by translating multiple exception handlers into a single exception handler with an `if` statement which distinguishes the different cases. See Snippet 4.7 for an example. For each exception handler with a restriction we check whether the currently active exception is compatible with the restriction as defined by the PLR by calling into the runtime. We then execute the first exception handler we find that is compatible with the currently active exception and bind this exception to the specified target if and only if a target has been specified. After executing the handler the target is deleted as required by the PLR. If no exception handler is found, then the exception is simply re-raised. For further details we refer to our implementation [35].

```python
1  try:
2      ...  # try body
3  except RESTRICTION0 as TARGET0:
4      ...  # body of the first exception handler
5  except RESTRICTION1:
6      ...  # body of the second exception handler
```

$$\Downarrow$$

```python
1  if is_exception_compatible(GET_ACTIVE_EXC(), RESTRICTION0):
2      TARGET0 = GET_ACTIVE_EXC()
3      try:
4          ...  # body of the first exception handler with a target
5      finally:
6          del TARGET0
7  elif is_exception_compatible(GET_ACTIVE_EXC(), RESTRICTION1):
8      ...  # body of the second exception handler without a target
9  else:
10     raise  # or the handler without an restriction if present
```

Snippet 4.7: Desugaring of multiple exception handlers.

**The `assert` Statement**

The `assert` statement [20, §7.3] is primarily used to insert checkpoints for debugging at which certain conditions have to hold. The semantics of the assert statement is defined by the PLR by desugaring it into an `if` statement checking the condition and then raising an `AssertionError` if the condition turns out false. Whether assertions are enabled is determined by the built-in variable `__debug__` to which assignments are illegal [20, §7.3.]. Like the reference implementation, we implement this behavior in the translator where code can be translated with assertions enabled or disabled.

### The `for` Statement

The `for` statement [20, §8.3] is used to iterate over an *iterable*. Iterables are Python objects implementing the *iterable protocol*, i. e., implement a method `__iter__` which returns an *iterator*. An iterator is a Python object implementing the *iterator protocol*, i. e., implement a method `__next__` which either returns the next element or throws a `StopIteration` exception if there is no such element. Unlike many other languages where there is an explicit method to check whether there is a next element, e. g., `hasNext` in Java[7], iterators in Python have no explicit method to check whether there is a next element. For our formal semantics, we desugar `for` loops as sketched in Snippet 4.8 by transforming them into an equivalent `while` loop.

```python
1  for TARGET in ITERABLE:
2      ...  # for body
3  else:
4      ...  # else body
```

$$\Downarrow$$

```python
1   __iterator0__ = iter(ITERABLE)
2   __check_next0__ = True
3   while __check_next0__:
4       try:
5           TARGET = next(__iterator0__)
6       except StopIteration:
7           __check_next0__ = False
8       else:
9           ...  # for body
10  else:
11      ...  # else body
```

Snippet 4.8: Desugaring of `for` statements.

For the translation, two fresh local variables are introduced, `__iterator__` and `__check_next__`, which are numbered to make them unique for the particular loop and thereby allow for nested loops. At the beginning, `__iterator__` is set to the iterator obtained from the iterable the `for` loop should iterate over. The variable `__check_next__` is a boolean flag indicating whether there should be a next cycle where the iterator is checked for the next element. It is initially set to `True`. In each iteration of the `while` loop, in a `try-except` statement, it is tried to assign the next element of the iterator to the target of the `for` loop. In case this succeeds, i. e., there is a next element, the body of the `for` loop is executed. In case this throws a `StopIteration` exception indicating that the iterator is exhausted, `__check_next__` is set to `False`. As a result the `while` loop is left without the body of the `for` loop executing and instead the `else` body of the `for` loop is executed.

---
[7] https://docs.oracle.com/javase/8/docs/api/java/util/Iterator.html

When desugaring a `for` loop, special attention has to be payed to `break` and `continue` statements having the desired effect. If a `break` or `continue` statement appears in the body of the for loop, it directly interacts with the `while` loop. If a `break` statement is hit, the `else` case of the for loop should not be executed and this is what happens here. Also, if there is a `break` or `continue` in the `else` case, this does not interact with the `while` emulating the desugared `for` loop but instead with the enclosing outer loop.

### Function Definitions

The body of a function definition is translated to a term and then put into a code object (cf. Section 4.2.3). Functions in Python are runtime objects, i. e., they are constructed at runtime when a function definition is executed. To this end, the runtime function `build_function` takes the code object generated from a function definition, a global namespace in terms of a dictionary-like object, a mapping from cell names to cells, and a mapping from parameter names to default values for those parameters, and constructs a function object based on these objects. Before binding the resulting function object to the name of the function in the scope of the definition, the decorator expressions are evaluated and applied to the created function. Afterwards, the resulting object is bound to the name of the function.

The global namespace and cells are extracted from the local namespace of the scope the function definition is executed in. Recall that the global namespace is always bound to the local variable `__globals__` and a mapping from cell names to cells is available via the local variable `__cells__`. The name and docstring for the function are taken from the code object.

### Class Definitions

Analogously to function definitions, the body of a class definition is translated to a term and put into a code object. The signature of this code object consists of one parameter, named `__dict__`, which is bound to the namespace of the class when the class definition is executed. At runtime, a class definition is translated by first constructing a function from the code object. By construing a function from the code object, the global namespace and closure of the class definition is set. The class definition itself is implemented by calling the built-in function `__build_class__` with the constructed function as first argument, the name of the class as second argument, and passing on the remaining arguments of the class definition. Snippet 4.9 shows the code of this built-in function as suggested by van Rossum [48].

The built-in function `__build_class__` first extracts the meta class from the provided base classes in case no meta class has been provided. It then checks whether the meta class provides a method named `__prepare__`. In case the meta class provides a method `__prepare__`, it invokes this method to obtain a namespace for the class. The namespace has to be a dictionary-like object and will be bound to the parameter `__dict__` of the code object generated from the class definition. Recall how accesses to variables with the access mechanisms `CLASS_CELL` and `CLASS_GLOBAL` have been handled by retrieving a class's namespace bound to `__dict__` (cf. Section 4.3.1). After the namespace has been created, the constructed function is invoked with the namespace as first and only argument. Invoking this function, will populate the namespace with the functions and other entities defined within the body of the class definition. Now, to construct the actual class object, an object of the meta class is constructed by invoking it with the namespace.

```python
1  def __build_class__(func, name, *bases, metaclass=None, **kwargs):
2      if metaclass is None:
3          metaclass = extract_metaclass(bases)
4      prepare = getattr(metaclass, "__prepare__", None)
5      if prepare:
6          namespace = prepare(name, bases, **kwargs)
7      else:
8          namespace = {}
9      func(namespace)
10     return metaclass(name, bases, namespace, **kwargs)
```

Snippet 4.9: Built-in function for constructing a class [48].

Analogously to how decorators are handled for function definitions, the decorators are evaluated after the class has been constructed and are then applied to the class object returned by `__build_function__`. Finally, the object returned by the other decorator is bound to the name of the class in the scope of the class definition.

```python
1  class X:
2      def f(self):
3          print("X.f")
4
5  class Y(X):
6      def f(self):
7          print("Y.f")
8          super().f()
9
10 >>> Y().f()
11 Y.f
12 X.f
```

Snippet 4.10: Example of calling super without any arguments.

During translation, the built-in function super undergoes a special treatment if it appears without arguments within a function definition within a class body. PEP 3135 specifies the semantics of super without arguments within a function definition within a class body [54]. The built-in function super is used to delegate a method call or attribute access to a super or sibling class of the class the function has been defined in. Snippet 4.10 shows an example of calling `super` without any arguments. With the help of `super` it is possible to traverse complex class hierarchies due to multiple inheritance by delegating upwards in the C3 linearization [54, 7]. For this purpose, `super` needs to know the object the method has been invoked on as well as the class the method being executed is defined in. Based on this information it decides to which class to delegate method calls and attribute accesses.

PEP 3135 explicitly allows for handling calls to `super` without arguments syntactically during a compilation or translation process. However, the reference implementation does not do that but instead inspects the call stack to retrieve the necessary information required by `super`. Following PEP 3135, we make the class a method has been defined in available via a special cell named `__class__`. Calls to `super` without arguments within a function definition in a class body are then translated to `super(__class__, x)` where $x$ is the identifer of the first argument of the function, usually `self` [54]. To mimic the quirky behavior of the reference implementation, a few tricks are necessary during this translation. We spare the reader these rather cumbersome details and refer to our implementation for further details on how class definitions and `super` are translated [35].

## 4.4 Semantic Layers

In Section 4.1.2, we introduced the idea of decomposing the semantics into multiple layers. Hit hereto, our focus has been on the intraprocedural layer covering the semantics of expressions and statements. We saw that some rules require non-local information, e. g., loading a local variable or a reference, have side effects, e. g., storing a value in memory, or induce non-local control flow, e. g., calling a function or returning a value. All of these *non-local intents* have in common that they cannot be handled locally on the level of the respective syntactic construct and, hence, are explicitly signaled to the surrounding execution context via actions. The idea is to carry out the respective intents within their own layer by providing the necessary information, effectuating a side effect, or inducing non-local control flow. In this section, we present and describe the remaining inference rules complementing the inference rules introduced so far by handling non-local intents.

### 4.4.1 The Frame Layer

The purpose of the frame layer is to manage a single execution frame. To this end, it stores the local variables and execution state of a frame. The terms, the frame layer deals with, have the form (`frame` $\mathfrak{F}$) where $\mathfrak{F}$ is a frame descriptor. Recall that a frame descriptor is a record with a field `body` containing a term representing the execution state and a field `locals` containing a mapping from identifiers to values storing the local variables (cf. Section 4.2.3).

The rules of the frame layer execute the body of a frame as stored in the field `body`. They provide information about local variables to the intraprocedural layer and handle side effects concerning local variables.

If the action produced by the body of a frame is not a *frame action*, it is not handled by the frame layer and passed through:

$$\frac{\mathfrak{F}.\texttt{body} \xrightarrow{\alpha} t \quad \neg\texttt{is\_frame}(\alpha)}{(\texttt{frame } \mathfrak{F}) \xrightarrow{\alpha} (\texttt{frame } (\mathfrak{F}.\texttt{body} := t))} \text{ \scriptsize FRAME-NOP}$$

To load the value of a local variable, the rule small caps LOAD-LOCAL uses a `LOAD_LOCAL` action with the identifier of the local variable and a default value. The rule small caps FRAME-LOAD-LOCAL handles these actions by providing the value based on the provided identifier and default value. It looks up the identifier among the local variables stored in the frame and returns the default value in case the identifier is not bound to any value:

$$\frac{\mathfrak{F}.\texttt{body} \xrightarrow{(\texttt{LOAD\_LOCAL } id \; \mathfrak{F}.\texttt{locals}[id]?d \; d)} t}{(\texttt{frame } \mathfrak{F}) \xrightarrow{\tau} (\texttt{frame } (\mathfrak{F}.\texttt{body} := t))} \text{ \scriptsize FRAME-LOAD-LOCAL}$$

The rule small caps FRAME-STORE-LOCAL complements the rule small caps STORE-LOCAL-EXEC by storing a value for a local variable as requested via a `STORE_LOCAL` action:

$$\frac{\mathfrak{F}.\texttt{body} \xrightarrow{(\texttt{STORE\_LOCAL } id \; v)} t}{(\texttt{frame } \mathfrak{F}) \xrightarrow{\tau} (\texttt{frame } ((\mathfrak{F}.\texttt{locals} := \mathfrak{F}.\texttt{locals}\,[id \mapsto v]).\texttt{body} := t))} \text{ \scriptsize FRAME-STORE-LOCAL}$$

Finally, the rule small caps FRAME-DELETE-LOCAL deletes a local variable by removing it from the mapping of local variables. It complements the rule small caps DELETE-LOCAL which signals deletion intents via `DELETE_LOCAL` actions:

$$\frac{\mathfrak{F}.\texttt{body} \xrightarrow{(\texttt{DELETE\_LOCAL } id \; \mathfrak{F}.\texttt{locals}[id]?\bot)} t}{(\texttt{frame } \mathfrak{F}) \xrightarrow{\tau} (\texttt{frame } ((\mathfrak{F}.\texttt{locals} := \texttt{delitem}(\mathfrak{F}.\texttt{locals}, \; id)).\texttt{body} := t))} \text{ \scriptsize FRAME-DELETE-LOCAL}$$

Recall how we translated `raise` statements without an expression by retrieving the active exception using a `GET_ACTIVE_EXC` action (cf. Section 4.3.2). If there is no enclosing `try-finally` or `try-except` statement, then there is no active exception and the action is handled by the frame layer:

$$\frac{\mathfrak{F}.\texttt{body} \xrightarrow{(\texttt{GET\_ACTIVE\_EXC ref(None)})} t}{(\texttt{frame } \mathfrak{F}) \xrightarrow{\tau} (\texttt{frame } (\mathfrak{F}.\texttt{body} := t))} \text{ \scriptsize FRAME-NO-ACTIVE-EXC}$$

By returning `ref(None)` as the active exception, the required `RuntimeError` will be thrown by the runtime function `check_active_exception`.

## 4.4.2   The Stack Layer

The *stack layer* is responsible for all interprocedural control flow. Analogously to the frame layer, there is a rule small caps STACK-NOP passing through actions that are not handled by the stack layer:

$$\frac{s \xrightarrow{\alpha} t \quad \neg \mathtt{is\_stack}(\alpha)}{(\mathtt{stack}\ (xs :: s)) \xrightarrow{\alpha} (\mathtt{stack}\ (xs :: t))}\ \textsc{stack-nop}$$

The terms of the stack layer have the following form (cf. Figure 4.2)

$$(\mathtt{stack}\ (((\mathtt{(nil}\ :: t_n)\ \ldots\ )\ :: t_2)\ :: t_1))$$

where $t_i$ for $1 \leq i \leq n$ represent the frames on the call stack. The stack is encoded like the stack we used for proving Turing-completeness (cf. Section 3.3.1). Technically, valid terms representing a stack are defined inductively. The term $\mathtt{nil}$ is a stack and if a term $xs$ is a stack then so is $(xs :: t)$ for a frame $t$. All rules for the stack layer, with the exception of $\textsc{stack-nop}$, modify the call stack in some way or another.

The rule $\textsc{stack-call}$ handles $\mathtt{CALL}$ actions as produced by the rules $\textsc{call-exec}$ and $\textsc{runtime-exec}$. These actions instruct the stack layer to push a frame described by the provided frame descriptor onto the stack:

$$\frac{s \xrightarrow{(\textsc{call}\ \mathfrak{F})} t}{(\mathtt{stack}\ (xs :: s)) \xrightarrow{\tau} (\mathtt{stack}\ ((xs :: t)\ ::\ (\mathtt{frame}\ \mathfrak{F})))}\ \textsc{stack-call}$$

Whenever a frame returns as the result of executing a $\mathtt{return}$ statement, the result is passed to its predecessor on the call stack or, if no predecessor exists, is passed through to the surrounding execution context. This intuition is captured by the following two inference rules:

$$\frac{s \xrightarrow{(\textsc{return}\ v)} t \quad s' \xrightarrow{(\textsc{result}\ v)} t'}{(\mathtt{stack}\ ((xs :: s')\ :: s)) \xrightarrow{\tau} (\mathtt{stack}\ (xs :: t'))}\ \textsc{stack-return}$$

$$\frac{s \xrightarrow{(\textsc{return}\ v)} t}{(\mathtt{stack}\ (\mathtt{nil} :: s)) \xrightarrow{(\textsc{return}\ v)} (\mathtt{stack}\ \mathtt{nil})}\ \textsc{stack-result}$$

Both rules handle $\mathtt{RETURN}$ actions carrying a value to return as produced by the rule $\textsc{return-exec}$ in response to a $\mathtt{return}$ statement (cf. Section 4.3.2). The value is passed to the predecessor on the call stack by means of a $\mathtt{RESULT}$ action which is handled by the rule $\textsc{entry-result}$ for normal entry points and $\textsc{send-res-result}$ for send entry points.

As the result of executing a $\mathtt{raise}$ statement, an exception may be thrown in a frame. If this exception is not handled intraprocedurally by a $\mathtt{try\text{-}except}$ statement, it is passed to the caller by the stack layer. Analogously to the two rules for returning a value, there are two rules for throwing an exception up the call stack to the caller:

$$\frac{s \xrightarrow{(\textsc{throw}\ v)} t \quad s' \xrightarrow{(\textsc{error}\ v)} t'}{(\mathtt{stack}\ ((xs :: s')\ :: s)) \xrightarrow{\tau} (\mathtt{stack}\ (xs :: t'))}\ \textsc{stack-throw}$$

$$\frac{s \xrightarrow{(\textsc{throw}\ v)} t}{(\mathtt{stack}\ (\mathtt{nil} :: s)) \xrightarrow{(\textsc{throw}\ v)} (\mathtt{stack}\ \mathtt{nil})}\ \textsc{stack-error}$$

Here, both rules handle `THROW` actions carrying exceptions as produced by the rule RAISE-EXEC (cf. Section 4.3.2). If not already handled intraprocedurally, this exception is passed to the caller by means of an `ERROR` action which is handled by the rule ENTRY-ERROR for normal entry points and by the rule SEND-RES-ERROR for send entry points.

**Generators** The rule STACK-YIELD handles `YIELD` actions carrying a value to yield as produced by the rule YIELD-EXEC when executing a `yield` expression (cf. Section 4.3.1). The value is passed back to the caller together with the frame descriptor of the frame to later resume the execution of the generator. To this end, a `VALUE` action is used which, in turn, is handled by the rule SEND-RES-VALUE for send entry points:

$$\frac{s \xrightarrow{\text{(YIELD } v)} (\text{frame } t) \quad s' \xrightarrow{\text{(VALUE } t\ v)} t'}{(\text{stack } ((xs :: s') :: s)) \xrightarrow{\tau} (\text{stack } (xs :: t'))} \text{ STACK-YIELD}$$

A `yield` expression results in a normal entry point. Sending a value or throwing an exception into a generator, hence, is achieved by `RESULT` and `ERROR` actions, respectively. A `send_value` $\mathfrak{F}$ $v$ expression sends the value $v$ into the frame $\mathfrak{F}$ extracted from a generator by producing an action `SEND_VALUE` $\mathfrak{F}$ $v$ (cf. SEND-VALUE). Being interprocedural control flow, this action is in turn handled by the rule STACK-SEND-VALUE within the stack layer:

$$\frac{s \xrightarrow{\text{(SEND\_VALUE } \mathfrak{F}\ v)} t \quad (\text{frame } \mathfrak{F}) \xrightarrow{\text{(RESULT } v)} t'}{(\text{stack } (xs :: s)) \xrightarrow{\tau} (\text{stack } ((xs :: t) :: t'))} \text{ STACK-SEND-VALUE}$$

Analogously, the rule STACK-SEND-ERROR throws an exception inside the frame of the generator by means of an `ERROR` action handled by ENTRY-ERROR:

$$\frac{s \xrightarrow{\text{(SEND\_THROW } \mathfrak{F}\ v)} t \quad (\text{frame } \mathfrak{F}) \xrightarrow{\text{(ERROR } v)} t'}{(\text{stack } (xs :: s)) \xrightarrow{\tau} (\text{stack } ((xs :: t) :: t'))} \text{ STACK-SEND-ERROR}$$

### 4.4.3 The Memory Layer

The memory layer is the outermost layer of our semantics and is responsible for managing the heap and handling memory intents.

Like the other two layers, the memory layer comes with a rule MEM-NOP passing through actions that are not memory intents:

$$\frac{s \xrightarrow{\alpha} t \quad \neg \texttt{is\_memory}(\alpha)}{(\text{memory } \mathcal{H} : s) \xrightarrow{\alpha} (\text{memory } \mathcal{H} : t)} \text{ MEM-NOP}$$

The rule MEM-LOAD-EXEC produces `MEM_LOAD` actions in order to load the value stored for a specific reference. The rule MEM-LOAD handles those actions by loading the value from the heap:

$$\frac{s \xrightarrow{\text{(MEM\_LOAD } r\ \mathcal{H}[r])} t}{(\text{memory } \mathcal{H} : s) \xrightarrow{\tau} (\text{memory } \mathcal{H} : t)} \text{ MEM-LOAD}$$

The rule MEM-STORE-EXEC produces MEM_STORE actions in order to store a value for a specific reference on the heap. The rule MEM-STORE handles those actions by storing the value on the heap:

$$\frac{s \xrightarrow{\texttt{(MEM\_STORE } r \ v\texttt{)}} t}{(\texttt{memory } \mathcal{H} : s) \xrightarrow{\tau} (\texttt{memory store}(\mathcal{H}, \ r, \ v) : t)} \ \text{\small MEM-STORE}$$

Finally, the rule MEM-NEW-EXEC produces MEM_NEW actions in order to create and initialize fresh references. The rule MEM-NEW handles those actions:

$$\frac{s \xrightarrow{\texttt{(MEM\_NEW } r \ v\texttt{)}} t \quad (\mathcal{H}', \ r) = \texttt{new}(\mathcal{H}, \ v)}{(\texttt{memory } \mathcal{H} : s) \xrightarrow{\tau} (\texttt{memory } \mathcal{H}' : t)} \ \text{\small MEM-NEW}$$

The primitive new operator takes two arguments, a heap $\mathcal{H}$ and a value $\nu$, and returns a sequence term of the form $(\mathcal{H}', \ r)$ where $r$ is a fresh numeric reference that has no value assigned to it according to the previous heap $\mathcal{H}$, i.e., $r \notin \mathrm{dom}(\mathcal{H})$, and $\mathcal{H}'(r) = \nu$. Note that the identity of a reference might make a semantic difference via the built-in function id and the hash of an object. Our implementation concretizes the function new. By ranging over all functions satisfying the given requirements, our semantics allows capturing all possible behaviors induced by different identities.

## 4.5 Evaluation

To evaluate the proposed semantics with respect to its completeness and correctness, we cash in on the advantages of having it formalized in our meta-theoretic framework. The inference rules presented in Section 4.3 and Section 4.4 have been exported to LATEX for the purpose of presentation. From the same definitions, our tool automatically derives an interpreter applying the inference rules to execute a given program. Enabled by the derived interpreter, we evaluate our semantics by executing test cases.

In this section, we (i) describe the process of assembling a test suite, (ii) compare our semantics to other strains of research on formal semantics for Python, and (iii), based on the thereby obtained insights, discuss the limitations of our semantics. In addition, we discuss a few improvements made to Algorithm 1 to speed up the execution of test cases.

### 4.5.1 Assembling a Test Suite

Prior work on formal semantics for Python already established a quite extensive suite of test cases. We base our evaluation on the test suites developed by Guth [27] and Politz et al. [44]. Together their test cases already cover a broad range of different language features. For a few corner cases, where we felt their test suites were lacking, we developed additional tests.

The test suite of Politz et al. has been developed for Python 3.2 [44] while the test suite of Guth has been developed for Python 3.3 [27]. Prior to their work, there has been work on formal semantics for Python 2 [53]. Our test suite does not include those tests for Python 2 because there have been quite a few changes to the core language since then. For the same reason, we do not compare our semantics to the semantics for Python 2.

The test cases due to Guth [27] and Politz et al. [44] have in common that they rely on assertions. Snippet 4.11 shows an example of a test due to Guth for testing `for` loops. To asses whether a semantics passes such a test, one needs a means to decide whether the program executed completely and threw an exception or returned normally. Because of the incompleteness of all existing semantics with respect to the built-ins they support [27, 44, 22] it is expected that some tests throw exceptions that do not indicate an actually erroneous semantics but rather an incomplete one. For the purpose of this thesis, we adopt the following criteria for evaluating a semantics based on a test suite based on assertions. A semantics *fails* a test if it throws an `AssertionError` or does not execute the test completely. A semantics *passes* a test if it executes the test completely without throwing an exception. In case of another exception than `AssertionError`, we assume the semantics to be incomplete with regard to the language features utilized by the test.

```python
1  y = 0
2  for x in (5,6):
3      y = y + 1
4  else:
5      assert x == 6 and y == 2
6      z = 1
7  assert z == 1
```

Snippet 4.11: Test `testfor1.py` of the test suite by Guth [27].

The test cases have been sourced from the public GitHub repositories of $\mathbb{K}$ Python[8] and $\lambda_\pi$[9]. In total, the test suite comprises 536 test cases, 277 due to Guth, 253 due to Politz et al., and 6 developed by ourselves.

### 4.5.2  Comparison to Related Work

There are three main strains of related work on formal semantics for Python 3: Guth [27] developed a formal semantics formalized in the $\mathbb{K}$ framework [49], Politz et al. developed a formal semantics, coined $\lambda_\pi$, using Racket[10] [44], and Fromherz et al. [22] and Monat et al. [39] proposed a big-step formal semantics developed as part of the analysis tool Mopsa [33]. We compare our own semantics to their efforts by means of the test suite we assembled. Appendix B.4 contains a table showing the test results for various versions of CPython and the different strains of existing work. These test results are the baseline for the comparison and evaluation of our semantics.

---

[8] https://github.com/kframework/python-semantics (Accessed: 2020/12/08)
[9] https://github.com/brownplt/lambda-py (Accessed: 2020/12/08)
[10] https://racket-lang.org/ (Accessed: 2020/12/14)

**Comparison to $\mathbb{K}$ Python**

The formal semantics due to Guth in the $\mathbb{K}$ framework seems to be the most complete and developed semantics [27]. Unfortunately, we were unable to reproduce the experiments conducted by Guth. The required version of the $\mathbb{K}$ framework, the formal semantics has been developed with, is no longer available. We contacted the authors of the $\mathbb{K}$ framework as well as Guth himself for help, however, parts of the required version of the $\mathbb{K}$ framework have apparently been removed from the repository as part of their migration from SVN to Git [40]. After trying to source the missing parts from other sources, we finally gave up. According to Guth, the semantics probably also needs to be completely rewritten in order to work with the newest version of the $\mathbb{K}$ framework [40]. Being unable to conduct tests ourselves, we can only base our comparison on the results presented by Guth.

Apparently, the semantics by Guth passes all the tests of their own test suite and 186 out of the 224 tests[11] [27] of the test suite of $\lambda_\pi$. In comparison, our semantics passes 159 tests of the suite by Guth and 128 tests of the suite of $\lambda_\pi$ (cf. Appendix B.4). As there seem to be no detailed records on which tests the semantics passes, a more detailed comparison is impossible.

Taking just the number of passed tests into account, the semantics by Guth within the $\mathbb{K}$ framework seems to be more complete than our own semantics. However, the semantics has been abandoned and without rewriting it from scratch based on the latest version of $\mathbb{K}$ there seems to be little hope that one can derive any benefit from the semantics.

A built-in function the semantics due to Guth does not support[12] is `print`. Thus, a program run with their semantics cannot print anything. In contrast, we support the built-in function `print` and our semantics emits a visible side effect in terms of a PRINT action (cf. Section 4.3.1).

**Comparison to $\lambda_\pi$**

The $\lambda_\pi$ semantics due to Politz et al. using Racket [44] has been available in their GitHub repository. To asses their semantics, we compiled it with Racket version 7.2 on Ubuntu 20.04. As already shown by Guth their semantics is often incorrect [27]. Out of the 536 tests of our test suite, the semantics actually failed 29 tests. In comparison, our own semantics fails only 4 tests that, in our opinion, test implementation details.

---

[11] According to our counting there are 277 test cases. We counted all Python files in the directory `tests` of the repository as test cases.

[12] This assessment is based on the `README.md` of their GitHub repository https://github.com/kframework/python-semantics (Accessed: 2020/12/08).

Regarding completeness, $\lambda_\pi$ passes 329 tests of our test suite while our semantics passes 293 tests. Neither semantics subsumes the other because there are certain language features only $\lambda_\pi$ or only our semantics covers. For instance, we do neither support augmented assignments nor unpacking assignments, while $\lambda_\pi$ supports them to some extent. In contrast, the test results suggest that our implementation of Python's data model is more complete and, more importantly, correct (cf. Appendix B.4).

Guth attributes the incorrectness of $\lambda_\pi$ to the attempt to specify large parts of the semantics of built-ins by implementing them in pure Python:

> *"The blame for this [incorrect behavior] once again is solely at the feet of the design of the semantics: it implements much of the standard library, where it does at all, using pure Python code, despite the fact that many features of many built-in functions are simply not expressible in this fashion, because they rely on underlying primitives that Python does not expose."*
>
> — Dwight Guth, 2013 [27, p. 43]

The solution of Guth is to specify the behavior of built-in functions and types mostly on the meta-theoretical level using the primitives provided by the $\mathbb{K}$ framework. Our semantics constitutes a middle ground between those two extremes. We spent a considerable effort on figuring out what primitives one has to choose in order to arrive at a semantics that is correct but still specifies the behavior of most of the built-in functions and types using Python code carefully extended with the necessary primitives.

Guth bases his evaluation on the ground that "60 tests failed [on $\lambda_\pi$] with incorrect behavior or failed assertions" [27, p. 43]. Note that we only count failed assertions as incorrect behavior because the tests should be designed in a way such that they check the behavior with assertions. We believe that the discrepancy between the 29 failed tests we counted and the 60 failed tests Guth counted is in part caused by counting "incorrect behavior" besides assertions as failures and in part because we based our evaluation on the newest version of $\lambda_\pi$ taken from the Git repository where most likely some of these problems have already been fixed.

For generators, $\lambda_\pi$ relies on desugaring functions into classes [44]. While we also desugar some language constructs, the desugaring we perform is not as involved as what is necessary for $\lambda_\pi$. We take it to be an advantage of our semantics, that it does not perform complex translations whose correctness is questionable and cannot be independently established.

**Comparison to MOPSA**

The work of Fromherz et al. [22] on the static analysis of Python utilizing abstract interpretation has been continued by Monat et al. [39]. The fruits of their work have been implemented in the MOPSA static analysis framework [33]. Unfortunately, for the time being, there is no interpreter based on the concrete semantics underlying the abstract semantics available. Nevertheless, Monat suggested to run MOPSA on the test cases and check whether it complains about an assertion [41]. Clearly, comparing a static analysis tool with a concrete semantics in that way is not ideal.

MOPSA complains about an assertion if it cannot verify that this assertion will not lead to an `AssertionError` being thrown at runtime. If MOPSA complains about an assertion this does, however, not necessarily imply that this assertion will be thrown by the concrete semantics. Hence, a failed test does not imply that the concrete semantics the analysis is based on is erroneous in any way. Nevertheless, a failed test is at least an indication that the analysis could be improved. In addition, tests that neither fail nor succeed allow us to gain insights about the completeness of MOPSA.

Based on our experiments, MOPSA passes 138 tests, fails 106 tests, and is incomplete with respect to the remaining 292 tests. We would like to emphasize again that the amount of failed tests does not mean that the underlying semantics is erroneous, although it might very well be in some of these cases. The amount of tests MOPSA does not support indicates that our semantics is more complete than the semantics underlying MOPSA. Again some of these tests might not succeed due to the over-approximative nature of the analysis MOPSA performs. However, out of the 292 tests that neither fail nor succeed, 153 tests cause an internal error in MOPSA and 4 tests contain unsupported syntax. A further investigation of why certain tests fail or do not succeed is out of scope of this thesis.

MOPSA is based on a big-step semantics [22, 39]. In contrast, our semantics is a small-step semantics naturally allowing for concurrency by interleaving and interaction with an environment. In general, the aim of MOPSA is to be an analysis tool, while our focus has been entirely on the semantics.

### 4.5.3 Limitations

Recapitulate that out of the 536 tests our semantics passes 293 tests, fails 4 tests, and does not support the remaining 239 tests. We consider those tests our semantics does fail implementation details of CPython.

Out of the 239 tests our semantics does not support, 145 tests are outside of the fragment we identified in Section 2.1.2. Most of these tests utilize augmented assignments, unpacking assignments, or the `import` statement. The other 94 tests utilize built-ins we do not fully support.

We expect that extending our semantics with the missing built-in functions and types is mostly straightforward. Instead of touching the inference rules, additional built-in functions and types can be added by implementing them with the primitives we extended Python with. Introducing new primitives,

e. g., operations on primitive strings, in the sense of the `apply` mechanism, is fairly easy. Therefore, we are confident that a significant part of the missing built-ins can be covered with some additional effort.

Adding support for augmented assignments, unpacking assignments, and the `import` statement should also be possible. As we argued in Section 2.1.2, augmented assignments and unpacking assignments are mostly syntactic sugar. Supporting the `import` statement is possible in different ways. One approach would be to add an additional *import layer* to the semantics which stores code of modules that can be imported. Another approach would be to signal that the executing code would like to import a module by means of actions to the environment similar to how printing works and then supply the module via another action. We consider the exact mechanism how imports are resolved not to be a part of the semantic core of Python.

An area where our semantics currently lacks is the semantics of garbage collection. The reference implementation uses reference counting and an additional garbage collector for reference cycles [20, §3.3.1.]. The PLR does not seem to stipulate any particular mechanism for garbage collection and other implementations use other strategies. For instance, PyPy does not use reference counting.[13] Our semantics does not cover garbage collection and as a result destructors are never called. We did not investigate the possibility to add garbage collection to our semantics in further detail. Existing work on formal semantics for Python seems to leave aside the formalization of garbage collection as well [27, 44, 22].

Another area where our semantics currently lacks is the possibility to generate stack traces for exceptions. In principle, it should be possible to generate a stack trace based on the information available at the stack layer, however, we did not investigate this in further detail. For now, our semantics simply does not assign stack traces to exceptions.

There are a few other rather minor limitations our semantics currently has. However, discussing all of them in detail, would not be insightful. Despite its limitations, we take our semantics to be a step forward with respect to the formalization of Python. The presented semantics at least serves as an elaborate case study of an SOS-style semantics for Python and as a starting point for future endeavours in that direction.

### 4.5.4 Algorithmic Improvements

While testing our semantics, we noticed that the running time of test cases increased with the complexity of the semantics. The average running time over all test cases is $90\,s$ per test case.[14] The average number of transitions is $108\,664$ per test case (cf. Appendix B.4). To bring down the time required to execute a test case, we modified Algorithm 1 for our purposes. Take the simple program x = 6 as an example. Without these modifications, our tools needs 17.46 min to execute the 1 462 transitions of this simple program. In contrast, with our modifications, the tool only needs 2.6 s.

---

[13] https://doc.pypy.org/en/latest/gc_info.html (Accessed: 2020/12/08)
[14] Executed on a machine with an Intel i7-6700K at $4\,GHz$ and $32\,GB$ RAM.

Algorithm 1 constructs inferences trees starting at the bottom of the trees it constructs. For our semantics, at the bottom, all rules for the memory layer apply giving rise to respective branches in the search graph. The algorithm cannot know which rule to choose until it constructed almost the whole tree and found out whether the program state can perform any of the memory actions or not. The subtrees above the memory layer are almost identical, however, no information is reused and the algorithm quite ineffectively constructs the same subtrees over and over again. Analogously, the stack layer, the frame layer, and other rules handling actions blow up the search space because which action the inner-most syntactic construct is able to perform is usually only decided at the leaves of the inference tree.

Instead of using Algorithm 1 for the execution of test cases, we developed another algorithm which constructs inference trees starting at their leaves. The algorithm exploits some properties of the inference rules of our semantics. For instance, the meta variables occurring on the left-hand side of a transition within a premise are usually bound by mere syntactical decomposition of the left-hand side of the inference question asked. Intuitively, this allows us to destruct the program state for which we would like to infer a transition until a rule is applicable fixing the kind of action to be performed. Working our way backwards through the applicable inference rules, an algorithm constructing trees starting at their leaves is obtained. The actual algorithm used for the execution of test cases is a bit more involved because not all inference rules satisfy this property. For instance, the rule STACK-SEND-VALUE contains the meta variable $\tilde{\mathfrak{s}}$ at the left-hand side of the transition of the second premise. However, this variable is not bound by syntactical decomposition of the program state but instead is provided by the action associated with the first premise. For details we refer to our implementation [35].

Another modification improving the execution speed has been to dynamically collapse multiple $\tau$ transitions into a single $\tau$ transition. As our semantics is deterministic, this optimization in essences applies a minimization of the resulting LTS semantics regarding its observable behavior.

# Chapter 5

# Epilog

Aiming at an executable structural operational formal semantics for Python, we developed a meta-theoretic framework for the formalization of structural operational semantics and demonstrated its feasibility by developing a semantics for Python. In this chapter, we conclude this thesis by (1) comparing our meta-theoretic framework to existing work on the mechanization of formal semantics, (2) discussing the contributions made by this thesis, and (3) giving an outlook on opportunities for future work.

## 5.1  Related Work

The idea to meta-theoretically capture definitions of formal semantics is not new. While work on *rule formats* focused on the theoretical investigation of properties of structural operational semantics [24], several approaches and tools have been proposed to aid the development of formal semantics within a meta-theoretic framework [49, 12, 10]. Like our framework, these tools allow deriving an interpreter automatically. Some of these frameworks also aim to provide analysis tools fully automatically.

**The $\mathbb{K}$ Framework**   The $\mathbb{K}$ semantic framework due to Roșu and Șerbănuță is a rewriting-based framework for the definition of formal semantics for programming languages [49]. In their paper, Roșu and Șerbănuță discuss several desirable features for language definitional frameworks and also compare their approach to small-step SOS [49, p. 399].

In contrast to the $\mathbb{K}$ framework, where rewrite rules apply globally on the program state [49], SOS rules are *hierarchic* in nature. Roșu and Șerbănuță claim that "SOS transitions only apply at the top of the term" [49] which is strictly speaking not correct. As we have seen, inference rules can modify parts of the program state which are not "at the top." For instance, our rules for the `while` statement modify the program state based on the action produced by the loop's body and do not apply only "at the top" of the term. What is true, however, is that SOS inference rules are more limited in the scope where they can apply changes. Over a framework where rules may just "apply anywhere" [49] we believe SOS has the advantage of being more *structural* and, hence, allows for a more clear and organized formal semantics because the scope where rules can perform changes is limited. So, while we generally agree with the assessment of Roşu and Şerbănuţă, we draw a different conclusion and believe that the alleged disadvantage to limit the scope where rules can perform changes is in fact an advantage.

There are two more disadvantages Roşu and Şerbănuţă ascribe to small-step SOS: (a) a lack of modularity and (b) the complexity of computing an execution step [49]. Leveraging action labels for control-flow one can, however, achieve a certain degree of modularity as evidenced by our Python semantics. For instance, the intraprocedural layer can simply be replaced by rules for another language while keeping the rules of the other layers. With regard to computational complexity, we proved that the problem of finding execution steps is in general undecidable in our framework. With certain syntactical restrictions on the form of inference rules, however, the problem becomes decidable, for instance, if the terms always shrink by syntactical decomposition. While we did not investigate the computational complexity of our approach, with the modifications of the inference algorithm (cf. Algorithm 1) discussed in Section 4.5.4, the complexity of computing an execution step in our framework does not prevent us from evaluating our semantics with test cases. The Python semantics written in an old version of the $\mathbb{K}$ framework also has been too slow too be useful beyond mere evaluation [27, 40].

Our framework stays close to how SOS are used in existing research and thus semantic definitions in our framework are easy and intuitive to understand if one is in general familiar with SOS. Recapitulate the inference rules for CCS (cf. Section 2.3) which have been formalized in our framework exactly as they appear in the literature [38, 37]. There is no need to learn the details of the framework or even know that it exists. For the same reason, learning how to write semantic definitions using our framework should be straightforward for anyone familiar with SOS. We believe this to be an advantage over the $\mathbb{K}$ framework and hope that it fosters adoption by the community.

Clearly, the $\mathbb{K}$ framework is a much more mature and developed tool. It also offers features beyond the mere specification of formal semantics like deriving verification tools from formal semantics [56].

**Process Algebra Compiler**   The *Process Algebra Compiler* (PAC) is one of the earliest works aiming to mechanize structural operational semantics dating back to 1995 [12]. Like our tool, it produces an interpreter for state space exploration based on the provided inference rules. It does so by compiling the rules using one of two backends. It also allows adding custom data values by implementing them in a backend specific language.

Our meta theory can be seen as a conservative extension of the PAC. In particular, the PAC imposes certain restrictions on the form of inference rules, which we do not. According to the theory underlying the PAC, a relation, e. g., the transition relation, defined with inference rules is required to have *inputs*. The PAC then generates code for computing the relation based on its inputs. For instance, the source state of the transition relation may be the input and the action and target state are what should be computed. The PAC would then generate code from the inference rules to compute the action and target state based on the source state. In this sense, the PAC is very similar to our approach. For compilation, the PAC imposes several restrictions on the form of inference rules which restricts the "flow of data through a rule: information flows from the inputs of the conclusion to the [inputs of the] premises, and the outputs of the premises flow (together with inputs of the conclusion) to the side condition and the outputs of the conclusion" [12, p. 166]. In contrast, our approach does not pose such restrictions on the inference rules. Take the rule MEM-LOAD-EXEC (cf. Section 4.3.1) as an example. The action is neither an input nor an "output" of this rule but the "flow of data" through the different inference rules involved in loading is more complex.

With our theory we are able to handle such complex flow and also allow operator terms to occur almost anywhere thanks to the theory of semantic term unification we developed (cf. Section 3.2). Certainly, we need restrictions on the inference rules as well to prevent not agnostically solvable problems from arising. The restrictions imposed by the PAC ensure that all problems arising from a formal semantics are agnostically solvable as it determines a fixed order in which meta variables are bound and it does not allow placing operator terms freely.[1] While these restrictions are sufficient, they are not necessary to ensure that problems are always agnostically solvable. In fact, our Python semantics violates some of these restrictions and still no problems arise which are not agnostically solvable. It remains for future work to syntactically characterize the agnostically solvable fragment.

**Skeletal Semantics**  Recent work by Bodin et al. proposed *sekletal semantics* [10] as a means to mechanize structural operational semantics and derive both, concrete and abstract interpretations, from a formal semantics. To this end, *skeletons* have to be specified which abstract over concrete and abstract inference rules. Skeletons add another layer of abstraction for the specification of semantics while our framework is based directly on inference rules. Whether our framework can be extended to derive an abstract interpreter from inference rules remains to be investigated.

## 5.2  Discussion

An explicit goal of this thesis was to assess whether the SOS framework is in general suitable for the formalization of Python (cf. Chapter 1). As the PLR is often vague or incomplete, we aimed at an executable semantics so that we can convince ourselves of its correctness by executing test cases as done in Section 4.5. To this end, and due to a lack of existing tools, we developed our meta-theoretic framework in tandem with the necessary tool support for the automated derivation of interpreters from a formal semantics. Let us now discuss whether we meet the goals set for the meta-theoretic framework (cf. Chapter 3) as well as the feasibility and adequacy of the overall approach to a formal semantics for Python taken in this thesis.

---

[1]  From the material on the PAC we surveyed, it is unclear whether there exists an analogon to operator terms and how it interfaces with the remainder of the PAC. A user can define *functions*, however, given how the PAC is described those cannot occur arbitrarily.

**Meta-Theoretic Framework**

The main goals for our meta-theoretic framework were (a) a small language-agnostic core, (b) the ability to formalize semantics in a way that closely resembles how SOS are used in research, and (c) the ability to automatically derive an interpreter from a formal semantics.

As evidenced by our implementation, the core of the meta-theoretic framework is indeed small. We managed to implement the proposed algorithms in roughly 300 lines of Python code. By staying agnostic on data values and operators thereon, it is easy to adapt our framework for a particular language. For CCS we were able to introduce the complement operator for actions with just a few lines of code (cf. Section 3.4.2). For Python, we introduced the `apply` operator and a few other operators like `compare`. Thus, it is safe to say, that the framework has a small language-agnostic core.

As evidenced in Section 2.3 and Section 3.4.2, our framework allows us to capture inference rules of structural operational semantics as they appear in the literature and introductory courses. Having used the framework for the quite involved Python semantics, we are confident that it is also fairly easy to capture other semantics and formalisms making use of inference rules. How far this applicability goes needs further investigation.

With the optimizations tailored to structural operational semantics discussed in Section 4.5.4, the tool is able to cope with complex and large semantics such as our Python semantics. For semantics with less complexity and inference rules, the algorithm provided in Section 3.3 is feasible as well. The derived interpreter is clearly not performant enough to compete with the reference implementation or to be used beyond evaluating the semantics. However, it is useful for the evaluation of the semantics.

**SOS for Python**

For the purpose of this thesis, we took the decision to develop a structural operational semantics as a bedrock. We expected to thereby establish a nice correspondence between syntax and semantics.

The process of translation is, however, more involved than initially anticipated. Nevertheless, while we deviate from the AST quite a bit for some constructs like `for` statements, binary expressions, or identifiers, these translations are still straightforward and easy to grasp. Especially, when compared with the quite complex transformations Politz et al. deploy for $\lambda_\pi$ [44]. Still, the question arises whether it would be better to define the formal semantics on a lower-level of abstraction like Python's bytecode as suggested by Mark Shannon [1]. To our knowledge, all existing formalizations for Python are based on the AST and not on the bytecode of Python. Investigating a formalization of Python's bytecode VM may be worthwhile.

Independent on how one formalizes the semantics of statements and expressions, one has to base the data model of Python on a rigorous foundation. By identifying a parsimonious set of primitives, we managed to succinctly define the semantic core of the language with just 98 inference rules. By using actions as a universal means to handle intra- and inter-procedural control flow, we believe to have defined a semantics that is comparably easy to comprehend and intelligible. Concerning the adequacy criteria defined in Section 2.2.1, we believe that our semantics satisfies most of them. The semantics is correct and covers a significant subset of Python as evidenced in Section 4.5. It is based on a parsimonious set of primitives enabling to write built-ins in an extended variant of Python. In particular, we were able to discuss each and every inference rule comprising the semantics in this thesis. Hence, the semantics is rather succinct and intelligible especially when compared to prior work. At the same time, we are confident that capturing the missing parts of Python is possible with some additional effort. Especially with regard to the built-in functions and types, primitives can be added as needed via the `apply` mechanism. These primitives can then be used to implement what is missing in the variant of Python extended with primitives.

## 5.3 Future Work

Future work on formalizing Python's semantics should focus on further carving out primitives and separating implementation details and built-ins from the *semantic core* of the language. To this end, we propose to develop a unified test suite focusing on a small semantic core where each test covers a distinctive feature of the language. Many tests of the existing test suites unnecessarily rely on language features such as augmented assignments in order to test other language features. We believe that a unified test suite endorsed by the *Python Software Foundation* (PSF) would help advance research on the formalization and analysis of Python. Formalizing Python should not be a one-way-street but results produced by academia should be fed back into the language specification itself.

While we explored the idea to formalize Python based on the general framework of structural operational semantics as pioneered by Gordon Plotkin [42], other ideas might be worth pursuing. In particular, other more low-level rep-resentations of Python programs might be more suitable for the formalization of Python. Mark Shannon recently proposed to formalize the Python bytecode VM [1]. Existing research, including this thesis, does not make any attempts in that direction except providing general insights.

With regard to the meta-theoretic framework developed in this thesis, a syntactic characterization of the agnostically solvable fragment is highly desirable but has been postponed to future work to stay within scope. We believe that the key to characterizing this fragment is to investigate the order in which variables can and should be bound in order to prevent not agnostically solvable problems from arising. Syntactic restrictions on inference rules such as the ones imposed by the PAC [12] (cf. Section 5.1) allow defining a fixed order in which solutions for variables are obtained. Thereby, such restrictions guarantee that all problems arising from a formal semantics will be agnostically solvable proving completeness of the approach.

While we deployed the meta-theoretic framework for structural operational semantics specifically, we believe it to be useful beyond defining formal semantics. Future work should focus on investigating the limits of the framework and unveil other potential areas of application. In particular, it might turn out valuable for teaching purposes allowing students to play with and explore formalisms based on inference rules.

Concerning the semantics put forward in this thesis, there is ample room for improvements as described in Section 4.5.3. For once, the missing syntactic features such as augmented assignments or support for `import` statements can be added. In addition, the missing built-ins can be implemented. However, we believe those improvements to be of little interest from an academic perspective. With our semantics, we already capture most if not all "hard parts" [44] of the semantics like generators and scoping.

## 5.4 Conclusion

With the vision of harvesting static analysis techniques like abstract interpretation for Python, we developed an executable structural operational formal semantics covering a significant subset of Python. To our best knowledge, we are the first who investigated the general feasibility of the SOS framework for the development of a formal semantics for Python. We innovated by developing a flexible and language-agnostic meta-theoretic framework for the formalization of structural operational semantics. To this end, we established the theory of semantic term unification. The meta-theoretic framework enabled the automated derivation of an interpreter from the inference rules of our Python semantics. We cashed in on this advantage and evaluated our semantics with respect to the reference implementation and pre-existing work by executing a suite of test cases. We thereby evidenced the correctness of the semantics and evaluated its completeness.

Defining a formal semantics for Python should not be a one-way street. To be successful, insights gained from the formalization process should flow back into the language specification itself. This thesis explored the possibility to define the semantics of Python as a structural operational semantics which turned out to be feasible although the semantics is still limited. However, these limitations should be fairly easy to overcome with additional effort because we formalized most if not all "hard parts" [44] of Python. The approach taken in this thesis, i.e., to define the semantics within a meta-theoretic framework enabling the generation of an interpreter, turned out fruitful to assess the quality of the semantics. We conclude that the SOS framework is feasible for the formalization of Python and that the formalization project should, independently of the semantic framework, further focus on entangling the intricacies of the Python language by separating the semantic core of the language from implementation details and built-ins.

# Appendix A

# Implementation

**rigorous/core/unification.py**

```python
1   # -*- coding:utf-8 -*-
2   #
3   # Copyright (C) 2020, Maximilian Köhl <mail@koehlma.de>
4   #
5   # fmt: off
6
7   from __future__ import annotations
8
9   import dataclasses as d
10  import typing as t
11
12  import collections
13
14  from . import terms
15
16
17  Equation = t.Tuple[terms.Term, terms.Term]
18
19
20  class NoSolutionError(Exception):
21      pass
22
23
24  def _substitute(
25      term: terms.Term, substitution: terms.Substitution
26  ) -> terms.Term:
27      for variable in term.variables:
28          if variable in substitution:
29              return term.substitute(substitution)
30      return term
31
32
33  @d.dataclass(eq=False)
34  class Solver:
35      _failure: bool = False
36      _deferred: t.Set[Equation] = d.field(default_factory=set)
37      _pending: t.Deque[Equation] = d.field(default_factory=collections.deque)
38      _solutions: t.Dict[terms.Variable, terms.Term] = d.field(
```

```
39          default_factory=dict
40      )
41
42      def _reintegrate_deferred(self) -> None:
43          self._pending.extend(self._deferred)
44          self._deferred.clear()
45
46      def add_equation(self, equation: Equation) -> None:
47          self._pending.append(equation)
48          self._reintegrate_deferred()
49
50      def add_equations(self, equations: t.Iterable[Equation]) -> None:
51          self._pending.extend(equations)
52          self._reintegrate_deferred()
53
54      @property
55      def has_no_solutions(self) -> bool:
56          self.solve()
57          return self._failure
58
59      @property
60      def is_solved(self) -> bool:
61          self.solve()
62          return not self._deferred and not self._failure
63
64      @property
65      def not_agnostically_solvable(self) -> bool:
66          self.solve()
67          return bool(self._deferred and not self._failure)
68
69      @property
70      def solution(self) -> t.Mapping[terms.Variable, terms.Term]:
71          self.solve()
72          return self._solutions
73
74      def merge(self, other: Solver) -> None:
75          assert not other._failure
76          assert self._solutions.keys().isdisjoint(other._solutions.keys())
77          self._solutions = {
78              variable: solution.substitute(other._solutions)
79              for variable, solution in self._solutions.items()
80          }
81          for variable, solution in other._solutions.items():
82              self._solutions[variable] = solution.substitute(self._solutions)
83          self._pending.extend(other._pending)
84          self._pending.extend(other._deferred)
85          self._reintegrate_deferred()
86
87      def solve(self) -> None:
88          did_discover_solutions: bool = False
89          while self._pending and not self._failure:
90              equation = self._pending.popleft()
91              left = _substitute(equation[0], self._solutions).evaluated
92              right = _substitute(equation[1], self._solutions).evaluated
93              if left is None or right is None:
94                  self._failure = True
95                  return
96              if isinstance(right, terms.Variable):
97                  if not isinstance(left, terms.Variable):
98                      left, right = right, left
```

```python
                if isinstance(left, terms.Variable):
                    if left in right.unguarded_variables:
                        self._failure = True
                        return
                    self._solutions, solutions = {}, self._solutions
                    for variable, solution in solutions.items():
                        new_solution = _substitute(
                            solution, {left: right}
                        ).evaluated
                        if new_solution is None:
                            self._failure = True
                            return
                        self._solutions[variable] = new_solution
                    self._solutions[left] = right
                    did_discover_solutions = True
                elif isinstance(left, terms.Sequence):
                    if isinstance(right, terms.Sequence):
                        if left.length != right.length:
                            self._failure = True
                            return
                        self._pending.extend(zip(left.elements, right.elements))
                    elif not isinstance(right, terms.Apply):
                        self._failure = True
                        return
                    else:
                        self._deferred.add((left, right))
                elif left == right:
                    if left.guarded_variables:
                        self._deferred.add((left, right))
                else:
                    if not left.is_operator and not right.is_operator:
                        self._failure = True
                        return
                    self._deferred.add((left, right))
            if (
                not self._pending
                and not self._failure
                and self._deferred
                and did_discover_solutions
            ):
                did_discover_solutions = False
                self._reintegrate_deferred()

    def clone(self) -> Solver:
        return Solver(
            _failure=self._failure,
            _deferred=set(self._deferred),
            _pending=collections.deque(self._pending),
            _solutions=self._solutions,
        )

def get_solution(
    solutions: terms.Substitution, variable: terms.Variable
) -> terms.Term:
    try:
        return solutions[variable]
    except KeyError:
        for key, value in solutions.items():
            if value == variable:
```

```
159                    return key
160            raise NoSolutionError(f"no solution for variable {variable}")
161
162
163    def match(
164        pattern: terms.Term, term: terms.Term
165    ) -> t.Optional[terms.Substitution]:
166        solver = Solver()
167        solver.add_equation((pattern, term))
168        if not solver.is_solved or solver.solution.keys() != pattern.variables:
169            return None
170        return solver.solution
```

### rigorous/core/inference.py

```
1    # -*- coding:utf-8 -*-
2    #
3    # Copyright (C) 2020, Maximilian Köhl <mail@koehlma.de>
4    #
5    # fmt: off
6
7    from __future__ import annotations
8
9    import dataclasses as d
10   import typing as t
11
12   import collections
13   import enum
14   import functools
15   import itertools
16
17   from . import terms, unification
18
19
20   class Verdict(enum.Enum):
21       SATISFIABLE = "satisfiable"
22       VIOLATED = "violated"
23       SATISFIED = "satisfied"
24
25
26   class Condition(t.Protocol):
27       def get_verdict(self, substititon: terms.Substitution) -> Verdict:
28           pass
29
30
31   Premises = t.Tuple[terms.Term, ...]
32   Constraints = t.Tuple[unification.Equation, ...]
33   Conditions = t.Tuple[Condition, ...]
34
35
36   @d.dataclass(frozen=True)
37   class Rule:
38       conclusion: terms.Term
39       premises: Premises = ()
40       constraints: Constraints = ()
41       conditions: Conditions = ()
42       name: t.Optional[str] = None
```

```
43
44        @functools.cached_property
45        def variables(self) -> t.AbstractSet[terms.Variable]:
46            variables = set(self.conclusion.variables)
47            for premise in self.premises:
48                variables |= premise.variables
49            for left, right in self.constraints:
50                variables |= left.variables | right.variables
51            return frozenset(variables)
52
53
54    @d.dataclass(frozen=True)
55    class Instance:
56        rule: Rule
57        substitution: terms.Substitution
58
59        def __post_init__(self) -> None:
60            assert self.rule.variables == self.substitution.keys()
61
62        def _instantiate(
63            self, term: terms.Term, *, evaluate: bool = True
64        ) -> terms.Term:
65            result = term.substitute(self.substitution)
66            if evaluate:
67                assert result.evaluated is not None
68                result = result.evaluated
69            return result
70
71        @functools.cached_property
72        def conclusion(self) -> terms.Term:
73            return self._instantiate(self.rule.conclusion)
74
75        @functools.cached_property
76        def premises(self) -> t.Sequence[terms.Term]:
77            return tuple(
78                self._instantiate(premise) for premise in self.rule.premises
79            )
80
81        @functools.cached_property
82        def constraints(self) -> t.Sequence[unification.Equation]:
83            return tuple(
84                (self._instantiate(left), self._instantiate(right, evaluate=False))
85                for left, right in self.rule.constraints
86            )
87
88
89    @d.dataclass(frozen=True)
90    class Tree:
91        instance: Instance
92        premises: t.Tuple[Tree, ...]
93
94
95    Question = terms.Term
96
97
98    @d.dataclass(frozen=True)
99    class Answer:
100        substitution: terms.Substitution
101        tree: Tree
102
```

```
103
104  @d.dataclass(eq=False)
105  class System:
106      _rules: t.List[Rule]
107
108      def __init__(self, rules: t.Optional[t.Iterable[Rule]] = None) -> None:
109          self._rules = []
110          if rules is not None:
111              for rule in rules:
112                  self.add_rule(rule)
113
114      def add_rule(self, rule: Rule) -> None:
115          self._rules.append(rule)
116
117      @property
118      def rules(self) -> t.Sequence[Rule]:
119          return self._rules
120
121      def iter_answers(
122          self, question: Question, *, depth_first: bool = False
123      ) -> t.Iterator[Answer]:
124          variables = question.variables
125          queue = collections.deque([_Node.create_root(self, question)])
126          while queue:
127              head = queue.popleft()
128              if depth_first:
129                  queue.extendleft(head.expand())
130              else:
131                  queue.extend(head.expand())
132              if head.is_solved and variables <= head.solver.solution.keys():
133                  assert head.solver.is_solved, "not agnostically solvable"
134                  substitution = head.solver.solution
135                  yield Answer(
136                      {
137                          variable: substitution[variable]
138                          for variable in variables
139                      },
140                      head.construct_tree(question),
141                  )
142
143
144  @d.dataclass(frozen=True, eq=False)
145  class _RenamedRule:
146      original: Rule
147      renaming: terms.Renaming
148      conclusion: terms.Term
149      premises: Premises = ()
150      constraints: Constraints = ()
151
152
153  def _create_renamed_rule(rule: Rule) -> _RenamedRule:
154      renaming = {variable: variable.clone() for variable in rule.variables}
155      return _RenamedRule(
156          original=rule,
157          renaming=renaming,
158          conclusion=rule.conclusion.substitute(renaming),
159          premises=tuple(
160              premise.substitute(renaming) for premise in rule.premises
161          ),
162          constraints=tuple(
```

```python
                   (left.substitute(renaming), right.substitute(renaming))
                   for left, right in rule.constraints
               ),
       )

@d.dataclass(frozen=True, eq=False)
class _Node:
    system: System

    solver: unification.Solver

    renamed_rules: t.Mapping[terms.Term, _RenamedRule]

    pending_terms: t.Tuple[terms.Term, ...]
    pending_conditions: t.FrozenSet[t.Tuple[_RenamedRule, Condition]]

    @classmethod
    def create_root(cls, system: System, term: terms.Term) -> _Node:
        return cls(system, unification.Solver(), {}, (term,), frozenset())

    @property
    def is_solved(self) -> bool:
        return (
            not self.pending_terms
            and not self.solver.has_no_solutions
            and not self.pending_conditions
        )

    def construct_tree(self, term: terms.Term) -> Tree:
        renamed_rule = self.renamed_rules[term]
        return Tree(
            Instance(
                renamed_rule.original,
                {
                    variable: self.solver.solution.get(
                        renamed_variable, terms.symbol("FATAL_ERROR")
                    )
                    for variable, renamed_variable
                    in renamed_rule.renaming.items()
                },
            ),
            tuple(
                self.construct_tree(premise)
                for premise in renamed_rule.premises
            ),
        )

    def expand(self) -> t.Iterator[_Node]:
        if not self.pending_terms:
            return
        term = self.pending_terms[0]
        for rule in self.system.rules:
            renamed_rule = _create_renamed_rule(rule)
            solver = self.solver.clone()
            solver.add_equation((term, renamed_rule.conclusion))
            if solver.has_no_solutions:
                continue
            solver.add_equations(renamed_rule.constraints)
            if solver.has_no_solutions:
```

```
223                    continue
224            pending_conditions: t.Set[t.Tuple[_RenamedRule, Condition]] = set()
225            conditions_iterator = itertools.chain(
226                self.pending_conditions,
227                ((renamed_rule, condition) for condition in rule.conditions),
228            )
229            for condition_instance, condition in conditions_iterator:
230                verdict = condition.get_verdict(
231                    {
232                        variable: solver.solution[renamed_variable]
233                        for variable, renamed_variable
234                        in condition_instance.renaming.items()
235                        if renamed_variable in solver.solution
236                    }
237                )
238                if verdict is Verdict.VIOLATED:
239                    break
240                elif verdict is Verdict.SATISFIABLE:
241                    pending_conditions.add((condition_instance, condition))
242            else:
243                yield _Node(
244                    self.system,
245                    solver,
246                    {term: renamed_rule, **self.renamed_rules},
247                    renamed_rule.premises + self.pending_terms[1:],
248                    frozenset(pending_conditions),
249                )
```

## A.1   Simple Arithmetic Expressions

**rigorous/semantics/arithmetic/semantics.py**

```python
1   # -*- coding:utf-8 -*-
2   #
3   # Copyright (C) 2020, Maximilian Köhl <mail@koehlma.de>
4   #
5   # fmt: off
6
7   from __future__ import annotations
8
9   from ...core import inference, terms
10  from ...data import booleans, numbers, sets
11  from ...pretty import define
12
13  from .. import sos
14
15
16  BINARY_ADD = terms.symbol("+")
17  BINARY_SUB = terms.symbol("-")
18  BINARY_MUL = terms.symbol("*")
19  BINARY_DIV = terms.symbol("/")
20
21  BINARY_OPERATORS = sets.create({BINARY_ADD, BINARY_SUB, BINARY_MUL, BINARY_DIV})
22
23
24  def is_binary_operator(operator: terms.Term) -> inference.Condition:
```

```python
25          return booleans.check(sets.contains(BINARY_OPERATORS, operator))
26
27
28  left_expr = define.variable("left_expr", text="el", math="e_l")
29  right_expr = define.variable("right_expr", text="er", math="e_r")
30
31  left_int = define.variable("left_int", text="zl", math="z_l")
32  right_int = define.variable("right_int", text="zr", math="z_r")
33
34  some_result = define.variable("result", text="u", math="u")
35
36  some_operator = define.variable("operator", text="()", math="\\circ")
37
38
39  def binary_expr(
40      left: terms.Term, operator: terms.Term, right: terms.Term
41  ) -> terms.Term:
42      return terms.sequence(left, operator, right)
43
44
45  l_eval_rule = define.rule(
46      name="l-eval",
47      premises=(
48          sos.transition(
49              source=left_expr, action=sos.some_action, target=some_result
50          ),
51      ),
52      conditions=(is_binary_operator(some_operator),),
53      conclusion=sos.transition(
54          source=binary_expr(left_expr, some_operator, right_expr),
55          action=sos.some_action,
56          target=binary_expr(some_result, some_operator, right_expr),
57      ),
58  )
59
60  r_eval_rule = define.rule(
61      name="r-eval",
62      premises=(
63          sos.transition(
64              source=right_expr, action=sos.some_action, target=some_result
65          ),
66      ),
67      conditions=(is_binary_operator(some_operator),),
68      conclusion=sos.transition(
69          source=binary_expr(left_expr, some_operator, right_expr),
70          action=sos.some_action,
71          target=binary_expr(left_expr, some_operator, some_result),
72      ),
73  )
74
75  add_eval_rule = define.rule(
76      name="add-eval",
77      conclusion=sos.transition(
78          source=binary_expr(left_int, BINARY_ADD, right_int),
79          action=sos.ACTION_TAU,
80          target=numbers.add(left_int, right_int),
81      ),
82  )
83
84  sub_eval_rule = define.rule(
```

```python
85          name="sub-eval",
86          conclusion=sos.transition(
87              source=binary_expr(left_int, BINARY_SUB, right_int),
88              action=sos.ACTION_TAU,
89              target=numbers.sub(left_int, right_int),
90          ),
91  )
92
93  mul_eval_rule = define.rule(
94          name="mul-eval",
95          conclusion=sos.transition(
96              source=binary_expr(left_int, BINARY_MUL, right_int),
97              action=sos.ACTION_TAU,
98              target=numbers.mul(left_int, right_int),
99          ),
100 )
101
102 div_eval_rule = define.rule(
103         name="div-eval",
104         conclusion=sos.transition(
105             source=binary_expr(left_int, BINARY_DIV, right_int),
106             action=sos.ACTION_TAU,
107             target=numbers.floor_div(left_int, right_int),
108         ),
109 )
110
111 system = inference.System(
112         [
113             l_eval_rule,
114             r_eval_rule,
115             add_eval_rule,
116             sub_eval_rule,
117             mul_eval_rule,
118             div_eval_rule,
119         ]
120 )
```

## A.2   Calculus of Communicating Systems

**rigorous/semantics/ccs/semantics.py**

```python
1   # -*- coding:utf-8 -*-
2   #
3   # Copyright (C) 2020, Maximilian Köhl <mail@koehlma.de>
4   #
5   # fmt: off
6
7   from __future__ import annotations
8
9   import dataclasses as d
10  import typing as t
11
12  from ...core import terms, inference
13  from ...data import mappings, sets, booleans
14  from ...pretty import define
15  from ...semantics import sos
```

```
16
17
18   # Let's introduce process variables and actions as a new data type.
19   @d.dataclass(frozen=True)
20   class ProcessVariable(terms.Value):
21       identifier: str
22
23
24   # The process that cannot do anything.
25   DEAD_PROCESS = terms.symbol("0")
26
27
28   @terms.function_operator
29   def complement(action: terms.Sequence) -> t.Optional[terms.Term]:
30       if len(action.elements) != 2:
31           return None
32       name, modifier = action.elements
33       if isinstance(modifier, terms.Symbol):
34           if modifier.symbol == "!":
35               return terms.sequence(name, "?")
36           elif modifier.symbol == "?":
37               return terms.sequence(name, "!")
38       return None
39
40
41   # Next, we define some auxillary functions to build process terms.
42
43
44   def generative_action(action: terms.Term) -> terms.Term:
45       return terms.sequence(action, "!")
46
47
48   def reactive_action(action: terms.Term) -> terms.Term:
49       return terms.sequence(action, "?")
50
51
52   def prefix(action: terms.Term, process: terms.Term) -> terms.Term:
53       return terms.sequence(action, ".", process)
54
55
56   def choice(left: terms.Term, right: terms.Term) -> terms.Term:
57       return terms.sequence(left, "+", right)
58
59
60   def parallel(left: terms.Term, right: terms.Term) -> terms.Term:
61       return terms.sequence(left, "⎕", right)
62
63
64   def restrict(process: terms.Term, actions: terms.Term) -> terms.Term:
65       return terms.sequence(process, terms.symbol("\\"), actions)
66
67
68   def fix(process_variable: terms.Term, process: terms.Term) -> terms.Term:
69       return terms.sequence(
70           terms.symbol("fix"), process_variable, terms.symbol("="), process
71       )
72
73
74   def create_environment(
75       binding: t.Mapping[terms.Term, terms.Term]
```

```
76  ) -> terms.Term:
77      return mappings.create(binding)
78
79
80  # We are going to need some meta variables for our inference rules.
81
82  some_environment = define.variable("the_env", text="Γ", math="\\Gamma")
83
84  some_process = define.variable("some_process", text="P", math="P")
85  some_successor = define.variable("some_successor", text="P'", math="P'")
86
87  other_process = define.variable("other_process", text="Q", math="Q")
88  other_successor = define.variable("other_successor", text="Q'", math="Q'")
89
90  process_variable = define.variable("process_variable", text="X", math="X")
91
92  action_set = define.variable("action_set", text="H", math="H")
93
94
95  prefix_rule = define.rule(
96      "prefix",
97      conclusion=sos.transition(
98          environment=some_environment,
99          source=prefix(sos.some_action, some_process),
100         action=sos.some_action,
101         target=some_process,
102     ),
103 )
104
105 choice_l_rule = define.rule(
106     "choice-l",
107     premises=(
108         sos.transition(
109             environment=some_environment,
110             source=some_process,
111             action=sos.some_action,
112             target=some_successor,
113         ),
114     ),
115     conclusion=sos.transition(
116         environment=some_environment,
117         source=choice(some_process, other_process),
118         action=sos.some_action,
119         target=some_successor,
120     ),
121 )
122
123 choice_r_rule = define.rule(
124     "choice-r",
125     premises=(
126         sos.transition(
127             environment=some_environment,
128             source=other_process,
129             action=sos.some_action,
130             target=other_successor,
131         ),
132     ),
133     conclusion=sos.transition(
134         environment=some_environment,
135         source=choice(some_process, other_process),
```

```
136              action=sos.some_action,
137              target=other_successor,
138          ),
139      )
140
141      par_l_rule = define.rule(
142          "par-l",
143          premises=(
144              sos.transition(
145                  environment=some_environment,
146                  source=some_process,
147                  action=sos.some_action,
148                  target=some_successor,
149              ),
150          ),
151          conclusion=sos.transition(
152              environment=some_environment,
153              source=parallel(some_process, other_process),
154              action=sos.some_action,
155              target=parallel(some_successor, other_process),
156          ),
157      )
158
159      par_r_rule = define.rule(
160          "par-r",
161          premises=(
162              sos.transition(
163                  environment=some_environment,
164                  source=other_process,
165                  action=sos.some_action,
166                  target=other_successor,
167              ),
168          ),
169          conclusion=sos.transition(
170              environment=some_environment,
171              source=parallel(some_process, other_process),
172              action=sos.some_action,
173              target=parallel(some_process, other_successor),
174          ),
175      )
176
177      sync_rule = define.rule(
178          name="sync",
179          premises=(
180              sos.transition(
181                  environment=some_environment,
182                  source=some_process,
183                  action=sos.some_action,
184                  target=some_successor,
185              ),
186              sos.transition(
187                  environment=some_environment,
188                  source=other_process,
189                  action=complement(sos.some_action),
190                  target=other_successor,
191              ),
192          ),
193          conclusion=sos.transition(
194              environment=some_environment,
195              source=parallel(some_process, other_process),
```

```
196            action=sos.ACTION_TAU,
197            target=parallel(some_successor, other_successor),
198        ),
199    )
200
201    rec_rule = define.rule(
202        "rec",
203        premises=(
204            sos.transition(
205                environment=some_environment,
206                source=some_process,
207                action=sos.some_action,
208                target=some_successor,
209            ),
210        ),
211        constraints=(
212            (some_process, mappings.getitem(some_environment, process_variable)),
213        ),
214        conclusion=sos.transition(
215            environment=some_environment,
216            source=process_variable,
217            action=sos.some_action,
218            target=some_successor,
219        ),
220    )
221
222    res_rule = define.rule(
223        "res",
224        premises=(
225            sos.transition(
226                environment=some_environment,
227                source=some_process,
228                action=sos.some_action,
229                target=some_successor,
230            ),
231        ),
232        conditions=(
233            booleans.check(sets.not_contains(action_set, sos.some_action)),
234        ),
235        conclusion=sos.transition(
236            environment=some_environment,
237            source=restrict(some_process, action_set),
238            action=sos.some_action,
239            target=restrict(some_successor, action_set),
240        ),
241    )
242
243    fix_rule = define.rule(
244        "fix",
245        premises=(
246            sos.transition(
247                environment=some_environment,
248                source=terms.replace(
249                    some_process,
250                    process_variable,
251                    fix(process_variable, some_process),
252                ),
253                action=sos.some_action,
254                target=some_successor,
255            ),
```

```
256        ),
257        conclusion=sos.transition(
258            environment=some_environment,
259            source=fix(process_variable, some_process),
260            action=sos.some_action,
261            target=some_successor,
262        ),
263    )
264
265
266    system = inference.System(
267        [
268            prefix_rule,
269            choice_l_rule,
270            choice_r_rule,
271            par_l_rule,
272            par_r_rule,
273            sync_rule,
274            rec_rule,
275            res_rule,
276            fix_rule,
277        ]
278    )
```

# Appendix B

# Python Semantics

## B.1 Primitive Functions

The following listing of primitive functions for the apply mechanism (cf. Section 4.3.1) has been automatically generated with the command:

```
rigorous-python latexify primitives $LATEX_FILE
```

| | |
|---|---|
| `reference_id` : *Ref → Num* | `primitives.py:150` |

Returns a unique numeric identifier for the given reference.

We use this primitive to implement Python's id function.

| | |
|---|---|
| `reference_hash` : *Ref → Num* | `primitives.py:169` |

Returns a hash for the given reference.

In Python most objects are hashable based on their identity. This function returns a hash value based on the identity of the reference.

| | |
|---|---|
| `number_add` : *Num × Num → Num* | `primitives.py:186` |

Returns the sum of both numbers.

| | |
|---|---|
| `number_sub` : *Num × Num → Num* | `primitives.py:194` |

Returns the difference of both numbers.

| | |
|---|---|
| `number_mul` : *Num × Num → Num* | `primitives.py:202` |

Returns the product of both numbers.

---

`number_str` : $Num \rightarrow Str$        `primitives.py:210`

Converts a number into a primitive string.

---

`number_neg` : $Num \rightarrow Num$        `primitives.py:218`

Flips the sign of the number.

---

`number_hash` : $Num \rightarrow Num$        `primitives.py:226`

Computes the hash of the number.

Numbers are hashed in Python by treating them as rationals. This allows hashing numbers independent of their representation, e.g., the float $1.0$ and the integer $1$ have the same hash.

---

`make_record` : $Vec \rightarrow Rec$        `primitives.py:244`

Turns a vector of pairs $(s, t)$ into a record.

The first component of each pair is required to be a string representing a field name. The resulting record then maps those fields to the respective terms given by the second component of each pair.

---

`record_get` : $Rec \times Str \rightarrow \mathcal{T}$        `primitives.py:264`

Retrives the value of the given field from a record.

---

`record_get_default` : $Rec \times Str \times \mathcal{T} \rightarrow \mathcal{T}$        `primitives.py:277`

Retrives the value of the given field from a record and returns the default value in case no such field does exist.

---

`record_set` : $Rec \times Str \times \mathcal{T} \rightarrow Rec$        `primitives.py:288`

Sets the specified field to the provided term.

Returns a record that is identical to the first argument except that the field represented by the second argument is set to the value provided as third argument.

---

`sequence_length` : $Vec \rightarrow Num$        `primitives.py:308`

Returns the length of the vector.

---

`sequence_get` : $Vec \times Int \rightarrow \mathcal{T}$        `primitives.py:316`

Retrieves the element at the provided index of the vector.

---

| `sequence_set` : $Vec \times Int \times \mathcal{T} \rightarrow Vec$ | `primitives.py:329` |
|---|---|

Sets the value at the specified index of the vector.

| `sequence_push` : $Vec \times \mathcal{T} \rightarrow Vec$ | `primitives.py:344` |
|---|---|

Appends a value to the right side of the vector.

| `sequence_push_left` : $Vec \times \mathcal{T} \rightarrow Vec$ | `primitives.py:354` |
|---|---|

Appends a value to the left side of the vector.

| `sequence_pop` : $Vec \rightarrow Vec$ | `primitives.py:364` |
|---|---|

Removes a value from the right side of the vector.

| `sequence_pop_left` : $Vec \rightarrow Vec$ | `primitives.py:372` |
|---|---|

Removes a value from the left side of the vector.

| `sequence_concat` : $Vec \times Vec \rightarrow Vec$ | `primitives.py:380` |
|---|---|

Concatenates both vectors.

| `sequence_delete` : $Vec \times Int \rightarrow Vec$ | `primitives.py:390` |
|---|---|

Removes the value at the specified index from the vector.

| `sequence_slice` : $Vec \times Int \times Int \rightarrow Vec$ | `primitives.py:402` |
|---|---|

Returns the specified slice of the vector.

| `string_hash` : $Str \rightarrow Num$ | `primitives.py:418` |
|---|---|

Returns the hash for the string.

String hashing in Python is randomized. The value returned by this primitive depends on the environment variable `PYTHONHASHSEED`.

| `string_equals` : $Str \times Str \rightarrow Boolean$ | `primitives.py:429` |
|---|---|

Checks equality of two strings.

| `string_join` : $Str \times Vec \rightarrow Str$ | `primitives.py:439` |
|---|---|

Joins a vector of strings with the provided seperator.

| `string_concat` : $Str \times Str \rightarrow Str$ | `primitives.py:454` |
|---|---|

Concatenates two strings.

| | |
|---|---|
| `string_rpartition` : $Str \times Str \to Vec$ | `primitives.py:464` |
| Partitions a string from the right. | |
| `string_repr` : $Str \to Str$ | `primitives.py:474` |
| Returns the Python `repr` of the string. | |
| `string_length` : $Str \to Num$ | `primitives.py:482` |
| Returns the length of the string. | |
| `mapping_get` : $Map \times \mathcal{T} \to \mathcal{T}$ | `primitives.py:496` |
| Retrieves a value from the mapping using the provided key. | |
| `mapping_get_default` : $Map \times \mathcal{T} \times \mathcal{T} \to \mathcal{T}$ | `primitives.py:504` |
| Retrives a value from the mapping using the provided key. In case no such value exists, the provided default value is returned. | |
| `mapping_set` : $Map \times \mathcal{T} \times \mathcal{T} \to Map$ | `primitives.py:515` |
| Puts a key-value pair into the mapping. | |
| `mapping_contains` : $Map \times \mathcal{T} \to Boolean$ | `primitives.py:525` |
| Checks whether the mapping contains the provided key. | |
| `mapping_delete` : $Map \times \mathcal{T} \to Map$ | `primitives.py:535` |
| Deletes a key from the mapping. | |
| `mapping_update` : $Map \times Map \to Map$ | `primitives.py:545` |
| Updates a mapping with the key-value pairs of the other mapping. | |
| `mapping_keys` : $Map \to Vec$ | `primitives.py:555` |
| Returns the sequence of keys of the mapping. | |
| `mapping_size` : $Map \to Num$ | `primitives.py:563` |
| Returns the number of entries of the mapping. | |
| `send_value` : $Rec \times \mathcal{T} \to \mathcal{T}$ | `primitives.py:577` |
| Constructs a `send_value` term for sending a value to a generator. The first argument is the frame descriptor to resume execution from and the second is the value to send. | |

---

```
send_throw : Rec × 𝒯 → 𝒯                        primitives.py:587
```

Constructs a `send_throw` term for throwing an exception into a generator. The first argument is the frame descriptor to resume execution from and the second is the exception to throw.

---

```
make_frame : Rec × Map → Rec                     primitives.py:598
```

Creates a frame descriptor from the given code object and namespace.

---

## B.2   Runtime Functions

The following listing of runtime functions for the runtime mechanism has (cf. Section 4.3.1) been automatically generated with the command:

```
rigorous-python latexify runtime $LATEX_FILE
```

---

```
store_cell(cells, identifier, value)
runtime.py:38
```

Stores a value in the cell of the closure `cells`.

---

```
load_cell(cells, identifier)
runtime.py:45
```

Loads a value from the cell of the closure `cells`.

---

```
load__class__(cells)
runtime.py:52
```

Loads the special cell `__class__` from the closure `cells`.

---

```
delete_cell(cells, identifier)
runtime.py:65
```

Deletes the value in the cell of the closure `cells`.

---

```
class_super(cls, instance)
runtime.py:72
```

Used to implement `super` without any arguments.

---

```
make_list(elements)
runtime.py:124
```

Constructs a list from the provided vector of elements.

---

```
make_tuple(elements)
runtime.py:131
```

Constructs a tuple from the provided vector of arguments.

```
make_dict(entries)
runtime.py:172
```

Constructs a dict from the provided vector of entries.

```
get_cls_slot(cls, name)
runtime.py:311
```

Retrieves the dunder method `name` from the provided class `cls`.

The argument `cls` is required to be a Python type object and `name` is required to be a primitive string.

```
lowlevel_isinstance(obj, cls)
runtime.py:333
```

Checks whether `obj` is an instance of `cls`.

```
lowlevel_issubclass(cls, other)
runtime.py:347
```

Checks whether `cls` is a subclass of `other`.

```
store_global(namespace, identifier, value)
runtime.py:369
```

Stores `value` in the provided global namespace.

```
load_global(namespace, identifier)
runtime.py:376
```

Loads a value from the provided global namespace.

```
load_global_default(namespace, identifier, default)
runtime.py:391
```

Loads a value from the provided global namespace. In case no value has been bound to `identifier`, the provided default value is returned.

```
delete_global(namespace, identifier)
runtime.py:403
```

Deletes an identifier from the global namespace.

```
store_class(namespace, identifier, value)
runtime.py:413
```

Stores a value in the namespace of a class.

```
load_class_global(class_namespace, global_namespace, identifier)
runtime.py:420
```

Loads a value from the namespace of a class and reverts to the provided global namespace in case no value exists in the namespace of the class.

```
load_class_cell(class_namespace, cells, identifier)
runtime.py:432
```

Loads a value from the namespace of a class and reverts to the provided closure in case no value exists in the namespace of the class.

```
get_attribute(obj, name)
runtime.py:450
```

Retrieves an attribute from an object.

```
set_attribute(obj, name, value)
runtime.py:463
```

Sets an attribute on an object.

```
delete_attribute(obj, name)
runtime.py:473
```

Deletes an attribute from an object.

```
set_item(mapping, key, value)
runtime.py:486
```

Sets an item on an object.

```
get_item(obj, key)
runtime.py:496
```

Retrieves an item from an object.

```
delete_item(mapping, key)
runtime.py:506
```

Deletes an item from an object.

---

`ensure_exception(obj_or_cls, context)`
`runtime.py:522`

---

Constructs an exception from `obj_or_cls`.

---

`check_active_exception(cls)`
`runtime.py:536`

---

Checks whether there is an active exception.

---

`is_exception_compatible(exception, pattern)`
`runtime.py:545`

---

Implements the compatibility check specified in Section 8.4. of the PLR.

---

`call(positional_arguments, keyword_arguments, target)`
`runtime.py:750`

---

Calls a callable object with the provided arguments.

---

`build_function(code, global_namespace, cells, defaults)`
`runtime.py:800`

---

Constructs a function object from the provided code, global namespace, closure, and default values for parameters.

---

`rich_cmp(left, right, normal, swapped)`
`runtime.py:855`

---

Implements rich comparisons between two objects.

---

`binary_operator(left, right, left_slot, right_slot)`
`runtime.py:927`

---

Applies a binary operator to the provided operands.

---

`unary_operator(operand, slot_name)`
`runtime.py:949`

---

Applies a unary operator to the provided operand.

---

`unpack_iterable(iterable)`
`runtime.py:966`

---

Takes an iterable and returns a primitive vector of its elemens.

---

`unpack_str_mapping(mapping)`
`runtime.py:979`

---

Takes a python mapping and returns a primitive mapping.

---

| `compute_cls_layout(mro)` |
|---|
| `runtime.py:1145` |
| Computes the class layout based on the MRO. |

| `compute_mro(cls, bases)` |
|---|
| `runtime.py:1165` |
| Computes the method resolution order (MRO) for `cls`. |

| `create_assertion_error(message)` |
|---|
| `runtime.py:1263` |
| Creates an `AssertionError`. |

| `unbound_local_error(identifier)` |
|---|
| `runtime.py:1270` |
| Creates an `UnboundLocalError`. |

| `convert_bool(obj)` |
|---|
| `runtime.py:1277` |
| Converts an object to a boolean. |

## B.3   Translation Macros

The following listing of translation macros (cf. Section 4.1) has been automatically generated with the command:

```
rigorous-python latexify macros $LATEX_FILE
```

| `GET_CLS_SLOT` | `macros.py:59` |
|---|---|
| Retrieves a dunder method from the provided class. | |

| `GET_SLOT` | `macros.py:73` |
|---|---|
| Retrieves a dunder method from the provided object. | |

| `CALL_SLOT` | `macros.py:91` |
|---|---|
| Calls the specified dunder method on the provided object. | |

| `CLS_OF` | `macros.py:124` |
|---|---|
| Retrieves the class of the provided object. | |

| VALUE_OF | `macros.py:136` |
|---|---|
| Retrieves the primitive value of the provided object. | |

| SET_VALUE | `macros.py:148` |
|---|---|
| Sets the primitive value of the provided object. | |

| LOAD | `macros.py:172` |
|---|---|
| Creates a `mem_load` term for loading from a reference. | |

| STORE | `macros.py:180` |
|---|---|
| Creates a `mem_store` term for storing a value. | |

| NEW | `macros.py:193` |
|---|---|
| Creates a `mem_new` term for creating a new reference. | |

| LITERAL | `macros.py:207` |
|---|---|
| Casts a Python literal into a primitive literal. | |

| RECORD | `macros.py:233` |
|---|---|
| Constructs a record with the provided fields. | |

| NEW_FROM_VALUE | `macros.py:251` |
|---|---|
| Constructs an expression utilizing `mem_new` to create a new object of the given class with the given value. | |

| CALL | `macros.py:284` |
|---|---|
| Creates a `call` term from the given frame. | |
| This macro is used by the runtime function `call` for creating a `call` term which delegates control to the constructed frame. | |

| PRINT | `macros.py:296` |
|---|---|
| Creates a `print` term. | |

| HALT | `macros.py:304` |
|---|---|
| Creates a `HALT` term useful for debugging. | |

## B.4 Evaluation Table

The following table contains the results of evaluating the available tools based on the assembled test suite (cf. Section 4.5). The evaluation has been conducted on a fairly modern machine with an Intel i7-6700K at $4\,\mathrm{GHz}$ and $32\,\mathrm{GB}$ of RAM.

| | |
|---|---|
| ✓ | Test has successfully executed without any exceptions. |
| X | Test has failed due to an `AssertionError`. |
| – | Test has not succeeded due to an exception other than `AssertionError`. |
| T | Test has taken too long to execute (timeout after $2$ hours). |
| U | Test contains unsupported syntax. |
| P | Test caused an internal error (panic) in the tool (MOPSA only). |

| | CPython | | | $\lambda_\pi$ | MOPSA | SOS Python | |
|---|---|---|---|---|---|---|---|
| | 3.2.5 | 3.3.3 | 3.7.3 | | | Time $[s]$ | Transitions |
| k-python/testall.py | ✓ | ✓ | ✓ | – | U | – | – |
| k-python/testany.py | ✓ | ✓ | ✓ | X | P | U | – | – |
| k-python/testassert1.py | ✓ | ✓ | ✓ | ✓ | ✓ | 17.20 | 11 734 |
| k-python/testassert2.py | ✓ | ✓ | ✓ | – | P | U | – | – |
| k-python/testassert3.py | ✓ | ✓ | ✓ | – | P | – | – | – |
| k-python/testassign1.py | ✓ | ✓ | ✓ | ✓ | ✓ | 4.08 | 2 345 |
| k-python/testassign10.py | ✓ | ✓ | ✓ | – | U | U | – | – |
| k-python/testassign2.py | ✓ | ✓ | ✓ | ✓ | ✓ | U | – | – |
| k-python/testassign3.py | ✓ | ✓ | ✓ | – | ✓ | U | – | – |
| k-python/testassign4.py | ✓ | ✓ | ✓ | – | X | U | – | – |
| k-python/testassign5.py | ✓ | ✓ | ✓ | – | ✓ | U | – | – |
| k-python/testassign6.py | ✓ | ✓ | ✓ | ✓ | ✓ | U | – | – |
| k-python/testassign7.py | ✓ | ✓ | ✓ | ✓ | ✓ | U | – | – |
| k-python/testassign8.py | ✓ | ✓ | ✓ | – | U | U | – | – |
| k-python/testassign9.py | ✓ | ✓ | ✓ | – | U | U | – | – |
| k-python/testaugassign1.py | ✓ | ✓ | ✓ | ✓ | ✓ | U | – | – |

| | CPython | | | $\lambda_\pi$ | Mopsa | SOS Python | | |
|---|---|---|---|---|---|---|---|---|
| | 3.2.5 | 3.3.3 | 3.7.3 | | | | Time [$s$] | Transitions |
| k-python/testaugassign10.py | ✓ | ✓ | ✓ | ✓ | ✓ | U | – | – |
| k-python/testaugassign11.py | ✓ | ✓ | ✓ | ✓ | ✓ | U | – | – |
| k-python/testaugassign12.py | ✓ | ✓ | ✓ | X | X | U | – | – |
| k-python/testaugassign2.py | ✓ | ✓ | ✓ | ✓ | ✓ | U | – | – |
| k-python/testaugassign3.py | ✓ | ✓ | ✓ | ✓ | ✓ | U | – | – |
| k-python/testaugassign4.py | ✓ | ✓ | ✓ | ✓ | ✓ | U | – | – |
| k-python/testaugassign5.py | ✓ | ✓ | ✓ | ✓ | ✓ | U | – | – |
| k-python/testaugassign6.py | ✓ | ✓ | ✓ | ✓ | ✓ | U | – | – |
| k-python/testaugassign7.py | ✓ | ✓ | ✓ | ✓ | ✓ | U | – | – |
| k-python/testaugassign8.py | ✓ | ✓ | ✓ | ✓ | ✓ | U | – | – |
| k-python/testaugassign9.py | ✓ | ✓ | ✓ | ✓ | ✓ | U | – | – |
| k-python/testbools1.py | ✓ | ✓ | ✓ | ✓ | ✓ | ✓ | 1.76 | 1 053 |
| k-python/testbools10.py | ✓ | ✓ | ✓ | ✓ | ✓ | ✓ | 74.91 | 72 598 |
| k-python/testbools11.py | ✓ | ✓ | ✓ | – | X | ✓ | 39.99 | 37 241 |
| k-python/testbools2.py | ✓ | ✓ | ✓ | – | P | ✓ | 79.14 | 97 547 |
| k-python/testbools3.py | ✓ | ✓ | ✓ | ✓ | P | ✓ | 2.66 | 1 644 |
| k-python/testbools4.py | ✓ | ✓ | ✓ | ✓ | ✓ | ✓ | 5.52 | 4 139 |
| k-python/testbools5.py | ✓ | ✓ | ✓ | ✓ | X | ✓ | 24.03 | 20 863 |
| k-python/testbools6.py | ✓ | ✓ | ✓ | ✓ | X | ✓ | 108.05 | 118 979 |
| k-python/testbools7.py | ✓ | ✓ | ✓ | X | ✓ | ✓ | 68.09 | 62 864 |
| k-python/testbools8.py | ✓ | ✓ | ✓ | ✓ | ✓ | ✓ | 57.42 | 49 096 |
| k-python/testbools9.py | ✓ | ✓ | ✓ | ✓ | ✓ | ✓ | 5.84 | 4 181 |
| k-python/testbytes1.py | ✓ | ✓ | ✓ | – | – | U | – | – |
| k-python/testbytes2.py | ✓ | ✓ | ✓ | – | – | U | – | – |
| k-python/testbytes3.py | ✓ | ✓ | ✓ | – | X | – | – | – |
| k-python/testbytes4.py | ✓ | ✓ | ✓ | – | ✓ | U | – | – |
| k-python/testbytes5.py | ✓ | ✓ | ✓ | – | X | U | – | – |
| k-python/testcallable.py | ✓ | ✓ | ✓ | ✓ | P | ✓ | 72.94 | 72 820 |
| k-python/testclasses1.py | ✓ | ✓ | ✓ | – | P | ✓ | 301.31 | 394 444 |
| k-python/testclasses2.py | ✓ | ✓ | ✓ | X | X | ✓ | 147.44 | 176 571 |
| k-python/testclasses3.py | ✓ | ✓ | ✓ | X | P | ✓ | 173.11 | 198 270 |
| k-python/testclasses4.py | ✓ | ✓ | ✓ | – | ✓ | – | – | – |
| k-python/testclasses5.py | ✓ | ✓ | ✓ | – | P | U | – | – |

| | CPython | | | $\lambda_\pi$ | Mopsa | SOS Python | |
|---|---|---|---|---|---|---|---|
| | 3.2.5 | 3.3.3 | 3.7.3 | | | Time [$s$] | Transitions |
| k-python/testclasses6.py | ✓ | ✓ | ✓ | – | – | U | – | – |
| k-python/testclasses7.py | ✓ | ✓ | ✓ | ✓ | ✓ | ✓ | 83.83 | 77 895 |
| k-python/testclasses8.py | ✓ | ✓ | ✓ | ✓ | ✓ | ✓ | 70.97 | 66 381 |
| k-python/testclasses9.py | ✓ | ✓ | ✓ | ✓ | ✓ | ✓ | 68.98 | 64 492 |
| k-python/testclassmethod.py | ✓ | ✓ | ✓ | X | P | ✓ | 107.98 | 145 228 |
| k-python/testcoercion1.py | ✓ | ✓ | ✓ | – | X | ✓ | 122.94 | 136 776 |
| k-python/testcoercion2.py | ✓ | ✓ | ✓ | ✓ | X | ✓ | 133.97 | 154 815 |
| k-python/testcompile1.py | ✓ | ✓ | – | – | X | – | – | – |
| k-python/testcompile2.py | ✓ | ✓ | ✓ | – | P | – | – | – |
| k-python/testcomprehensions.py | ✓ | ✓ | ✓ | X | P | U | – | – |
| k-python/testdecorator.py | ✓ | ✓ | ✓ | ✓ | – | ✓ | 122.08 | 174 980 |
| k-python/testdel.py | ✓ | ✓ | ✓ | – | X | U | – | – |
| k-python/testdicts1.py | ✓ | ✓ | ✓ | – | X | ✓ | 175.47 | 228 329 |
| k-python/testdicts2.py | ✓ | ✓ | ✓ | ✓ | X | ✓ | 57.56 | 61 028 |
| k-python/testdocstring.py | ✓ | ✓ | ✓ | – | – | U | – | – |
| k-python/testeval1.py | ✓ | ✓ | ✓ | – | X | – | – | – |
| k-python/testeval2.py | ✓ | ✓ | ✓ | – | P | – | – | – |
| k-python/testeval3.py | ✓ | ✓ | ✓ | – | – | – | – | – |
| k-python/testeval4.py | ✓ | ✓ | ✓ | – | X | – | – | – |
| k-python/testeval5.py | ✓ | ✓ | ✓ | – | X | – | – | – |
| k-python/testexceptions1.py | – | ✓ | ✓ | – | P | ✓ | 87.66 | 104 289 |
| k-python/testexec1.py | ✓ | ✓ | ✓ | – | X | – | – | – |
| k-python/testexec2.py | ✓ | ✓ | ✓ | – | X | – | – | – |
| k-python/testexec3.py | ✓ | ✓ | ✓ | – | X | – | – | – |
| k-python/testexec4.py | ✓ | ✓ | ✓ | – | X | – | – | – |
| k-python/testfloat1.py | ✓ | ✓ | ✓ | – | ✓ | – | – | – |
| k-python/testfor1.py | ✓ | ✓ | ✓ | ✓ | ✓ | ✓ | 20.78 | 16 341 |
| k-python/testfor2.py | ✓ | ✓ | ✓ | ✓ | X | ✓ | 17.36 | 18 948 |
| k-python/testfor3.py | ✓ | ✓ | ✓ | ✓ | X | ✓ | 30.83 | 42 376 |
| k-python/testfor4.py | ✓ | ✓ | ✓ | ✓ | X | ✓ | 55.28 | 104 572 |
| k-python/testformat1.py | ✓ | ✓ | ✓ | – | P | – | – | – |
| k-python/testformat2.py | ✓ | ✓ | ✓ | – | X | – | – | – |
| k-python/testformat3.py | ✓ | ✓ | ✓ | – | P | – | – | – |

| | CPython | | | $\lambda_\pi$ | Mopsa | SOS Python | | |
|---|---|---|---|---|---|---|---|---|
| | 3.2.5 | 3.3.3 | 3.7.3 | | | | Time $[s]$ | Transitions |
| k-python/testformat4.py | ✓ | ✓ | ✓ | − | X | − | − | − |
| k-python/testformat5.py | ✓ | ✓ | ✓ | − | − | − | − | − |
| k-python/testfuncfinally1.py | ✓ | ✓ | ✓ | ✓ | ✓ | ✓ | 12.55 | 8 486 |
| k-python/testfuncfinally2.py | ✓ | ✓ | ✓ | ✓ | ✓ | ✓ | 9.33 | 5 744 |
| k-python/testfuncfinally3.py | ✓ | ✓ | ✓ | X | ✓ | ✓ | 8.44 | 5 745 |
| k-python/testfuncfinally4.py | ✓ | ✓ | ✓ | − | ✓ | ✓ | 27.22 | 23 280 |
| k-python/testfuncraise1.py | ✓ | ✓ | ✓ | X | ✓ | ✓ | 23.31 | 17 497 |
| k-python/testfuncraise2.py | ✓ | ✓ | ✓ | − | − | − | − | − |
| k-python/testfunctions1.py | ✓ | ✓ | ✓ | − | − | ✓ | 68.17 | 77 690 |
| k-python/testfunctions10.py | ✓ | ✓ | ✓ | ✓ | P | ✓ | 4.84 | 3 089 |
| k-python/testfunctions11.py | ✓ | ✓ | ✓ | ✓ | − | ✓ | 8.02 | 4 531 |
| k-python/testfunctions12.py | ✓ | ✓ | ✓ | ✓ | P | ✓ | 6.34 | 3 796 |
| k-python/testfunctions13.py | ✓ | ✓ | ✓ | − | − | ✓ | 17.77 | 12 121 |
| k-python/testfunctions14.py | ✓ | ✓ | ✓ | − | − | U | − | − |
| k-python/testfunctions2.py | ✓ | ✓ | ✓ | ✓ | P | ✓ | 18.50 | 13 124 |
| k-python/testfunctions3.py | ✓ | ✓ | ✓ | ✓ | X | ✓ | 27.47 | 24 647 |
| k-python/testfunctions4.py | ✓ | ✓ | ✓ | − | P | ✓ | 32.23 | 27 019 |
| k-python/testfunctions5.py | ✓ | ✓ | ✓ | ✓ | P | ✓ | 20.50 | 15 033 |
| k-python/testfunctions6.py | ✓ | ✓ | ✓ | X | X | ✓ | 101.19 | 139 602 |
| k-python/testfunctions7.py | ✓ | ✓ | ✓ | ✓ | − | ✓ | 163.74 | 257 434 |
| k-python/testfunctions8.py | ✓ | ✓ | ✓ | ✓ | − | ✓ | 10.94 | 8 554 |
| k-python/testfunctions9.py | ✓ | ✓ | ✓ | ✓ | P | ✓ | 17.58 | 13 499 |
| k-python/testgetattr1.py | ✓ | ✓ | ✓ | X | ✓ | ✓ | 276.38 | 388 936 |
| k-python/testgetattr2.py | ✓ | ✓ | ✓ | ✓ | ✓ | ✓ | 143.25 | 172 530 |
| k-python/testgetattr3.py | ✓ | ✓ | ✓ | − | P | ✓ | 235.36 | 292 957 |
| k-python/testgetattr4.py | ✓ | ✓ | ✓ | ✓ | X | ✓ | 23.94 | 23 567 |
| k-python/testgetattr5.py | ✓ | ✓ | ✓ | − | P | ✓ | 132.64 | 139 846 |
| k-python/testgetattr6.py | ✓ | ✓ | ✓ | − | P | U | − | − |
| k-python/testhash.py | ✓ | ✓ | ✓ | X | P | ✓ | 83.27 | 79 395 |
| k-python/testif.py | ✓ | ✓ | ✓ | ✓ | ✓ | ✓ | 1.72 | 1 007 |
| k-python/testimport1.py | ✓ | ✓ | ✓ | − | P | U | − | − |
| k-python/testimport2.py | ✓ | ✓ | ✓ | − | − | U | − | − |
| k-python/testimport3.py | ✓ | ✓ | ✓ | − | − | U | − | − |

| | CPython | | | $\lambda_\pi$ | Mopsa | SOS Python | |
|---|---|---|---|---|---|---|---|
| | 3.2.5 | 3.3.3 | 3.7.3 | | | Time [$s$] | Transitions |
| k-python/testimport4.py | – | ✓ | – | – | – | U – | – |
| k-python/testin.py | ✓ | ✓ | ✓ | ✓ | ✓ | ✓ 15.61 | 13 596 |
| k-python/testintegers1.py | ✓ | ✓ | ✓ | ✓ | ✓ | ✓ 5.38 | 1 937 |
| k-python/testintegers10.py | ✓ | ✓ | ✓ | ✓ | P | ✓ 90.27 | 94 804 |
| k-python/testintegers2.py | ✓ | ✓ | ✓ | – | P | – – | – |
| k-python/testintegers3.py | ✓ | ✓ | ✓ | ✓ | ✓ | – – | – |
| k-python/testintegers4.py | ✓ | ✓ | ✓ | ✓ | ✓ | ✓ 10.74 | 7 229 |
| k-python/testintegers5.py | ✓ | ✓ | ✓ | ✓ | ✓ | ✓ 2.76 | 1 609 |
| k-python/testintegers6.py | ✓ | ✓ | ✓ | – | P | ✓ 80.16 | 96 177 |
| k-python/testintegers7.py | ✓ | ✓ | ✓ | ✓ | P | ✓ 16.56 | 11 953 |
| k-python/testintegers8.py | ✓ | ✓ | ✓ | – | – | – – | – |
| k-python/testintegers9.py | ✓ | ✓ | ✓ | – | P | ✓ 49.44 | 58 782 |
| k-python/testio1.py | ✓ | ✓ | ✓ | – | – | U – | – |
| k-python/testio2.py | ✓ | ✓ | ✓ | – | – | U – | – |
| k-python/testio3.py | ✓ | ✓ | – | – | – | U – | – |
| k-python/testio4.py | ✓ | ✓ | ✓ | – | – | U – | – |
| k-python/testio5.py | ✓ | ✓ | ✓ | – | – | U – | – |
| k-python/testio6.py | ✓ | ✓ | ✓ | – | – | U – | – |
| k-python/testio7.py | – | ✓ | ✓ | – | – | U – | – |
| k-python/testio8.py | ✓ | ✓ | ✓ | – | – | U – | – |
| k-python/testis.py | ✓ | ✓ | ✓ | ✓ | X | ✓ 1.78 | 1 038 |
| k-python/testisinstance1.py | ✓ | ✓ | ✓ | X | X | ✓ 126.12 | 147 475 |
| k-python/testisinstance2.py | ✓ | ✓ | ✓ | ✓ | P | ✓ 55.83 | 63 375 |
| k-python/testissubclass1.py | ✓ | ✓ | ✓ | X | X | ✓ 142.52 | 167 939 |
| k-python/testissubclass2.py | ✓ | ✓ | ✓ | ✓ | P | ✓ 68.05 | 78 724 |
| k-python/testiter1.py | ✓ | ✓ | ✓ | – | P | ✓ 85.44 | 108 528 |
| k-python/testiter2.py | ✓ | ✓ | ✓ | ✓ | ✓ | U – | – |
| k-python/testlambda.py | ✓ | ✓ | ✓ | ✓ | ✓ | ✓ 4.49 | 2 632 |
| k-python/testlen.py | ✓ | ✓ | ✓ | – | X | ✓ 207.97 | 249 050 |
| k-python/testlists1.py | ✓ | ✓ | ✓ | ✓ | ✓ | ✓ 19.48 | 15 996 |
| k-python/testlists10.py | ✓ | ✓ | ✓ | ✓ | X | ✓ 38.92 | 38 689 |
| k-python/testlists11.py | ✓ | ✓ | ✓ | – | X | ✓ 33.61 | 30 754 |
| k-python/testlists2.py | ✓ | ✓ | ✓ | X | X | ✓ 9.28 | 7 071 |

| | CPython | | | $\lambda_\pi$ | Mopsa | SOS Python | | |
|---|---|---|---|---|---|---|---|---|
| | 3.2.5 | 3.3.3 | 3.7.3 | | | | Time [$s$] | Transitions |
| k-python/testlists3.py | ✓ | ✓ | ✓ | − | ✓ | ✓ | 17.50 | 13 107 |
| k-python/testlists4.py | ✓ | ✓ | ✓ | − | ✓ | ✓ | 17.89 | 13 103 |
| k-python/testlists5.py | ✓ | ✓ | ✓ | ✓ | − | ✓ | 114.91 | 194 453 |
| k-python/testlists6.py | ✓ | ✓ | ✓ | ✓ | X | ✓ | 99.64 | 124 797 |
| k-python/testlists7.py | ✓ | ✓ | ✓ | ✓ | X | ✓ | 7.00 | 5 044 |
| k-python/testlists8.py | ✓ | ✓ | ✓ | X | X | ✓ | 66.34 | 60 428 |
| k-python/testlists9.py | ✓ | ✓ | ✓ | − | − | ✓ | 49.69 | 46 424 |
| k-python/testloopfinally1.py | ✓ | ✓ | ✓ | − | ✓ | ✓ | 20.97 | 15 414 |
| k-python/testloopfinally2.py | ✓ | ✓ | ✓ | − | ✓ | ✓ | 7.38 | 4 554 |
| k-python/testloopfinally3.py | ✓ | ✓ | ✓ | ✓ | X | ✓ | 15.01 | 14 233 |
| k-python/testloopfinally4.py | ✓ | ✓ | ✓ | ✓ | X | ✓ | 6.94 | 4 527 |
| k-python/testloopfinally5.py | ✓ | ✓ | ✓ | − | ✓ | ✓ | 21.66 | 16 240 |
| k-python/testloopfinally6.py | ✓ | ✓ | ✓ | ✓ | ✓ | ✓ | 8.20 | 5 380 |
| k-python/testloopfinally7.py | ✓ | ✓ | ✓ | ✓ | X | ✓ | 19.39 | 16 915 |
| k-python/testloopfinally8.py | ✓ | ✓ | ✓ | ✓ | ✓ | ✓ | 8.58 | 5 353 |
| k-python/testloopfinally9.py | ✓ | ✓ | ✓ | − | ✓ | ✓ | 12.66 | 8 039 |
| k-python/testmap.py | ✓ | ✓ | ✓ | − | P | − | − | − |
| k-python/testmockimport1.py | ✓ | ✓ | ✓ | − | − | U | − | − |
| k-python/testmockimport2.py | ✓ | ✓ | ✓ | − | − | U | − | − |
| k-python/testmockimport3.py | ✓ | ✓ | ✓ | X | − | U | − | − |
| k-python/testmockimport4.py | ✓ | ✓ | ✓ | X | P | U | − | − |
| k-python/testmockimport5.py | ✓ | ✓ | ✓ | − | P | U | − | − |
| k-python/testmodule1.py | ✓ | ✓ | ✓ | − | P | U | − | − |
| k-python/testnone.py | − | ✓ | ✓ | − | P | ✓ | 69.23 | 69 283 |
| k-python/testobject1.py | ✓ | ✓ | ✓ | X | P | ✓ | 71.91 | 70 051 |
| k-python/testobject2.py | ✓ | ✓ | ✓ | − | P | ✓ | 111.25 | 119 433 |
| k-python/testobject3.py | ✓ | ✓ | ✓ | − | P | ✓ | 37.50 | 36 384 |
| k-python/testobject4.py | ✓ | ✓ | ✓ | − | X | ✓ | 134.36 | 154 685 |
| k-python/testord.py | ✓ | ✓ | ✓ | − | X | − | − | − |
| k-python/testpass.py | ✓ | ✓ | ✓ | ✓ | ✓ | ✓ | 1.70 | 977 |
| k-python/testposix1.py | − | − | − | − | − | U | − | − |
| k-python/testposix2.py | − | − | − | − | − | U | − | − |
| k-python/testposix3.py | − | − | − | − | − | U | − | − |

| | CPython | | | $\lambda_\pi$ | Mopsa | SOS Python | |
|---|---|---|---|---|---|---|---|
| | 3.2.5 | 3.3.3 | 3.7.3 | | | Time [s] | Transitions |
| k-python/testrange.py | ✓ | ✓ | ✓ | ✓ | X | – | – |
| k-python/testrepr1.py | ✓ | ✓ | ✓ | – | – | ✓ 49.92 | 58 971 |
| k-python/testrepr2.py | ✓ | ✓ | ✓ | – | – | ✓ 92.42 | 93 895 |
| k-python/testreversed1.py | ✓ | ✓ | ✓ | – | X | – | – |
| k-python/testreversed2.py | ✓ | ✓ | ✓ | – | – | – | – |
| k-python/testscope1.py | ✓ | ✓ | ✓ | ✓ | – | ✓ 60.69 | 60 989 |
| k-python/testscope10.py | ✓ | ✓ | ✓ | – | – | U – | – |
| k-python/testscope11.py | ✓ | ✓ | ✓ | – | X | ✓ 97.26 | 106 523 |
| k-python/testscope12.py | ✓ | ✓ | ✓ | – | – | U – | – |
| k-python/testscope2.py | ✓ | ✓ | ✓ | – | – | – | – |
| k-python/testscope3.py | ✓ | ✓ | ✓ | – | – | – | – |
| k-python/testscope4.py | ✓ | ✓ | ✓ | – | – | – | – |
| k-python/testscope5.py | ✓ | ✓ | ✓ | – | X | – | – |
| k-python/testscope6.py | ✓ | ✓ | ✓ | – | – | – | – |
| k-python/testscope7.py | ✓ | ✓ | ✓ | – | ✓ | ✓ 156.33 | 235 628 |
| k-python/testscope8.py | ✓ | ✓ | ✓ | – | – | ✓ 30.06 | 28 148 |
| k-python/testscope9.py | ✓ | ✓ | ✓ | – | – | – | – |
| k-python/testsetattr1.py | ✓ | ✓ | ✓ | ✓ | X | ✓ 99.11 | 95 639 |
| k-python/testsetattr2.py | ✓ | ✓ | ✓ | – | X | ✓ 125.19 | 133 915 |
| k-python/testsetattr3.py | ✓ | ✓ | ✓ | – | P | ✓ 53.05 | 49 182 |
| k-python/testsetattr4.py | ✓ | ✓ | ✓ | ✓ | ✓ | ✓ 124.59 | 137 781 |
| k-python/testsetattr5.py | ✓ | ✓ | ✓ | ✓ | ✓ | ✓ 94.05 | 98 211 |
| k-python/testsetattr6.py | ✓ | ✓ | ✓ | X | X | U – | – |
| k-python/testsets.py | ✓ | ✓ | ✓ | X | – | U – | – |
| k-python/testslice.py | ✓ | ✓ | ✓ | – | X | U – | – |
| k-python/teststaticmethod.py | ✓ | ✓ | ✓ | ✓ | ✓ | ✓ 89.94 | 92 930 |
| k-python/teststrings1.py | ✓ | ✓ | ✓ | ✓ | ✓ | ✓ 31.97 | 25 389 |
| k-python/teststrings10.py | ✓ | ✓ | ✓ | ✓ | P | ✓ 86.27 | 86 672 |
| k-python/teststrings11.py | ✓ | ✓ | ✓ | ✓ | ✓ | ✓ 90.25 | 94 520 |
| k-python/teststrings2.py | ✓ | ✓ | ✓ | ✓ | X | ✓ 32.58 | 26 458 |
| k-python/teststrings3.py | X | X | X | – | P | X – | – |
| k-python/teststrings4.py | ✓ | ✓ | ✓ | – | P | ✓ 60.08 | 64 228 |
| k-python/teststrings5.py | ✓ | ✓ | ✓ | – | P | – | – |

| | CPython | | | $\lambda_\pi$ | Mopsa | SOS Python | | |
|---|---|---|---|---|---|---|---|---|
| | 3.2.5 | 3.3.3 | 3.7.3 | | | | Time [$s$] | Transitions |
| k-python/teststrings6.py | ✓ | ✓ | ✓ | – | X | – | – | – |
| k-python/teststrings7.py | ✓ | ✓ | ✓ | – | X | U | – | – |
| k-python/teststrings8.py | ✓ | ✓ | ✓ | – | ✓ | – | – | – |
| k-python/teststrings9.py | ✓ | ✓ | ✓ | – | P | – | – | – |
| k-python/testsuper1.py | ✓ | ✓ | ✓ | T | – | ✓ | 187.88 | 264 009 |
| k-python/testsuper2.py | ✓ | ✓ | ✓ | – | P | ✓ | 38.78 | 40 257 |
| k-python/testsuper3.py | ✓ | ✓ | – | – | – | ✓ | 475.83 | 677 115 |
| k-python/testsuper4.py | ✓ | – | – | – | – | ✓ | 323.62 | 425 585 |
| k-python/testsuper5.py | ✓ | ✓ | ✓ | X | P | ✓ | 389.16 | 501 481 |
| k-python/testsys.py | X | X | X | X | X | U | – | – |
| k-python/testtry1.py | ✓ | ✓ | ✓ | ✓ | ✓ | ✓ | 7.70 | 5 085 |
| k-python/testtry10.py | ✓ | ✓ | ✓ | – | ✓ | ✓ | 18.67 | 13 207 |
| k-python/testtry11.py | – | | | | | – | – | – |
| k-python/testtry12.py | ✓ | ✓ | ✓ | ✓ | ✓ | ✓ | 42.84 | 36 558 |
| k-python/testtry13.py | – | | | – | | U | – | – |
| k-python/testtry14.py | ✓ | ✓ | ✓ | ✓ | ✓ | ✓ | 40.83 | 43 082 |
| k-python/testtry15.py | ✓ | ✓ | ✓ | ✓ | ✓ | ✓ | 26.91 | 25 872 |
| k-python/testtry2.py | ✓ | ✓ | ✓ | – | ✓ | ✓ | 17.94 | 13 195 |
| k-python/testtry3.py | ✓ | ✓ | ✓ | – | ✓ | ✓ | 21.72 | 18 415 |
| k-python/testtry4.py | ✓ | ✓ | ✓ | ✓ | ✓ | ✓ | 37.70 | 40 429 |
| k-python/testtry5.py | ✓ | ✓ | ✓ | ✓ | ✓ | ✓ | 38.17 | 40 413 |
| k-python/testtry6.py | ✓ | ✓ | ✓ | ✓ | X | ✓ | 37.88 | 39 751 |
| k-python/testtry7.py | ✓ | ✓ | ✓ | – | – | ✓ | 23.94 | 21 242 |
| k-python/testtry8.py | ✓ | ✓ | ✓ | – | – | – | – | – |
| k-python/testtry9.py | ✓ | ✓ | ✓ | – | – | U | – | – |
| k-python/testtuples1.py | ✓ | ✓ | ✓ | ✓ | X | ✓ | 4.12 | 2 545 |
| k-python/testtuples10.py | ✓ | ✓ | ✓ | – | ✓ | ✓ | 31.89 | 25 235 |
| k-python/testtuples11.py | ✓ | ✓ | ✓ | ✓ | X | ✓ | 15.20 | 9 726 |
| k-python/testtuples12.py | ✓ | ✓ | ✓ | – | P | – | – | – |
| k-python/testtuples2.py | ✓ | ✓ | ✓ | ✓ | ✓ | ✓ | 24.22 | 23 505 |
| k-python/testtuples3.py | ✓ | ✓ | ✓ | ✓ | – | ✓ | 34.80 | 27 814 |
| k-python/testtuples4.py | ✓ | ✓ | ✓ | – | – | – | – | – |
| k-python/testtuples5.py | ✓ | ✓ | ✓ | – | – | – | – | – |

| | CPython | | | $\lambda_\pi$ | Mopsa | SOS Python | |
|---|---|---|---|---|---|---|---|
| | 3.2.5 | 3.3.3 | 3.7.3 | | | Time [$s$] | Transitions |
| k-python/testtuples6.py | ✓ | ✓ | ✓ | ✓ | ✓ | 21.26 | 17 312 |
| k-python/testtuples7.py | ✓ | ✓ | ✓ | – | P | X | – | – |
| k-python/testtuples8.py | ✓ | ✓ | ✓ | – | – | ✓ | 31.61 | 30 106 |
| k-python/testtuples9.py | ✓ | ✓ | ✓ | X | ✓ | ✓ | 9.33 | 7 071 |
| k-python/testweakref.py | ✓ | ✓ | ✓ | – | – | U | – | – |
| k-python/testwhile1.py | ✓ | ✓ | ✓ | ✓ | – | ✓ | 12.02 | 11 088 |
| k-python/testwhile2.py | ✓ | ✓ | ✓ | ✓ | X | ✓ | 13.69 | 11 595 |
| k-python/testwhile3.py | ✓ | ✓ | ✓ | ✓ | X | ✓ | 19.69 | 23 031 |
| k-python/testwhile4.py | ✓ | ✓ | ✓ | ✓ | X | ✓ | 52.05 | 71 793 |
| k-python/testwith1.py | ✓ | ✓ | ✓ | X | – | U | – | – |
| k-python/testwith2.py | ✓ | ✓ | ✓ | X | X | U | – | – |
| k-python/testwith3.py | ✓ | ✓ | ✓ | – | X | U | – | – |
| k-python/testwith4.py | ✓ | ✓ | ✓ | – | ✓ | U | – | – |
| k-python/testyield1.py | – | ✓ | ✓ | – | P | ✓ | 123.88 | 154 532 |
| k-python/testyield10.py | ✓ | ✓ | ✓ | – | – | ✓ | 77.58 | 87 568 |
| k-python/testyield11.py | ✓ | ✓ | ✓ | ✓ | ✓ | ✓ | 91.52 | 117 398 |
| k-python/testyield12.py | ✓ | ✓ | ✓ | ✓ | ✓ | ✓ | 87.12 | 112 347 |
| k-python/testyield13.py | ✓ | ✓ | ✓ | – | P | ✓ | 101.12 | 105 998 |
| k-python/testyield14.py | ✓ | ✓ | ✓ | – | X | ✓ | 57.50 | 58 910 |
| k-python/testyield2.py | ✓ | ✓ | ✓ | – | P | ✓ | 89.48 | 107 040 |
| k-python/testyield3.py | ✓ | ✓ | ✓ | ✓ | ✓ | ✓ | 46.41 | 43 052 |
| k-python/testyield4.py | ✓ | ✓ | ✓ | ✓ | ✓ | ✓ | 67.12 | 78 459 |
| k-python/testyield5.py | ✓ | ✓ | ✓ | ✓ | – | ✓ | 75.36 | 93 010 |
| k-python/testyield6.py | ✓ | ✓ | ✓ | ✓ | – | ✓ | 55.38 | 62 376 |
| k-python/testyield7.py | ✓ | ✓ | ✓ | – | P | ✓ | 79.41 | 93 965 |
| k-python/testyield8.py | ✓ | ✓ | ✓ | – | – | ✓ | 75.44 | 82 515 |
| k-python/testyield9.py | ✓ | ✓ | – | – | X | ✓ | 98.41 | 121 293 |
| lambda-py/memofib-decorator.py | ✓ | ✓ | ✓ | T | P | – | – | – |
| lambda-py/methodinstance.py | ✓ | ✓ | ✓ | X | P | X | – | – |
| lambda-py/setitem.py | ✓ | ✓ | ✓ | – | P | U | – | – |
| lambda-py/gen-comp.py | ✓ | ✓ | ✓ | – | P | U | – | – |
| lambda-py/gen-except.py | ✓ | ✓ | ✓ | – | P | – | – | – |
| lambda-py/gen-method.py | ✓ | ✓ | ✓ | – | – | ✓ | 132.27 | 169 887 |

| | CPython | | | $\lambda_\pi$ | Mopsa | | SOS Python | |
|---|---|---|---|---|---|---|---|---|
| | 3.2.5 | 3.3.3 | 3.7.3 | | | | Time [$s$] | Transitions |
| lambda-py/send-gen.py | ✓ | ✓ | ✓ | – | – | U | – | – |
| lambda-py/support_exception.py | – | – | – | – | | U | – | – |
| lambda-py/test_exception.py | ✓ | ✓ | ✓ | ✓ | P | U | – | – |
| lambda-py/test_import_name_binding.py | ✓ | ✓ | ✓ | – | P | U | – | – |
| lambda-py/test_meta_path.py | ✓ | – | – | – | – | U | – | – |
| lambda-py/test_unary.py | ✓ | ✓ | ✓ | – | P | U | – | – |
| lambda-py/test_with_extension.py | – | – | – | – | | U | – | – |
| lambda-py/empty.py | ✓ | ✓ | ✓ | ✓ | ✓ | ✓ | 2.02 | 977 |
| lambda-py/__init__.py | ✓ | ✓ | ✓ | ✓ | ✓ | ✓ | 1.80 | 977 |
| lambda-py/test_augassign.py | ✓ | ✓ | ✓ | – | P | U | – | – |
| lambda-py/test_binop.py | ✓ | ✓ | ✓ | – | P | U | – | – |
| lambda-py/test_bool.py | ✓ | ✓ | ✓ | – | P | U | – | – |
| lambda-py/test_builtin.py | ✓ | – | – | – | – | U | – | – |
| lambda-py/test_dict.py | ✓ | ✓ | ✓ | – | P | U | – | – |
| lambda-py/test_dictviews.py | ✓ | ✓ | ✓ | – | P | U | – | – |
| lambda-py/test_file.py | ✓ | ✓ | ✓ | – | – | U | – | – |
| lambda-py/test_float.py | – | – | T | – | – | U | – | – |
| lambda-py/test_int.py | ✓ | ✓ | ✓ | – | P | U | – | – |
| lambda-py/test_isinstance.py | ✓ | ✓ | ✓ | – | – | U | – | – |
| lambda-py/test_list.py | ✓ | ✓ | ✓ | – | P | U | – | – |
| lambda-py/test_raise.py | ✓ | ✓ | ✓ | – | – | U | – | – |
| lambda-py/test_scope.py | ✓ | ✓ | ✓ | – | – | U | – | – |
| lambda-py/test_string.py | ✓ | ✓ | ✓ | – | – | U | – | – |
| lambda-py/test_types.py | ✓ | ✓ | ✓ | – | P | U | – | – |
| lambda-py/test_unary.py | ✓ | ✓ | ✓ | – | P | U | – | – |
| lambda-py/super-external-self.py | ✓ | – | – | – | P | ✓ | 131.16 | 153 904 |
| lambda-py/super-external.py | ✓ | – | – | – | P | ✓ | 138.11 | 153 835 |
| lambda-py/super-nested-noarg.py | ✓ | – | – | – | P | ✓ | 137.22 | 152 948 |
| lambda-py/super-nested-not-self.py | ✓ | ✓ | ✓ | X | P | – | – | – |
| lambda-py/super-nested-self.py | ✓ | ✓ | ✓ | ✓ | P | – | – | – |
| lambda-py/parser-regress.py | ✓ | ✓ | ✓ | – | | U | – | – |
| lambda-py/base.py | ✓ | ✓ | ✓ | – | | – | – | – |
| lambda-py/lambda_getter.py | ✓ | ✓ | ✓ | – | | – | – | – |

| | CPython | | | $\lambda_\pi$ | Mopsa | SOS Python | |
| --- | --- | --- | --- | --- | --- | --- | --- |
| | 3.2.5 | 3.3.3 | 3.7.3 | | | Time [$s$] | Transitions |
| lambda-py/bool-callable.py | ✓ | ✓ | ✓ | ✓ | P | ✓ 25.84 | 27 592 |
| lambda-py/bool-compare-list-dict.py | ✓ | ✓ | ✓ | ✓ | X | – – | – |
| lambda-py/bool-compare.py | ✓ | ✓ | ✓ | ✓ | ✓ | ✓ 62.14 | 70 887 |
| lambda-py/bool-convert.py | ✓ | ✓ | ✓ | ✓ | P | ✓ 69.86 | 85 380 |
| lambda-py/bool-float.py | ✓ | ✓ | ✓ | ✓ | ✓ | – – | – |
| lambda-py/bool-int.py | ✓ | ✓ | ✓ | ✓ | X | ✓ 45.22 | 52 943 |
| lambda-py/bool-isinstance.py | ✓ | ✓ | ✓ | ✓ | ✓ | ✓ 78.11 | 100 945 |
| lambda-py/bool-math.py | ✓ | ✓ | ✓ | ✓ | X | – – | – |
| lambda-py/bool-str.py | ✓ | ✓ | ✓ | ✓ | X | ✓ 28.27 | 28 715 |
| lambda-py/all.py | ✓ | ✓ | ✓ | ✓ | P | U – | – |
| lambda-py/any.py | ✓ | ✓ | ✓ | ✓ | P | U – | – |
| lambda-py/callable.py | ✓ | ✓ | ✓ | ✓ | P | ✓ 141.77 | 168 512 |
| lambda-py/filter.py | ✓ | ✓ | ✓ | ✓ | – | – – | – |
| lambda-py/get_set_attribute.py | ✓ | ✓ | ✓ | ✓ | – | – – | – |
| lambda-py/isinstance-tuple.py | ✓ | ✓ | ✓ | ✓ | – | ✓ 270.89 | 342 206 |
| lambda-py/isinstance.py | ✓ | ✓ | ✓ | ✓ | P | ✓ 210.22 | 254 769 |
| lambda-py/len.py | ✓ | ✓ | ✓ | ✓ | X | ✓ 70.67 | 81 193 |
| lambda-py/class-decorators.py | ✓ | ✓ | ✓ | ✓ | P | ✓ 218.83 | 265 652 |
| lambda-py/metaclass-class1.py | ✓ | ✓ | ✓ | ✓ | P | ✓ 193.72 | 233 943 |
| lambda-py/metaclass-class2.py | ✓ | ✓ | ✓ | ✓ | P | ✓ 193.47 | 234 934 |
| lambda-py/metaclass-class3.py | ✓ | ✓ | ✓ | ✓ | P | ✓ 254.92 | 323 588 |
| lambda-py/metaclass-function.py | ✓ | ✓ | ✓ | ✓ | P | ✓ 152.48 | 175 441 |
| lambda-py/test_class_dict.py | ✓ | ✓ | ✓ | ✓ | P | X – | – |
| lambda-py/test_metaclass.py | ✓ | ✓ | ✓ | ✓ | P | ✓ 345.91 | 456 082 |
| lambda-py/dict-attribute.py | ✓ | ✓ | ✓ | ✓ | P | ✓ 81.39 | 80 300 |
| lambda-py/dict-bool.py | ✓ | ✓ | ✓ | ✓ | P | ✓ 37.56 | 39 477 |
| lambda-py/dict-clear.py | ✓ | ✓ | ✓ | ✓ | P | – – | – |
| lambda-py/dict-contains.py | ✓ | ✓ | ✓ | ✓ | X | ✓ 45.36 | 50 161 |
| lambda-py/dict-get.py | ✓ | ✓ | ✓ | ✓ | P | ✓ 141.91 | 186 649 |
| lambda-py/dict-getitem.py | ✓ | ✓ | ✓ | ✓ | P | ✓ 82.80 | 84 571 |
| lambda-py/dict-hash-effects.py | ✓ | ✓ | ✓ | ✓ | X | ✓ 191.17 | 230 498 |
| lambda-py/dict-items.py | ✓ | ✓ | ✓ | ✓ | – | U – | – |
| lambda-py/dict-keys-effects.py | ✓ | ✓ | ✓ | ✓ | X | U – | – |

| | CPython | | | $\lambda_\pi$ | Mopsa | SOS Python | |
|---|---|---|---|---|---|---|---|
| | 3.2.5 | 3.3.3 | 3.7.3 | | | Time [$s$] | Transitions |
| lambda-py/dict-keys.py | ✓ | ✓ | ✓ | ✓ | – | – | |
| lambda-py/dict-list.py | ✓ | ✓ | ✓ | ✓ | – | 44.53 | 47 065 |
| lambda-py/dict-set-comprehensions.py | ✓ | ✓ | ✓ | ✓ | X | U | – | – |
| lambda-py/dict-set-keys.py | ✓ | ✓ | ✓ | ✓ | – | U | – |
| lambda-py/dict-update-iterable.py | ✓ | ✓ | ✓ | ✓ | – | – | – |
| lambda-py/dict-update.py | ✓ | ✓ | ✓ | ✓ | X | – | – |
| lambda-py/dict-values-set.py | ✓ | ✓ | ✓ | ✓ | – | U | – |
| lambda-py/dict-values.py | ✓ | ✓ | ✓ | ✓ | – | – | – |
| lambda-py/test-dict-aug-assign.py | ✓ | ✓ | ✓ | ✓ | X | U | – | – |
| lambda-py/assert-silent.py | ✓ | ✓ | ✓ | ✓ | P | ✓ | 20.47 | 17 534 |
| lambda-py/assertion-error-msg.py | ✓ | ✓ | ✓ | ✓ | – | ✓ | 33.70 | 28 637 |
| lambda-py/assertion-error.py | ✓ | ✓ | ✓ | ✓ | P | ✓ | 22.91 | 21 846 |
| lambda-py/assertRaises.py | ✓ | ✓ | ✓ | ✓ | P | ✓ | 21.52 | 19 469 |
| lambda-py/elseReturn.py | ✓ | ✓ | ✓ | ✓ | ✓ | | 5.08 | 3 006 |
| lambda-py/except-as.py | ✓ | ✓ | ✓ | ✓ | ✓ | | 31.41 | 33 158 |
| lambda-py/except-reraise.py | ✓ | ✓ | ✓ | ✓ | P | ✓ | 34.94 | 37 507 |
| lambda-py/invalid-reraise.py | ✓ | ✓ | ✓ | ✓ | X | – | – | – |
| lambda-py/nested-else.py | ✓ | ✓ | ✓ | ✓ | ✓ | | 39.72 | 42 620 |
| lambda-py/nested-reraise.py | ✓ | ✓ | ✓ | ✓ | P | – | – |
| lambda-py/nested.py | ✓ | ✓ | ✓ | ✓ | ✓ | | 30.22 | 30 214 |
| lambda-py/reraise.py | ✓ | ✓ | ✓ | ✓ | ✓ | | 29.94 | 29 739 |
| lambda-py/test-finally-reraise.py | ✓ | ✓ | ✓ | ✓ | P | | 31.42 | 32 393 |
| lambda-py/test-try-except-else.py | ✓ | ✓ | ✓ | ✓ | ✓ | | 24.56 | 23 039 |
| lambda-py/test-try-except-no-exception.py | ✓ | ✓ | ✓ | ✓ | ✓ | | 11.48 | 7 182 |
| lambda-py/test-try-except.py | ✓ | ✓ | ✓ | ✓ | ✓ | | 19.72 | 16 836 |
| lambda-py/test_with.py | ✓ | ✓ | ✓ | ✓ | P | U | – | |
| lambda-py/try-except-else-finally-no-exception.py | ✓ | ✓ | ✓ | ✓ | ✓ | | 21.92 | 20 556 |
| lambda-py/try-except-else-finally.py | ✓ | ✓ | ✓ | ✓ | ✓ | | 29.64 | 29 727 |
| lambda-py/try-except-else-no-exception.py | ✓ | ✓ | ✓ | ✓ | ✓ | | 16.88 | 13 868 |
| lambda-py/try-except-finally-no-exception.py | ✓ | ✓ | ✓ | ✓ | ✓ | | 16.94 | 13 870 |
| lambda-py/try-except-finally.py | ✓ | ✓ | ✓ | ✓ | ✓ | | 25.17 | 23 524 |
| lambda-py/try-finally-no-exception.py | ✓ | ✓ | ✓ | ✓ | ✓ | | 12.11 | 7 667 |
| lambda-py/while-break-finally.py | ✓ | ✓ | ✓ | ✓ | ✓ | | 14.53 | 9 550 |

| | CPython | | | $\lambda_\pi$ | Mopsa | SOS Python | | |
|---|---|---|---|---|---|---|---|---|
| | 3.2.5 | 3.3.3 | 3.7.3 | | | | Time [$s$] | Transitions |
| lambda-py/while-continue-finally.py | ✓ | ✓ | ✓ | ✓ | ✓ | ✓ | 16.56 | 11 750 |
| lambda-py/func_attr.py | ✓ | ✓ | ✓ | ✓ | — | ✓ | 17.11 | 11 930 |
| lambda-py/func_attr2.py | ✓ | ✓ | ✓ | ✓ | — | ✓ | 16.62 | 11 691 |
| lambda-py/func_defaults.py | ✓ | ✓ | ✓ | ✓ | P | ✓ | 151.61 | 175 925 |
| lambda-py/func_stararg.py | ✓ | ✓ | ✓ | ✓ | X | ✓ | 152.16 | 176 354 |
| lambda-py/keywords_args1.py | ✓ | ✓ | ✓ | ✓ | X | ✓ | 53.30 | 51 419 |
| lambda-py/keywords_args2.py | ✓ | ✓ | ✓ | ✓ | P | ✓ | 109.94 | 131 107 |
| lambda-py/keywords_args3.py | ✓ | ✓ | ✓ | ✓ | P | ✓ | 214.22 | 302 225 |
| lambda-py/keywords_args4.py | ✓ | ✓ | ✓ | ✓ | P | ✓ | 98.16 | 111 335 |
| lambda-py/kwarg.py | ✓ | ✓ | ✓ | ✓ | P | ✓ | 175.39 | 195 000 |
| lambda-py/kwonlyargs.py | ✓ | ✓ | ✓ | ✓ | P | ✓ | 86.20 | 90 870 |
| lambda-py/lambda_defaults.py | ✓ | ✓ | ✓ | ✓ | P | ✓ | 153.99 | 175 925 |
| lambda-py/lambda_stararg.py | ✓ | ✓ | ✓ | ✓ | P | ✓ | 154.86 | 176 354 |
| lambda-py/memofib-function-attributes.py | ✓ | ✓ | ✓ | ✓ | P | ✓ | 154.25 | 284 638 |
| lambda-py/non_tuple_starag.py | ✓ | ✓ | ✓ | ✓ | — | — | – | – |
| lambda-py/varargs.py | ✓ | ✓ | ✓ | ✓ | P | ✓ | 29.62 | 28 921 |
| lambda-py/basic-gen.py | ✓ | ✓ | ✓ | ✓ | P | U | – | – |
| lambda-py/gen-arg.py | ✓ | ✓ | ✓ | ✓ | P | ✓ | 41.28 | 40 061 |
| lambda-py/gen-exception.py | ✓ | ✓ | ✓ | ✓ | ✓ | ✓ | 73.30 | 85 794 |
| lambda-py/gen-in-gen.py | ✓ | ✓ | ✓ | ✓ | — | — | – | – |
| lambda-py/gen-list.py | ✓ | ✓ | ✓ | ✓ | X | U | – | – |
| lambda-py/gen-multiyield.py | ✓ | ✓ | ✓ | ✓ | P | U | – | – |
| lambda-py/gen-return-uncatchable.py | ✓ | ✓ | ✓ | ✓ | ✓ | ✓ | 61.39 | 69 670 |
| lambda-py/gen-return.py | ✓ | ✓ | ✓ | ✓ | ✓ | ✓ | 69.81 | 85 655 |
| lambda-py/simple-gen.py | ✓ | ✓ | ✓ | ✓ | ✓ | ✓ | 61.08 | 72 038 |
| lambda-py/continue.py | ✓ | ✓ | ✓ | ✓ | X | U | – | – |
| lambda-py/filter-comprehension.py | ✓ | ✓ | ✓ | ✓ | X | U | – | – |
| lambda-py/iter-classes.py | ✓ | ✓ | ✓ | ✓ | X | — | – | – |
| lambda-py/iter-comprehensions.py | ✓ | ✓ | ✓ | ✓ | X | U | – | – |
| lambda-py/iter-misc.py | ✓ | ✓ | ✓ | ✓ | P | — | – | – |
| lambda-py/iter-simple.py | ✓ | ✓ | ✓ | ✓ | X | — | – | – |
| lambda-py/iter-stop.py | ✓ | ✓ | ✓ | ✓ | — | — | – | – |
| lambda-py/test-for-else.py | ✓ | ✓ | ✓ | ✓ | X | U | – | – |

|  | CPython | | | $\lambda_\pi$ | Mopsa | SOS Python | |
|---|---|---|---|---|---|---|---|
|  | 3.2.5 | 3.3.3 | 3.7.3 | | | Time [s] | Transitions |
| lambda-py/test-genexp-lazyness.py | ✓ | ✓ | ✓ | ✓ | P | U | – | – |
| lambda-py/simple_append.py | ✓ | ✓ | ✓ | ✓ | X | ✓ | 44.27 | 45 570 |
| lambda-py/simple-extend.py | ✓ | ✓ | ✓ | ✓ | X | ✓ | 130.23 | 177 196 |
| lambda-py/test_list_assign.py | ✓ | ✓ | ✓ | ✓ | X | ✓ | 6.47 | 4 316 |
| lambda-py/test_list_identity.py | ✓ | ✓ | ✓ | ✓ | – | ✓ | 1.80 | 1 006 |
| lambda-py/test_list_simple.py | ✓ | ✓ | ✓ | ✓ | – | – | – | – |
| lambda-py/test_list_truth.py | ✓ | ✓ | ✓ | ✓ | P | ✓ | 13.94 | 12 244 |
| lambda-py/support.py | ✓ | ✓ | ✓ | ✓ | ✓ | ✓ | 59.81 | 54 621 |
| lambda-py/test_basic.py | ✓ | ✓ | ✓ | ✓ | – | U | – | – |
| lambda-py/test_failing_support_sticks.py | ✓ | ✓ | ✓ | ✓ | P | U | – | – |
| lambda-py/test_file.py | ✓ | ✓ | ✓ | ✓ | P | – | – | – |
| lambda-py/test_in_modules.py | ✓ | ✓ | ✓ | ✓ | – | U | – | – |
| lambda-py/test_scope.py | ✓ | ✓ | ✓ | ✓ | – | U | – | – |
| lambda-py/dunder-methods.py | ✓ | ✓ | ✓ | ✓ | P | ✓ | 87.03 | 89 231 |
| lambda-py/methods.py | ✓ | ✓ | ✓ | ✓ | P | ✓ | 129.75 | 148 834 |
| lambda-py/test-classmethod-desugar.py | ✓ | ✓ | ✓ | ✓ | ✓ | ✓ | 160.55 | 202 049 |
| lambda-py/test-staticmethod-decorator.py | ✓ | ✓ | ✓ | ✓ | ✓ | ✓ | 94.92 | 98 636 |
| lambda-py/test-staticmethod-oldstyle.py | ✓ | ✓ | ✓ | ✓ | ✓ | ✓ | 101.27 | 102 448 |
| lambda-py/test_consistency_with_epg.py | ✓ | ✓ | ✓ | ✓ | P | ✓ | 303.80 | 395 347 |
| lambda-py/test_diamond_inheritance.py | ✓ | ✓ | ✓ | ✓ | P | ✓ | 568.72 | 761 661 |
| lambda-py/test_ex5_from_c3_switch.py | ✓ | ✓ | ✓ | ✓ | P | ✓ | 308.34 | 407 765 |
| lambda-py/test_monotonicity.py | ✓ | ✓ | ✓ | ✓ | P | ✓ | 454.25 | 615 033 |
| lambda-py/test_mro_disagreement.py | ✓ | ✓ | ✓ | ✓ | X | ✓ | 439.53 | 580 478 |
| lambda-py/test_mro_disagreement_type.py | ✓ | ✓ | ✓ | ✓ | P | ✓ | 430.49 | 565 909 |
| lambda-py/test_multiple_inheritance.py | ✓ | ✓ | ✓ | ✓ | – | – | – | – |
| lambda-py/test_super_multiple_inheritance.py | ✓ | ✓ | ✓ | ✓ | P | ✓ | 1389.69 | 2 095 709 |
| lambda-py/type-dict.py | ✓ | ✓ | ✓ | ✓ | P | ✓ | 72.99 | 64 956 |
| lambda-py/override_getattr_descr.py | ✓ | ✓ | ✓ | ✓ | – | – | – | – |
| lambda-py/shadow.py | ✓ | ✓ | ✓ | ✓ | P | ✓ | 76.55 | 74 757 |
| lambda-py/simple_getter.py | ✓ | ✓ | ✓ | ✓ | ✓ | U | – | – |
| lambda-py/simple_getter_exn.py | ✓ | ✓ | ✓ | ✓ | P | U | – | – |
| lambda-py/simple_property.py | ✓ | ✓ | ✓ | ✓ | X | U | – | – |
| lambda-py/simple_property_decorator.py | ✓ | ✓ | ✓ | ✓ | – | U | – | – |

| | CPython | | | $\lambda_\pi$ | Mopsa | SOS Python | | |
|---|---|---|---|---|---|---|---|---|
| | 3.2.5 | 3.3.3 | 3.7.3 | | | | Time [$s$] | Transitions |
| lambda-py/test_property_decorator_baseclass.py | ✓ | ✓ | ✓ | ✓ | — | ✓ | 269.59 | 344 737 |
| lambda-py/test_property_decorator_subclass.py | ✓ | ✓ | ✓ | ✓ | — | — | – | – |
| lambda-py/test_property_getter_doc_override.py | ✓ | ✓ | ✓ | ✓ | | — | – | – |
| lambda-py/range-errs.py | ✓ | ✓ | ✓ | ✓ | P | — | – | – |
| lambda-py/range-list.py | ✓ | ✓ | ✓ | ✓ | X | — | – | – |
| lambda-py/range-vars.py | ✓ | ✓ | ✓ | ✓ | X | — | – | – |
| lambda-py/slice1.py | ✓ | ✓ | ✓ | ✓ | X | U | – | – |
| lambda-py/test-range-lazyness.py | ✓ | ✓ | ✓ | ✓ | X | U | – | – |
| lambda-py/bound-and-free.py | ✓ | ✓ | ✓ | ✓ | — | ✓ | 149.72 | 191 910 |
| lambda-py/class-bases-scope.py | ✓ | ✓ | ✓ | ✓ | — | ✓ | 112.36 | 134 590 |
| lambda-py/comprehensions-leak.py | ✓ | ✓ | ✓ | ✓ | ✓ | U | – | – |
| lambda-py/destructuring-bind.py | ✓ | ✓ | ✓ | ✓ | ✓ | U | – | – |
| lambda-py/extra-nesting.py | ✓ | ✓ | ✓ | ✓ | — | ✓ | 130.06 | 130 308 |
| lambda-py/for-loops.py | ✓ | ✓ | ✓ | ✓ | X | ✓ | 173.56 | 242 720 |
| lambda-py/freevar-in-method.py | ✓ | ✓ | ✓ | ✓ | X | ✓ | 294.25 | 380 061 |
| lambda-py/global-in-class-body.py | ✓ | ✓ | ✓ | ✓ | ✓ | ✓ | 119.73 | 136 669 |
| lambda-py/global-in-parallel-nested-functions.py | ✓ | ✓ | ✓ | ✓ | — | U | – | – |
| lambda-py/global_scope.py | ✓ | ✓ | ✓ | ✓ | — | | – | – |
| lambda-py/immutable-alias.py | ✓ | ✓ | ✓ | ✓ | X | ✓ | 70.12 | 75 277 |
| lambda-py/lambda1.py | ✓ | ✓ | ✓ | ✓ | — | ✓ | 107.06 | 128 974 |
| lambda-py/lambda2.py | ✓ | ✓ | ✓ | ✓ | — | ✓ | 111.94 | 130 292 |
| lambda-py/lambda3.py | ✓ | ✓ | ✓ | ✓ | ✓ | ✓ | 18.00 | 11 405 |
| lambda-py/lambda4.py | ✓ | ✓ | ✓ | ✓ | — | ✓ | 138.47 | 183 245 |
| lambda-py/locals-class.py | ✓ | ✓ | ✓ | ✓ | P | — | – | – |
| lambda-py/locals-function-renamed.py | ✓ | ✓ | ✓ | ✓ | — | U | – | – |
| lambda-py/locals-function.py | ✓ | ✓ | ✓ | ✓ | P | U | – | – |
| lambda-py/mixed-freevars-and-cellvars.py | ✓ | ✓ | ✓ | ✓ | — | ✓ | 193.39 | 268 648 |
| lambda-py/mutable-alias.py | ✓ | ✓ | ✓ | ✓ | X | ✓ | 167.27 | 176 823 |
| lambda-py/nearest-enclosing-scope.py | ✓ | ✓ | ✓ | ✓ | — | ✓ | 67.75 | 68 084 |
| lambda-py/nested-classes.py | ✓ | ✓ | ✓ | ✓ | X | ✓ | 51.94 | 51 285 |
| lambda-py/nested-nonlocal.py | ✓ | ✓ | ✓ | ✓ | — | U | – | – |
| lambda-py/nesting-global-no-free.py | ✓ | ✓ | ✓ | ✓ | — | ✓ | 24.08 | 20 493 |
| lambda-py/nesting-global-under-local.py | ✓ | ✓ | ✓ | ✓ | X | ✓ | 20.95 | 17 516 |

|  | CPython | | | $\lambda_\pi$ | Mopsa | SOS Python | |
|---|---|---|---|---|---|---|---|
|  | 3.2.5 | 3.3.3 | 3.7.3 | | | Time [$s$] | Transitions |
| lambda-py/nesting-plus-free-ref-to-global.py | ✓ | ✓ | ✓ | ✓ | ✓ | ✓ 24.48 | 20 859 |
| lambda-py/nesting-through-class.py | ✓ | ✓ | ✓ | ✓ | — | ✓ 195.31 | 246 355 |
| lambda-py/nonlocal-class.py | ✓ | ✓ | ✓ | ✓ | — | U | — |
| lambda-py/nonlocal-function-vardef-shadow.py | ✓ | ✓ | ✓ | ✓ | — | U | — |
| lambda-py/nonlocal-function-vardef-unbound.py | — | — | — | — | — | U | — |
| lambda-py/nonlocal-function-vardef.py | ✓ | ✓ | ✓ | ✓ | — | U | — |
| lambda-py/nonlocal-function.py | ✓ | ✓ | ✓ | ✓ | — | U | — |
| lambda-py/nonlocal-method.py | ✓ | ✓ | ✓ | ✓ | — | U | — |
| lambda-py/recursion.py | ✓ | ✓ | ✓ | ✓ | P | ✓ 149.52 | 232 872 |
| lambda-py/simple-and-rebinding.py | ✓ | ✓ | ✓ | ✓ | — | ✓ 164.05 | 228 516 |
| lambda-py/simple-nesting.py | ✓ | ✓ | ✓ | ✓ | — | ✓ 109.16 | 128 982 |
| lambda-py/test-assign-field.py | ✓ | ✓ | ✓ | ✓ | ✓ | ✓ 71.62 | 64 264 |
| lambda-py/test-builtin-scope-iter.py | ✓ | ✓ | ✓ | ✓ | X | U | — |
| lambda-py/test-builtin-scope.py | ✓ | ✓ | ✓ | ✓ | — | ✓ 21.62 | 20 378 |
| lambda-py/unbound-local.py | ✓ | ✓ | ✓ | ✓ | P | — | — |
| lambda-py/unboundlocal-after-del.py | ✓ | ✓ | ✓ | ✓ | P | — | — |
| lambda-py/unboundlocal-augassign.py | ✓ | ✓ | ✓ | ✓ | P | U | — |
| lambda-py/super-not-self.py | ✓ | ✓ | ✓ | ✓ | P | ✓ 440.72 | 582 864 |
| lambda-py/test-super.py | ✓ | ✓ | ✓ | ✓ | P | ✓ 436.77 | 582 864 |
| lambda-py/tuple-add.py | ✓ | ✓ | ✓ | ✓ | — | U | — |
| lambda-py/tuple-constructors.py | ✓ | ✓ | ✓ | ✓ | X | — | — |
| lambda-py/tuple-in.py | ✓ | ✓ | ✓ | ✓ | X | ✓ 53.94 | 54 897 |
| lambda-py/tuple-length.py | ✓ | ✓ | ✓ | ✓ | ✓ | ✓ 36.75 | 39 625 |
| lambda-py/tuple-mul.py | ✓ | ✓ | ✓ | ✓ | — | U | — |
| lambda-py/tuple-to-list.py | ✓ | ✓ | ✓ | ✓ | — | — | — |
| lambda-py/tuple-truth.py | ✓ | ✓ | ✓ | ✓ | P | ✓ 16.39 | 12 244 |
| lambda-py/address-of-none.py | ✓ | ✓ | ✓ | ✓ | ✓ | ✓ 6.16 | 3 735 |
| lambda-py/assign-to-self.py | ✓ | ✓ | ✓ | ✓ | ✓ | — | — |
| lambda-py/bitwise_test.py | ✓ | ✓ | ✓ | ✓ | ✓ | — | — |
| lambda-py/test-add-with-funcs.py | ✓ | ✓ | ✓ | ✓ | ✓ | U | — |
| lambda-py/test-aliasing.py | ✓ | ✓ | ✓ | ✓ | X | ✓ 96.25 | 96 855 |
| lambda-py/test_booleans.py | ✓ | ✓ | ✓ | ✓ | ✓ | ✓ 3.55 | 2 539 |
| lambda-py/test_comparisons.py | ✓ | ✓ | ✓ | ✓ | ✓ | — | — |

| | CPython | | | $\lambda_\pi$ | Mopsa | SOS Python | |
|---|---|---|---|---|---|---|---|
| | 3.2.5 | 3.3.3 | 3.7.3 | | | Time $[s]$ | Transitions |
| lambda-py/test_div_zero.py | ✓ | ✓ | ✓ | ✓ | ✓ | – | – |
| lambda-py/test_dynamics.py | ✓ | ✓ | ✓ | ✓ | P | – | – |
| lambda-py/test_floats.py | ✓ | ✓ | ✓ | ✓ | – | – | – |
| lambda-py/test_int_float_primitives.py | ✓ | ✓ | ✓ | ✓ | X | – | – |
| lambda-py/test_object_class.py | ✓ | ✓ | ✓ | ✓ | P | – | – |
| lambda-py/test_pow.py | ✓ | ✓ | ✓ | ✓ | P | – | – |
| lambda-py/test_simple_strings.py | ✓ | ✓ | ✓ | ✓ | ✓ | 24.67 | 23 466 |
| lambda-py/test_simple_string_ops.py | ✓ | ✓ | ✓ | ✓ | P | – | – |
| lambda-py/test_string_slices.py | ✓ | ✓ | ✓ | ✓ | – | U | – |
| lambda-py/types_truthy1.py | ✓ | ✓ | ✓ | ✓ | – | – | – |
| lambda-py/types_truthy2.py | ✓ | ✓ | ✓ | ✓ | ✓ | 60.00 | 53 431 |
| lambda-py/complex-defs.py | ✓ | ✓ | ✓ | ✓ | P | 113.59 | 135 747 |
| lambda-py/complex-defs2.py | ✓ | ✓ | ✓ | ✓ | P | 114.33 | 136 751 |
| lambda-py/multiple-locals-calls.py | ✓ | ✓ | ✓ | X | P | U | – |
| lambda-py/nonlocal-from-class-body.py | – | – | – | – | ✓ | 66.45 | 60 120 |
| lambda-py/unboundlocal-augassign.py | ✓ | ✓ | ✓ | ✓ | P | U | – |
| k-python/testdocstring.py | ✓ | ✓ | ✓ | – | – | ✓ | 78.84 | 66 462 |
| k-python/testsetattr4.py | ✓ | ✓ | ✓ | ✓ | – | ✓ | 125.62 | 137 785 |
| k-python/testtuples7.py | ✓ | ✓ | ✓ | – | P | – | – |
| koehl/do-not-run-init.py | ✓ | ✓ | ✓ | ✓ | ✓ | 69.30 | 65 700 |
| koehl/except-target-delete.py | ✓ | ✓ | ✓ | ✓ | ✓ | 52.95 | 57 312 |
| koehl/mcs-attribute-lookup.py | ✓ | ✓ | ✓ | ✓ | P | 214.84 | 262 433 |
| koehl/non-string-keywords.py | ✓ | ✓ | ✓ | – | P | 22.33 | 16 031 |
| koehl/nonlocal-in-class.py | ✓ | ✓ | ✓ | ✓ | ✓ | 117.42 | 139 349 |
| koehl/simple-property.py | ✓ | ✓ | ✓ | ✓ | – | ✓ | 182.03 | 218 529 |
| ✓ | 519 | 520 | 516 | 329 | 138 | 293 | |
| X | 2 | 2 | 2 | 29 | 106 | 4 | |
| – | 15 | 14 | 17 | 176 | 135 | 94 | |
| T | 0 | 0 | 1 | 2 | 0 | 0 | |
| U | 0 | 0 | 0 | 0 | 4 | 145 | |
| P | 0 | 0 | 0 | 0 | 153 | 0 | |

Generated by `evaluation/latexify.py`.